\documentclass[11pt,a4paper]{article}
\pdfoutput=1
\usepackage{jheppub}
\usepackage{gensymb}
\DeclareSymbolFont{matha}{OML}{txmi}{m}{it}% txfonts
\DeclareMathSymbol{\varv}{\mathord}{matha}{118}
\usepackage{amsmath}
\usepackage{amssymb}
\usepackage{graphicx}
\usepackage{subfigure}
\usepackage{pstricks}
\usepackage{bm}
\usepackage{pbox}
\usepackage{placeins}
\usepackage{booktabs}
\usepackage[T1]{fontenc}
\usepackage{footnote}
\usepackage{pdfpages}
\usepackage{hhline}
\usepackage{multirow}
\usepackage{multicol}
\usepackage{enumitem}
\usepackage[toc,page]{appendix}
\allowdisplaybreaks
\usepackage{comment}
\usepackage{float}                                                                      

\usepackage{textcomp}
\usepackage{gensymb}

\usepackage[numbers]{natbib}
\usepackage{notoccite}

\usepackage{rotating}
\usepackage{tabularx}
\usepackage[labelfont=bf]{caption} % optional

\FloatBarrier
\usepackage[many]{tcolorbox}

\renewcommand{\thetable}{\Roman{table}}

\newcommand{\beq}{\begin {equation}}  
\newcommand{\eeq}{\end   {equation}} 
\newcommand{\bea}{\begin {eqnarray}} 
\newcommand{\eea}{\end   {eqnarray}}  
\newcommand{\baa}{\begin {array}   } 
\newcommand{\eaa}{\end   {array}   }     
\newcommand{\bit}{\begin {itemize} }
\newcommand{\eit}{\end   {itemize} }
\newcommand{\be }{\begin {equation}} 
\newcommand{\ee }{\end   {equation}}
\newcommand{\mwt}{m_{\omega^{2/3}}}        
\newcommand{\mwo}{m_{\omega^{-1/3}}}   
\newcommand{\mw}{m_W}
\definecolor{MyDarkBlue}{rgb}{0.1, 0.1, 0.8} %defining the color 'MyDarkBlue'
\definecolor{SBlue}{rgb}{0.2, 0.4, 0.7} %defining the color 'MyDarkBlue'
\definecolor{MyLightBlue}{rgb}{0.22,0.51,0.9}
\definecolor{MyGreen}{rgb}{0.0, 0.5, 0.0}
\definecolor{BrickRed}{rgb}{0.8, 0.25, 0.33}
\RequirePackage{hyperref}
\hypersetup{colorlinks, citecolor=BrickRed,linkcolor=MyDarkBlue, urlcolor=MyLightBlue}

\begin{document}
%\vspace*{-0.2in}
\preprint{FERMILAB-PUB-19-304-T, OSU-HEP-19-04}
%\color{SBlue}{\bf OSU-HEP-19-04}
%\end{flushright}
%\vspace{0.5cm}
%\begin{center}
%{\Large\bf 
\title{
 Non-Standard Interactions in Radiative Neutrino Mass Models
 }
 \author[a,b]{K. S. Babu,}
 \author[c,b]{P. S. Bhupal Dev,}
 \author[a,b]{Sudip Jana,}
 \author[a,b]{Anil Thapa}
 
 \affiliation[a]{Department of Physics, Oklahoma State University, Stillwater, OK 74078, USA}
 \affiliation[b]{Theoretical Physics Department, Fermi National Accelerator Laboratory, \\ 
 P.O. Box 500, Batavia,
IL 60510, USA}
\affiliation[c]{Department of Physics and McDonnell Center for the Space Sciences,\\ Washington University, St. Louis, MO 63130, USA}
\emailAdd{kaladi.babu@okstate.edu, bdev@wustl.edu, sudip.jana@okstate.edu, thapaa@okstate.edu}
%%%%%%%%%%%%%%%%%%%%%%%%%%%%%
\abstract{
Models of radiative Majorana neutrino masses require new scalars and/or fermions to induce lepton-number-violating interactions.  We show that these new particles also generate observable neutrino non-standard interactions (NSI) with matter. We classify radiative models as type-I or II, with type-I models containing at least one Standard Model (SM) particle inside the loop diagram generating neutrino mass, and type-II models having no SM particle inside the loop.  While type-II radiative models do not generate NSI at tree-level, popular models which fall under the type-I category are shown, somewhat surprisingly, to generate observable NSI at tree-level, while being consistent with direct and indirect constraints from colliders, electroweak precision data and charged-lepton flavor violation (cLFV).  We survey such models where neutrino masses arise at one, two and three loops.  In the  prototypical Zee model which generates neutrino masses via one-loop diagrams involving  charged scalars, we find that diagonal NSI can be as large as ($8\%, 3.8\%, 9.3\%$) for ($\varepsilon_{ee},\varepsilon_{\mu \mu}, \varepsilon_{\tau \tau}$), while off-diagonal NSI can be at most ($10^{-3}\%, 0.56\%, 0.34\%$) for ($\varepsilon_{e\mu},\varepsilon_{e \tau}, \varepsilon_{\mu \tau}$). In one-loop neutrino mass models  using leptoquarks (LQs),  
$(\varepsilon_{\mu\mu},\, \varepsilon_{\tau\tau})$ can be as large as $(21.6\%,\,51.7\%)$, while $\varepsilon_{ee}$ and  $(\varepsilon_{e\mu},\, \varepsilon_{e\tau},\,\varepsilon_{\mu\tau})$ can at most be 0.6\%. Other two- and three-loop LQ models are found to give NSI of similar strength. The most stringent constraints on the diagonal NSI are found to come from neutrino oscillation and scattering experiments, while the off-diagonal NSI are mostly constrained by low-energy processes, such as atomic parity violation and 
cLFV. We also comment on the future sensitivity of these radiative models in long-baseline neutrino experiments, such as DUNE. While our analysis is focused on radiative neutrino mass models, it essentially covers 
all NSI possibilities with heavy mediators.    
}


\arxivnumber{1907.09498}

\maketitle

%%%%%%%%%%%%%%%%%%%%%%%%%%%%%%%%%%%%%%%%%%%%%%%
%%%%%%%%%%%%%%%%%%%%%%%%%%%%%%%%%%%%%%%%%%%%%%%
\section{Introduction}\label{sec:Intro}
%%%%%%%%%%%%%%%%%%%%%%%%%%%%%%%%%%%%%%%%%%%%%
%%%%%%%%%%%%%%%%%%%%%%%%%%%%%%%%%%%%%%%%%%%%%

The origin of tiny neutrino masses needed to explain the observed neutrino oscillation data is of fundamental importance in particle physics.  Most attempts that explain the smallness of these masses assume the neutrinos to be Majorana particles, in which case their masses could arise from effective higher dimensional operators, suppressed by a high energy scale that characterizes lepton number violation.  This is the case with the seesaw mechanism, where the dimension-five operator \cite{Weinberg:1979sa}
\begin{equation}
    {\cal O}_1 \ = \ L^iL^j H^kH^l \epsilon_{ik} \epsilon_{jl}
    \label{O1}
\end{equation}
suppressed by an inverse mass scale $\Lambda$ is induced by integrating out Standard Model (SM) singlet fermions~\cite{Minkowski:1977sc,Mohapatra:1979ia,Yanagida:1979as,GellMann:1980vs,Glashow:1979nm},  $SU(2)_L$ triplet scalars~\cite{Schechter:1980gr,Cheng:1980qt,Mohapatra:1980yp,Lazarides:1980nt}, or  $SU(2)_L$ triplet fermions~\cite{Foot:1988aq} with mass of order $\Lambda$.\footnote{For a clear discussion of the classification of seesaw types see Ref. \cite{Ma:1998dn}.} In Eq.~\eqref{O1}, $L$ stands for the lepton doublet, and $H$ for the Higgs doublet, with $i,j,k,l$ denoting $SU(2)_L$ indices, and $\epsilon_{ik}$ is the $SU(2)_L$ antisymmetric tensor.  Once the vacuum expectation value (VEV) of the Higgs field, $\langle H^0 \rangle \simeq 246$ GeV is inserted in Eq.~(\ref{O1}), Majorana masses for the neutrinos given by $m_\nu = v^2/\Lambda$ will be induced. For light neutrino masses in the observed range, $m_\nu \sim (10^{-3}-10^{-1})$ eV, the scale $\Lambda$ should be around $10^{14}$ GeV.  The mass of the new particle that is integrated out need not be $\Lambda$, since it is parametrically different, involving a combination of Yukawa couplings and $\Lambda$.  For example, in the type-I seesaw model the heavy right-handed neutrino mass goes as $M_R \sim y_D^2 \Lambda$, which can be near the TeV scale, if the Dirac Yukawa coupling $y_D \sim 10^{-6}$.  However, it is also possible that $y_D \sim {\cal O}(1)$, in which case the new physics involved in neutrino mass generation could not be probed directly in experiments.\footnote{This is strictly true for one generation case. For more than one generation, the scale could be lower~\cite{Kersten:2007vk}.}

An alternative explanation for small neutrino masses is that they arise only as quantum corrections \cite{Zee:1980ai,Zee:1985id,Babu:1988ki} (for a review, see Ref.~\cite{Cai:2017jrq}).  In these radiative neutrino mass models, the tree-level Lagrangian does not generate ${\cal O}_1$ of Eq.~(\ref{O1}), owing to the particle content or symmetries present in the model.  If such a model has lepton number violation, then small Majorana masses for neutrinos will be induced at the loop level. The leading diagram may arise at one, two, or three loop level, depending on the model details, which will have an appropriate loop suppression factor, and typically a chiral suppression factor involving a light fermion mass as well.\footnote{The magnitude of $m_\nu$ would be too small if it is induced at four or higher loops, assuming that the diagrams have chiral suppression factors proportional to the SM charged fermion masses; see Sec.~\ref{sec:four-loop}.}  For example, in the two-loop neutrino mass model of Refs.~\cite{Zee:1985id,Babu:1988ki}, small and calculable $m_\nu$ arises from the diagram shown in Fig. \ref{zee_babu}, which is estimated to be of order
\begin{equation}
m_\nu \ \approx \ \frac{f^2 h}{(16\pi^2)^2} \frac{m_\mu^2}{M} \, ,
\end{equation}
assuming normal ordering of neutrino masses and requiring large $\mu-\tau$ mixing.  Here
$f,h$ are Yukawa couplings involving new charged scalars with mass of order $M$.  Even with $f \sim h \sim 1$, to obtain $m_\nu \sim 0.1$ eV, one would require the scalar mass $M \sim$ TeV.  This type of new physics can be directly probed at colliders, enabling direct tests of the origin of neutrino mass.

When the mediators of neutrino mass generation have masses around or below the TeV scale, they can also induce other non-standard processes.  The focus of this paper is neutrino non-standard interactions (NSI)~\cite{Wolfenstein:1977ue} induced by these mediators.  These NSI are of great phenomenological interest, as their presence would modify the standard three-neutrino oscillation picture.  The NSI will modify scattering experiments, as the production and detection vertices are corrected; they would also modify neutrino oscillations, primarily through new contributions to matter effects. There have been a variety of phenomenological studies of NSI in the context of oscillations, but relatively lesser effort has gone into the ultraviolet (UV) completion of models that yield such NSI (for a recent update, see Ref.~\cite{Dev:2019anc}). A major challenge in generating observable NSI in any UV-complete model is that there are severe constraints arising from charged-lepton flavor violation (cLFV) \cite{Gavela:2008ra}.  One possible way to avoid such constraints is to have light mediators for NSI~\cite{Farzan:2015hkd,Babu:2017olk,Denton:2018xmq}. In contrast to these attempts, in this paper we focus on heavy mediators, and study the range of NSI allowed in a class of radiative neutrino mass models.\footnote{Analysis of Ref.~\cite{Forero:2016ghr, Dey:2018yht} of neutrino NSI in a model with charged singlet and/or doublet scalars, although not in the context of a neutrino mass model, is analogous to one model we analyze.}  Apart from being consistent with cLFV constraints, these models should also be consistent with direct collider searches for new particles and precision electroweak constraints.  We find, somewhat surprisingly, that the strengths of the diagonal NSI can be (20-50)\%  of the weak interaction strength for the flavor diagonal components in a class of popular models that we term as type-I radiative neutrino mass models, while they are absent at tree-level in another class, termed type-II radiative models.

\subsection{Type-I and type-II radiative neutrino mass models} \label{sec:intro1}

We propose a nomenclature that greatly helps the classification of various radiative models of neutrino mass generation.  One class of models can be described by lepton number violating effective higher dimensional operators, similar to Eq.~(\ref{O1}).  A prototypical example is the Zee model \cite{Zee:1980ai} which introduces a second Higgs doublet and a charged $SU(2)_L$-singlet scalar to the SM.  Interactions of these fields violate lepton number, and would lead to the effective lepton number violating $(\Delta L = 2)$ dimension 7 operator
\begin{equation}
    {\cal O}_2 \ = \ L^i L^j L^k e^c H^l \epsilon_{ij} \epsilon_{kl}
    \label{O2}
\end{equation}
with indices $i,j,..$ referring to $SU(2)_L$, and $e^c$ standing for the $SU(2)_L$ singlet let-handed positron state.  Neutrino masses arise via the one-loop diagram shown in Fig.~\ref{loopzee}.  The induced neutrino mass has an explicit chiral suppression factor, proportional to the charged lepton mass inside the loop.  Operator ${\cal O}_2$ can be obtained by cutting the diagram of Fig.~\ref{loopzee}.  We call radiative neutrino mass models of this type, having a loop suppression and a chirality suppression proportional to a light charged fermion mass, and expressible in terms of an effective higher dimensional operator as in Eq.~(\ref{O2}) as type-I radiative models. A classification of low dimensional operators that violate lepton number by two units has been worked out in Ref.~\cite{Babu:2001ex}. Each of these operators can generate a finite set of type-I radiative neutrino mass models in a well-defined manner. Lepton number violating phenomenology of these operators has been studied in Ref. \cite{deGouvea:2007qla}.  

Another well known example in this category is the two-loop neutrino mass model of Refs.~\cite{Zee:1985id,Babu:1988ki}, which induces an effective $d=9$ operator
\begin{equation}
{\cal O}_9 \ = \ L^i L^j L^k e^c L^l e^c \epsilon_{ij} \epsilon_{kl} \, .
\label{O9}
\end{equation}
Neutrino masses arise in this model via the two-loop diagrams shown in Fig.~\ref{zee_babu}, which has a chiral suppression factor proportional to $m_\ell^2$, with $\ell$ standing for the charged leptons of the SM.  

This category of type-I radiative neutrino mass models is populated by one-loop, two-loop, and three-loop models. Popular one-loop type-I models include the Zee model \cite{Zee:1980ai} (cf.~Sec.~\ref{sec:Zee}), and its variant with LQs replacing the charged scalars (cf.~Sec.~\ref{sec:LQ}). This variant is realized in supersymmetric models with $R$-parity violation \cite{Hall:1983id}.  Other one-loop models include $SU(2)_L$-triplet LQ models (cf.~Sec.~\ref{subsec:1loopLQ}) wherein the neutrino mass is proportional to the up-type quark masses \cite{Dorsner:2017wwn,AristizabalSierra:2007nf}. Ref.~\cite{Cai:2014kra} has classified simple realizations of all models leading to $d=7$ lepton number violating operators, which is summarized in Sec. \ref{sec:operator}. Popular type-I two-loop models include the Zee-Babu model \cite{Zee:1985id,Babu:1988ki} (cf.~Sec.~\ref{sec:zeebabu}), a  variant of it using LQs and a diquark (DQ)~\cite{Kohda:2012sr} (cf.~Sec.~\ref{subsec:colorzeebabu}), a pure LQ extension~\cite{Babu:2010vp} (cf.~Sec.~\ref{subsec:color2}), a model with LQs and vector-like fermions~\cite{Babu:2011vb} (cf.~Sec.~\ref{subsec:VQLQ}), and the Angelic model~\cite{Angel:2013hla} (cf.~Sec.~\ref{subsec:angelic}).  We also present here a new two-loop model (cf.~Sec.~\ref{subsec:newcolor}) with LQs wherein the neutrino masses are proportional to the up-type quark masses.  Type-I three-loop models include the KNT model \cite{Krauss:2002px} (cf.~Sec.~\ref{sec:KNT}), an LQ variant of the KNT model~\cite{Nomura:2016ezz} (cf.~Sec.~\ref{sec:3loopLQ}), the AKS model \cite{Aoki:2008av} (cf.~Sec.~\ref{sec:AKS}), and the cocktail model \cite{Gustafsson:2012vj} (cf.~Sec.~\ref{sec:cocktail}).  For a review of this class of models, see Ref.~\cite{Cai:2017jrq}.

A systematic approach to identify type-I radiative models is to start from a given $\Delta L =2$ effective operators of the type ${\cal O}_2$ of Eq.~(\ref{O2}), open the operator in all possible ways, and identify the mediators that would be needed to generate the operator.  Such a study was initiated in Ref. \cite{Babu:2001ex}, and further developed in Refs.~\cite{Angel:2012ug,Cai:2014kra}.  We shall rely on these techniques.  In particular, the many models suggested in Ref.~\cite{Cai:2014kra} have been elaborated on in Sec. \ref{sec:other_type1}, and their implications for NSI have been identified. This method has been applied to uncover new models in Ref. \cite{Klein:2019jgb}.  

\begin{table}[!t]
    \centering
    \begin{tabular}{|c|c|}
    \hline \hline
      \textbf{  Particle Content}  & \textbf{Lagrangian term} \\ \hline \hline
   $\eta^+({\bf 1},{\bf 1},1)$ or $h^+({\bf 1},{\bf 1},1)$ & $f_{\alpha\beta}L_\alpha L_\beta\, \eta^+$  or $f_{\alpha\beta}L_\alpha L_\beta\, h^+$\\ \hline 
   $\Phi\left({\bf 1},{\bf 2},\frac{1}{2}\right) = \left(\phi^+,\phi^0\right)$ & $Y_{\alpha\beta}L_\alpha \ell^c_\beta \widetilde{\Phi}$ \\ \hline
   $\Omega\left({\bf 3},{\bf 2},\frac{1}{6}\right) =\begin{pmatrix} \omega^{2/3} , \omega^{-1/3} \end{pmatrix}$
   & $\lambda_{\alpha\beta}L_\alpha d^c_\beta \Omega$ \\ \hline
   $\chi\left({\bf 3},{\bf 1},-\frac{1}{3}\right)$ & $\lambda'_{\alpha\beta}L_\alpha Q_\beta \chi^\star$ \\ \hline
   $\bar{\rho}\left(\bar{\bf 3},{\bf 3},\frac{1}{3}\right)=\left(\bar{\rho}^{4/3},\bar{\rho}^{1/3},\bar{\rho}^{-2/3}\right)$ & $\lambda''_{\alpha\beta}L_\alpha Q_\beta \bar{\rho}$ \\ \hline
   $\delta\left({\bf 3},{\bf 2},\frac{7}{6}\right)=\left(\delta^{5/3},\delta^{2/3}\right)$ & $\lambda'''_{\alpha\beta}L_\alpha u^c_\beta \delta$ \\ \hline
   $\Delta({\bf 1},{\bf 3},1)=\left(\Delta^{++},\Delta^{+},\Delta^0\right)$ & $f'_{\alpha\beta}L_\alpha L_\beta \Delta$ \\ \hline 
   \hline
    \end{tabular}
    \caption{Summary of new particles, their $SU(3)_c\times SU(2)_L\times U(1)_Y$ quantum numbers (with the non-Abelian charges in boldface), field components and electric charges (in superscript), and corresponding Lagrangian terms responsible for NSI in various type-I radiative neutrino mass models discussed in Secs.~\ref{sec:Zee}, \ref{sec:LQ} and \ref{sec:other_type1}. Here $\widetilde{\Phi}=i\tau_2\Phi^\star$, with $\tau_2$ being the second Pauli matrix. For a singly charged scalar, $\eta^+$ and $h^+$ are used interchangeably, to be consistent with literature. }
    \label{tab:glossory}
    \end{table}
    
In all these models there are new scalar bosons, which are almost always necessary for neutrino mass generation in type-I radiative models using effective higher dimensional operators. For future reference, we list in Table~\ref{tab:glossory} all possible new scalar mediators in type-I radiative models that can couple to neutrinos, along with their $SU(3)_c\times SU(2)_L\times U(1)_Y$ quantum numbers, field components and electric charges (in superscript), and corresponding Lagrangian terms responsible for NSI. We will discuss them in detail in~\ref{sec:Zee},~\ref{sec:LQ} and~\ref{sec:other_type1}. The models discussed in Sec.~\ref{sec:other_type1} contain other particles as well, which are however not relevant for the NSI discussion, so are not shown in Table~\ref{tab:glossory}.  Note that the scalar triplet $\Delta({\bf 1},{\bf 3},1)$ could induce neutrino mass at tree-level via type-II seesaw mechanism~\cite{Schechter:1980gr,Cheng:1980qt,Mohapatra:1980yp,Lazarides:1980nt}, which makes radiative models involving $\Delta$ field somewhat unattractive, and therefore, is not included in our subsequent discussion.  

There is one exception to the need for having new scalars for type-I radiative models (see Sec.~\ref{subsec:mrism}). The Higgs boson and the $W,Z$ bosons of the SM can be the mediators for radiative neutrino mass generation, with the new particles being fermions.  In this case, however, there would be tree-level neutrino mass \'{a} la type-I seesaw mechanism~\cite{Minkowski:1977sc,Mohapatra:1979ia,Yanagida:1979as,GellMann:1980vs,Glashow:1979nm}, which should be suppressed by some mechanism or symmetry.  Such a model has been analyzed in Refs.~\cite{Pilaftsis:1991ug,Dev:2012sg}, which leads to interesting phenomenology, see Sec.~\ref{sec:other_1loop}.

From the perspective of neutrino NSI, these type-I radiative models are the most interesting, as the neutrino couples to a SM fermion and a new scalar directly, with the scalar mass near the TeV scale. We have analyzed the ranges of NSI possible in all these type-I radiative models.  Our results are summarized in Fig.~\ref{fig:summaryplot} and Table~\ref{Table_Models}.

A second class of radiative neutrino mass models has entirely new (i.e., non-SM) particles inside the loop diagrams generating the mass.  These models cannot be derived from effective $\Delta L = 2$ higher-dimensional operators, as there is no way to cut the loop diagram and generate such operators.  We term this class of models type-II radiative neutrino mass models (cf.~Sec.~\ref{sec:type2}).  The induced neutrino mass may have a chiral suppression, but this is not proportional to any light fermion mass.  Effectively, these models generate operator ${\cal O}_1$ of Eq.~(\ref{O1}), but with some loop suppression. From a purely neutrino mass perspective, the scale of new physics could be of order $10^{10}$ GeV in these models.  However, there are often other considerations which make the scale near a TeV, a prime example being the identification of a WIMP dark matter with a particle that circulates in the loop diagram generating neutrino mass.

A well-known example of the type-II radiative neutrino mass model is the scotogenic model~\cite{Ma:2006km}  which assumes a second Higgs doublet and right-handed neutrinos $N$ beyond the SM.  A discrete $Z_2$ symmetry is assumed under which $N$ and the second Higgs doublet are odd.  If this $Z_2$ remains unbroken, the lightest of the $Z_2$-odd particles can serve as a dark matter candidate.  Neutrino mass arises through the diagram of Fig.~\ref{scotogenic}.  Note that this diagram cannot be cut in any way to generate an effective higher dimensional operator of the SM.  While the neutrino mass is chirally suppressed by $M_N$, this need not be small, except for the desire for it (or the neutral component of the scalar) to be TeV-scale dark matter. There are a variety of other models that fall into the type-II category \cite{Kubo:2006yx,FileviezPerez:2009ud,Law:2013saa,Restrepo:2013aga,Baek:2015mna,Dutta:2018qei}.

The type-II radiative neutrino mass models will have negligible neutrino NSI, as the neutrino always couples to non-SM fermions and scalars.  Any NSI would be induced at the loop level, which would be too small to be observable in experiments. As a result, in our comprehensive analysis of radiative neutrino mass models for NSI, we can safely ignore type-II models.

One remark is warranted here.  Consider an effective operator of the type
\begin{equation}
{\cal O}'_1 \ = \ L^i L^j H^k H^l \epsilon_{ik} \epsilon_{jl} (u^c u^c) (u^c u^c)^\star.
\label{Op}
\end{equation}
Such an operator would lead to neutrino masses at the two-loop level, as can be seen in an explicit model shown in Fig.~\ref{type0}.  Although this model can be described as arising from an effective $\Delta L = 2$ operator, the neutrino mass has no chiral suppression here.
The mass scale of the new scalars could be as large as $10^{10}$ GeV. Such models do belong to type-I radiative models; however, they are more like type-II models due to the lack of a chiral suppression.  In any case, the NSI induced by the LQs that go inside the loop diagram for neutrino masses is already covered in other type-I radiative models that we have analyzed.  Another example of this type of operator is $L^i L^j H^k H^l \epsilon_{ik} \epsilon_{jl} (H^\dagger H)$, which is realized for instance in the minimal radiative inverse seesaw model (MRISM) of Ref.~\cite{Dev:2012sg} (see Sec.~\ref{subsec:mrism}). Such effective operators, which appear as products of lower operators,  were treated as trivial in the classification of Ref.~\cite{Babu:2001ex}.

\subsection{Summary of results} \label{sec:sum}

We have mapped out in this paper the allowed ranges for the neutrino NSI parameters $\varepsilon_{\alpha\beta}$ (cf.~Sec.~\ref{sec:NSI}) in radiative neutrino mass models.  We present a detailed analysis of the Zee model \cite{Zee:1980ai} with light  charged scalar bosons (cf.~Sec.~\ref{sec:Zee}).  To map out the allowed values of $\varepsilon_{\alpha\beta}$, we have analyzed constraints arising from the following experimental and theoretical considerations:
    i) Contact interaction limits from LEP (cf.~Sec.~\ref{sec:contact}); 
    ii) Monophoton constraints from LEP (cf.~Sec.~\ref{sec:monop}); 
    iii) Direct searches for charged scalar pair and single production at LEP (cf.~Sec.~\ref{sec:LEPZee}); 
    iv) Pair production of charged scalars at LHC (cf.~Sec.~\ref{sec:LHCZee}); 
    v) Higgs physics constraints from LHC (cf.~Sec.~\ref{sec:HiggsOb}); 
    vi) Lepton universality in $W^\pm$ decays (cf.~Sec.~\ref{sec:Wuniv}); 
    vii) Lepton universality in $\tau$ decays (cf.~Sec.~\ref{sec:taudecay}); 
    viii) Electroweak precision data (cf.~Sec.~\ref{sec:ewpt}); 
    ix) charged-lepton flavor violation (cf.~Sec.~\ref{sec:lfv}); 
    x) Perturbative unitarity of Yukawa and quartic couplings;  and
    xi) charge-breaking minima of the Higgs potential (cf.~Sec.~\ref{sec:CBM}).

Imposing these constraints, we find that light charged scalars, arising either from the $SU(2)_L$-singlet or doublet field or an admixture, can have a mass near 100 GeV.  Neutrino NSI obtained from the pure $SU(2)_L$-singlet component turns out to be unobservably small.  However, the $SU(2)_L$-doublet component in the light scalar can have significant Yukawa couplings to the electron and the neutrinos, thus inducing potentially large NSI. The maximum allowed NSI in this model is summarized below (cf.~Table~\ref{tab:Zee}):
\begin{center}
\begin{tabular}{|llll|}\hline
{\bf Zee } & $\varepsilon_{ee}^{\rm max} \  = \ 8\%$ \, ,  & 
    $\varepsilon_{\mu\mu}^{\rm max} \ = \ 3.8\%$ \, , &
     $ \varepsilon_{\tau\tau}^{\rm max} \ = \ 9.3\%$ \, , \\
 {\bf model:} & $\varepsilon_{e\mu}^{\rm max} \ = \ 0.0015\%$ \, , &
   $\varepsilon_{e\tau}^{\rm max} \ = \ 0.56\%$ \, , &
   $\varepsilon_{\mu\tau}^{\rm max} \ = \ 0.34\%$ \, . \\ \hline
\end{tabular}
\end{center}
These values are significantly larger than the ones obtained in Ref.~\cite{Herrero-Garcia:2017xdu}, where the contributions from the doublet Yukawa couplings of the light charged Higgs were ignored. 

We have also analyzed in detail LQ models of radiative neutrino mass generation.  As the base model we analyze the LQ version of the Zee model (cf.~Sec.~\ref{sec:LQ}), the results of which can also be applied to other LQ models with minimal modifications.  This analysis took into account the following experimental constraints: i) Direct searches for LQ pair and single production at LHC (cf.~Sec.~\ref{sec:highconstraints});  ii) APV (cf.~Sec.~\ref{sec:APV}); iii) charged-lepton flavor violation (cf.~Secs.~\ref{sec:llgLQ} and \ref{sec:tauLQ}); and iv) rare meson decays (cf.~Sec.~\ref{sec:Dmeson}).  Including all these constraints we found the maximum possible NSI induced by the singlet and doublet LQ components, as given below (cf.~Table~\ref{tab:LQ}):
\begin{center}
\begin{tabular}{|llll|}\hline
{\bf $SU(2)_L$-singlet} & $\varepsilon_{ee}^{\rm max} \  = \ 0.69\%$,  & 
    $\varepsilon_{\mu\mu}^{\rm max} \ = \ 0.17\%$, &
     $ \varepsilon_{\tau\tau}^{\rm max} \ = \ 34.3\%$, \\
{\bf LQ model:}  & $\varepsilon_{e\mu}^{\rm max} \ = \ 1.5\times 10^{-5}\%$, &
   $\varepsilon_{e\tau}^{\rm max} \ = \ 0.36\%$, &
   $\varepsilon_{\mu\tau}^{\rm max} \ = \ 0.43\%$. \\ \hline
\end{tabular}
\end{center}

\begin{center}
\begin{tabular}{|llll|}\hline
{\bf $SU(2)_L$-doublet} & $\varepsilon_{ee}^{\rm max} \  = \ 0.4\%$,  & 
    $\varepsilon_{\mu\mu}^{\rm max} \ = \ 21.6\%$, &
     $ \varepsilon_{\tau\tau}^{\rm max} \ = \ 34.3\%$, \\
{\bf LQ model:}  & $\varepsilon_{e\mu}^{\rm max} \ = \ 1.5\times 10^{-5}\%$, &
   $\varepsilon_{e\tau}^{\rm max} \ = \ 0.36\%$, &
   $\varepsilon_{\mu\tau}^{\rm max} \ = \ 0.43\%$. \\ \hline
\end{tabular}
\end{center}
 Our results yield somewhat larger NSI compared to the results of Ref.~\cite{Wise:2014oea} which analyzed, in part,  effective interactions obtained by integrating out the LQ fields. 
 %Also note that the maximum $\varepsilon_{\tau\tau}$ values allowed in these LQ models (and the maximum $\varepsilon_{\mu\mu}$ in the $SU(2)_L$-doublet LQ models) are larger than the Zee model predictions, since these are not subject to the 
 
 We also analyzed a variant of the LQ model with $SU(2)_L$-triplet LQs, which have couplings to both up and down quarks simultaneously. The maximum NSI in this case are found to be as follows (cf.~Eq.~\eqref{eq:NSI-rho}): 
\begin{center}
\begin{tabular}{|llll|}\hline
{\bf $SU(2)_L$-triplet} & $\varepsilon_{ee}^{\rm max} \  = \ 0.59\%$,  & 
    $\varepsilon_{\mu\mu}^{\rm max} \ = \ 2.49\%$, &
     $ \varepsilon_{\tau\tau}^{\rm max} \ = \ 51.7\%$, \\
{\bf LQ model:}  & $\varepsilon_{e\mu}^{\rm max} \ = \ 1.9\times 10^{-6}\%$, &
   $\varepsilon_{e\tau}^{\rm max} \ = \ 0.50\%$, &
   $\varepsilon_{\mu\tau}^{\rm max} \ = \ 0.38\%$. \\ \hline
\end{tabular}
\end{center}

For completeness, we also list here the maximum possible tree-level NSI in the two-loop Zee-Babu model (cf.~Eq.~\eqref{eq:maxZB}):
\begin{center}
\begin{tabular}{|llll|}\hline
{\bf Zee-Babu} & $\varepsilon_{ee}^{\rm max} \  = \ 0\%$,  & 
    $\varepsilon_{\mu\mu}^{\rm max} \ = \ 0.9\%$, &
     $ \varepsilon_{\tau\tau}^{\rm max} \ = \ 0.3\%$  , \\
{\bf model:}  & $\varepsilon_{e\mu}^{\rm max} \ = \ 0\%$, &
   $\varepsilon_{e\tau}^{\rm max} \ = \ 0\%$, &
   $\varepsilon_{\mu\tau}^{\rm max} \ = \ 0.3\%$. \\ \hline
\end{tabular}
\end{center}

The NSI predictions in all other models analyzed here will fall into one of the above categories (except for the MRISM discussed in Sec.~\ref{subsec:mrism}). Our results for the base models mentioned above are summarized in Fig.~\ref{fig:summaryplot}, and the results for all the models analyzed in this paper are tabulated in Table~\ref{Table_Models}. We emphasize that while our analysis is focused on radiative neutrino mass models, it essentially covers 
all NSI possibilities with heavy mediators, and thus is more general.

The rest of the paper is structured as follows. In Sec.~\ref{sec:operator}, we discuss the classification of low-dimensional lepton-number violating operators and their UV completions. In Sec.~\ref{sec:NSI}, we briefly review neutrino NSI and establish our notation.  Sec.~\ref{sec:Zee} discusses the Zee model of neutrino masses and derives the various experimental and theoretical constraints on the model.  Applying these constraints, we derive the allowed range for the NSI parameters.  Here we also show how neutrino oscillation data may be consistently explained with large NSI.  In Sec.~\ref{sec:LQ} we turn to the one-loop radiative model for neutrino mass with LQs.  Here we delineate the collider and low energy constraints on the model and derive the ranges for neutrino NSI. In Sec.~\ref{sec:CCSVO39}, we discuss a variant of the one-loop LQ model with triplet LQ. In Sec.~\ref{sec:other_type1} we discuss other type-I models of radiative neutrino mass and obtain the allowed values of $\varepsilon_{\alpha\beta}$. We briefly discuss NSI in type-II models in Sec.~\ref{sec:type2}. In Sec.~\ref{sec:con} we conclude. Our results are tabulated in Table~\ref{Table_Models} and summarized in Fig.~\ref{fig:summaryplot}. In Appendix~\ref{app:A}, we present the analytic expressions for the charged-scalar production cross sections in electron-positron collisions.
%%%%%%%%%%%%%%%%%%

\section{Classification of \texorpdfstring{$\Delta L = 2$}{delta} operators and their UV completions} \label{sec:operator}
It is instructive to write down low-dimensional effective operators that carry lepton number of two units ($\Delta L = 2$), since all type-I radiative models can be constructed systematically from these operators.  Here we present a summary of such operators through $d=7$ \cite{Babu:2001ex}. We use  two component Weyl notation for SM fermions and denote them as
\begin{equation}
L\left({\bf 1},{\bf 2},-\frac{1}{2}\right),~~ e^{c}({\bf 1},{\bf 1},1),~~ Q\left({\bf 3},{\bf 2}, \frac{1}{6}\right), ~~d^{c}\left(\overline{\bf 3}, {\bf 1}, \frac{1}{3}\right),~~ u^{c}\left(\overline{\bf 3}, {\bf 1},-\frac{2}{3}\right)~.
\end{equation}
The Higgs field of the SM is denoted as $H\left({\bf 1},{\bf 2},\frac{1}{2}\right)$.  
The $\Delta L =2$ operators in the SM are all odd-dimensional. The full list of operators through $d=7$ is given by \cite{Babu:2001ex}:
\begin{subequations}
\begin{eqnarray}
\mathcal{O}_{1}&\ = \ &L^{i} L^{j} H^{k} H^{l} \epsilon_{i k} \epsilon_{j l} \, , \label{eq:O1} \\
\mathcal{O}_{2} &\ = \ &L^{i} L^{j} L^{k} e^{c} H^{l} \epsilon_{i j} \epsilon_{k l} \, ,\label{eq:O2}  \\ 
\mathcal{O}_{3} & \ = \ & \left\{ L^{i} L^{j} Q^{k} d^{c} H^{l} \epsilon_{i j} \epsilon_{k l}, ~~  L^{i} L^{j} Q^{k} d^{c} H^{l} \epsilon_{i k} \epsilon_{j l}\right\} \ \equiv \ \{\mathcal {O}_{3a},~\mathcal{O}_{3b} \} \, , \label{eq:O3} \\
\mathcal{O}_{4} & \ = \ & \left\{L^{i} L^{j} \overline{Q}_{i} \overline{u^c} H^{k} \epsilon_{j k}, ~~ L^{i} L^{j} \overline{Q}_{k} \overline{u^c} H^{k} \epsilon_{i j}\right\} \ \equiv \  \{\mathcal {O}_{4a}, ~\mathcal{O}_{4b} \} \, , \label{eq:O4} \\ 
%\mathcal{O}_{5} &\ = \ &L^{i} L^{j} Q^{k} d^{c} H^{l} H^{m} \overline{H}_{i} \epsilon_{j l} \epsilon_{k m} \, ,\label{eq:O5} \\ 
%\mathcal{O}_{6} & \ = \ &L^{i} L^{j} \overline{Q}_{k} \overline{u^c} H^{l} H^{k} \overline{H}_{i} \epsilon_{j l} \, , \label{eq:O6} \\
%\mathcal{O}_{7} &\ = \ &L^{i} Q^{j} \overline{e^c} \overline{Q}_{k} H^{k} H^{l} H^{m} \epsilon_{i l} \epsilon_{j m} \, , \label{eq:O7} \\
\mathcal{O}_{8} & \ = \ &L^{i} \overline{e^c}\, \overline{u^c} d^{c} H^{j} \epsilon_{i j} \, \label{eq:O8} ~.
\end{eqnarray}
\end{subequations}
Not listed here are products of lower-dimensional operators, such as $\mathcal{O}_1 \times \overline{H} H$, with the $SU(2)_L$ contraction of $\overline{H} H$ being a singlet.  Here  $\mathcal{O}_1$ is the Weinberg operator \cite{Weinberg:1979sa}, while the remaining operators are all $d=7$.\footnote{In the naming convention of Ref. \cite{Babu:2001ex}, operators were organized based on how many fermion fields are in them. Operators $\mathcal{O}_5-\mathcal{O}_7$, which are $d=9$ operators, appeared ahead of the $d=7$ operator $\mathcal{O}_8$.} In this paper, we shall analyze all models of neutrino mass arising from these $d=7$ operators for their NSI, as well as the two-loop Zee-Babu model arising from $\mathcal{O}_9$ of Eq. (\ref{O9}). A few other models that have been proposed in the literature with higher dimensional operators will also be studied.  The full list of $d=9$ models is expected to contain a large number, which has not been done to date.

%%%%%%%%%%%%%%%%%%%%%%%%%%%%%%%%%%%%%%%%%
%%%%%%%%%%%%%%%%%%%%%%%%%%%%%%%%%%%%%%%%%
\begin{figure}[t!]
         \centering
         \subfigure[]{
        \includegraphics[scale=0.5]{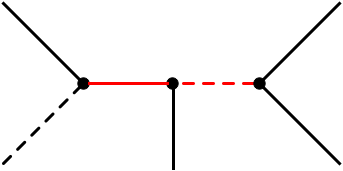}}
          \hspace{0.15in}
          \subfigure[]{
         \includegraphics[scale=0.5]{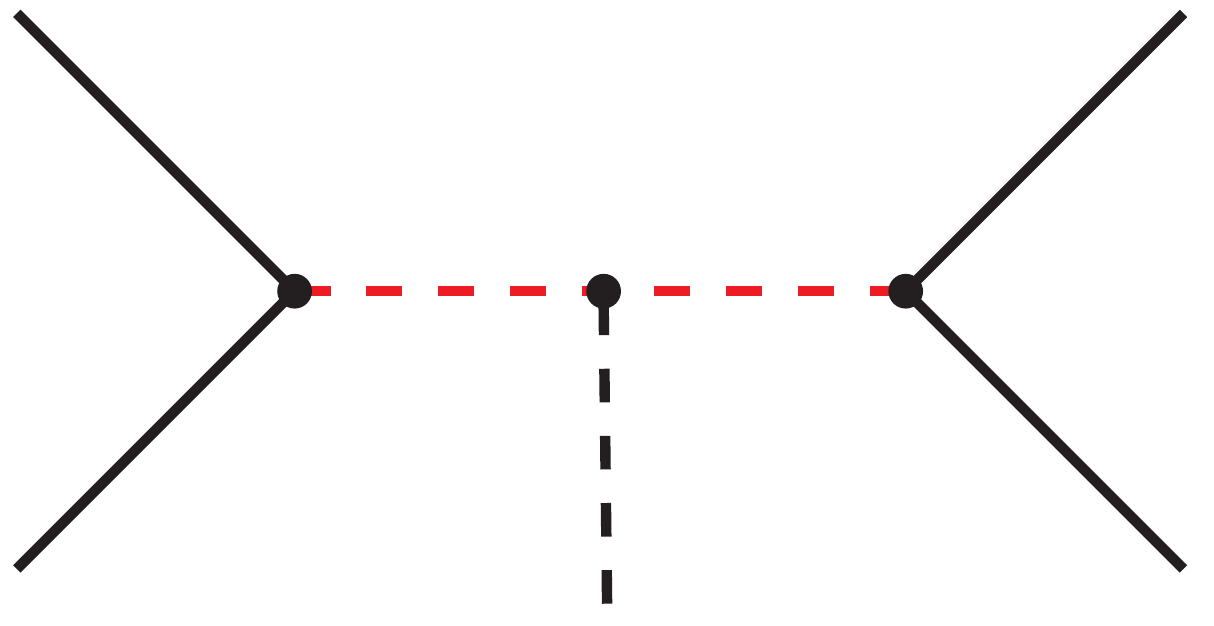}}
         \caption{Diagrams that generate operators of dimension 7 via (a) scalar and vectorlike fermion exchange, and  (b) by pure scalar exchange. }
         \label{fig:Topology}
    \end{figure}
%%%%%%%%%%%%%%%%%%%%%%%%%%%%%%%%%%%%%%%%%%%%%%%%%%%%%%%%%%%%%%%%%%%%%%%%%%%%%%%%%%

Each of these $d=7$ operators can lead to finite number of UV complete neutrino mass models. The generic diagrams that induce all of the $d=7$ operators are shown in Fig.~\ref{fig:Topology}.  Take for example the operator $\mathcal{O}_2$ in Eq.~\eqref{eq:O2}. There are two classes of models that can generate this operator with the respective mediators obtained from the following contractions (see Table~\ref{tab:O2}):
\begin{equation}
\mathcal{O}_{2}^1 \ = \ L (LL)(e^c H)\, ,\qquad \mathcal{O}_{2}^2 \ = \ H (LL)(L e^c)~.
\end{equation}
Here the pairing of fields suggests the mediator necessary.  The $(LL)$ contraction would require a scalar that can be either an $SU(2)_L$ singlet, or a triplet. The $(e^c H)$ contraction would require a new fermion, which is typically a vectorlike fermion.\footnote{There is a third contraction allowed in principle, $e^c(LL)(LH)$. However, the mediator needed to realize this would generate $d=5$ operator $LLLH$ either via type-I or type-II seesaw at tree-level, and hence this contraction is not used in radiative neutrino mass models.} Thus,  $\mathcal{O}_{2}^1$ has two UV completions, with the addition of a vectorlike lepton $\psi\left({\bf 1},{\bf 2},-\frac{3}{2}\right)$ to the SM, along with a scalar which is either a singlet $\eta^+({\bf 1},{\bf 1},1)$, or a triplet $\Delta({\bf 1},{\bf 3},1)$. The choice of $\Delta({\bf 1},{\bf 3},1)$ can lead to the generation of the lower $d=5$ operator at tree level via type-II seesaw, and therefore, is usually not employed in radiative models.  The model realizing $\mathcal{O}_{2}^1$ with  $\psi\left({\bf 1},{\bf 2},-\frac{3}{2}\right)$ vectorlike lepton and $\eta^+({\bf 1},{\bf 1},1)$ scalar is discussed in Sec.~\ref{subsec:CCSVO21}.  Similarly operator $\mathcal{O}_{2}^2$ has a unique UV completion, with two scalars added to the SM -- one  $\eta^+({\bf 1},{\bf 1},1)$ and one $\Phi\left({\bf 1},{\bf 2},\frac{1}{2}\right)$.  This is the Zee model of neutrino mass, discussed at length in Sec.~\ref{sec:Zee}. 

\begin{table}[t!]
\begin{center}
\begin{tabular}{cc}
\toprule
& $\mathcal{O}_2^1$  \\
& $L (LL) (e^c H)$ \\
\midrule
$\phi$ &   $({\bf 1}, {\bf 1}, 1)$ \\[1ex]
$\psi$ &   $({\bf 1}, {\bf 2}, -\frac{3}{2})$ \\
\bottomrule
\end{tabular}
\hspace{5ex}
\begin{tabular}{cc}
\toprule
& $\mathcal{O}_2^2$\\
 & $ H (LL) (L e^c)$\\
\midrule
     $\phi$ & $({\bf 1}, {\bf 1}, 1)$ \\[1ex]
  $\eta$ & $({\bf 1}, {\bf 2}, \frac{1}{2})$ \\
\bottomrule
\end{tabular}
\end{center}
\caption{Minimal UV completions of operator $\mathcal{O}_{2}$~\cite{Cai:2014kra}. Here $\phi$ and $\eta$ generically denote scalars and $\psi$ is a generic vectorlike fermion. The SM quantum numbers of these new fields are as indicated.}
\label{tab:O2}
\end{table}

\begin{table}[t!]
\begin{center}
   \begin{tabular}{ccccccc}
   \toprule
& $\mathcal{O}_{3}^{1}$ &  $\mathcal{O}_{3}^{2}$ & $\mathcal{O}_{3}^{3}$ & $\mathcal{O}_{3}^{4}$ & $\mathcal{O}_{3}^{5}$ & $\mathcal{O}_{3}^{6}$\\
     & $Q(LL) ({d^c}H)$ & ${d^c} (LL) (Q H)$ & $L (L{d^c}) (QH)$ & $L(LQ) ({d^c} H)$  & $L(LQ) ({d^c} H)$ & $L (L{d^c}) (QH)$\\
      \midrule
      $\phi$    & $({\bf 1},{\bf  1}, 1)$ &   $({\bf 1}, {\bf 1}, 1)$& $\left({\bf 3},{\bf 2},\frac{1}{6}\right)$ & $\left({\bf 3}, {\bf 1}, -\frac{1}{3}\right)$  & $\left({\bf 3}, {\bf 3}, -\frac{1}{3}\right)$ & $\left({\bf 3},{\bf 2},\frac{1}{6}\right)$  \\[1ex]
      $\psi$  &  $\left({\bf 3}, {\bf 2}, -\frac{5}{6}\right)$& $\left({\bf 3}, {\bf 1}, \frac{2}{3}\right)$ & $\left({\bf 3}, {\bf 1}, \frac{2}{3}\right)$ & $\left({\bf 3}, {\bf 2}, -\frac{5}{6}\right)$  & $\left({\bf 3}, {\bf 2}, -\frac{5}{6}\right)$  & $\left({\bf 3}, {\bf 3}, \frac{2}{3}\right)$ \\
      \midrule
&$\mathcal{O}_{3a}$&$\mathcal{O}_{3a}$&$\mathcal{O}_{3a}$&$\mathcal{O}_{3b}$&$\mathcal{O}_{3a},\mathcal{O}_{3b}$&$\mathcal{O}_{3a},\mathcal{O}_{3b}$\\
\bottomrule
   \end{tabular}

\vspace{2ex}
   \begin{tabular}{cccc}
   \toprule
 & $\mathcal{O}_{3}^{7}$ &  $\mathcal{O}_{3}^{8}$ &  $\mathcal{O}_{3}^{9}$\\
   & $H (LL) (Q {d^c})$& $H (LQ) (L{d^c})$& $H (LQ) (L{d^c})$\\
   \midrule
    $\phi$ & $({\bf 1},{\bf 1},{\bf 1})$&  $\left({\bf 3}, {\bf 1}, -\frac{1}{3}\right)$&  $\left({\bf 3}, {\bf 3}, -\frac{1}{3}\right)$\\[1ex]
    $\eta$  & $\left({\bf 1}, {\bf 2}, \frac{1}{2}\right) $   &  $\left({\bf 3},{\bf 2},\frac{1}{6}\right)$  &  $\left({\bf 3},{\bf 2},\frac{1}{6}\right)$\\ %/ 8
    \midrule
&$\mathcal{O}_{3a}$&$\mathcal{O}_{3b}$&$\mathcal{O}_{3a},\mathcal{O}_{3b}$\\
\bottomrule
   \end{tabular}

\end{center}
\caption{Minimal UV completions of operators $\mathcal{O}_{3a}$ and $\mathcal{O}_{3b}$ \cite{Cai:2014kra}. Here the models in the top segment require a new scalar $\phi$ and a vectorlike fermion $\psi$, while those in the lower segment require two scalar fields $\phi$ and $\eta$.}
\label{tab:O3}
\end{table}

\begin{table}[t!]
\centering
\begin{tabular}{ccc}
\toprule
 & $\mathcal{O}_{4}^{1}$ &  $\mathcal{O}_{4}^{2}$\\
&   $\overline{Q} (LL)(\overline{u^c} H)$ & $\overline{u^c} (LL)(\overline{Q} H)$\\
\midrule
$\phi$ &  $({\bf 1},{\bf 1},1)$ & $({\bf 1},{\bf 1},1)$\\
$\psi$ &  $\left({\bf 3},{\bf 2},\frac{7}{6}\right)$ & $\left({\bf 3},{\bf 1},-\frac{1}{3}\right)$\\[1ex]
\midrule
 & $\mathcal{O}_{4b}$& $\mathcal{O}_{4b}$\\
\bottomrule
\end{tabular}
\hspace{5ex}
\begin{tabular}{cc}
\toprule
& $\mathcal{O}_{4}^{3}$ \\
& $H(LL)(\overline{Q} \overline{u^c}) $ \\
\midrule
$\phi$ &  $({\bf 1},{\bf 1},1)$\\[1ex]
$\eta$ &  $\left({\bf 1},{\bf 2},\frac{1}{2}\right)$\\ %/8
\midrule
& $\mathcal{O}_{4b}$\\
\bottomrule
\end{tabular}
\caption{Minimal UV completions of the operators $\mathcal{O}_{4a}$ and $\mathcal{O}_{4b}$. Note that only the operator $\mathcal{O}_{4b}$ is generated. Fields $\phi$ and $\eta$ are scalars, while the $\psi$ fields are vectorlike fermions.}
\label{tab:O4}
\end{table}

\begin{table}[htb]
\centering
\begin{tabular}{cccc}
\toprule
 &$\mathcal{O}_{8}^{1}$ &$\mathcal{O}_{8}^{2}$ &  $\mathcal{O}_{8}^{3}$ \\
 & $L(\overline{e^c} \,\overline{u^c}) (d^c H)$  & $\overline{u^c} (L{d^c})(\overline{e^c} H)$ 
& $\overline{e^c} (L{d^c})(\overline{u^c} H)$  \\
\midrule
$\phi$          & $\left({\bf 3},{\bf 1},-\frac{1}{3}\right)$  &  $\left({\bf 3},{\bf 2},\frac{1}{6}\right)$    & $\left({\bf 3},{\bf 2},\frac{1}{6}\right)$   \\[1ex]
$\psi$    & $\left( {\bf 3},{\bf 2},-\frac56\right)$  &  $\left({\bf 1},{\bf 2},-\frac{1}{2}\right)$ &  $\left({\bf 3},{\bf 2},\frac{7}{6}\right)$ \\
\bottomrule
\end{tabular}
\hspace{5ex}
\begin{tabular}{cccc}
\toprule
 & $\mathcal{O}_{8}^{4}$ \\
& $(L{d^c})(\overline{u^c} \overline{e^c})H$\\
\midrule
$\phi$     & $\left({\bf 3},{\bf 1},-\frac{1}{3}\right)$\\[1ex]
$\eta$ & $\left({\bf 3},{\bf 2},\frac{1}{6}\right)$      \\
\bottomrule
\end{tabular}
\caption{Minimal UV completions of operator $\mathcal{O}_8$. Fields $\phi$ and $\eta$ are scalars, while the $\psi$ fields are vectorlike fermions.}
\label{tab:O8}
\end{table}

Operators $\mathcal{O}_{3a}$ and $\mathcal{O}_{3b}$ in Eq.~\eqref{eq:O3} can be realized by the  UV complete models given in Table.~\ref{tab:O3}  \cite{Cai:2014kra}. Here all possible contraction among the fields are shown, along with the required mediators to achieve these contractions.  Fields denoted as $\phi$ and $\eta$ are scalars, while $\psi$ is a vectorlike fermion.  The SM quantum numbers for each field are also indicated in the Table. We shall analyze neutrino NSI arising from each of these models in Sec. \ref{sec:other_type1}.

The UV completions of operators $\mathcal{O}_4$ and $\mathcal{O}_8$ are shown in Tables \ref{tab:O4} and \ref{tab:O8} respectively \cite{Cai:2014kra}.  These models will all be analyzed in Sec.~\ref{sec:other_type1} for neutrino NSI.  Note that in both $\mathcal{O}_4$ and $\mathcal {O}_8$, pairing of un-barred and barred fermion fields is not included, as the mediators for such an UV completion will have to be vector bosons which would make such models difficult to realize.  As a result, only ${\cal O}_{4b}$ can be realized with scalar and fermionic exchange.

%%%%%%%%%%%%%%%%%%%%%%%%%%%%%%%%%%%%%%%%%%%%
\subsection*{Classification based on topology of diagrams} %\label{sec:topo}
\begin{figure}[!t]
         \centering
         \includegraphics[scale=0.6]{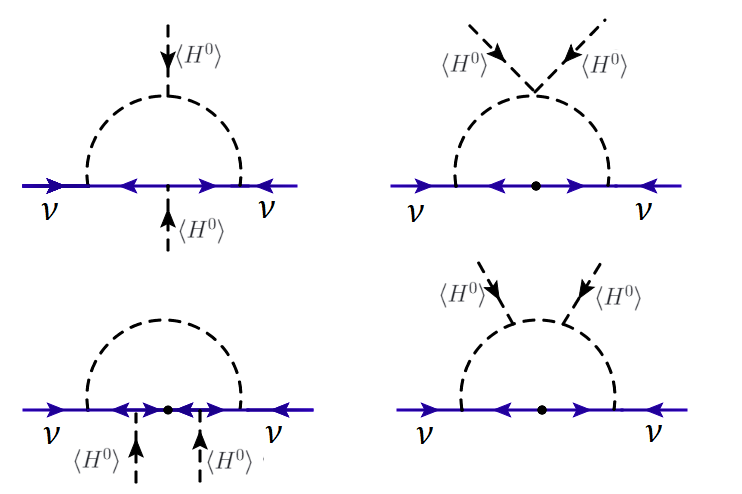}
         \caption{Topologies of one-loop radiative neutrino mass diagrams. } 
         \label{loop1}
    \end{figure}

\begin{figure}[!t]
         \centering
         \includegraphics[scale=0.5]{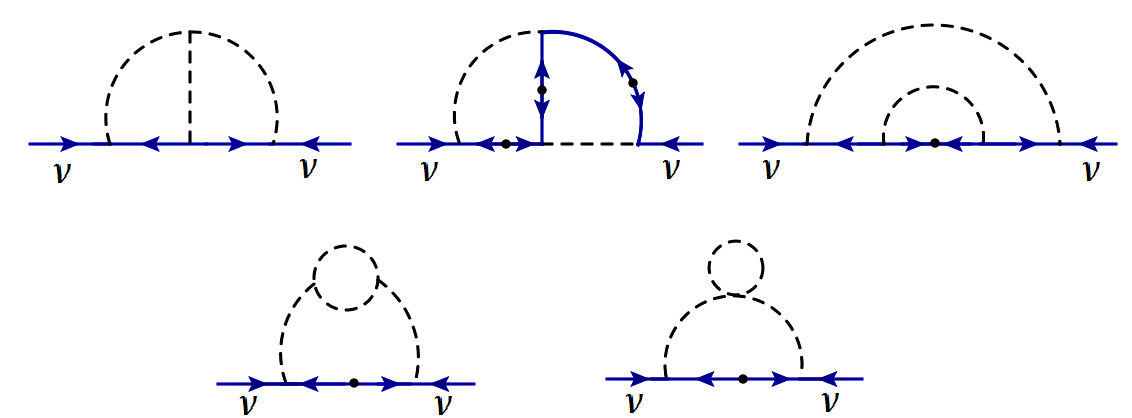}
         \caption{Topologies of two-loop neutrino mass diagrams. Two Higgs bosons should be attached to internal lines in all possible ways.} 
         \label{loop2}
    \end{figure}

Rather than classifying radiative neutrino mass models in terms of effective $\Delta L = 2$ operators, one could also organize them in terms of the topology of the loop diagrams \cite{Ma:1998dn,Bonnet:2012kz,Sierra:2014rxa}.  Possible one-loop topologies are shown in Fig.~\ref{loop1} \cite{Ma:1998dn,Bonnet:2012kz}, and the two-loop topologies are shown in Fig.~\ref{loop2} \cite{Sierra:2014rxa}.  Note that in the two-loop diagrams, two Higgs particles that are connected to two internal lines in possible ways are not shown. Recently the three-loop topologies that generate operator ${\cal O}_1$ has been classified in Ref.~\cite{Cepedello:2018rfh}.

%%%%%%%%%%%%%%%%%%%%%%%%%%%%%%%%%%%%%%%%%%%%%%%%%%%%%%%%%%%%%%%%%%%%%%%%%%%%%%%%%%
%%%%%%%%%%%%%%%%%%%%%%%%

%%%%%%%%%%%%%%%%%%%%%%%%

For the purpose of NSI, we find the classification based on type-I and type-II suggested here more convenient. The classification based on the diagram topology does not specify whether the internal particles are SM fermions or not, and the NSI effects arise only when neutrino couples to the SM fermions.  Let us also note that the first diagram of Fig.~\ref{loop1} and the first two diagrams of Fig.~\ref{loop2} are the ones that appear most frequently in the explicit type-I radiative models that we discuss in subsequent sections.

%%%%%%%%%%%%%%%%%
\section{Neutrino non-standard interactions}
\label{sec:NSI}
%%%%%%%%%%%%%%%%%%
Neutrino NSI can be of two types: Neutral Current (NC) and Charged Current (CC). The CC NSI of neutrinos with the matter fields in general affects the production and detection of neutrinos, while the NC NSI affects the neutrino propagation in matter. In the low-energy regime, neutrino NSI with matter fields can be formulated in terms of an effective four-fermion Lagrangian as follows~\cite{Wolfenstein:1977ue}:  
\begin{align}
  \mathcal{L}_{\rm NSI}^{\rm NC} \ & = \ -2\sqrt{2} G_F \sum_{f,X,\alpha,\beta} \varepsilon_{\alpha \beta}^{fX}\left(\bar{\nu}_{\alpha}\gamma^\mu P_L\nu_{\beta}\right)\left(\bar{f}\gamma_{\mu}P_Xf \right)\, , \label{NSI-NC}\\
  \mathcal{L}_{\rm NSI}^{\rm CC} \ & = \ -2\sqrt{2}G_F\, \sum_{f,f',X,\alpha,\beta}\varepsilon_{\alpha \beta}^{ff'X} \left(\bar{\nu}_{\alpha}\gamma^\mu P_L\ell_{\beta}\right)\left(\bar{f}'\gamma_{\mu}P_X f \right) \, , \label{NSI-CC} 
\end{align}
where $G_F$ is the Fermi coupling constant, and  $P_X$ (with $X=L,R$) denotes the chirality projection operators $P_{L,R}=(1\mp \gamma^5)/2$. These projection operators can also be reparametrized into vector and axial components of the interaction. The dimensionless coefficients $\varepsilon_{\alpha \beta}$ are the NSI parameters that quantify the strength of the NSI between neutrinos of flavors $\alpha$ and $\beta$ and the matter fields $f, f'\in \{e,u,d\}$. If $\varepsilon_{\alpha \beta}\neq 0$ for $\alpha\neq \beta$, the NSI violates lepton flavor, while for $\varepsilon_{\alpha \alpha}\neq \varepsilon_{\beta \beta}$, it violates lepton flavor universality. 

The vector component of NSI, $\varepsilon_{\alpha\beta}^{fV}=\varepsilon_{\alpha\beta}^{fL}+\varepsilon_{\alpha\beta}^{fR}$, affects neutrino  oscillations by providing a new flavor-dependent matter effect.\footnote{The axial-vector part of the weak interaction gives a nuclear spin-dependent contribution that averages to zero in the non-relativistic limit for the nucleus.}
The effective Hamiltonian for the matter effect is given by 
\begin{align}
H \ = \ \frac{1}{2E}U_{\rm PMNS}
\begin{pmatrix}
0 & 0 & 0\\ 0 &\Delta m^2_{21} & 0\\ 0 & 0 &\Delta m^2_{31}
\end{pmatrix}U_{\rm PMNS}^\dagger+ \sqrt{2}G_F N_e(x)
\begin{pmatrix}
1+\varepsilon_{ee}&\varepsilon_{e\mu}&\varepsilon_{e\tau}\\
\varepsilon_{e\mu}^\star&\varepsilon_{\mu\mu}&\varepsilon_{\mu\tau}\\
\varepsilon_{e\tau}^\star&\varepsilon_{\mu\tau}^\star&\varepsilon_{\tau\tau}
\end{pmatrix}\,,
\label{eq:nsi hamiltonian}
\end{align}
where $U_{\rm PMNS}$ is the standard $3\times 3$ lepton mixing matrix, $E$ is the neutrino energy, $N_e(x)$ is the electron number density as a function of the distance $x$ traveled by the neutrino in matter, and the 1 in the $1+\varepsilon_{ee}$ term is due to the standard CC matter potential. The Hamiltonian level NSI in Eq.~\eqref{eq:nsi hamiltonian} is related to the Lagrangian level NSI in Eq.~\eqref{NSI-NC} as follows:
\begin{align}
\varepsilon_{\alpha\beta} \ & = \ \sum_{f\in\{e,u,d\}} \left\langle \frac{N_f(x)}{N_e(x)} \right\rangle \varepsilon_{\alpha\beta}^{fV} \nonumber \\
 \ & = \ \varepsilon_{\alpha \beta}^{eV}+\left\langle \frac{N_p(x)}{N_e(x)}\right\rangle (2\varepsilon_{\alpha \beta}^{uV}+\varepsilon_{\alpha \beta}^{dV})+\left\langle \frac{N_n(x)}{N_e(x)}\right\rangle (\varepsilon_{\alpha \beta}^{uV}+2\varepsilon_{\alpha \beta}^{dV}) \, ,
 \label{eq:eps1}
\end{align}
where $N_f(x)$ is the number density of fermion $f$ at position $x$, and $\langle N_{p(n)}/N_e \rangle$ is the average ratio of the density of protons (neutrons) to the density of electrons along the neutrino propagation path. Note that the  coherent forward scattering of neutrinos with nucleons can be thought of as the incoherent sum of the neutrino scattering with the constituent quarks, because  the nucleon form factors are equal to one in the limit of zero momentum transfer. Assuming electric charge neutrality of the medium, we can set $\langle N_p(x)/N_e(x)\rangle=1$ and define the ratio $Y_n(x)\equiv \langle N_n(x)/N_e(x)\rangle$ to rewrite Eq.~\eqref{eq:eps1} as
\begin{align}
 \varepsilon_{\alpha\beta} \  = \ \varepsilon_{\alpha \beta}^{eV}+\left[2+Y_n(x)\right]\varepsilon_{\alpha \beta}^{uV}+\left[1+2Y_n(x)\right]\varepsilon_{\alpha \beta}^{dV} \, .
 \label{eq:eps2}
\end{align}
In the Earth, the ratio $Y_n$ which characterizes the matter chemical composition can be taken to be constant to very good approximation. According to the Preliminary Reference Earth Model (PREM)~\cite{Dziewonski:1981xy}, $Y_n=1.012$ in the mantle and 1.137 in the core, with an average value $Y_n = 1.051$ all over the Earth. On the other hand, for solar neutrinos, $Y_n(x)$ depends on the distance to the center of the Sun and drops from about 1/2 in the
center to about 1/6 at the border of the solar core~\cite{Serenelli:2009yc, Coloma:2016gei}. 

In the following sections, we will derive the predictions for the NSI parameters $\varepsilon_{\alpha \beta}$ in various radiative neutrino mass models, which should then be compared with the experimental and/or global-fit constraints~\cite{Coloma:2017egw, Farzan:2017xzy, Esteban:2018ppq, Esteban:2019lfo} on $\varepsilon_{\alpha\beta}$ using Eq.~\eqref{eq:eps2}. We would like to emphasize two points in this connection: 
\begin{enumerate}
\item [(i)] Depending on the model, we might have NSI induced only in the neutrino-electron or neutrino-nucleon interactions, or involving only left- or right-chirality of the matter fields. In such cases, only the relevant terms in Eq.~\eqref{eq:eps2} should be considered, while comparing with the experimental or global-fit constraints. 

\item [(ii)] Most of the experimental constraints~\cite{Farzan:2017xzy} are derived assuming only one NSI parameter at a time, whereas within the framework of a given model, there might exist some non-trivial correlation between NSI involving different neutrino flavors, as we will see below. On the other hand, the global-fits~\cite{Esteban:2018ppq, Esteban:2019lfo} usually  perform a scan over all NSI parameters switched on at the same time in their analyses, whereas for a given model, the cLFV constraints usually force the NSI involving some flavor combinations to be small, in order to allow for those involving some other flavor combination to be sizable.  To make a  conservative comparison with our model predictions, we will quote the most stringent values from the set of experimental and global-fit constraints both, as well as the future DUNE sensitivities~\cite{deGouvea:2015ndi, Coloma:2015kiu, Blennow:2016etl, dev_pondd} (cf.~Tables~\ref{tab:Zee} and \ref{tab:LQ}). 
\end{enumerate}

%%%%%%%%%%%%%%%%%%%%%%%%
\section{Observable NSI in the Zee model}\label{sec:Zee}
%%%%%%%%%%%%%%%%%%%%%%%%%%%%%%%%%%%%%%%%%%%%%%%

One of the simplest extensions of the SM that can generate neutrino mass radiatively is the
Zee Model \cite{Zee:1980ai}, wherein small Majorana masses arise through one-loop diagrams.  This is a type-I radiative model, as it can be realized by opening up the $\Delta L = 2$ effective $d=7$ operator ${\cal O}_2 = L^i L^j L^k e^c H^l \epsilon_{ij} \epsilon_{kl}$, and since the induced neutrino mass has a chiral suppression factor proportional to the charged lepton mass.  Due to the loop and the chiral suppression factors, the new physics scale responsible for neutrino mass can be at the TeV scale.  The model belongs to the classification $\mathcal{O}_2^2$ of Table \ref{tab:O2}.

The model assumes the SM gauge symmetry
 $SU(3)_{c} \times SU(2)_{L} \times U(1)_{Y} $, with an extended scalar sector.  Two Higgs doublets $\Phi_{1,2}({\bf 1},{\bf 2},1/2)$, and a charged scalar singlet $\eta^+ ({\bf 1},{\bf 1},1)$ are introduced to facilitate lepton number violating interactions and thus nonzero neutrino mass. The leptonic Yukawa Lagrangian of the model is given by:
 \begin{align}
       -  \mathcal{L}_Y \ \supset \ f_{\alpha \beta} L_{\alpha}^{i}    L_{\beta}^j \epsilon_{ij}  \eta^+ +(y_1)_{\alpha \beta} \widetilde{\Phi}_1^{i} L_\alpha^{j}  \ell_\beta^c\epsilon_{ij}  + (y_2)_{\alpha \beta} \widetilde{\Phi}_2^i L_\alpha^{j}  \ell_\beta^c \epsilon_{ij} + {\rm H.c.} \, ,
         \label{Lfab}
    \end{align}
where $\{\alpha,\beta\}$ are generation indices, $\{i,j\}$ are $SU(2)_L$ indices,  $\widetilde{\Phi}_a \equiv i \tau_2 \Phi_a^ \star$ ($a=1,2$) and  $\ell^c$ denotes the left-handed antilepton fields.  Here and in what follows, a transposition and charge conjugation between two fermion fields is to be understood. Note that due to Fermi statistics, $f_{\alpha \beta} = -f_{\beta\alpha}$.  Expanding the first term of the Lagrangian Eq.~(\ref{Lfab}) leads to the following couplings of $\eta^+$:
\begin{equation}
       -  \mathcal{L}_Y \ \supset \ 2 \eta^+\left[f_{e\mu} (\nu_e  \mu -\nu_\mu  e  )+f_{e\tau} (\nu_e  \tau -\nu_\tau  e)+f_{\mu\tau} (\nu_\mu  \tau -\nu_\tau  \mu)\right] + {\rm H.c.} \, 
       \label{eq:Zeefex}
\end{equation}

%%%%%%%%%%%%%%%%%%%%%%

The presence of two Higgs doublets $\Phi_{1,2}$ allows for a  cubic coupling in the Higgs potential,
    \begin{equation}
         V \ \supset \ \mu \,  \Phi_1^i \,  \Phi_2^j \epsilon_{i j} \, \eta^- + {\rm H.c.} \, ,
         \label{eq:V}
    \end{equation}
which, along with the Yukawa couplings of Eq. (\ref{Lfab}), would lead to lepton number violation. The magnitude of the parameter $\mu$ in Eq. (\ref{eq:V}) will determine the range of NSI allowed in the model.   Interestingly, $\mu$ cannot be arbitrarily large, as it would lead to charge-breaking minima of the Higgs potential which are deeper than the charge conserving minimum ~\cite{Barroso:2005hc, Babu:2014kca} (see Sec.~\ref{sec:CBM}). 
%%%%%%%%%%%%%%%%%%%%%%%%%%%%%%%%%%%%%%%%%%

%%%%%%%%%%%%%%%%%%%%%%%%%%%%%
\subsection{Scalar sector} \label{sec:scalar}
%%%%%%%%%%%%%%%%%%%%%%%%%%%%%%
We can start with a general basis, where both $\Phi_1$ and $\Phi_2$ acquire vacuum expectation values (VEVs): 
\begin{equation}
    \langle \Phi_1 \rangle \ = \ \frac{1}{\sqrt 2}\left(
          \begin{array}{c}
           0 \\
             v_1 \\
          \end{array}
         \right), \hspace{5mm}
      \langle \Phi_2 \rangle\ = \  \frac{1}{\sqrt 2}\left(
        \begin{array}{c}
         0 \\
         v_2  e^{i \xi} \\
        \end{array}
      \right).
      \label{vev}
\end{equation}
%\begin{equation}
 %   \tan{\beta}=\frac{v_2}{v_1}
%\end{equation}
However, without loss of generality, we can choose to work in the Higgs basis~\cite{Davidson:2005cw} where only one of the doublet fields gets a VEV $v$ given by $v = \sqrt{v_1^2 + v_2^2}\simeq 246$ GeV. The transformation to the new basis $\{ H_1, H_2 \}$ is given by:
 \begin{equation}
     \left(
        \begin{array}{c}
            H_1 \\
            H_2 \\
        \end{array}
    \right) \ = \ \left(
        \begin{array}{cc}
            c_{\beta } & e^{-i\xi} s_{\beta } \\
            - e^{i\xi} s_{\beta } & c_{\beta } \\
        \end{array}
        \right)      \left(
        \begin{array}{c}
            \Phi_1 \\
            \Phi_2 \\
        \end{array}
    \right)\,,
 \end{equation}
where $s_\beta \equiv \sin{\beta}$ and $c_\beta \equiv \cos{\beta}$, with $\tan\beta = v_2/v_1$. In this new basis, we can parametrize the two doublets as 
    \begin{equation}
         H_1 = \left(
            \begin{array}{c}
                G^+ \\
                \frac{1}{\sqrt{2}} (v + H_1^0 + i G^0) \\
            \end{array}
          \right), \hspace{10mm} 
         H_2 \ = \ \left(
           \begin{array}{c}
                 H_2^+ \\
                \frac{1}{\sqrt{2}} ( H_2^0 + i A) \\
            \end{array}
         \right) \, ,
    \end{equation}
where $(G^+, G^0)$ are the Goldstone bosons,  $(H_1^0$, $H_2^0)$, $A$, and $H_2^+$ are the neutral $\mathcal{CP}$-even and odd, and charged scalar fields, respectively. We shall work in the $\mathcal{CP}$ conserving limit, since phases such as $\xi$ in Eq. (\ref{vev}) will not have a significant impact on  NSI  phenomenology which is our main focus here. 

The most general renormalizable scalar potential involving the doublet fields $H_1, H_2$ and the singlet field $\eta^+$ can be written as 
\begin{align}
     V(H_1, H_2, \eta)  \ = \ & -\mu_1^2 H_1^{\dagger} H_1 + \mu_2^2 H_2^{\dagger} H_2 - (\mu_3^2 H_2^{\dagger} H_1 + {\rm H.c.} )  \nonumber \\
     &+ \frac{1}{2} \lambda_1 (H_1^{\dagger} H_1)^2+\frac{1}{2} \lambda_2 (H_2^{\dagger} H_2)^2 + \lambda_3 (H_1^{\dagger} H_1) (H_2^{\dagger} H_2) + \lambda_4 ( H_1^{\dagger} H_2) (H_2^{\dagger} H_1) \nonumber \\
     &+ \left[ \frac{1}{2} \lambda_5 (H_1^\dagger H_2)^2 +  \left\{\lambda_6 ( H_1^\dagger H_1 )+ \lambda_7 (H_2^\dagger H_2)\right\}H_1^\dagger H_2 + {\rm H.c.} \right] \nonumber \\
     &+ \mu_\eta^2 |\eta|^2 + \lambda_\eta |\eta|^4 +  \lambda_8  |\eta|^2 H_1^\dagger H_1 + \lambda_9  |\eta|^2 H_2^\dagger H_2 \nonumber \\
     &+ (\lambda_{10}  |\eta|^2 H_1^\dagger H_2 + {\rm H.c.}) + ( \mu \, \epsilon_{ij} H_1^i H_2^j \eta^- + {\rm H.c.} ) \, 
     \label{eq:pot}
\end{align}
Differentiating $V$ with respect to $H_1$ and $H_2$, we obtain the following minimization conditions:
    \begin{equation}
         \mu_1^2 \ = \ \frac{1}{2} \lambda_1 v^2, \hspace{10 mm} \mu_3^2 \ = \ \frac{1}{2} \lambda_6 v^2,
         \label{eq:mu12}
    \end{equation}
where, for simplicity, we have chosen $\mu_3^2$  to be real. The mass matrix for the charged scalars in the basis $\{ H_2^+ , \eta^+ \}$ becomes
   \begin{align}
       M_{\rm charged}^2 \ & = \ \left(
            \begin{array}{cc}
             M_2^2 & -\mu v/\sqrt{2} \\
            -\mu v/\sqrt{2} & M_3^2 \\
            \end{array}
        \right)  \, ,
        \end{align}
where
\begin{align}
    \quad M_2^2 \ & = \ \mu_2^2 + \frac{1}{2} \lambda_3 v^2, \hspace{8mm} M_3^2 \ = \ \mu_\eta^2 + \frac{1}{2}  \lambda_8 v^2 \, .
   \end{align}
The physical masses of the charged scalars $\{ h^+, H^+ \}$ are given by:
    \begin{equation}
         m_{h^+ , H^+}^2 \ = \ \frac{1}{2} \, \left\{ M_2^2 + M_3^2 \mp \sqrt{(M_2^2 - M_3^2)^2 + 2 \,v^2 \mu^2} \right\} \, ,
         \label{eq:cHiggs}
    \end{equation}
    where
    \begin{eqnarray}
        h^+ & \ = \ & \cos\varphi\, \eta^+ + \sin\varphi\, H_2^+ \, , \nonumber \\
        H^+ &\ = \ & -\sin\varphi \,\eta^+ + \cos\varphi \,H_2^+ \, ,
        \label{eq:charged}
    \end{eqnarray}
with the mixing angle $\varphi$  given by 
    \begin{equation}
        \sin{2\varphi} \ = \ \frac{-\sqrt{2} \ v \mu}{m_{H^+}^2-m_{h^+}^2}~.
        \label{mixphi}
    \end{equation}
As we shall see later, this mixing parameter $\varphi$, which is proportional to $\mu$ will play a crucial role in the NSI phenomenology of the model. 
%%%%%%%%%%%%%%%%%%%%%%%%%%

Similarly, the matrix for the $\mathcal{CP}$-even and odd neutral scalars in the basis $\{ H_1^0, H_2^0 , A \}$ can written as \cite{Babu:2018uik}: 
   \begin{equation}
    M^2_{\rm neutral} \ = \    \left(
            \begin{array}{ccc}
             \lambda_1 v^2 & \text{Re}(\lambda_6) v^2  & - \text{Im} (\lambda_6) v^2\\
            \lambda_6 v^2  & M_2^2 + \frac{1}{2} v^2 (\text{Re}(\lambda_5) + \lambda_4) & -\frac{1}{2} \text{Im}(\lambda_5) v^2 \\
            - \text{Im} (\lambda_6) v^2 & -\frac{1}{2} \text{Im}(\lambda_5) v^2 &  M_2^2 + \frac{1}{2} v^2 (-\text{Re}(\lambda_5) + \lambda_4)
            \end{array}
        \right)  \, .
        \label{eq:neut}
   \end{equation}
In the $\mathcal{CP}$-conserving limit where Im($\lambda_{5,6}) = 0$, the $\mathcal{CP}$-odd state will decouple from the $\mathcal{CP}$-even states. One can then rotate the $\mathcal{CP}$-even states into a physical basis $\{ h, H \}$ which would have masses given by \cite{Babu:2018uik}:
    \begin{equation}
        m_{h,H}^2 \ = \ \frac{1}{2} \left[ m_A^2 + (\lambda_1 + \lambda_5) v^2 \pm \sqrt{\{m_A^2 + (\lambda_5 - \lambda_1) v^2\}^2 + 4 \lambda_6^2 v^4} \right] \, ,
        \label{eq:nHiggs}
    \end{equation}
whereas the $\mathcal{CP}$-odd scalar  mass is given by 
    \begin{equation}
         m_A^2 \ = \ M_2^2 - \frac{1}{2} (\lambda_5 - \lambda_4) v^2 \, .
    \end{equation}
The mixing angle between the $\mathcal{CP}$-even eigenstates $\{ H_1^0, H_2^0 \}$, defined as
\begin{eqnarray}
 h & \ = \ & \cos(\alpha-\beta) \,H_1^0 + \sin(\alpha-\beta) \,H_2^0 \, , \nonumber \\
 H &\ = \ & -\sin(\alpha-\beta) \,H_1^0 + \cos(\alpha-\beta) \,H_2^0 \, ,
    \end{eqnarray}
is given by
    \begin{equation}
        \sin{2(\alpha-\beta)} \ = \ \frac{2 \lambda_6 v^2}{m_{H}^2 - m_{h}^2} \, .
    \end{equation}
We will identify the lightest $\mathcal{CP}$-even eigenstate $h$ as the observed 125 GeV SM-like Higgs and use the LHC Higgs data to obtain constraints on the heavy Higgs sector (see Sec.~\ref{sec:HiggsOb}). 
%This is similar to the constraints on the inert two-Higgs-doublet model (2HDM). 
We will work in the alignment/decoupling limit, where $\beta-\alpha \rightarrow 0$~\cite{Gunion:2002zf, Carena:2013ooa, Dev:2014yca, Das:2015mwa}, as suggested by the LHC Higgs data~\cite{Bernon:2015qea, Chowdhury:2017aav}.

%%%%%%%%%%%%%%%%%%%%%%%%%%%%%%%%%%%%%%%%%%
%%%%%%%%%%%%%%%%%%%%%%%%%%%%%%%%%%%%%%%%%%%%%%%%%%%%%%%%%%
%%%%%%%%%%%%%%%%%%%%%%%%%%%%%
\subsection{Neutrino mass} \label{sec:neutrino}
In the Higgs basis where only the neutral component of $H_1$ gets a VEV, the Yukawa interaction terms in Eq.~\eqref{Lfab} of fermions with the scalar doublets $H_1$ and $H_2$ become
    \begin{equation}
      - {\cal L}_Y \ \supset \ \widetilde{Y}_{\alpha \beta} \widetilde{H}_1^i L^{j}_\alpha   \ell_{\beta}^c \epsilon_{ij} + Y_{\alpha\beta} \widetilde{H}_2^i L^{j}_\alpha  \ell_{\beta}^c \epsilon_{ij}  + {\rm H.c.} \, ,
       \label{eq:Yuk1}
    \end{equation}
where $Y$ and $\widetilde{Y}$ are the redefined  couplings in terms of the original Yukawa couplings $y_1$ and $y_2$ given in Eq.~\eqref{Lfab} and where $\widetilde{H}_a = i \tau_2 H^ \star_a$ ($a=1,2)$ with $\tau_2$ being the second Pauli matrix.  After electroweak symmetry breaking, the charged lepton mass matrix reads as
\begin{align}
M_\ell \ = \ \widetilde{Y} \langle H_1^0 \rangle \ = \ \widetilde{Y} \frac{v}{\sqrt{2}} \, . 
\end{align}
Without loss of generality, one can work in a basis where $M_\ell$ is diagonal, i.e., $M_\ell = \text{diag}\, (m_e,\, m_\mu,\, m_\tau)$. The Yukawa coupling matrix $f$ involving the $\eta^+$ field in Eq.~\eqref{Lfab} is taken to be defined in this basis.

\begin{figure}[!t]
         \centering
         \includegraphics[scale=0.5]{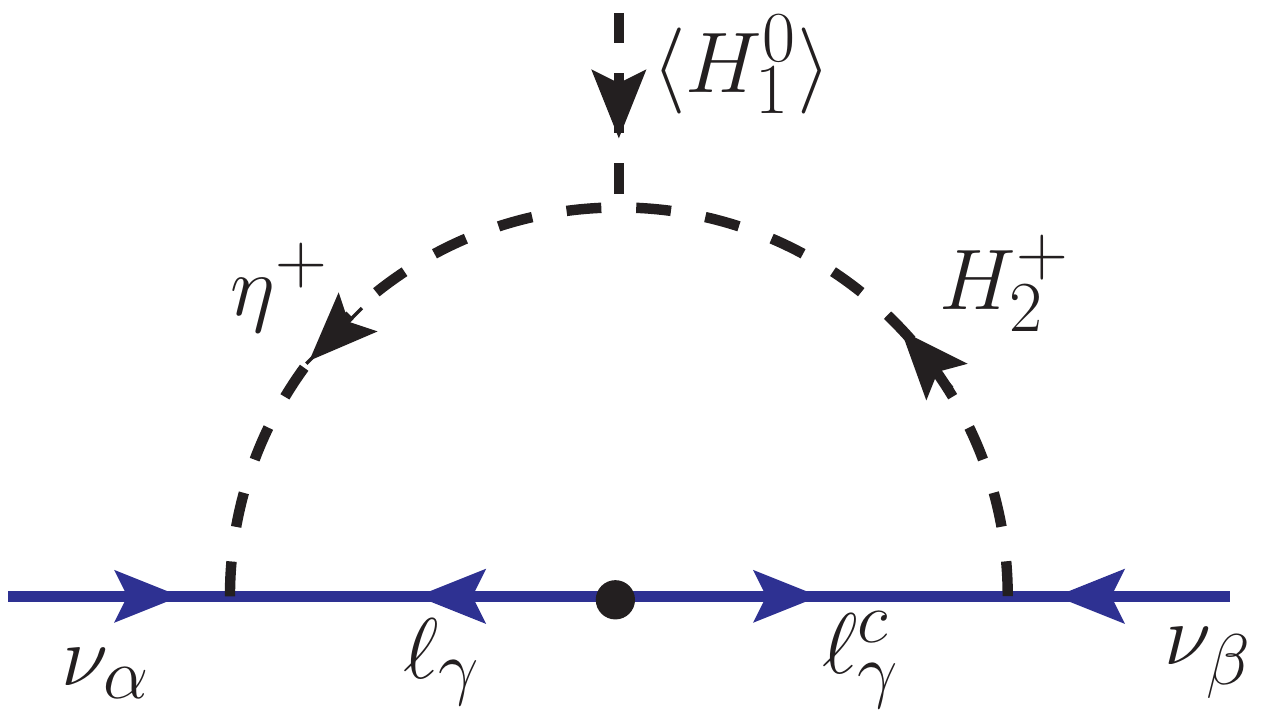}
         \caption{Neutrino mass generation at one-loop level in the Zee model~\cite{Zee:1980ai}. The dot ($\bullet$) on the SM fermion line indicates mass insertion due to the SM Higgs VEV.} 
         \label{loopzee}
    \end{figure}

The Yukawa couplings in Eq.~\eqref{Lfab}, together with the trilinear term in the scalar potential Eq.~\eqref{eq:V}, generate neutrino mass at the one-loop level, as shown in Fig.~\ref{loopzee}. Here the dot ($\bullet$) on the SM fermion line indicates mass insertion due to the SM Higgs VEV.  There is a second diagram obtained by reversing the arrows on the internal particles. Thus, we have a symmetric neutrino mass matrix given by 
%%%%%%%%%%%%%%%%%%%%%%%%
    \begin{equation}
         M_\nu \ = \ \kappa \, (f M_\ell Y + Y^T M_\ell f^T) \, ,
         \label{nuMass}
    \end{equation}
%%%%%%%%%%%%%%%%%%%%%%%%%
where  $\kappa$ is the one-loop factor given by
%%%%%%%%%%%%%%%%%%%%%%%%%%%%
    \begin{equation}
          \kappa \ = \ \frac{1}{16 \pi^2} \sin{2 \varphi} \log\left(\frac{m_{h^+}^2}{m_{H^+}^2}\right) \, ,
          \label{kfactor}
    \end{equation}
%%%%%%%%%%%%%%%%%%%%%%%%%%%%
with $\varphi$ given in Eq.~\eqref{mixphi}. From Eq.~\eqref{nuMass} it is clear that only the product of the  Yukawa couplings $f$ and $Y$ is constrained by the neutrino oscillation data. Therefore, by taking some of the $Y$ couplings to be of $\sim {\cal O}(1)$ and all $f$ couplings very small in the neutrino mass matrix of Eq.~\eqref{nuMass}, we can correctly reproduce the neutrino oscillation parameters (see Sec.~\ref{sec:neu}).  This choice maximizes the neutrino NSI in the model. We shall adopt this choice. With the other possibility, namely, $Y\ll 1$, the stringent cLFV constraints on $f$ couplings  (cf.~Table~\ref{tab:zee-babu}) restrict the maximum NSI to $\lesssim 10^{-8}$~\cite{Herrero-Garcia:2017xdu}, well below any foreseeable future experimental sensitivity.

The matrix $f$ that couples the  left-handed lepton doublets to the charged scalar $\eta^+$ can be made real by a phase redefinition $\hat{P} f \hat{P}$, where $\hat{P}$ is a diagonal phase matrix, while the Yukawa coupling $Y$ in Eq.~\eqref{eq:Yuk1} is in general a complex asymmetric matrix:   
    \begin{equation}
         f \ = \ \left(
    \begin{array}{ccc}
          0  & f_{e\mu}  &  f_{e\tau}  \\
          -f_{e\mu}  & 0  & f_{\mu \tau}  \\
         -f_{e\tau}  &  -f_{\mu \tau}  & 0  \\
    \end{array}
        \right), \hspace{10mm}
         Y \ = \ \left(
         \begin{array}{ccc}
             Y_{ee}  & Y_{e\mu}  &  Y_{e\tau}  \\
              Y_{\mu e}  & Y_{\mu \mu}  & Y_{\mu \tau}  \\
              Y_{\tau e}  &  Y_{\tau \mu}  & Y_{\tau \tau}  \\
        \end{array}
        \right) \, .
        \label{lepyuk}
    \end{equation}
Here the matrix $Y$ is multiplied by $(\bar{\nu}_e, \bar{\nu}_\mu,\,\bar{\nu}_\tau)$ from the left and $(e_R,\,\mu_R,\, \tau_R)^T$ from the right, in the interaction with the charged scalar $H^+$. 
Thus the neutrino NSI will be governed by the matrix elements $(Y_{ee},\,Y_{\mu e},\,Y_{\tau e})$, which parametrize the couplings of $\nu_\alpha$ with electrons in matter.  

Since the model has two Higgs doublets, in general both doublets will couple to up and down quarks.  If some of the leptonic Yukawa couplings $Y_{\alpha e}$ of Eq. (\ref{lepyuk}) are of order unity, so that significant neutrino NSI can be generated, then the quark Yukawa couplings of the second Higgs doublet $H_2$ will have to be small.  Otherwise chirality enhanced meson decays, such as $\pi^+ \rightarrow e^+ \nu$ will occur with unacceptably large rates.  Therefore, we assume that the second Higgs doublet $H_2$ is leptophilic in our analysis.

Note that in the limit $Y\propto M_l$, as was suggested by Wolfenstein~\cite{Wolfenstein:1980sy} by imposing a discrete $Z_2$ symmetry to forbid the tree-level flavor changing neutral currents (FCNC) mediated by the neutral Higgs bosons, the diagonal elements of $M_\nu$ would vanish, yielding neutrino mixing angles that are not compatible with observations \cite{Koide:2001xy,He:2003ih}.  For a variant of the Zee-Wolfenstein model with a family-dependent $Z_4$ symmetry which is consistent with neutrino oscillation data, see Ref. \cite{Babu:2013pma}.

%%%%%%%%%%%%%%%%%%%%%%%%%%%%%%%%%%%%%%%%%%%%%%%%%%%%%%%%%%
\subsection{Charge-breaking minima} \label{sec:CBM}
To have sizable NSI, we need a large mixing $\varphi$ between the singlet and doublet charged scalar fields $\eta^+$ and $H_2^+$. From Eq.~\eqref{mixphi}, this means that we need a large trilinear $\mu$-term. But $\mu$ cannot be arbitrarily large, as it leads to charge-breaking minima (CBM) of the potential~\cite{Barroso:2005hc, Babu:2014kca}. We numerically analyze the scalar potential given by  Eq.~\eqref{eq:pot} to ensure that it does not develop any CBM deeper than the charge-conserving minimum (CCM).

We take $\mu_2^2,\, \mu_\eta^2 >0$. The field $H_1$ is identified approximately as the SM Higgs doublet, and therefore, the value of $\lambda_1$ is fixed by the Higgs mass (cf.~Eq.~\eqref{eq:mu12}), and the corresponding mass-squared term is chosen to be negative to facilitate electroweak symmetry breaking ($\mu_1^2 > 0$ in Eq. (\ref{eq:pot})). Note that the cubic scalar coupling $\mu$ can be made real as any phase in it can be absorbed in  $\eta^-$ by a field redefinition.

In order to calculate the most general minima of the potential, we assign the following VEVs to the scalar fields:
    \begin{equation}
       \langle H_1 \rangle \ = \ \left(
          \begin{array}{c}
           0 \\
             v_1 \\
          \end{array}
         \right), \hspace{5mm}
      \langle H_2 \rangle\ = \  v_2\left(
        \begin{array}{c}
         \sin{\gamma} \,  e^{i \delta} \\
         \cos{\gamma}\,  e^{i \delta'} \\
        \end{array}
      \right), \hspace{5mm}
     \langle \eta^- \rangle \ = \ v_\eta \, ,
     \label{VEV}
    \end{equation}
where $v_\eta$  and $v_1$ can be made real and positive by $SU(2)_L \times U(1)_Y$ rotations.  A non-vanishing VEV $v_\eta$ would break electric charge conservation, as does a nonzero value of $\sin\gamma$. Thus, we must ensure that the CBM of the potential lie above the CCM.  The Higgs potential, after inserting  Eq.~\eqref{VEV} in   Eq.~\eqref{eq:pot}, reads as
%%%%%%%%%%%%%%%%%%%%
    \begin{eqnarray}
         V &\ = \ & -\mu_1^2 v_1^2 + \frac{ \lambda_1
         v_1^4 }{2} + (\mu_2^2 + \lambda_3) v_2^2 + \frac{\lambda_2 v_2^4}{2} + (\mu_\eta^2 + \lambda_8 v_1^2 + \lambda_9 v_2^2) v_\eta^2 + \lambda_\eta v_\eta^4 \nonumber\\
         && + v_1 v_2 \{ 2 \cos\gamma [-\mu_3^2 \cos{\delta'} + \lambda_6 v_1^2 \cos{(\theta_2 + \delta')}+ \lambda_7 v_2^3 \cos{(\theta_3 + \delta')} + \lambda_{10} v_\eta^2 \cos{(\theta_4 + \delta')}] \nonumber\\
         &&   + v_1 v_2 \cos{\gamma}^2 [\lambda_4 + \lambda_5 \cos{(\theta_1 + 2 \delta')}] - 2 \mu v_\eta \cos{\delta} \sin{\gamma}\}.
         \label{eq:pot2}
    \end{eqnarray}
%%%%%%%%%%%%%%%%%%%%%
Here $\theta_1, \theta_2, \theta_3,$ and $\theta_4$ are respectively the phases of the quartic couplings $\lambda_5, \lambda_6, \lambda_7, $ and $\lambda_{10}$. For simplicity, we choose these quartic couplings, as well as $\lambda_9$ to be small. This choice does not lead to any run-away behavior of the potential.  We keep all diagonal quartic couplings to be nonzero, so that the potential remains bounded.  (All boundedness conditions are satisfied if we choose,  as we do for the most part, all the quartic couplings to be positive.) We also keep the off-diagonal couplings $\lambda_3$ and $\lambda_8$ nonzero, as these couplings help in satisfying constraints from the SM Higgs boson properties from the LHC.  

\begin{figure}[!t]
    \centering
    \includegraphics[height=5.5cm, width=0.48\textwidth]{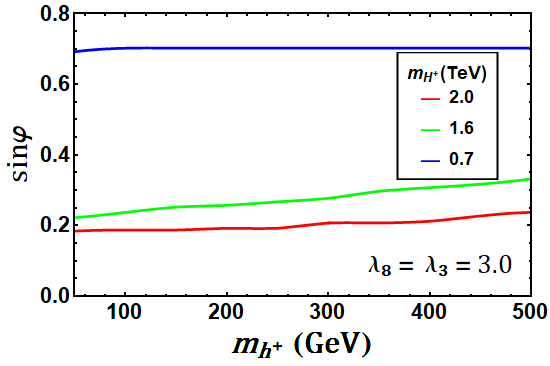}\hspace{2mm}
    \includegraphics[height=5.5cm, width=0.48\textwidth]{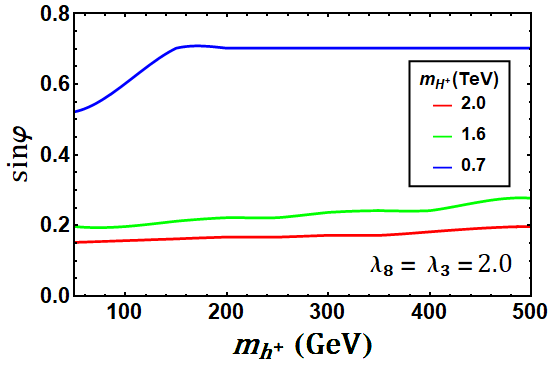}
    \caption{Maximum allowed value of the  mixing parameter $\sin\varphi$ from charge-breaking minima constraints as a function of the light charged Higgs mass $m_{h^+}$, for different values of the heavy charged Higgs mass $m_{H^+}=2$ TeV, 1.6 TeV and 0.7 TeV, shown by red, green and blue curves, respectively. We set the quartic couplings $\lambda_3= \lambda_8 = 3.0$ (left) and $\lambda_3= \lambda_8 = 2.0$ (right), and vary $\lambda_2, \lambda_\eta$ in the range [0.0,~3.0]. For a given $m_{H^+}$, the region above the corresponding curve leads to charge-breaking minima.}
    \label{CBM_mixing}
\end{figure}

Eq.~\eqref{eq:pot2} yields five minimization conditions from which  $\{v_1, v_2, v_\eta, \delta, \gamma \}$ can be solved numerically for any given set of masses and quartic couplings.  The mass parameters are derived from the physical masses of $h^+$, $H^+$ and $h$ in the CCM.  We vary $m_{h^+}$ from 50 to 500 GeV and choose three benchmark points for $m_{H^+}$: $\{0.7,\, 1.6,\,2.0\}$ TeV. To get an upper limit on the mixing  angle $\varphi$ (cf.~Eq.~\eqref{mixphi}] for our subsequent analysis,  we keep $\lambda_3 = \lambda_8$ fixed at two benchmark values (3.0 and 2.0) and vary the remaining nonzero quartic couplings  $\lambda_2$ and $\lambda_\eta$ in the range $[0.0,3.0]$. Our results on the maximum $\sin\varphi$ are shown in Fig.~\ref{CBM_mixing}. We do not consider values of the quartic couplings exceeding 3.0 to be consistent with perturbativity considerations~\cite{Kang:2013zba}. Each choice of mixing angle $\varphi$, and the parameters $\lambda_2$, $\lambda_\eta$, $m_{h^+} $, and $m_{H^+}$ yields different minimization conditions deploying different solutions to the VEVs. We compare the values of the potential for all cases of CBM and CCM. If any one of the CBM is deeper than CCM, we reject the solution and rerun the algorithm with different initial conditions until we meet the requirement of electroweak minimum being deeper than {\it all} CBM.

%%%%%%%%%%%%%%

For values of the mixing angle $\sin\varphi$ above the curves shown in Fig.~\ref{CBM_mixing} for a given $m_{H+}$, the potential develops CBM that are deeper than the electroweak minimum, which is unacceptable. This is mainly due to the fact that for these values of $\varphi$, the trilinear coupling $\mu$ becomes too large, which drives the potential to a deeper CBM~\cite{Barroso:2005hc}, even for positive $\mu_\eta^2$. From Fig.~\ref{CBM_mixing} it is found that $\sin\varphi < 0.23$ for $m_{H^+} = 2~{\rm TeV}$, while $\sin\varphi = 0.707$ is allowed when $m_{H^+} = 0.7$ TeV.  In all cases the maximum value of $|\mu|$ is found to be about 4.1 times the heavier mass $m_{H^+}$.  
Note that we have taken the maximum value of the mixing $\varphi_{\rm max}=\pi/4$ here, because for $\varphi>\pi/4$, the roles of $h^+$ and $H^+$ will be simply reversed, i.e.,~$H^+$ ($h^+$) will become the lighter (heavier) charged Higgs field.  The CBM limits from Fig.~\ref{CBM_mixing} will be applied when computing neutrino NSI in the model.

%In the limit where $\lambda_2$ and $\lambda_\eta$ are zero, the potential in Eq.~\eqref{eq:pot2} becomes much simpler and one can analytically show that the value of the potential for CBM is always larger than the CCM, because it leads to the condition
%\begin{equation}
%    \bigg( \mu^2 - \frac{\mu_2^2(\lambda_1  \mu_\eta^2+\lambda_8 \mu_1^2)}{\mu_1^2}\bigg)^2 \ > \  0 \, ,
%\end{equation}
%which is trivially satisfied for any choice of mass and mixing. However, this is not a realistic scenario, since the potential in the limit of $\lambda_{2,3}\to 0$ is unbounded below and one must have some diagonal quartic couplings nonzero to realize a bounded-from-below potential. Therefore, we have resorted to the numerical analysis above.  

%%%%%%%%%%%%%%%%%%%%%%%%%%%%%
\subsection{Electroweak precision constraints}\label{sec:ewpt}
%%%%%%%%%%%%%%%%%%%%%%%%%%%%%%%%%%%%%

The oblique parameters $S$, $T$  and $U$ can describe a variety of new physics in the electroweak sector parametrized arising through shifts in the gauge boson self-energies~\cite{Peskin:1990zt,Peskin:1991sw} and impose important constraints from precision data. These parameters have been calculated in the context of the Zee model in Ref. \cite{Grimus:2008nb}.
 We find that the $T$ parameter imposes the most stringent constraint, compared to the other oblique parameters. The $T$ parameter in the Zee model can be expressed as  \cite{Grimus:2008nb}:
\begin{align} 
T  \ = \ & \dfrac{1}{16\pi^2 \alpha_{\rm em} v^2} \left\lbrace  {\cos}^2\varphi \left[ {\sin}^2{(\beta-\alpha)} \mathcal{F}(m_{h^+}^2,m_{h}^2) + {\cos}^2{(\beta-\alpha)}\mathcal{F}(m_{h^+}^2,m_{H}^2) + \mathcal{F}(m_{h^+}^2,m_{A}^2) \right] \right. \nonumber\\
&+ {\sin}^2\varphi \left[ {\sin}^2{(\beta-\alpha)} \mathcal{F}(m_{H^+}^2,m_{h}^2) + {\cos}^2{(\beta-\alpha)}\mathcal{F}(m_{H^+}^2,m_{H}^2) + \mathcal{F}(m_{H^+}^2,m_{A}^2) \right] \nonumber\\ 
&- 2\,{\sin}^2\varphi {\cos}^2\varphi\, \mathcal{F}(m_{h^+}^2,m_{H^+}^2) - {\sin}^2{(\beta-\alpha)} \mathcal{F}(m_{h}^2,m_{A}^2) - {\cos}^2{(\beta-\alpha)} \mathcal{F}(m_{H}^2,m_{A}^2) \nonumber\\
&+ \left. 3 {\sin}^2{(\beta-\alpha)} \left[ \mathcal{F}(m_{Z}^2,m_{H}^2) - \mathcal{F}(m_{W}^2,m_{H}^2) - \mathcal{F}(m_{Z}^2,m_{h}^2) + \mathcal{F}(m_{W}^2,m_{h}^2) \right] \right\rbrace \,, \label{eq:T}
\end{align}
where the symmetric function $\mathcal{F}$ is given by
\begin{equation} \label{Fdef}
\mathcal{F}(m_1^2,m_2^2) \ = \ \mathcal{F}(m_2^2,m_1^2)\ \equiv \  \frac{1}{2}(m_1^2+m_2^2) -\frac{m_1^2m_2^2}{m_1^2-m_2^2}\ln\left(\frac{m_1^2}{m_2^2}\right)\,.
\end{equation}

\begin{figure}[!t]
\centering
\includegraphics[width=0.7\textwidth]{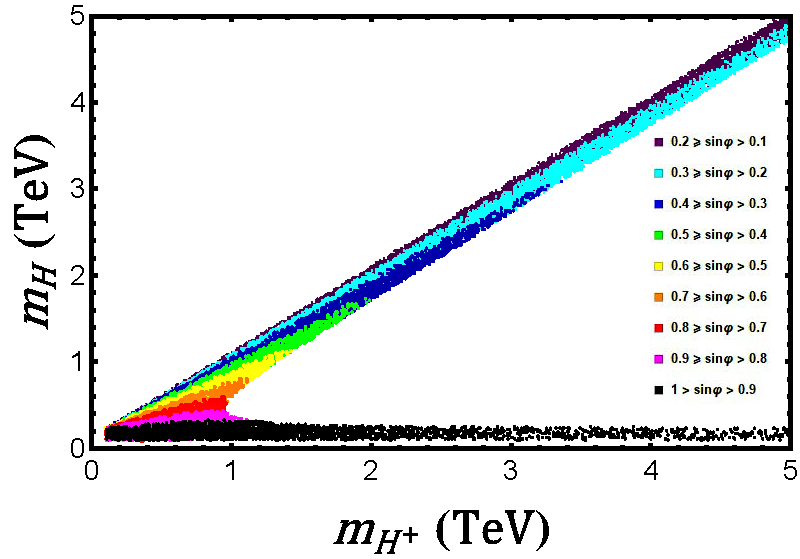}
 \caption{$T$-parameter constraint at the $2\sigma$ confidence level in the heavy charged and neutral Higgs mass plane in the Zee model. Here we have set the light charged scalar mass $m_{h^+}=100$ GeV. Different colored regions correspond to different values of the mixing angle $\sin\varphi$ between the charged Higgs bosons.}
 %Red : $\sin \phi < 0.3$, Blue : $0.3 <\sin \phi < 0.5$, Cyan : $0.5< \sin \phi < 0.7$, Green : $0.7< \sin \phi < 0.8$, Grey : $\sin \phi > 0.8$. }
\label{ewpt1}
\end{figure}

In order to generate large NSI effects in the Zee model, the mixing between the singlet and the doublet charged scalar, parametrized by the angle $\varphi$, should be significant. This mixing contributes to the gauge boson self-energies and will therefore  be bounded from the experimental value of the $T$ parameter: $T=0.01 \pm 0.12$~\cite{Tanabashi:2018oca}. For simplicity, we assume no mixing between the neutral $\mathcal{CP}$-even scalars $h$ and $H$.  Furthermore, we take the heavy neutral $\mathcal{CP}$-even ($H$) and odd ($A$) scalars to be degenerate in mass.  In Fig.~\ref{ewpt1}, we have shown our results from the $T$ parameter constraint, allowing for two standard deviation error bar, in the heavy neutral and charged Higgs mass plane. Here we have fixed the light charged scalar mass $m_{h^+}=100$ GeV.
 As shown in the figure, when the masses $m_H$ and $m_{H^{\pm}}$ are nearly equal (along the diagonal), the $T$ parameter constraint is easily satisfied. 
%%%%%%%%%%%%%%%%%%%%%%%%%%%%%%%%%%%%%%%%%%
%%%%%%%%%%%%%%%%%%%%%%%%%%%%%
%%%%%%%%%%%%%%%%%%%%%%%%%%%%%%
 \begin{figure}[t!]
$$
\includegraphics[height=5.5cm,width=0.7\textwidth]{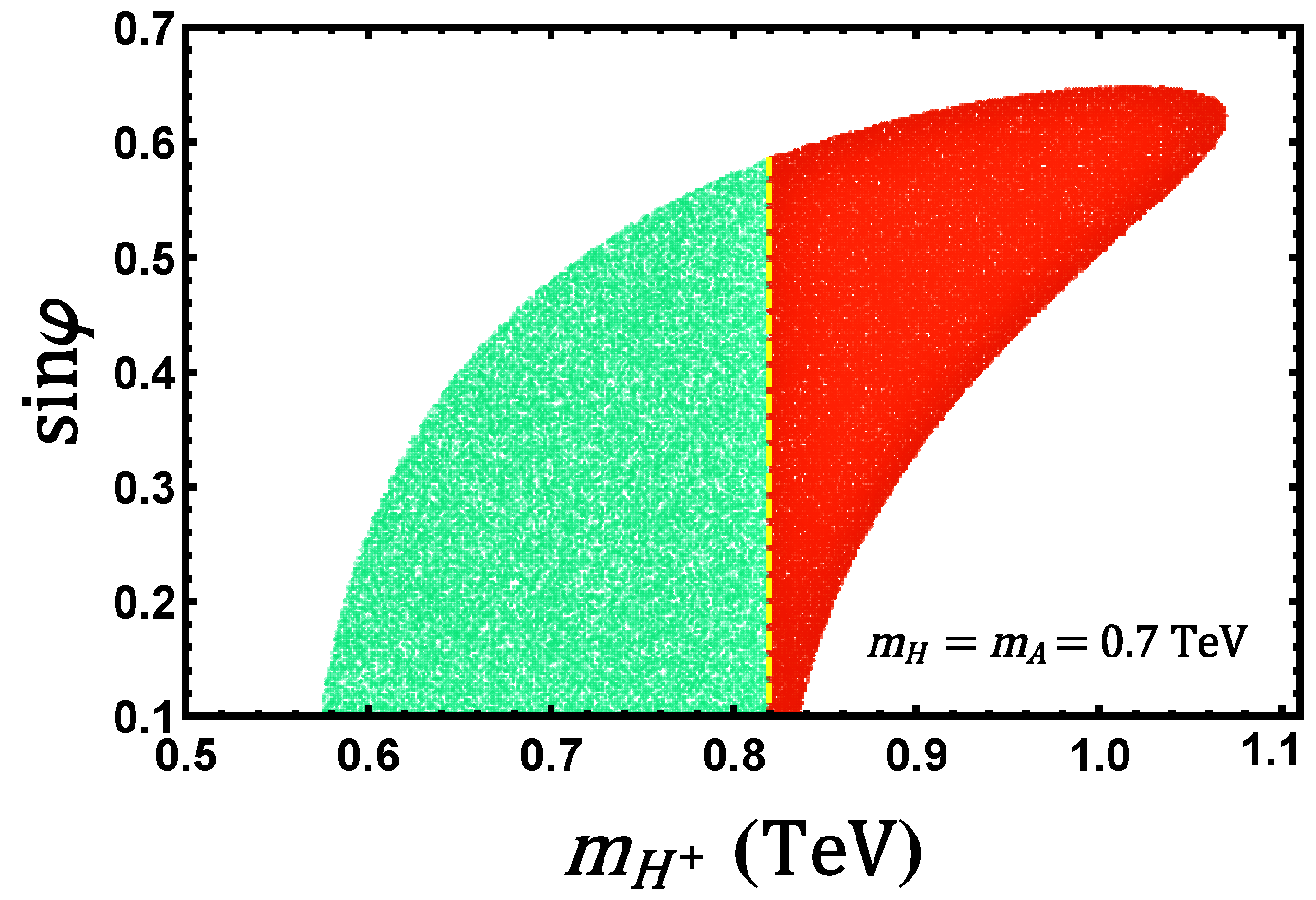}
 \hspace{0.1 in}
 $$
 \caption{$T$-parameter constraint in the mixing and heavy charged scalar mass plane in the Zee model for heavy neutral scalar masses $m_H=m_A=0.7$ TeV. The colored regions (both green and red) are allowed by the $T$-parameter constraint, while in the red-shaded region, $|\lambda_{4}|, |\lambda_{5}| >3.0 $, which we discard from perturbativity requirements.}
\label{fig:ewp2}
\end{figure}
%%%%%%%%%%%%%%%%%%%%%%%%%%%%%%%%%%

From Fig.~\ref{ewpt1}, we also find that for  specific values of $m_H$ and $m_{H^{\pm}}$, there is an upper limit on the mixing $\sin{\varphi}$. This is further illustrated in Fig.~\ref{fig:ewp2}. Here,  the colored regions (both green and red) depict the allowed parameter space in $m_H^+ - \sin{\varphi}$ plane resulting from the $T$ parameter constraint.  For example, if we set $m_H=0.7$ TeV, the maximum mixing  that is allowed by $T$ parameter is $(\sin{\varphi})_{\rm max}=0.63$. The mass splitting between the heavy neutral and the charged Higgs bosons is governed by the  relation (cf.~Eqs.~\eqref{eq:cHiggs} and \eqref{eq:nHiggs}): 
 \begin{align}
    m^2_{H^{\pm}} - m^2_{H} \ = \ \frac{1}{2} (\lambda_5 -\lambda_4) v^2 \, .
    \end{align}
We choose  $\lambda_5 = -\lambda_4$, which would maximize the mass splitting, as long as the quartic couplings remain perturbative. The red region in  Fig.~\ref{fig:ewp2} depicts the scenario where $|\lambda_{4}|, |\lambda_{5}| >3.0 $, which we discard from perturbativity requirements in a conservative approach.  Satisfying this additional requirement that these couplings be less than $3.0$, we get an upper limit on   $\sin{\varphi}<0.59$. For the degenerate case $m_{H^{\pm}} = m_{H}$ with $\lambda_4=\lambda_5$, the upper limit is stronger:  $\sin{\varphi}< 0.49$.
%On the other hand, the mass of the neutral partner $m_H$  from the same doublet has to be heavy to satisfy the LEP constraints as discussed earlier.     

%%%%%%%%%%%%%%%%%%%%%%%%%%%%%%
\subsection{Charged-lepton flavor violation constraints}\label{sec:lfv}
%%%%%%%%%%%%%%%%%%%%%%%%%%%%%%%%%

Charged-lepton flavor violation is an integral feature of the Lagrangian Eq.~\eqref{Lfab} of the model. We can safely ignore cLFV processes involving the $f_{\alpha\beta}$ couplings which are assumed to be of the order of $10^{-8}$ or so to satisfy the neutrino mass constraint, with $Y_{\alpha\beta}$ couplings being  order one.   Thus,  we focus on cLFV proportional to $Y_{\alpha\beta}$. 
Furthermore, as noted before, NSI arise proportional to $(Y_{ee},\,Y_{\mu e},\,Y_{\tau e})$, where the first index refers to the neutrino flavor and the second to the charged-lepton flavor in the coupling of charged scalars $h^+$ and $H^+$.  After briefly discussing the cLFV constraints arising from other $Y_{\alpha\beta}$, we shall focus on the set $(Y_{ee},\,Y_{\mu e},\,Y_{\tau e})$ relevant for NSI.  The neutral scalar bosons $H$ and $A$ will mediate cLFV of the type $\mu \rightarrow 3e$ and $\tau \rightarrow \mu ee$ at tree-level, while these neutral scalars and the charged scalars $(h^+,\,H^+)$ mediate processes of the type $\mu \rightarrow e \gamma$ via one-loop diagrams.  Both of these processes will be analyzed below.  We derive limits on the couplings $Y_{\alpha\beta}$ as functions of the scalar masses.  These limits need to be satisfied in the neutrino oscillation fit, see Sec.~\ref{sec:neu} for details. The constraints derived here will also be used to set upper limits of possible off-diagonal NSI. The various processes considered and the limits derived are summarized in Tables~\ref{lgdecay} and~\ref{3ldecay}. We now turn to the derivation of these bounds.   
%%%%%%%%%%%%%%%%%
%%%%%%%%%%%%%%%
%%%%%%%%%%%%%%%%%%%%%%%%%%%%%%%%%%%%%%%%%%%%%%%%
\subsubsection{\texorpdfstring{$\ell_\alpha \to \ell_\beta + \gamma$}{ell} decays} \label{sec:llg}
%%%%%%%%%%%%%%%%%%%%%%%%%%%%%%%%%%%%%%%%%%%%%%%%%\
\begin{figure}[!t]
   $$
    \includegraphics[scale=0.5]{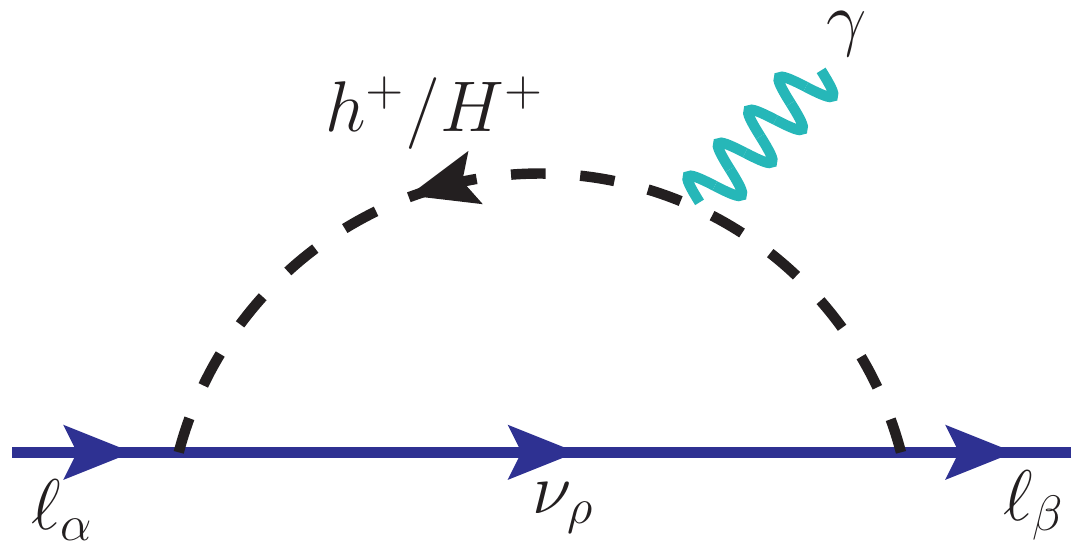} \hspace{10mm}
    \includegraphics[scale=0.5]{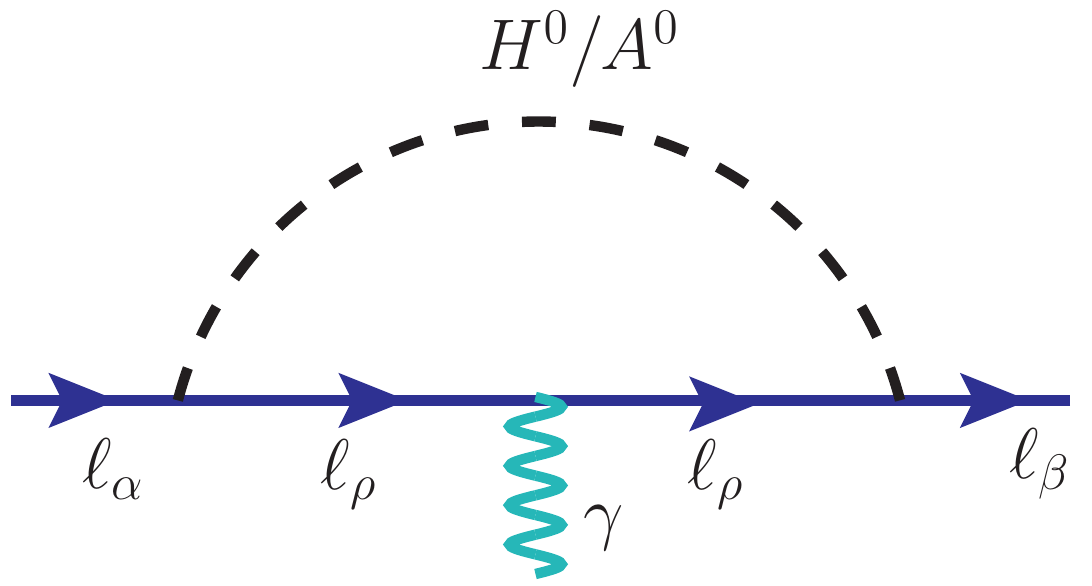}
   $$
    \caption{One-loop Feynman diagrams contributing to $\ell_\alpha \to \ell_\beta + \gamma$ process mediated by charged scalar (left) and neutral scalar (right) in the Zee model.}
    \label{mutoeg}
\end{figure}

%%%%%%%%%%%%%%%%%%%%%%%%%%%%%%%%%%%%%%%%%%%%%%%
The decay $\ell_\alpha \to \ell_\beta + \gamma$ arises from one-loop diagrams shown in Fig.~\ref{mutoeg}.
The general expression for this decay rate can be found in Ref.~\cite{Lavoura:2003xp}.  Let us focus on the special case where the FCNC coupling matrix $Y$ of Eq.~(\ref{lepyuk}) has nonzero entries either in a single row, or in a single column only.  In this case, the chirality flip necessary for the radiative decay will occur on the external fermion leg.  Suppose that only the right-handed component of fermion $f_\alpha$ has nonzero Yukawa couplings  with a scalar boson $B$ and fermion $F$, parametrized as
\begin{equation}
-{\cal L}_Y \ \supset \    B\,\sum_{\alpha =1,2}  Y_{\alpha \beta} \overline{F}_\beta P_R f_\alpha + {\rm H.c.} \, 
\end{equation}
The electric charges of fermions $F$ and $f$ are $Q_F$ and $Q_f$ respectively, while that of the boson $B$ is $Q_B$, which obey the relation $Q_f = Q_F-Q_B$.  The decay rate for $f_\alpha \rightarrow f_\beta+ \gamma$ is then given by
\begin{equation}
\Gamma(f_\alpha \rightarrow f_\beta + \gamma) \ = \ \frac{\alpha}{4}\frac{|Y_{\alpha \gamma}Y^\star_{\beta\gamma}|^2}{(16\pi^2)^2}\frac{m_\alpha^5}{m_B^4} \left[Q_F f_F(t) + Q_B f_B(t)\right]^2~.
\label{rate_rad}
\end{equation}
Here $\alpha=e^2/4\pi$ is the fine-structure constant, $t= m_F^2/m_B^2$, and the function $f_F(t)$ and $f_B(t)$ are given by
\begin{eqnarray}
  f_F(t)  &\ = \ & \frac{t^2-5t-2}{12(t-1)^3} + \frac{t \ {\rm log}t}{2(t-1)^4} \, , \nonumber\\
  f_B(t) &\ = \ & \frac{2t^2+5t-1}{12(t-1)^3} - \frac{t^2 \ {\rm log}t}{2(t-1)^4}.
\end{eqnarray}
These expressions are obtained in the approximation $m_\beta \ll m_\alpha$. % and $m_\alpha^2/m_B^2 \ll 1$.  

Let us apply these results to $\ell_\alpha \rightarrow \ell_\beta + \gamma$ mediated by the charged scalars ($h^+,\,H^+)$ in the Zee model where the couplings have the form $Y_{\alpha \beta} \bar{\nu}_\alpha P_R \ell_\beta h^+\sin\varphi$, etc.  Here $Q_F = 0$, while $Q_B = +1$.  Eq.~(\ref{rate_rad}) then reduces to (with $t \ll 1$)
\begin{equation}
\Gamma^{(h^+,H^+)}(\ell_\alpha \rightarrow \ell_\beta + \gamma) \ = \ \frac{\alpha}{4}\frac{|Y_{\gamma \alpha}Y^\star_{\gamma \beta}|^2}{(16\pi^2)^2}\frac{m_\alpha^5}{144} \left(\frac{\sin^2\varphi}{m_{h^+}^2} + \frac{\cos^2\varphi}{ m_{H^+}^2}  \right)^2.
\label{rate_rad1}
\end{equation}
If we set $m_{h^+} = 100$ GeV, $m_{H^+} = 700$ GeV and $\sin\varphi=0.7$,  then the experimental limit BR($\mu \rightarrow e \gamma) \leq 4.2 \times 10^{-13}$ \cite{Cei:2017puj} implies $|Y_{\alpha e} Y_{\alpha \mu}^\star| \leq 6 \times 10^{-5}$.  Similarly, the limit BR($\tau \rightarrow e \gamma)\leq 3.3 \times 10^{-8}$ \cite{Aubert:2009ag} implies $|Y_{\alpha \tau} Y_{\alpha e}^\star| \leq 4 \times 10^{-2}$, and the limit BR($\tau \rightarrow \mu \gamma)\leq 4.4 \times 10^{-8}$ \cite{Aubert:2009ag} implies $|Y_{\alpha \tau} Y_{\alpha \mu}^\star| \leq 4.6 \times 10^{-2}$.  These are rather stringent constraints, which suggest that no more than one entry in a given row of $Y$ can be large. Such a choice does not however affect the maximum NSI, as the elements of $Y$ that generate them are in the first column of $Y$. Keeping only the entries $(Y_{ee}, \, Y_{\mu e}, \, Y_{\tau e})$ nonzero does not lead to $\ell_\alpha \rightarrow \ell_\beta + \gamma$ decay mediated by the charged scalars ($h^+,\,H^+)$.  

However, nonzero values of $(Y_{ee}, \, Y_{\mu e}, \, Y_{\tau e})$, needed for NSI, would lead to $\ell_\alpha \rightarrow \ell_\beta + \gamma$ mediated by the heavy neutral scalars.  Taking $H$ and $A$ to be degenerate, the Yukawa couplings are of the form $\bar{\ell}_\alpha P_R \ell_\beta H$.  Thus, $Q_F = -1$ and $Q_B = 0$ in this case, leading to the decay width
\begin{equation}
\Gamma^{(H,A)}(\ell_\alpha \rightarrow \ell_\beta + \gamma) \ = \ \frac{\alpha}{144}\frac{|Y_{\alpha \gamma}Y^\star_{\beta\gamma}|^2}{(16\pi^2)^2}\frac{m_\alpha^5}{m_H^4} \, .
\label{rate_rad2}
\end{equation}
 We show the constraints on these product of Yukawa couplings for a fixed mass of the neutral Higgs $m_H$ in Table~\ref{lgdecay}. The severe constraint coming from $\mu \to e \gamma$ process prevents the off-diagonal NSI parameter $\varepsilon_{e\mu}$ from being in the observable range. However, $\varepsilon_{e\tau}$ and  $\varepsilon_{\mu\tau}$ can be in the observable range, consistent with these constraints.
 %%%%%%%%%%%%%%%%%%%%%%%%%%%%%%%%%%%%%%%%%%%%%%%%%\
\begin{table}[!t]
    \centering
    \begin{tabular}{|c|c|c|}
    \hline \hline
      \textbf{  Process }  & \textbf{Exp. bound} & \textbf{Constraint} \\ \hline \hline
       \rule{0pt}{10pt}  $\mu \to e \gamma$    &  BR \textless \, 4.2 $\times 10^{-13}$ \cite{TheMEG:2016wtm} &  $|Y_{\mu e}^\star Y_{ee}| $ \textless \, $ 1.05 \times 10^{-3} \left(\frac{m_H}{700~\text{GeV}}\right)^2$\\
       \rule{0pt}{20pt}  $\tau \to e \gamma$   &  BR \textless \, 3.3 $\times 10^{-8}$ \cite{Aubert:2009ag} & $|Y_{\tau e}^{\star} Y_{ee}| < \, 0.69 \left(\frac{m_H}{700~\text{GeV}}\right)^2 $ \\
      \rule{0pt}{20pt}   $\tau \to \mu \gamma$ & BR \textless \, 4.4 $\times 10^{-8}$ \cite{Aubert:2009ag} &  $|Y_{\tau e}^{\star} Y_{\mu e}| < \, 0.79  \left(\frac{m_H}{700 ~\text{GeV}}\right)^2$     \\
   \hline \hline
    \end{tabular}
    \caption{Constraints on Yukawa couplings as a function of heavy neutral scalar mass from $\ell_\alpha \to \ell_\beta + \gamma$ processes.}
    \label{lgdecay}
\end{table}

 %%%%%%%%%%%%%%
 \subsubsection{Electron anomalous magnetic moment} \label{sec:g-2e}
 %%%%%%%%%%%%%%%
 Another potential constraint comes from anomalous magnetic moment of leptons $(g-2)_\alpha$, which could get contributions from both charged and neutral scalars in the Zee model. The heavy neutral scalar contribution can be ignored here. For the Yukawa couplings relevant for NSI, the charged scalar contribution to muon $g-2$ is also absent. The only non-negligible contribution is to the electron $g-2$, which can be written at one-loop level as~\cite{Babu:2002uu} 
 \begin{align}
     \Delta a_e \ = \ -\frac{m_e^2}{96\pi}(Y^\dag Y)_{ee}\left(\frac{\sin^2\varphi}{m_{h^+}^2}+\frac{\cos^2\varphi}{m_{H^+}^2}\right) \, .
     \label{eq:g-2e}
 \end{align}
 Comparing this with $\Delta a_e\equiv a_e^{\rm exp}-a_e^{\rm SM}=(-87\pm 36)\times 10^{-14}$ (where $a_e\equiv (g-2)_e/2$), based on the difference between the experimental measurements~\cite{Hanneke:2010au} and  SM calculations~\cite{Aoyama:2017uqe} with the   updated value of the fine-structure constant~\cite{Parker:2018vye}, we find that  the charged scalar contribution~\eqref{eq:g-2e} goes in the right direction. However, for the allowed parameter space in $m_{h^+}-Y_{ee}\sin\varphi$ plane (see Fig.~\ref{fig:dnsi}), it turns out to be too small to explain the $2.4\sigma$ discrepancy in $\Delta a_e$.  For example, with $|Y_{\tau e}|\sin\varphi = 0.75$ and $m_{h^+} = 150$ GeV, which is a consistent choice (cf. Fig. \ref{fig:dnsi}), we would get $\Delta a_e = -2.2 \times 10^{-14}$, an order of magnitude too small to be relevant for experiments.

%%%%%%%%%%%%
%%%%%%%%%%%%%%%%%%%%%%%%%%%%%%%%%%%%%%%%%%%%%%%%%
\subsubsection{\texorpdfstring{$\ell_\alpha \to \bar{\ell}_\beta \ell_\gamma \ell_\delta$}{li lj lk ll} decays} \label{sec:3l}
%%%%%%%%%%%%%%%%%%%%%%%%%%%%%%%%%%%%%%%%%%%%%%%%
\begin{figure}[!t]
    \centering
    \includegraphics[scale=0.5]{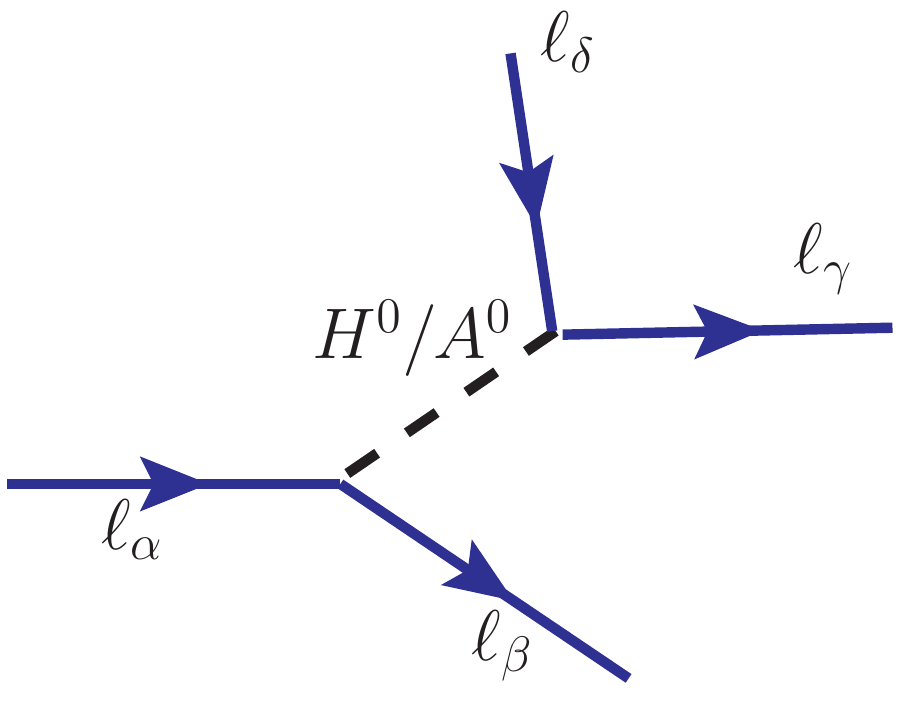}
    \caption{ Feynman diagram contributing to tree-level cLFV trilepton decay mediated by ${\cal CP}$-even and odd neutral scalars in the Zee model. At least two of the final state leptons must be of electron flavor to be relevant for NSI.}
    \label{tree_trilepton}
\end{figure}
%%%%%%%%%%%%%%%%%%%%%%

%%%%%%%%%%%%%%%%%%%%%%%%%%%%%%%%%%%%
The Yukawa coupling matrix $Y$ of the second Higgs doublet (cf. Eq. (\ref{lepyuk})) would lead to trilepton decay of charged leptons  mediated by the neutral scalars of the theory.  The tree-level Feynman diagrams for such decays are shown in Fig.~\ref{tree_trilepton}.  Partial rates for the trilepton decays are obtained in the limit when the masses of the decay products are neglected. The partial decay width for $\mu \rightarrow \bar{e} e e$ is given as follows:
%%%%%%%%%%%%%%%%%%%%%%%%%%%%%%%%%%%%%%%%%%
\begin{equation}
\Gamma(\mu^- \rightarrow e^+ e^- e^-) \ = \ \frac{1}{6144 \pi^3} |Y_{\mu e}^\star \, Y_{e e}|^{2} \frac{m_\mu^5}{m_H^4} \, .
\label{eq:mu3e}
\end{equation}
%%%%%%%%%%%%%%%%%%%%%%%%%%%%%%%%%%%%%%%%%%
The partial decay width for $\tau \rightarrow \bar{\ell}_{\alpha} \ell_{\beta} \ell_{\gamma}$ is given by 
%%%%%%%%%%%%%%%%%%%%%%%%%%
\begin{equation}
\Gamma\left(\tau \rightarrow \bar{\ell}_{\alpha} \ell_{\beta} \ell_{\gamma}\right) \ = \ \frac{1}{6144 \pi^3} \, S \, |Y_{\tau \alpha}^\star Y_{\beta\gamma}|^{2} \frac{m_\tau^5}{m_H^4} \, .
\label{eq:tau3l}
\end{equation}
%%%%%%%%%%%%%%%%%%%%%%%%%%%%%%%%%%%%%%%%%%
Here $S$ = 1 (2) for $\beta=\gamma \, (\beta \neq \gamma)$ is a symmetry factor. Using the total muon and tau decay widths, $\Gamma^{\rm tot}_\mu= 3.00\times 10^{-19}$ GeV and $\Gamma^{\rm tot}_\tau= 2.27\times 10^{-12}$ GeV respectively, we calculate the cLFV branching ratios for the processes $\mu^-\to e^+e^-e^-$, $\tau^-\to e^+e^-e^-$ and $\tau^-\to e^+e^-\mu^-$ using Eqs.~\eqref{eq:mu3e} and \eqref{eq:tau3l}. We summarize in Table~\ref{3ldecay} the current experimental bounds on these branching ratios and the constraints on the Yukawa couplings $Y_{\alpha \beta}$ as a function of mass of neutral Higgs boson $m_H = m_A$. It is clear from Table~\ref{3ldecay} that these trilepton decays put more stringent bounds on product of Yukawa couplings compared to the bounds arising from loop-level $\ell_\alpha \to \ell_\beta \gamma$ decays. This also implies that off-diagonal NSI are severely constrained. 

As already noted, the light charged Higgs $h^+$ would mediate $\ell_\alpha \rightarrow \ell_\beta + \gamma$ decay if more than one entry in a given row of $Y$ is large.  The heavy neutral Higgs bosons mediate trilepton decays of the leptons when there are more than one nonzero entry in the same column (or same row) of $Y$.  This last statement is however not valid for the third column of $Y$.  For example, nonzero $Y_{\tau \tau}$ and $Y_{\mu \tau}$ will not lead to tree-level trilepton decay of $\tau$.  Apart from the first column of $Y$, we shall allow nonzero entries in the third column as well.  In particular, for diagonal NSI $\varepsilon_{\alpha\alpha}$, we need one $Y_{\alpha e}$  entry for some $\alpha$  to be nonzero, and to avoid the trilepton constraints, the only other entry that can be allowed to be large is $Y_{\beta \tau}$ with $\beta\neq \alpha$. On the other hand, for off-diagonal NSI $\varepsilon_{\alpha\beta}$ (with $\alpha\neq \beta$), we must allow for both $Y_{\alpha e}$ and $Y_{\beta e}$ to be non-zero. In this case, however, the trilepton decay $\ell_\beta\to \ell_\alpha ee$ is unavoidable and severely restricts the NSI as we will see in Sec.~\ref{sec:NSIZee}. Also, the other entry that can be populated is $Y_{\gamma\tau}$ with $\gamma\neq \alpha,\beta$. This will lead to $\tau \rightarrow \ell+ \gamma$ decays, which, however, do not set stringent limits on the couplings (cf.~Table~\ref{lgdecay}). Some benchmark Yukawa textures satisfying all cLFV constraints are considered in 
Sec.~\ref{sec:neu} to show consistency with neutrino oscillation data. 

%%%%%%%%%%%%%%%%%%%%%%%%%%%%%%%%%%%%%%%%%%%%%%%%
%%%%%%%%%%%%%%%%%%%%%%%%%%%%%%%%%%%%%%%%%%%%%%%%
\begin{table}[!t]
    \centering
    \begin{tabular}{|c|c|c|}
    \hline \hline
       \textbf{ Process} & \textbf{Exp. bound}  & \textbf{Constraint} \\ \hline \hline
      \rule{0pt}{10pt}  $\mu^- \to e^+ e^- e^-  $      &  BR \textless \, 1.0 $\times 10^{-12}$~  \cite{Bertl:1985mw}      &    $|Y_{\mu e}^\star Y_{ee}| $ \textless \, $3.28 \times 10^{-5} \left(\frac{m_H}{700 ~\text{GeV}}\right)^2$ \\ 
      \rule{0pt}{20pt}  $\tau^- \to  e^+ e^- e^- $     &  BR \textless  \, 1.4 $\times 10^{-8}$~   \cite{Amhis:2016xyh}      & $|Y_{\tau e}^\star Y_{ee}| $ \textless \, $ 9.05 \times 10^{-3} \left(\frac{m_H}{700~ \text{GeV}}\right)^2$     \\
      \rule{0pt}{20pt}  $\tau^- \to e^+ e^- \mu^-    $ &  BR \textless   \, 1.1 $\times 10^{-8}$    ~ \cite{Amhis:2016xyh}   & $|Y_{\tau e}^\star Y_{\mu e}| $ \textless \, $5.68 \times 10^{-3} \left(\frac{m_H}{700 ~\text{GeV}}\right)^2$     \\
   \hline \hline
    \end{tabular}
    \caption{Constraints on Yukawa couplings as a function of heavy neutral scalar mass from $\ell_\alpha\to \bar{\ell}_\beta\ell_\gamma\ell_\delta$ decay (with at least two of the final state leptons of electron flavor to be relevant for NSI). }
    \label{3ldecay}
\end{table}
%%%%%%%%%%%%%%%%%%%%%%%%%%%%%%%%%%%%

\subsection{Collider constraints on neutral scalar mass} \label{sec:contact}
%%%%%%%%%%%%%%%%%%%%%%%%%%%%%%%%%%%%%%%}
In this section, we discuss the collider constraints on the neutral 
scalars $H$ and $A$ in the Zee model from various LEP and LHC searches.

\subsubsection{LEP contact interaction} \label{sec:lepcon}
%%%%%%%%%%%%%%%%
 Electron-positron collisions at center-of-mass energies above the $Z$-boson mass performed at LEP impose stringent constraints on contact interactions involving $e^+ e^-$ and a pair of fermions~\cite{LEP:2003aa}. Integrating out new particles in a theory one can express their effect via higher-dimensional (generally dimension-6) operators. An effective Lagrangian, $\mathcal{L}_{\rm eff}$,  can parametrize the contact interaction for the process $e^+ e^- \to f\bar{f}$  with the form~\cite{Eichten:1983hw}
\begin{equation}
{\cal L}_{\rm eff}\  = \ \frac{g^2}{{\Lambda}^2(1+ \delta_{ef})} \sum_{i,j=L,R} {\eta}_{ij}^{f}  (\bar e_i \gamma^\mu  e_i)(\bar f_j \gamma_\mu  f_j)\, ,  \label{eq:eff}
\end{equation}
where $\delta_{ef}$ is the Kronecker delta function,  $f$ refers to the final sate fermions, $g$ is the coupling strength, $\Lambda$ is the new physics scale and $\eta^f_{ij}=\pm 1$ or 0, depending on the chirality structure. LEP has put 95\% confidence level (CL) lower limits on the scale of the contact interaction $\Lambda$ assuming the coupling $g=\sqrt{4\pi}$~\cite{LEP:2003aa}. In the Zee model, the exchange of new neutral scalars ($H$ and $A$) emerging from the second Higgs doublet will affect the process $e^+ e^- \to \ell_\alpha^+\ell_\beta^-$ (with $\ell_{\alpha,\beta}=e,\mu,\tau$), and therefore, the LEP constraints on $\Lambda$ can be interpreted as a {\it lower} limit on the mass of the heavy neutral scalar, for a given set of Yukawa couplings. Here we assume that $H$ and $A$ are degenerate, and derive limits obtained by integrating out both fields.

In general, for $\ell_\alpha^+ \ell_\beta^- \to \ell_\gamma^+\ell_\delta^-$ via heavy neutral scalar exchange, the effective Lagrangian in the Zee model can be written as
\begin{equation}
{\cal L}^{\rm Zee}_{\rm eff}\  = \ \frac{Y_{\alpha\delta}Y_{\beta\gamma}^\star}{m_{H}^2} (\bar{\ell}_{\alpha L}\ell_{\delta R})(\bar{\ell}_{\beta R}\ell_{\gamma L}) \, .  \label{eq:effZ}
\end{equation}
By Fierz transformation, we can rewrite it in a form similar to Eq.~\eqref{eq:eff}: 
\begin{equation}
{\cal L}^{\rm Zee}_{\rm eff}\  = \ -\frac{1}{2}\frac{Y_{\alpha\delta}Y_{\beta\gamma}^\star}{m_{H}^2} (\bar{\ell}_{\alpha L}\gamma^\mu \ell_{\gamma L})(\bar{\ell}_{\beta R}\gamma_\mu \ell_{\gamma R}) \, .  \label{eq:effZee}
\end{equation}
Thus, the only relevant chirality structures in Eq.~\eqref{eq:eff} are $LR$ and $RL$, and the relevant process for deriving the LEP constraints is $e^+e^-\to \ell_\alpha^+\ell_\alpha^-$:
\begin{equation}
{\cal L}_{\rm eff}\  = \ \frac{g^2}{\Lambda^2(1+\delta_{e\alpha})}\left[\eta^{\ell}_{LR} (\bar{e}_{L}\gamma^\mu e_{L})(\bar{\ell}_{\alpha  R}\gamma_\mu \ell_{\alpha R})+\eta^\ell_{RL}(\bar{e}_R\gamma^\mu e_R) (\bar{\ell}_{\alpha L}\gamma_\mu \ell_{\alpha L})\right]\, ,  \label{eq:eff1}
\end{equation}
with $\eta^\ell_{LR}=\eta^\ell_{RL}=-1$. 
%The minus sign arises because the scalar exchange diagram has a destructive interference with the SM contributions via vector bosons. 

Now for $e^+e^-\to e^+e^-$, Eq.~\eqref{eq:effZee} becomes 
\begin{align}
{\cal L}_{\rm eff}^{\rm Zee}(e^+e^-\to e^+e^-) \ & = \ -\frac{|Y_{ee}|^2}{2m_{H}^2}(\bar e_L \gamma^\mu  e_L)(\bar e_R \gamma_\mu  e_R) \, .
\end{align}
Comparing this with Eq.~\eqref{eq:eff1}, we obtain 
\begin{align}
\frac{m_H}{|Y_{ee}|} \ & = \ \frac{\Lambda^-_{LR/RL}}{\sqrt 2 g} \, ,
     \label{eq:LEP1}
 \end{align}
 where $\Lambda^-$ corresponds to $\Lambda$ with $\eta^\ell_{LR}=\eta^\ell_{RL}=-1$.  The LEP constraints on $\Lambda$ were derived in Ref.~\cite{LEP:2003aa} for $g=\sqrt{4\pi}$, which can be translated into a lower limit on $m_H/|Y_{ee}|$ using Eq.~\eqref{eq:LEP1}, as shown in Table~\ref{tab:LEPcontact}. Similarly, for $e^+e^-\to \mu^+\mu^-$, Eq.~\eqref{eq:effZee} is   
\begin{align}
{\cal L}_{\rm eff}^{\rm Zee} (e^+e^-\to \mu^+\mu^-) \ = \    -\frac{1}{2m_H^2}\big[|Y_{e\mu}|^2(\bar e_L \gamma^\mu  e_L)(\bar \mu_R \gamma_\mu  \mu_R) +|Y_{\mu e}|^2(\bar e_R \gamma^\mu  e_R)(\bar \mu_L \gamma_\mu  \mu_L) \big] \, .
\label{eq:LEP2}
\end{align}
Since for NSI, only $Y_{\mu e}$ (neutrino interaction with electron) is relevant, we can set $Y_{e\mu}\to 0$, and compare Eq.~\eqref{eq:LEP2} with Eq.~\eqref{eq:eff1} to get a constraint on $m_H/|Y_{\mu e}|$, as shown in Table~\ref{tab:LEPcontact}. Similarly, for $e^+e^-\to \tau^+\tau^-$, we can set $Y_{e\tau}\to 0$ and translate the LEP limit on $\Lambda^-$ into a bound on $m_H/|Y_{\tau e}|$, as shown in Table~\ref{tab:LEPcontact}. 

\begin{table}[!t]
    \centering
    \begin{tabular}{|c|c|c|}
    \hline \hline
       \textbf{ Process} & \textbf{LEP bound}~\cite{LEP:2003aa} & \textbf{Constraint} \\ \hline \hline
      \rule{0pt}{10pt}  $e^+ e^- \to e^+ e^- $      &  $\Lambda^-_{LR/RL}>10$ TeV      &    $\frac{m_H}{|Y_{ee}|}>1.99$ TeV \\ [2pt]
      \rule{0pt}{10pt}  $e^+ e^- \to \mu^+ \mu^- $      &  $\Lambda^-_{LR/RL}>7.9$ TeV      &    $\frac{m_H}{|Y_{\mu e}|}>1.58$ TeV \\ [2pt]
      \rule{0pt}{10pt}  $e^+ e^- \to \tau^+ \tau^- $      &  $\Lambda^-_{LR/RL}>2.2$ TeV      &    $\frac{m_H}{|Y_{\tau e}|}>0.44$ TeV     \\
   \hline \hline
    \end{tabular}
    \caption{Constraints on the ratio of heavy neutral scalar mass and the Yukawa couplings from LEP contact interaction bounds.}
    \label{tab:LEPcontact}
\end{table}

The LEP constraints from the processes involving $q\bar{q}$ final states, such as $e^+ e^- \to c\bar{c}$ and $e^+ e^- \to b\bar{b}$, are not relevant in our case, since the neutral scalars are leptophilic. We will use the limits quoted in Table~\ref{tab:LEPcontact} while deriving the maximum NSI predictions in the Zee model. 

\subsubsection{LEP constraints on light neutral scalar} \label{sec:light_neutral}
The LEP contact interaction constraints discussed in Sec.~\ref{sec:contact} are not applicable if the  neutral scalars $H$ and $A$ are light. In this case, however, the cross section of $e^+e^-\to \ell_\alpha^+\ell_\alpha^-$ can still be modified, due to the $t$-channel contribution of $H/A$, which interferes with the SM processes. We implement our model file in {\tt FeynRules} package~\cite{Christensen:2008py} and compute the $e^+e^-\to \ell_\alpha^+\ell_\alpha^-$ cross-sections in the Zee model at the parton-level using {\tt MadGraph5} event generator~\cite{Alwall:2014hca}. These numbers are then compared with the measured cross sections~\cite{LEP:2003aa, Abbiendi:2003dh} to derive limits on $m_{H/A}$ as a function of the Yukawa couplings $Y_{\alpha e}$ (for $\alpha=e,\mu,\tau$). For a benchmark value of $m_H=m_A=130$ GeV, we find the following constraints on the Yukawa couplings $Y_{\alpha e}$ relevant for NSI: 
\begin{align}
    Y_{ee} \ < \ 0.80 \, , \qquad Y_{\mu e} \ < \ 0.74 \, , \qquad 
    Y_{\tau e} \ < \ 0.73 \, .
    \label{eq:LEP_light}
\end{align}
This implies that the second charged scalar $H^+$ can also be light, as long as it is allowed by other constraints (see Fig.~\ref{fig:col}). We will use this finding to maximize the NSI prediction for the Zee model (see Sec.~\ref{sec:Zee_light}).   

\subsubsection{LHC constraints}\label{sec:LHC_neutral}

Most of the LHC searches for heavy neutral scalars are done in the context of either MSSM or 2HDM, which are not directly applicable in our case because $H$ and $A$ do not couple to quarks, and therefore, cannot be produced via gluon fusion. The dominant channel to produce the  neutral scalars in our case at the LHC is via an off-shell $Z$ boson: $pp\to Z^\star\to HA \to \ell^+\ell^-\ell^+\ell^-$.\footnote{Only the $(H\overset{\leftrightarrow}{\partial}_\mu A)Z^\mu$ coupling is nonzero, while the $(H\overset{\leftrightarrow}{\partial}_\mu H)Z^\mu$ and $(A\overset{\leftrightarrow}{\partial}_\mu A)Z^\mu$ couplings vanish due to parity~\cite{Grzadkowski:2018ohf}.} Most of the LHC multilepton searches assume a heavy $ZZ^{(\star)}$ resonance~\cite{Sirunyan:2018qlb, Aaboud:2018puo}, which is not applicable in this case. The cross section limits from inclusive multilepton searches, mostly performed in the SUSY context with large missing transverse energy~\cite{Chatrchyan:2014aea, Aaboud:2018zeb}, turn out to be weaker than the LEP constraints derived above.

\subsection{Collider constraints on light charged scalar \label{sec:colliderZee}} 

  \begin{figure}[t!]
         \centering
         \subfigure[]{
         \includegraphics[scale=0.5]{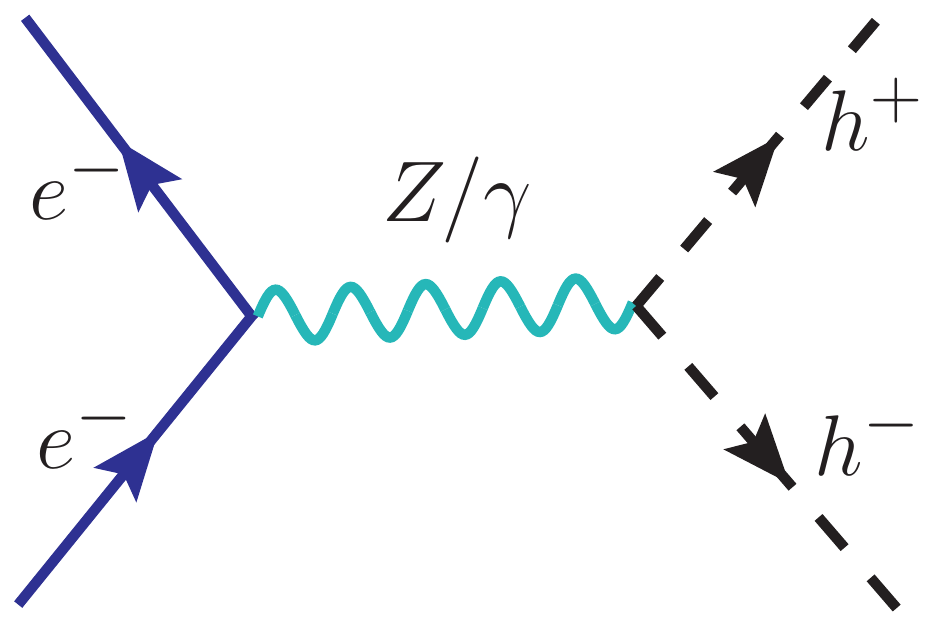}}
          \hspace{0.15in}
          \subfigure[]{
         \includegraphics[scale=0.6]{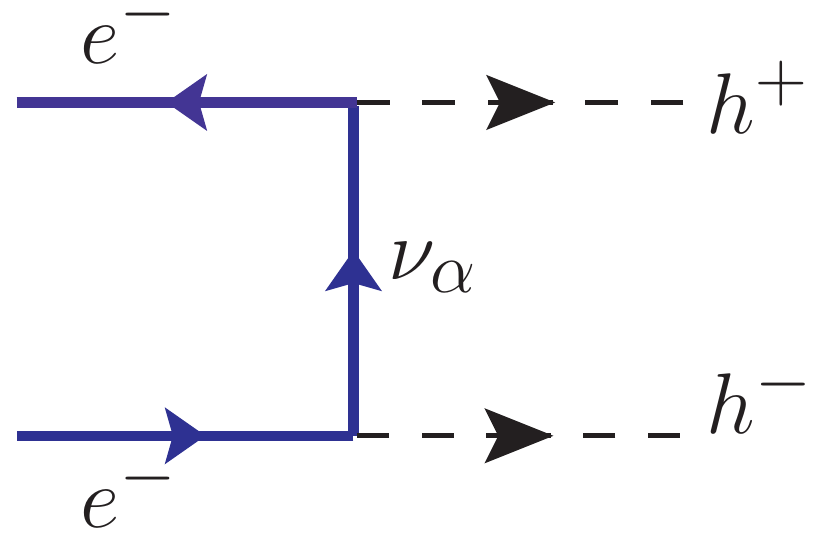}}
         \subfigure[]{  
            \includegraphics[scale=0.52]{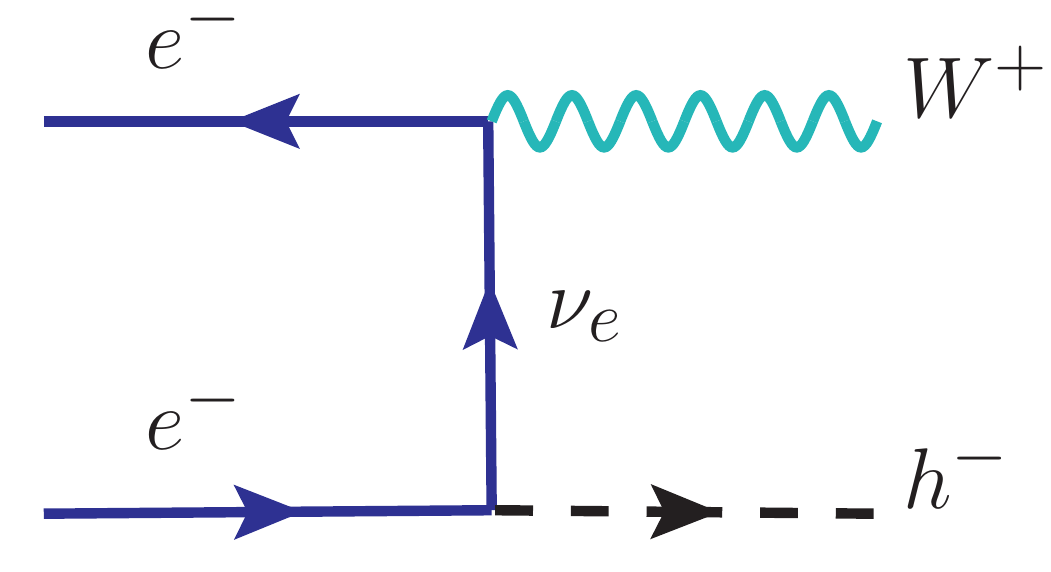}} \hspace{0.15in}
            \subfigure[]{
            \includegraphics[scale = 0.49]{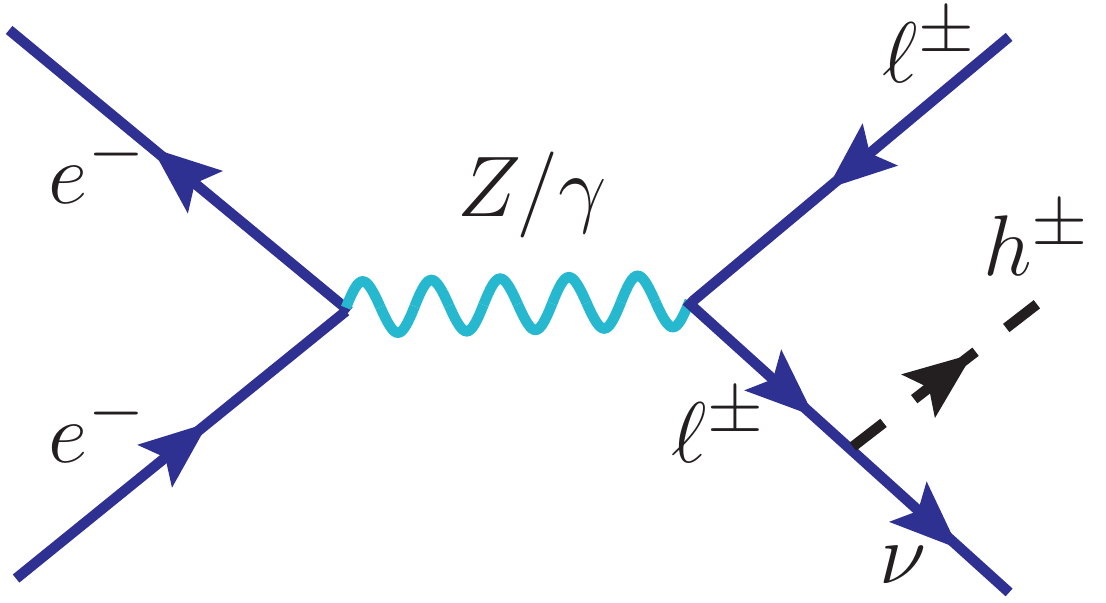}}
         \caption{Feynman diagrams for pair-  and single-production of singly-charged scalars $h^\pm$ at $e^+e^-$ collider.}
         \label{fig:feyn1}
    \end{figure}
%%%%%%%%%%%%%%%%%%%%%%%%%%%%%%%%%  
In this section, we discuss the collider constraints on the light charged scalar $h^\pm$ in the Zee model from various LEP and LHC searches.

\subsubsection{Constraints from LEP searches} \label{sec:LEPZee}
At LEP, $h^\pm$ can be pair-produced  through the $s$-channel Drell-Yan process mediated by either $\gamma$ or  $Z$ boson (see Fig.~\ref{fig:feyn1} (a)). It can also be  pair-produced through the $t$-channel processes mediated by a light neutrino (see Fig.~\ref{fig:feyn1} (b)). In addition, it can be singly-produced either in association with a $W$ boson (see Fig.~\ref{fig:feyn1} (c)) or via the Drell-Yan channel in association with leptons  (see Fig.~\ref{fig:feyn1} (d)). The analytic expressions for the relevant cross sections can be found in Appendix~\ref{app:A}. For our numerical study, we implement our model file in {\tt FeynRules} package~\cite{Christensen:2008py} and compute all the cross-sections at the parton-level using {\tt MadGraph5} event generator~\cite{Alwall:2014hca}. We  find good agreement between the numerically computed values and the analytic results presented in Appendix~\ref{app:A}.

Once produced on-shell, the charged scalar will decay into the leptonic final states $\nu_\alpha\ell_\beta$ through the  Yukawa coupling $Y_{\alpha\beta}$. Since we are interested in potentially large NSI effects, the charged scalar must couple to the electron. Due to stringent constraints from cLFV processes, especially the trilepton cLFV decays (see Table~\ref{3ldecay}), which is equally applicable for the product of two Yukawa entries either along a row or column, both $Y_{\alpha e}$ and $Y_{\alpha \mu}$ (or $Y_{\alpha e}$ and $Y_{\beta e}$) cannot be large simultaneously. So we consider the case where ${\rm BR}_{e\nu}+{\rm BR}_{\tau\nu}=100\%$ and ${\rm BR}_{\mu\nu}$ is negligible, in order to avoid more stringent limits from muon decay.\footnote{This choice is consistent with the  observed neutrino oscillation data (see Sec.~\ref{sec:neu}).}  
%%%%%%%%%%%%%%%%%%%%%%%%%%%%%%

%%%%%%%%%%%%%%%%%%%%%%%%%%%%%%

{\bf Electron channel:} For a given charged scalar decay branching ratio to electrons, ${\rm BR}_{e\nu}$, we can reinterpret the LEP selectron searches~\cite{lepsusy} to put a constraint on the charged scalar mass as a function of ${\rm BR}_{e\nu}$. In particular, the right-handed  selectron pair-production $e^+e^-\to \widetilde{e}_R\widetilde{e}_R$, followed by the decay of each selectron to electron and neutralino, $\widetilde{e}_R\to e_R+\widetilde{\chi}^0$, will mimic the $e^+ e^-\nu\bar{\nu}$ final state of our case in the massless neutralino limit. So we use the 95\% CL observed upper limits on the $\widetilde{e}_R\widetilde{e}_R$ production cross section~\cite{lepsusy} for $m_{\widetilde{\chi}}=0$ as an experimental upper limit on the quantity 
\begin{align}
\widetilde{\sigma}_{ee} \ \equiv \ \sigma(e^+e^-\to h^+h^-){\rm BR}^2_{e\nu}+\sigma(e^+e^-\to h^\pm W^\mp){\rm BR}_{e\nu}{\rm BR}_{W\to e\nu} \, ,
\end{align}
and derive the LEP exclusion region in the plane of charged scalar mass and ${\rm BR}_{e\nu}$, as shown in Fig.~\ref{fig:col} (a) by the orange-shaded region. Here we have chosen $Y_{ee}\sin\varphi=0.1$ and varied $Y_{\tau \alpha}$ (with $\alpha=\mu$ or $\tau$) to get the desired branching ratios. We find that for ${\rm BR}_{e\nu}=1$, charged scalar masses less than 100 GeV are excluded. For ${\rm BR}_{e\nu}<1$, these limits are weaker, as expected, and the charged scalar could be as light as 97 GeV (for ${\rm BR}_{e\nu}=0.33$), if we just consider the LEP selectron (as well as stau, see below) searches.

Fig.~\ref{fig:col} (b) shows the same constraints as in Fig.~\ref{fig:col} (a), but for the case of  $Y_{ee}\sin\varphi=0.2$. The LEP selectron constraints become stronger as we increase $Y_{ee}$ and extend to smaller ${\rm BR}_{e\nu}$. However, the mass limit of 100 GeV for ${\rm BR}_{e\nu}=1$ from Fig.~\ref{fig:col} (a) still holds here. This is because the charged scalar pair-production cross section drops rapidly for $m_{h^+}>100$ GeV due to the kinematic threshold of LEP II with $\sqrt{s}=209$ GeV and is already below the experimental cross section limit even for $Y_{ee}\sin\varphi=0.2$. In this regime, the single-production channel in Fig.~\ref{fig:feyn1} (d) starts becoming important, despite having a three-body phase space suppression. 

Figs.~\ref{fig:col} (c) and \ref{fig:col} (d) show the same constraints as in Fig.~\ref{fig:col} (a) and~\ref{fig:col} (b) respectively, but for the $Y_{ee}=0$ case. Here we have fixed $Y_{\tau e} \sin\varphi$ and varied $Y_{\tau\alpha}$ (with $\alpha=e$ or $\mu$) to get the desired branching ratios. In this case, the single-production channel in association with the $W$ boson (cf.~Fig.~\ref{fig:feyn1} (c)) goes away, and therefore, the limits from selectron and stau searches become slightly weaker. Note that for the NSI purpose, we must have a non-zero  $Y_{\alpha e}$ (for $\alpha=e,\mu$ or $\tau$). Therefore, the $t$-channel contribution to the pair-production (cf.~Fig.~\ref{fig:feyn1} (b)), as well as the Drell-Yan single-production channel are always present.\footnote{This might be the reason why the LEP limits derived here are somewhat more stringent than those reported in Ref.~\cite{Cao:2017ffm}, which presumably only considered the $s$-channel contribution.} 

{\bf Tau channel:} In the same way, we can also use the LEP stau  searches~\cite{lepsusy} to derive an upper limit on 
\begin{align}
\widetilde{\sigma}_{\tau\tau} \ \equiv \ \sigma(e^+e^-\to h^+h^-){\rm BR}^2_{\tau\nu}+\sigma(e^+e^-\to h^\pm W^\mp){\rm BR}_{\tau\nu}{\rm BR}_{W\to \tau\nu}
\end{align}
and the corresponding LEP exclusion region in the plane of charged scalar mass and ${\rm BR}_{\tau\nu}$, as shown in Fig.~\ref{fig:col} by the blue-shaded region. We find that for ${\rm BR}_{\tau\nu}=1$, charged scalar masses less than 104~(105) GeV are excluded for $Y_{ee}\sin\phi=0.1~(0.2)$. 

For ${\rm BR}_{\tau\nu}\neq 0$, a slightly stronger limit can be obtained from the LEP searches for the charged Higgs boson pairs in the 2HDM~\cite{Abbiendi:2013hk}. Their analysis focused on three kinds of final states, namely, $\tau\nu\tau\nu$, $c\bar{s}\tau\nu$ (or $\bar{c}s\tau\nu$) and $c\bar{s}\bar{c}s$, under the assumption that ${\rm BR}_{\tau\nu}+{\rm BR}_{c\bar{s}}=1$, which is valid in the 2HDM as the couplings of the charged Higgs boson to the SM fermions are proportional to the fermion masses. In our case, the observed LEP upper limit on 
$\sigma(e^+e^-\to h^+h^-){\rm BR}^2_{\tau\nu}$ for ${\rm BR}_{\tau\nu}=1$ can be recast into an upper limit on 
\begin{align}
\sigma^h_{\tau\tau} \ \equiv \ \sigma(e^+e^-\to h^+h^-){\rm BR}^2_{\tau\nu}+\sigma(e^+e^-\to h^\pm W^\mp){\rm BR}_{\tau\nu}{\rm BR}_{W\to \tau\nu}
\end{align}
and the corresponding exclusion region is shown in Fig.~\ref{fig:col} by the green-shaded region. We can also use the LEP cross section limit on $c\bar{s}\tau \nu$ for ${\rm BR}_{\tau\nu}\neq 1$ as an upper limit on $\sigma(e^+e^-\to h^\pm W^\mp){\rm BR}_{\tau\nu}{\rm BR}_{W\to c\bar{s}}$ and the corresponding exclusion region is shown in Fig.~\ref{fig:col} by the cyan-shaded region, which is found to be weaker than the $\tau\nu\tau\nu$ mode.  
\subsubsection{Constraints from LHC searches} \label{sec:LHCZee}
\begin{figure}[t!]
         \centering
         \subfigure[]{
         \includegraphics[scale=0.5]{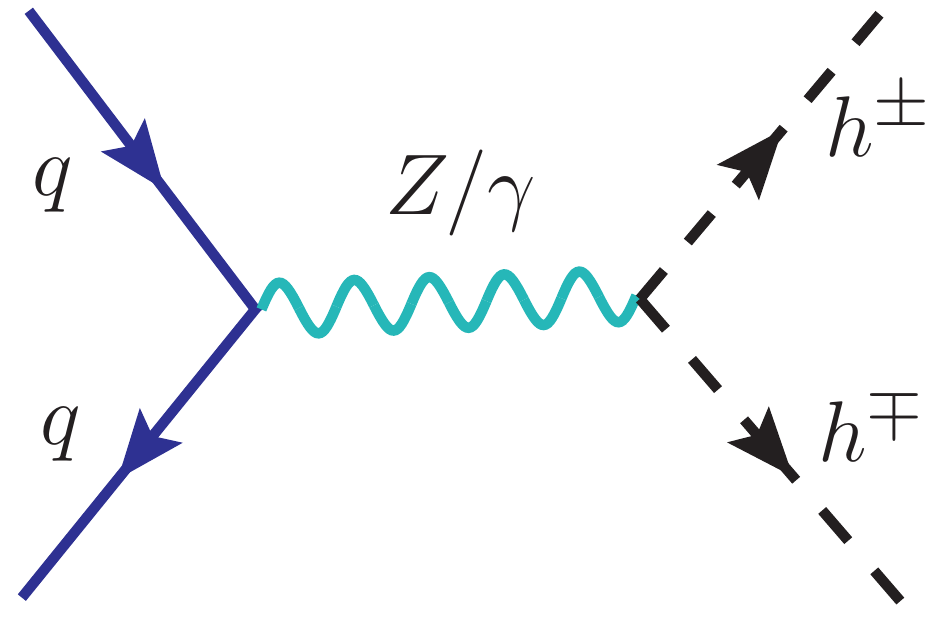}}
          \hspace{0.15in}
        \subfigure[]{ \includegraphics[scale=0.5]{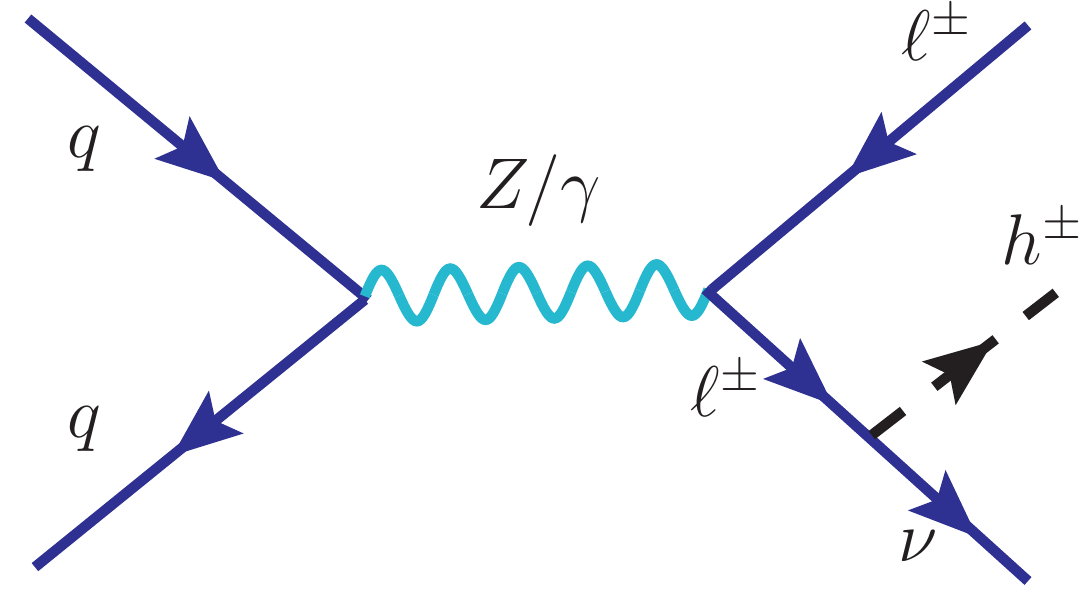}}
         \subfigure[]{ 
            \includegraphics[scale=0.5]{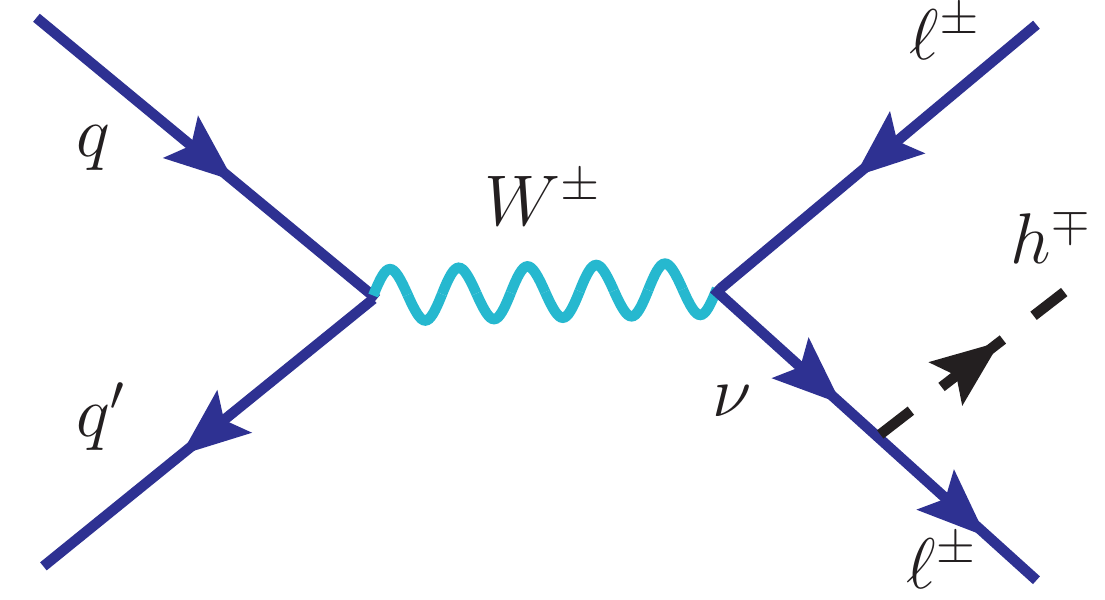}} \hspace{0.15in}
            \subfigure[]{
            \includegraphics[scale= 0.5]{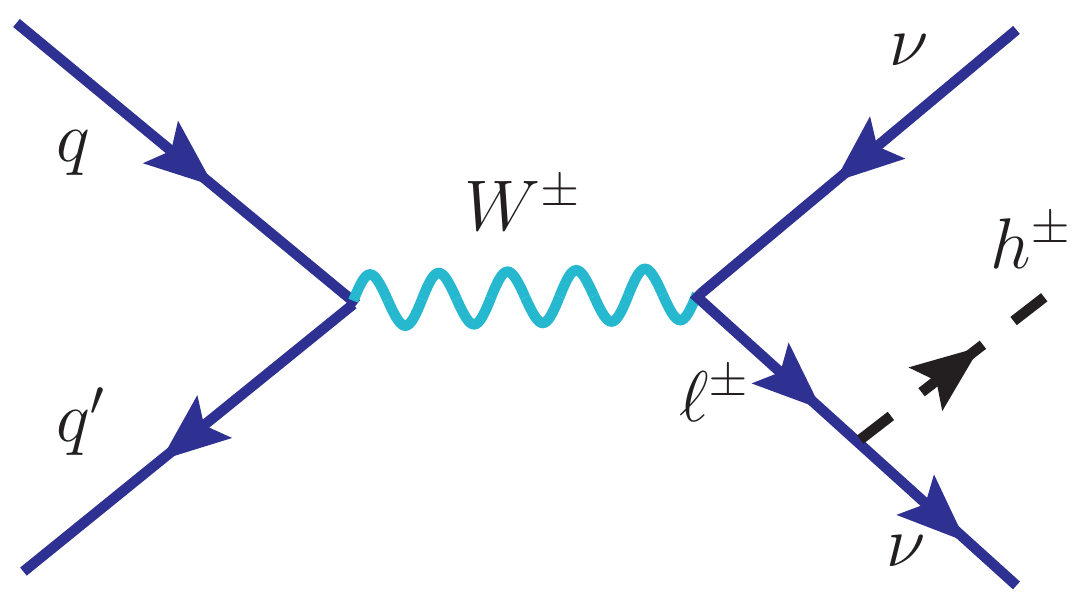}}
         \caption{Feynman diagrams for pair- and single-production of singly-charged scalars $h^\pm$ at LHC.}
         \label{fig:feyn2}
    \end{figure}

    As for the LHC constraints, there is no $t$-channel contribution to the singlet charged-scalar production. The only possible channel for pair-production is the $s$-channel Drell-Yan process $pp\to \gamma^\star/Z^\star\to h^+h^-$ (see Fig.~\ref{fig:feyn2} (a)), followed by the leptonic decay of $h^\pm\to \ell \nu$. There are also single-production processes as shown in Fig.~\ref{fig:feyn2} (b)-(d), which are less important. The relevant LHC searches are those for right-handed  selectrons/staus: $pp\to \widetilde{\ell}^+_R\widetilde{\ell}^-_R\to \ell^+_R\widetilde{\chi}^0\ell^-_R\widetilde{\chi}^0$, which will mimic the $\ell^+\nu\ell^-\nu$ final states from $h^+h^-$ decay in the massless neutralino limit. The $\sqrt s=13$ TeV LHC stau searches focus on the stau mass range above 100 GeV and it turns out that the current  limits~\cite{Sirunyan:2018vig} on the stau pair-production cross section are still a factor of five larger than the $h^+h^-$ pair-production cross section in our case; therefore, there are no LHC limits from the tau sector. A $\sqrt s=8$ TeV ATLAS analysis considered the mass range down to 80 GeV~\cite{Aad:2014yka}; however, the observed cross section is still found to be larger than the theoretical prediction in our case even for ${\rm BR}_{\tau\nu}=1$.  
    
\begin{figure}[t!]
\centering
\subfigure[]{
\includegraphics[height=0.52\textwidth, width=0.48\textwidth]{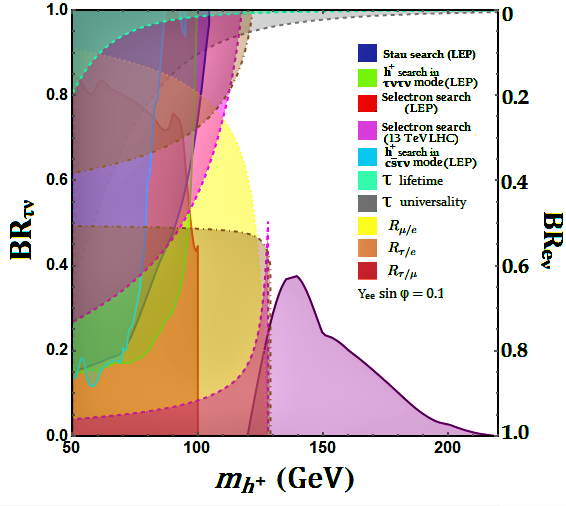}}
\subfigure[]{
\includegraphics[height=0.52\textwidth, width=0.48\textwidth]{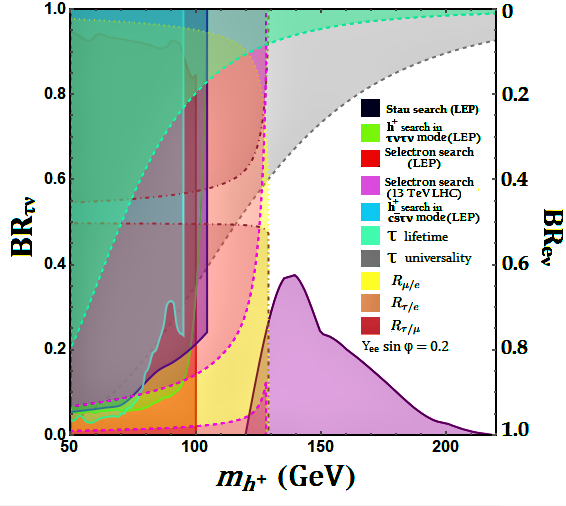}} \\
\subfigure[]{
\includegraphics[height=0.52\textwidth, width=0.48\textwidth]{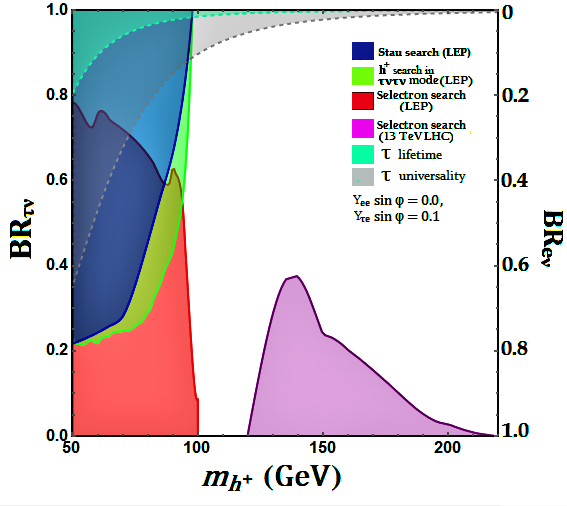}}
\subfigure[]{\includegraphics[height=0.52\textwidth, width=0.48\textwidth]{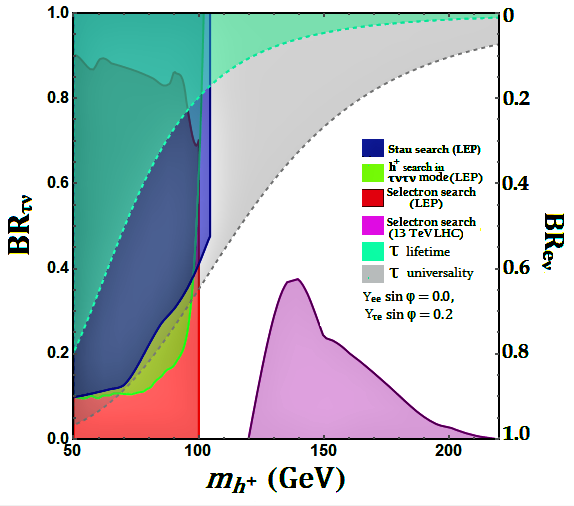}}
\caption{Collider constraints on light charged scalar $h^\pm$ in the Zee model for (a) $Y_{e e} \sin{\varphi}=0.1$, (b) $Y_{e e} \sin{\varphi}=0.2$, (c) $Y_{e e} \sin{\varphi}=0$,  $Y_{\tau e}\sin{\varphi}=0.1$, and (d) $Y_{e e} \sin{\varphi}=0$, $Y_{\tau e}\sin{\varphi}=0.2$. We plot the $h^\pm$ branching ratios to ${\tau \nu}$ and ${e\nu}$ (with the sum being equal to one) as a function of its mass. All shaded regions are excluded: Blue and orange regions from stau and selectron searches at LEP (see Sec.~\ref{sec:LEPZee}); purple region from selectron searches at LHC (see Sec.~\ref{sec:LHCZee}); yellow, brown, and pink regions from $W$ universality tests in LEP data for $\mu/e$, $\tau/e$, and $\tau/\mu$ sectors respectively (see Sec.~\ref{sec:Wuniv}); light green and gray  regions from tau decay universality and lifetime constraints respectively (see Sec.~\ref{sec:taudecay}). The $W$ universality constraints do not apply in panels (b) and (c), because the $h^\pm W^\mp$ production channel in Fig.~\ref{fig:feyn1} (c) vanishes in the $Y_{e e}\to 0$ limit.}
\label{fig:col}
\vspace{-1cm}
\end{figure}

 As for the selectron case, we take the $\sqrt s=13$ TeV CMS search~\cite{Sirunyan:2018nwe}, which focuses on the selectron masses above 120 GeV, and use the observed cross section limit on  $\sigma(pp\to e^+_R\widetilde{\chi}^0e^-_R\widetilde{\chi}^0)$ to derive an upper limit on $\sigma(pp\to h^+h^-){\rm BR}^2_{e\nu}$, which can be translated into  a bound on the charged scalar mass, as shown in Fig.~\ref{fig:col} by the purple-shaded regions. 
%%%%%%%%%%%%%%%%%%%%%%%%%%%%%%
\clearpage
%%%%%%%%%%%%%%%%%%%%%%%%%%%%%%
\noindent It is evident that the LHC limits can be evaded by going to larger ${\rm BR}_{\tau\nu}\gtrsim 0.4$, which can always be done for any given Yukawa coupling $Y_{\alpha e}$ by choosing an appropriate $Y_{\beta \tau}$. This however may not be the optimal choice for NSI, especially for $Y_{ee}\neq 0$, where the lepton universality constraints restrict us from having a larger ${\rm BR}_{\tau\nu}$. Thus, the LHC constraints will be most relevant for $\varepsilon_{ee}$, as we will see in Fig.~\ref{fig:dnsi} (a). 
 
 %%%%%%%%%%%%%%%%%%%%%%%%%%%%%%%%  

%%%%%%%%%%%%%%%%%%%%%%%
\subsection{Constraints from lepton universality in \texorpdfstring{$W^{\pm}$}{W} decays} \label{sec:Wuniv}

The presence of a light charged Higgs can also be constrained from precision measurements of $W$ boson decay rates. The topology of the charged Higgs pair production $h^+ h^-$ (Fig.~\ref{fig:feyn1} (a) and \ref{fig:feyn1} (b)) and the associated production $h^\pm W^\mp$ (Fig.~\ref{fig:feyn1} (c)) is very similar to the $W^+ W^-$ pair production at colliders, if the charged Higgs mass is within about 20 GeV of the $W$ boson mass. Thus, the leptonic decays of the charged Higgs which are not necessarily flavor-universal can be significantly constrained from the measurements of lepton universality in $W$ decays. From the combined LEP results~\cite{ALEPH:2004aa}, the constraints on the ratio of $W$ branching ratios to leptons of different flavors are as follows:  
\begin{align}
R_{\mu / e} & \ = \ \frac{\Gamma(W \rightarrow \mu \nu)}{\Gamma(W \rightarrow e \nu)} \ = \ 0.986 \pm 0.013 \, , \label{eq:Rmue} \\ 
R_{\tau / e} & \ = \ \frac{\Gamma(W \rightarrow \tau \nu)}{\Gamma(W \rightarrow e \nu)} \ = \ 1.043 \pm 0.024 \, ,\label{eq:Rtaue} \\ 
R_{\tau / \mu} & \ = \ \frac{\Gamma(W \rightarrow \tau \nu)}{\Gamma(W \rightarrow \mu \nu)} \ = \ 1.070 \pm 0.026 \, . \label{eq:Rtaumu}
\end{align}
Note that while the measured value of $R_{\mu/e}$ agrees with the lepton universality prediction of the SM, $R_{\mu/e}^{\rm SM}=1$, within $1.1\sigma$ CL, the $W$ branching ratio to tau with respect to electron is about $1.8\sigma$ and to muon is about $2.7\sigma$ away from the SM prediction: $R_{\tau/\ell}^{\rm SM}=0.9993$ (with $\ell=e,\mu$), using the one-loop calculation of Ref.~\cite{Kniehl:2000rb}. 

The best LEP limits on lepton universality in $W$ decays come from the $W^+W^-$ pair-production channel, where one $W$ decays leptonically, and the other $W$ hadronically, i.e.,~$e^+e^-\to W^+W^-\to \ell\nu q\bar{q}'$~\cite{ALEPH:2004aa}. However, due to the leptophilic nature of the charged Higgs $h^\pm$ in our model, neither the $e^+ e^-\to h^+ h^-$ channel (Figs.~\ref{fig:feyn1} (a) and \ref{fig:feyn1} (b)) nor the Drell-Yan single-production channel (Fig.~\ref{fig:feyn1} (d)) will lead to $\ell \nu q \bar{q}$ final state.  So the only relevant contribution to the $W$ universality violation could come from the $h^{\pm} W^{\mp}$ production channel (Fig.~\ref{fig:feyn1} (c)), with the $W$ decaying hadronically and $h^\pm$ decaying leptonically. The  pure leptonic channels ($e \nu e \nu$ and $\mu \nu \mu \nu$) have  $\sim 40\%$ uncertainties in the  measurement and are therefore not considered here. 

Including the $h^\pm W^\mp$ contribution, the modified ratios $R_{\ell/\ell'}$ can be calculated as follows:
\begin{align}
R_{\ell/\ell'} \ = \ \frac{\sigma(W^+W^-)  {\rm BR}^W_{q\bar{q}'}{\rm BR}^W_{\ell\nu} +\sigma(h^\pm W^\mp) {\rm 
BR}^W_{q\bar{q}'} {\rm BR}_{\ell\nu}}{\sigma(W^+W^-)  {\rm BR}^W_{q\bar{q}'}{\rm BR}^W_{\ell'\nu} +\sigma(h^\pm W^\mp) {\rm 
BR}^W_{q\bar{q}'} {\rm BR}_{\ell'\nu}} \, ,
\label{eq:Rll}
\end{align}
where $\sigma(W^+W^-)$ and $\sigma(h^\pm W^\mp)$ are the production cross sections for $e^+e^-\to W^+W^-$ and $e^+e^-\to h^\pm W^\mp$ respectively, ${\rm BR}^W_{\ell \nu}$ denotes the branching ratio of $W\to \ell \nu$ (with $\ell=e,\mu,\tau$), whereas ${\rm BR}_{\ell\nu}$ denotes the branching ratio of $h^\pm\to \ell\nu$ as before (with $\ell=e,\tau$). At LEP experiment, the $W^+W^-$ pair production cross section $\sigma_{W^+W^-}$  is computed to be 17.17 pb at  $\sqrt{s} = 209$ GeV~\cite{ALEPH:2004aa}.  Within the SM, $W^\pm$ decays equally to each generation of leptons with branching ratio of $10.83 \%$ and decays hadronically with branching ratio of $67.41 \%$~\cite{Tanabashi:2018oca}. We numerically compute using {\tt MadGraph5}~\cite{Alwall:2014hca} the $h^\pm W^\mp$ cross section at $\sqrt s=209$ GeV  as a function of $m_{h^\pm}$ and ${\rm BR}_{\ell\nu}$, and  compare Eq.~\eqref{eq:Rll} with the measured values given in Eqs.~\eqref{eq:Rmue}-\eqref{eq:Rtaumu} to derive the $2\sigma$ exclusion limits in the $m_{h^+}$-${\rm BR}_{\ell\nu}$ plane. This is shown in Figs.~\ref{fig:col} (a) and~\ref{fig:col} (b) by yellow, brown, and pink-shaded regions for $\mu/e$, $\tau/e$, and $\tau/\mu$ universality tests, respectively.  Note that these constraints are absent in Figs.~\ref{fig:col} (c) and \ref{fig:col} (d), because when $Y_{ee}=0$, there is no $W^\pm h^{\mp}$ production at LEP (cf.~Fig.~\ref{fig:feyn1} (c) in the Zee model. But when $Y_{ee}$ is relatively large, these constraints turn out to be some of the most stringent ones in the $m_{h^+}$-${\rm BR}_{\ell\nu}$ plane shown in Figs.~\ref{fig:col} (a) and \ref{fig:col} (b), and rule out charged scalars below 110 GeV (129 GeV) for $Y_{ee}\sin\varphi=0.1$ (0.2). These constraints are not  applicable for $m_{h^\pm}>129$ GeV, because $h^\pm W^\mp$ can no longer be produced on-shell at LEP II with maximum $\sqrt s=209$ GeV.

As mentioned before, the measured $W$ branching ratio to tau with respect to muon is $2.7\sigma$ above the SM prediction. 
Since in our case, $h^\pm$ decays to either $e\nu$ or $\tau\nu$, but not $\mu\nu$, this contributes to $R_{\tau\mu}$ only in the numerator, but not in the denominator. Therefore, the $2.7\sigma$ discrepancy can be explained in this model, as shown by the allowed region between the upper and lower pink-dashed curves in Fig.~\ref{fig:col} (a) with $Y_{ee}\sin\varphi=0.1$.\footnote{Light charged scalar has been used to address the lepton universality issue in $W$ decays in Ref.~\cite{Dermisek:2008dq}.} The upper  pink-shaded region with larger ${\rm BR}_{\tau\nu}$ gives $R_{\tau\mu}>1.122$, which is above the allowed $2\sigma$ range given in Eq.~\eqref{eq:Rtaumu}. On the other hand, the lower pink-shaded region with smaller ${\rm BR}_{\tau\nu}$ gives $R_{\tau\mu}<1.018$, which is below the allowed $2\sigma$ range given in Eq.~\eqref{eq:Rtaumu}.  For larger Yukawa coupling $Y_{ee}$, as illustrated in Fig.~\ref{fig:col} (b) with $Y_{ee}\sin\varphi=0.2$, the whole allowed range of parameter space from $R_{\tau/\mu}$ shifts to lower values of ${\rm BR}_{\tau\nu}$. This is because the $h^\pm W^\mp$ production cross section $\sigma(h^\pm W^\mp)$ in Eq.~\eqref{eq:Rll} is directly proportional to $|Y_{ee}|^2$, and therefore, for a large $Y_{ee}$, a smaller ${\rm BR}_{\tau\nu}$ would still be compatible with the $R_{\tau/\mu}$-preferred range.
%%%%%%%%%%%%%%%%%%%%%%%%%%%%%

%%%%%%%%%%%%%%%%%%%%%%%%%%%%%%%%%%%%%%%%%%%%
\subsection{Constraints from tau decay lifetime and universality} \label{sec:taudecay} 
%%%%%%%%%%%%%%%%%%%%%%%%%%%%%%%%%%%%%%%%%%%%
\begin{figure}[!t]
    \centering
    \includegraphics[scale=0.5]{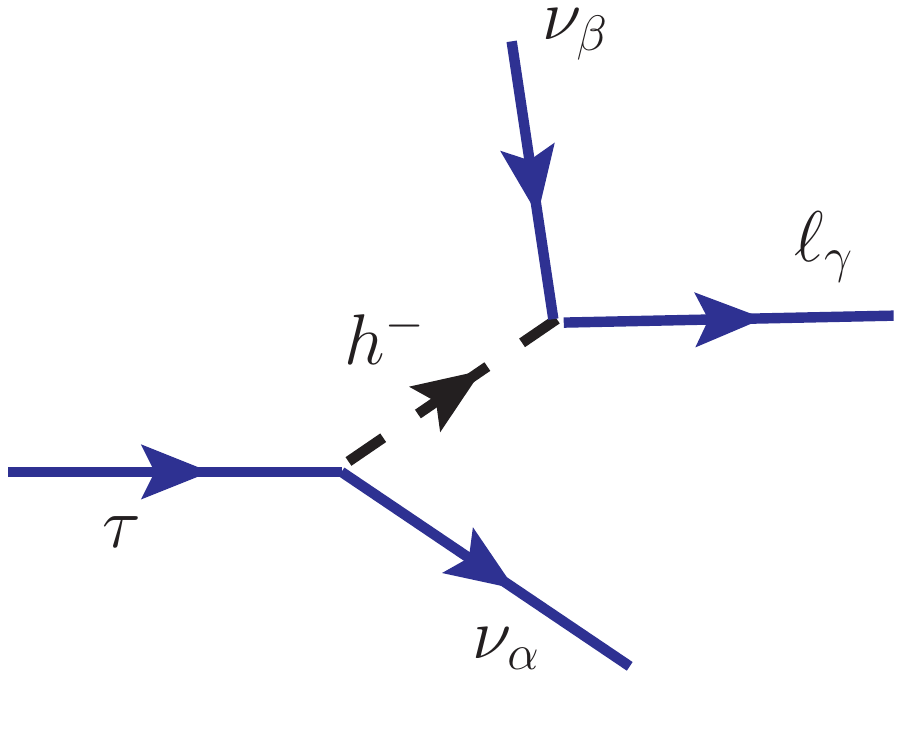}
    \caption{Feynman diagram for the new decay mode of the $\tau$ lepton mediated by light charged scalar in the Zee model.}
    \label{taudecay}
\end{figure}
%%%%%%%%%%%%%%%%%%%%%%%%%%%%%%%%%%%%%%%%%%%%%

In order to realize a light charged scalar $h^-$ consistent with LEP searches, we have assumed that the decay $h^- \rightarrow \tau \bar{\nu}_\beta$ proceeds with a significant branching ratio.  $h^-$ also has coupling with $e \bar{\nu}_\alpha$, so that non-negligible NSI is generated.  When these two channels are combined, we would get new decay modes for the $\tau$ lepton, as shown in Fig.~\ref{taudecay}.  This will lead to deviation in $\tau$-lifetime compared to the SM expectation.  The new decay modes will also lead to universality violation in $\tau$ decays, as the new modes preferentially lead to electron final states.  Here we analyze these constraints and evaluate the limitations these pose for NSI.

The effective four-fermion Lagrangian relevant for the new $\tau$ decay mode is given by
\begin{eqnarray}
            \mathcal{L}_{\rm eff} \ = \ (\bar{\nu}_{L \alpha} e_R) ( \bar{\tau}_R \nu_{L \beta}) Y_{\alpha e} Y_{\beta \tau}^{\star} \frac{\sin^2 \varphi}{m_{h^+}^2} \label{tauL}~.
            \end{eqnarray}
This can be recast, after a Fierz transformation, as
\begin{eqnarray}
           \mathcal{L}_{\rm eff} \ = \ -\frac{1}{2} (\bar{\nu}_{L \alpha} \gamma_\mu \nu_{L \beta}) (\bar{\tau}_R \gamma^\mu e_R) Y_{\alpha e} Y_{\beta \tau}^{\star} \frac{\sin^2 \varphi}{m_{h^+}^2} ~.
\end{eqnarray}
This can be directly compared with the SM $\tau$ decay Lagrangian, given by
\begin{eqnarray}
          {\cal L}^{\rm SM} \ = \ 2 \sqrt{2} G_F (\nu_{\tau L} \gamma_\mu \nu_{\tau L}) (\bar{\tau}_L \gamma^\mu e_L) ~.
          \label{SMfierz}
\end{eqnarray}
It is clear from here that the new decay mode will not interfere with the SM model (in the limit of ignoring the lepton mass), since the final state leptons have opposite helicity in the two decay channels. The width of the $\tau$ lepton is now increased from its SM value by a factor $1+\Delta$, with $\Delta$ given by \cite{Kuno:1999jp} 
\begin{equation}
\Delta \ = \ \frac{1}{4} |g_{RR}^s|^2~,
\end{equation}
where 
\begin{equation}
g_{RR}^s \ = \ - \frac{Y_{\alpha e} Y^\star_{\beta \tau} \sin^2\varphi}{2 \sqrt{2}G_F m_{h^+}^2}~.
\end{equation}
The global-fit result on $\tau$ lifetime is $\tau_{\tau} = (290.75 \pm 0.36)\times 10^{-15}$ s, while the SM prediction is $\tau_\tau^{\rm SM} = (290.39 \pm 2.17)\times 10^{-15}$ s~\cite{Tanabashi:2018oca}.  Allowing for 2$\sigma$ error, we find $\Delta \leq 1.5\%.$ If the only decay modes of $h^-$ are $h^- \to \bar{\nu}_\alpha e^-$ and $h^- \to \bar{\nu}_\beta \tau^-$, then we can express $|Y_{\beta \tau}|^2$ in terms of $|Y_{\alpha e}|^2$ as
\begin{equation}
    |Y_{\beta \tau}|^2 \ = \ |Y_{\alpha e}|^2 \frac{{\rm BR}(h^- \to \tau \nu)}{{\rm BR}(h^- \to e \nu)}~.
\end{equation}
Using this relation, we obtain
\begin{equation}
\Delta \ = \ |\varepsilon_{\alpha \alpha}|^2 \frac{{\rm BR (h^- \to \tau \nu})}{{\rm BR}(h^- \to e \nu)}~,
\end{equation}
where $\varepsilon_{\alpha\alpha}$ is the diagonal NSI parameter for which the expression is derived later in Eq.~\eqref{eq:nsi1}. Therefore, a constraint on $\Delta$ from the tau lifetime can be directly translated into a constraint on $\varepsilon_{\alpha\alpha}$:
\begin{equation}
|\varepsilon_{\alpha \alpha}| \ \leq \ 12.2\% \sqrt{\frac{{\rm BR}(h^- \to e \nu)}{{\rm BR} (h^- \to \tau \nu)}}~.
\end{equation}

An even stronger limit is obtained from $e-\mu$ universality in $\tau$ decays.  The experimental central value prefers a slightly larger width for $\tau \to \mu \nu \nu$ compared to $\tau \to e \nu \nu$. In our scenario, $h^-$ mediation enhances $\tau \to e \nu \nu$ relative to $\tau \to \mu \nu \nu$.  We have in this scenario
\begin{equation}
\frac{\Gamma(\tau \to \mu \nu \nu)}{  \Gamma(\tau \to e \nu \nu)} \ = \ 1- \Delta~,
\end{equation}
which constrains $\Delta \leq 0.002$, obtained by using the measured ratio $\frac{\Gamma(\tau \to \mu \nu \nu)}{\Gamma(\tau \to e \nu \nu)} = 0.9762 \pm 0.0028$ \cite{Tanabashi:2018oca}, and allowing 2$\sigma$ error.  This leads to a limit \begin{equation}
|\varepsilon_{\alpha \alpha}| \ \leq \ 4.5\% \sqrt{\frac{{\rm BR}(h^- \to e \nu)}{{\rm BR} (h^- \to \tau \nu)}}~.
\end{equation}
In deriving the limits on a light charged Higgs mass from LHC constraints, we have imposed the $\tau$ decay constraint as well as the universality constraint on $\Delta$, see Fig.~\ref{fig:col}. Avoiding the universality constraint by opening up the $\tau \to \mu \nu \nu$ channel will not work, since that will be in conflict with $\mu \to e \nu \nu$ constraints, which are more stringent.  

The Michel  parameters in $\tau$ decay will now be modified \cite{Babu:2016fdt}.  While the $\rho$ and $\delta$ parameters are unchanged compared to their SM value of 3/4, $\xi$ is modified from its SM value of 1 to 
\begin{equation}
\xi \ = \ 1- \frac{1}{2}|g_{RR}^s|^2~.
\end{equation}
However, the experimental value is $\xi = 0.985 \pm 0.030$ \cite{Tanabashi:2018oca}, which allows for significant room for the new decay. Again, our choice of Yukawa couplings does not modify the $\mu \to e \nu \nu$ decay, and is therefore, safe from the Michel parameter constraints in the muon sector, which are much more stringent.

%%%%%%%%%%%%%%%%%%%%%%%%%%%%%%%%%%
\subsection{Constraints from Higgs precision data}
\label{sec:HiggsOb}
%%%%%%%%%%%%%%%%%%%%%%%%%%%%%%%%%%
%%%%%%%%%%%%%%%%%%%%%%%%%%%%%%%%%%%%%%%%%%%%%%%%%%%%%%%%%%%%%%%%%%
\begin{figure}[!t]
\centering
\subfigure[]{
  \includegraphics[width=0.4\textwidth]{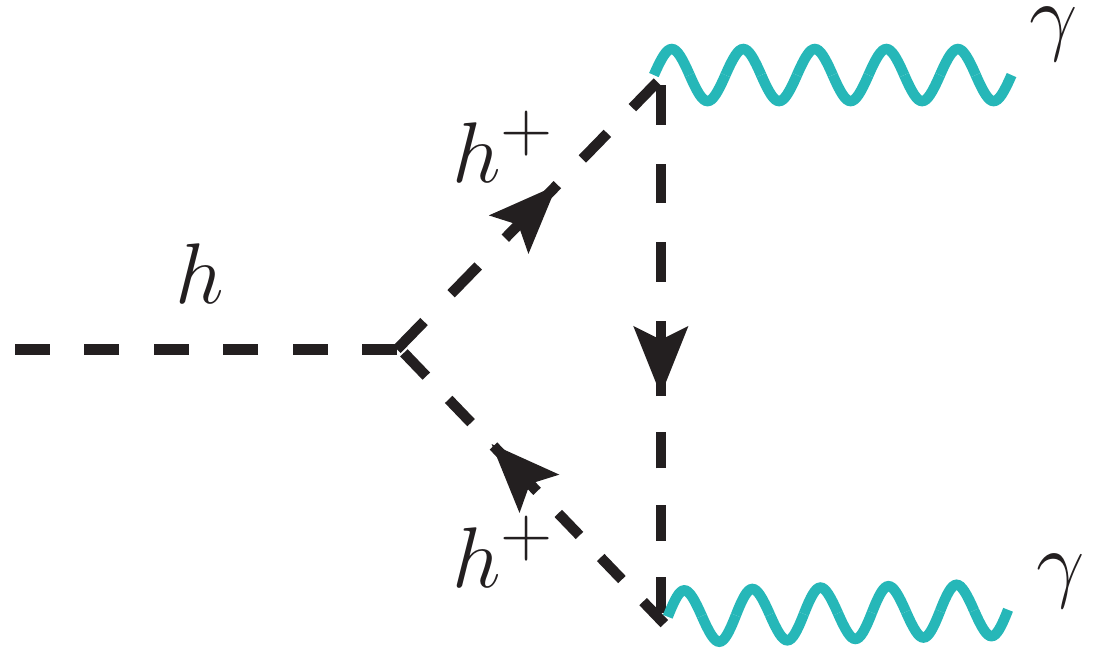}} 
  \hspace{10 mm}
  \subfigure[]{
  \includegraphics[width=0.4\textwidth]{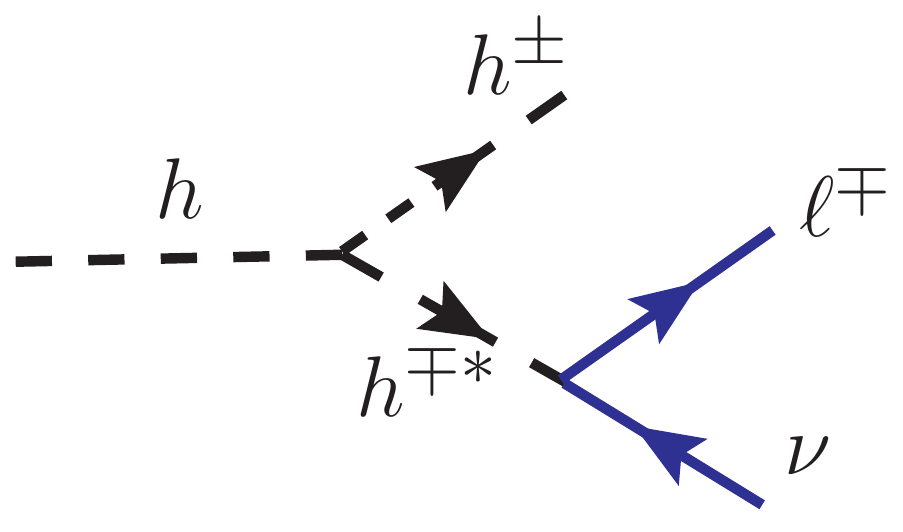}}
 \caption{(a) New contribution to $h \rightarrow \gamma
 \gamma$ decay mediated by charged scalar loop. (b) New contribution to $h\to 2\ell 2\nu$ via the exotic decay mode $h \to h^{\pm} h^{\mp \star}$.}
 \label{fig:hdiphoton}
\end{figure}
%%%%%%%%%%%%%%%%%%%%%%%%%%%%%%%%%%%%%%%%%%%%%%%%
In this subsection, we analyze the constraints on light charged scalar from LHC Higgs precision data. Both ATLAS and CMS collaborations have performed several measurements of the 125 GeV Higgs boson production cross sections and branching fractions at the LHC, both in Run I~\cite{Khachatryan:2016vau} and Run II~\cite{Sirunyan:2018koj, ATLAS:2019slw}. Since all the measurements are in good agreement with the SM expectations, any exotic contributions to either production or decay of the SM-like Higgs boson will be strongly constrained. In the Zee model, since the light charged scalar is leptophilic, it will not affect the production rate of the SM-like Higgs $h$ (which is dominated by gluon fusion via top-quark loop). However, it gives new contributions to the loop-induced $h\to \gamma \gamma$ decay (see Fig.~\ref{fig:hdiphoton} (a)) and mimics the tree-level $h\to WW^\star \to 2\ell 2\nu$ channel via the exotic decay mode $h\to h^\pm h^{\mp \star}\to h^\pm \ell \nu\to 2\ell 2\nu$ (see  Fig.~\ref{fig:hdiphoton} (b)). Both these contributions are governed by the effective $hh^+h^-$ coupling given by 
\begin{equation}
    \lambda_{h h^{+}h^{-}} \ = \  -\sqrt{2}\mu \sin{\varphi} \cos{\varphi}+\lambda_3 v \sin^2{\varphi} +\lambda_8 v \cos^2{\varphi} \, .
    \label{eq:effmu}
\end{equation}
Therefore, the Higgs precision data from the LHC can be used to set independent constraints on these Higgs potential parameters, as we show below.   

%Hence, these results can impose strong bounds on the free parameters of the Zee model affecting Higgs observables. It will also encourage us to venture any deviations in the Higgs observable making consistent within the uncertainty of these measurements, as well as if we can make predictions which can be tested at the LHC. 
The Higgs boson yield at the LHC is characterized by the signal strength,  defined as the ratio of the measured Higgs boson rate to its SM prediction. For a specific  production channel $i$ and decay into specific final states $f$, the signal strength of the Higgs boson $h$ can be expressed as
%%%%%%%%%%%%%%%%%%%%%%%%%%
\begin{equation}
\mu^i_f \ \equiv \ \frac{\sigma^i}{(\sigma^i)_{\rm SM}} \frac{ {\rm BR}_f}{{({\rm BR}_{f})_{\rm SM}}} \ \equiv \  \mu^i\cdot\mu_f \, ,
\label{eq:muif}
\end{equation}
%%%%%%%%%%%%%%%%%%
where $\mu^i$ (with $i=$ ggF, VBF, $Vh$, and $t\bar{t}h$) and $\mu_f$ (with $f = ZZ^{\star}, WW^{\star}, \gamma \gamma, \tau^+ \tau^-, b\bar{b}$) are the production and branching rates relative to the SM predictions in the relevant channels. As mentioned above, the production rate does not get modified in our case, so we will set $\mu^i=1$ in the following. As for the decay rates, the addition of the two new channels shown in Fig.~\ref{fig:hdiphoton} will increase the total Higgs decay width, and therefore, modify the partial widths in all the channels.

To derive the Higgs signal strength constraints on the model parameter space, we have followed the procedure outlined in Ref.~\cite{Babu:2018uik, Jana:2017hqg}, using the updated constraints on signal strengths reported by ATLAS and CMS collaboration for all individual production and decay modes at 95$\%$ CL,  based on the $\sqrt{s}=$ 13 TeV LHC data.  The individual analysis by each experiment examines a specific Higgs boson decay mode corresponding to various production processes. We use the measured signal strengths in the following dominant decay modes for our numerical analysis: $h \to \gamma\gamma$ \cite{CMS:1900lgv, CMS:2018rbc, ATLAS:2018uso, ATLAS:2019aqa}, $h \to ZZ^{\star}$ \cite{CMS:2019chr, ATLAS:2018bsg}, $ h \to WW^\star$ \cite{Aaboud:2018jqu, Aad:2019lpq,  Sirunyan:2018egh}, $h \to \tau \tau$ \cite{CMS:2019pyn, Aaboud:2018pen} and $h \to b \Bar{b}$ \cite{Sirunyan:2018kst, Aaboud:2018zhk, Aaboud:2018gay}.  

We formulate  the modified $h\to \gamma\gamma$ decay rate as 
\begin{align}
\Gamma({h\to \gamma\gamma}) \ = \ \kappa_\gamma^2  \Gamma({h\to \gamma\gamma})^{\rm SM} \, ,   
\end{align}
where the scaling factor $\kappa_\gamma$ is given by  
%In our study, the partial decay rate of $h\to \gamma \gamma$ is getting modified since the singly-charged scalar can mediate SM-like Higgs decay to a pair of photons in addition to $t$ and $W$ loops. A representative triangle loop diagram for the process is shown in Fig.~\ref{fig:hdiphoton}.  In fact, these processes can both augment or suppress the SM predicted $h\to \gamma \gamma$ rate at the LHC depending on the signs and relative strength of $\lambda_{h h^{\pm}h^{\mp}}$. 
%We formulate the scaling factor in the $h \gamma \gamma$  coupling as ~\cite{Djouadi:2005gi,Knapen:2015dap} 
\beq
\kappa_{\gamma} \ = \  \dfrac{\sum_f N^f_c Q^2_f A_{1/2}(\tau_f) +  A_{1}(\tau_W) + \dfrac{\lambda_{hh^+h^-}v}{2 m_{h^+}^2} A_0(\tau_{h^+})}{ \sum_f N_c^f Q^2_f A_{1/2}(\tau_f) +  A_{1}(\tau_W)}\, ,
\label{eq:kappag}
\eeq
where $N^f_c=3~(1)$ is the color factor for quark (lepton), $\sum_f$ is the sum over the SM fermions $f$ with charge $Q_f$, and the loop functions are given by~\cite{Djouadi:2005gi}
\bea
A_0(\tau) &\  = \ & - \tau + \tau^2 f(\tau),
\\ 
A_{1/2}(\tau) & \ = \ & 2 \tau [1+ (1-\tau) f(\tau)],
\\ 
A_1(\tau) & \ = \ & -2 - 3 \tau [1 + (2 - \tau) f(\tau) ], 
\label{loop_function_1}
\\
{\rm with}~~f(\tau) & \ = \ &   
	\begin{cases}
		{\rm arcsin}^2\left(\frac{1}{\sqrt{\tau}}\right),  & \mbox{if } \tau \geq 1 \\
		-\dfrac{1}{4}\left[\log \dfrac{1+\sqrt{1-\tau}}{1-\sqrt{1-\tau}} - i \pi\right]^2, & \mbox{if } \tau < 1 \,.
	\end{cases}
\label{loop_function_2}
\eea
The parameters $\tau_i=4 m^2_i/m^2_h$ are defined by the corresponding masses of the heavy particles in the loop. For the fermion loop, only the top quark contribution is significant, with the next leading contribution coming from the bottom quark which is an 8\% effect. Note that the new contribution in Eq.~\eqref{eq:kappag} due to the charged scalar  can interfere with the SM part either constructively or destructively, depending on the sign of the effective coupling $\lambda_{hh^+h^-}$ in Eq.~\eqref{eq:effmu}.

%%%%%%%%%%%%%%%%%%%%%%%%%%%%%%%%%%%%%%%%%%%%%
\begin{figure}[!t]
    $$
    \includegraphics[height=7cm,width=0.5\textwidth]{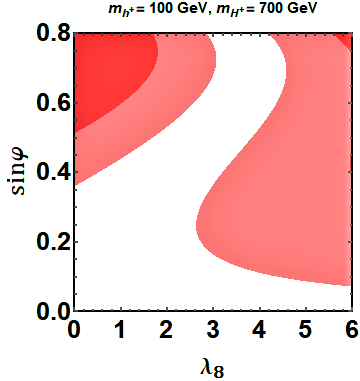}
     \includegraphics[height=7cm,width=0.5\textwidth]{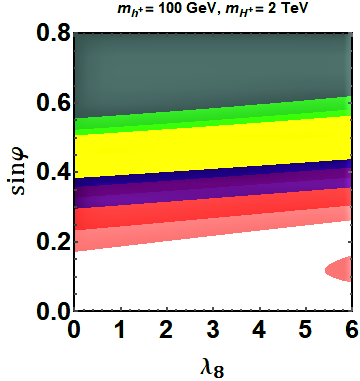}
    $$
       $$
    \includegraphics[height=7cm,width=0.5\textwidth]{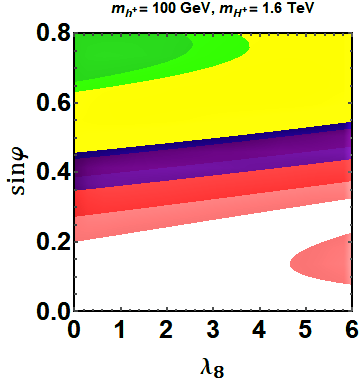}
     \includegraphics[height=7cm,width=0.5\textwidth]{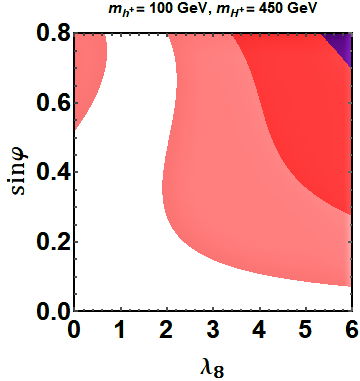}
    $$
    $$
     \includegraphics[height=1cm,width=0.8\textwidth]{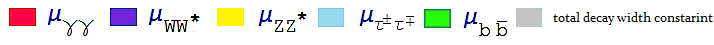}
    $$
    \caption{Constraints from the Higgs boson properties in $\lambda_8 - \sin{\varphi}$ plane in the Zee model (with $\lambda_3 = \lambda_8$). The red, cyan, green, yellow, and purple-shaded regions are excluded by the signal strength limits for various decay modes ($\gamma\gamma, \tau \tau, b\bar{b}, ZZ^\star, WW^\star$) respectively. The white unshaded region simultaneously satisfies all the experimental constraints. Grey-shaded region (only visible in the upper right panel) is excluded by total decay width constraint.}
    \label{fig:higgs}
\end{figure}

%%%%%%%%%%%%%%%%%%%%%%%%%%%%%%%%%%%%%%%%%%%%%%%

As for the new three-body decay mode $h \rightarrow h^{\pm} h^{\mp^{\star}}\to h^\pm \ell \nu $, the partial decay rate is given by 
\begin{align}
    \Gamma(h\to h^+ \ell^- \bar{\nu}) \  = \ \frac{|\lambda_{hh^+h^-}|^2}{64\pi^3m_h}{\rm Tr}(Y^\dag Y)\int_{\sqrt{r}}^{\frac{1}{2}(1+r)}dx \frac{(1-2x+r)\sqrt{x^2-r}}{(1-2x)^2+\frac{r^2\Gamma_{h^+}^2}{m_h^2}} \, ,
\end{align}
where $Y$ is the Yukawa coupling defined in Eq.~\eqref{eq:Yuk1}, $\Gamma_{h^+}={\rm Tr}(Y^\dag Y)m_{h^+}/8\pi$ is the total decay width of $h^+$, and $r=m_{h^+}^2/m_h^2$. With this new decay mode, the signal strength in the $h\to 2\ell 2\nu$ channel will be modified to include $\Gamma(h\to h^\pm \ell \nu\to 2\ell 2\nu)$ along with the SM contribution from $\Gamma(h\to WW^\star\to 2\ell 2\nu)$, and to some extent, from $\Gamma(h\to ZZ^\star\to 2\ell 2\nu)$.

The partial decay widths of $h$ in other channels will be the same as in the SM, but their partial widths will now be smaller, due to the enhancement of the total decay width. 
%In addition to the modification in the partial decay width for the process $h \to \gamma \gamma$, there will be significant modification in the total decay width for the SM Higgs boson since a new exotic decay mode opens up at tree level  $h \rightarrow h^{\pm} h^{\mp^{\star}} $ and which is shown in Fig. \ref{fig:hdiphoton}. This decay mode will exactly mimics the $h \to WW^{\star}$ process due to the same final states. Due to the enhancement of the total decay width, the branching ratio for the other decay modes will be modified significantly as the partial decay width remains almost unaltered other than the $h \to \gamma \gamma$ process.  
A comparison with the measured signal strengths therefore imposes an upper bound on the effective coupling $ \lambda_{h h^{\pm}h^{\mp}}$ which is a function of the cubic coupling $\mu$, quartic couplings $\lambda_3$ and $\lambda_8$,  and the mixing angle $\sin {\varphi}$ (cf.~Eq.~\eqref{eq:muif}). For suppressed effective coupling $ \lambda_{h h^{\pm}h^{\mp}}$ to be consistent with the Higgs observables, we need some cancellation between the cubic and quartic terms. In order to have large NSI effect, we need sufficiently large mixing $\sin {\varphi}$, which implies large value of $\mu$ (cf.~Eq.~\eqref{mixphi}). In order to find the maximum allowed value of $\sin\varphi$, we take $\lambda_3=\lambda_8$ in Eq.~\eqref{eq:muif} and show in Fig.~\ref{fig:higgs} the Higgs signal strength constraints in the $\lambda_8-\sin\varphi$ plane. The red, blue, yellow, cyan, and green-shaded regions are excluded by the signal strength limits $\gamma\gamma, WW^\star, ZZ^\star, \tau \tau$, and $b\bar{b}$ decay modes, respectively. We have fixed the light charged Higgs mass at 100 GeV, and the different panels are for different benchmark values of the heavy charged Higgs mass: $m_{H^+}=700$ GeV (upper left), 2 TeV (upper right), 1.6 TeV (lower left) and 450 GeV (lower right). The first choice is the benchmark value we will later use for NSI studies, while the other three values correspond to the minimum allowed values for the heavy neutral Higgs mass (assuming it to be degenerate with the heavy charged Higgs to easily satisfy the $T$-parameter constraint (cf.~Sec.~\ref{sec:ewpt})) consistent with the LEP contact interaction bounds for ${\cal O}(1)$ Yukawa couplings (cf.~Sec.~\ref{sec:contact}).  From  Fig.~\ref{fig:higgs}, we see that the $h\to \gamma\gamma$ signal strength gives the most stringent constraint. If we allow $\lambda_8$ to be as large as 3, then we can get maximum value of $\sin\varphi$ up to 0.67 (0.2) for $m_{H^+}=0.7$ (2) TeV. 

In addition to the modified signal strengths, the total Higgs width is enhanced due to the new decay modes. Both ATLAS~\cite{Aaboud:2018puo} and CMS~\cite{Sirunyan:2019twz} collaborations have put 95\% CL upper limits on the Higgs boson total width $\Gamma_h$ from measurement of off-shell production in the $ZZ\to 4\ell$ channel. Given the SM expectation $\Gamma_h^{\rm SM}\sim 4.1$ MeV, we use the CMS upper limit on $\Gamma_h<9.16$ MeV~\cite{Sirunyan:2019twz} to demand that the new contribution (mostly from $h\to h^\pm h^{\mp \star}$, because the $h\to \gamma\gamma$ branching fraction is much smaller) must be less than $5.1$ MeV. This is shown in Fig.~\ref{fig:higgs} by the grey-shaded region (only visible in the upper right panel), which turns out to be much weaker than the signal strength constraints in the individual channels.   
%%%%%%%%%%%%%%%%%%%%%%%%%%%%%%%%%%%%%%%%%%%%%%%%%%%%%%%
\subsection{Monophoton constraint from LEP}\label{sec:monop}
%%%%%%%%%%%%%%%%%%%%%%%%%%%%%%%%%%%%%%%%%%%%%%%%%%%%%%
\begin{figure}[!t]
    \centering
     \includegraphics[scale=0.5]{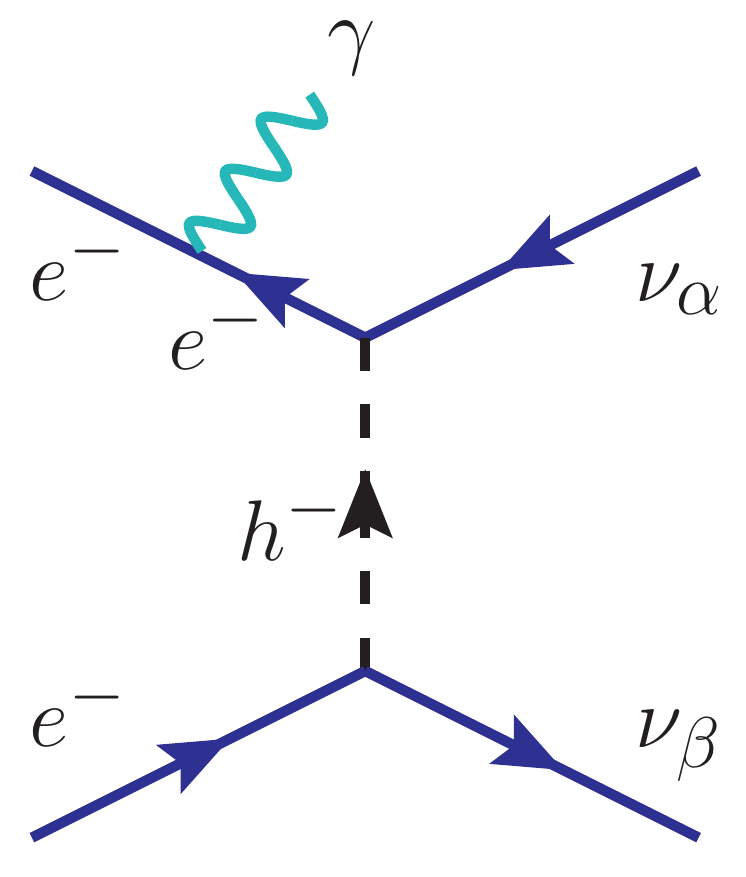}\hspace{10mm}
    \includegraphics[scale=0.5]{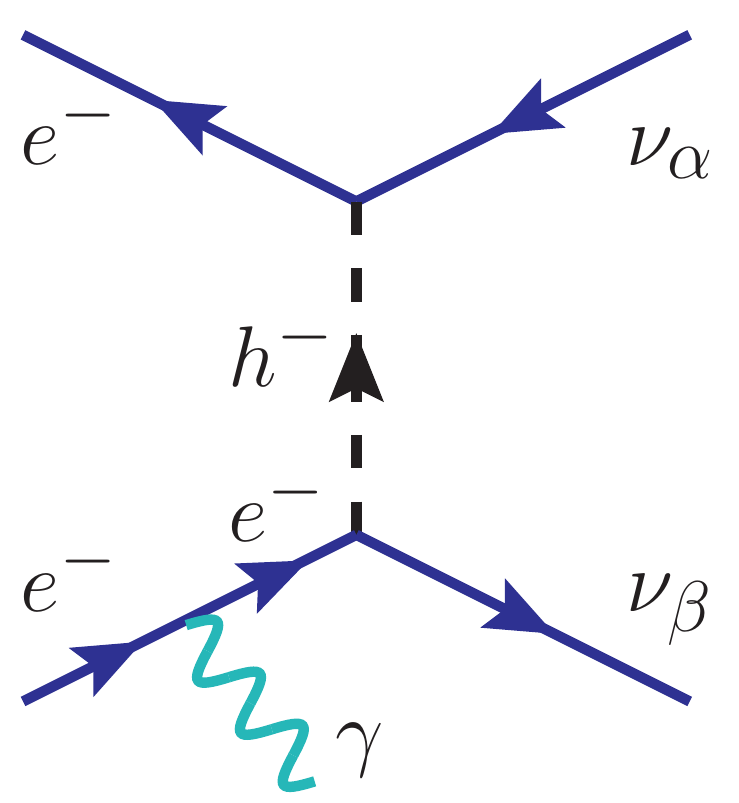} \hspace{10mm}
     \includegraphics[scale=0.5]{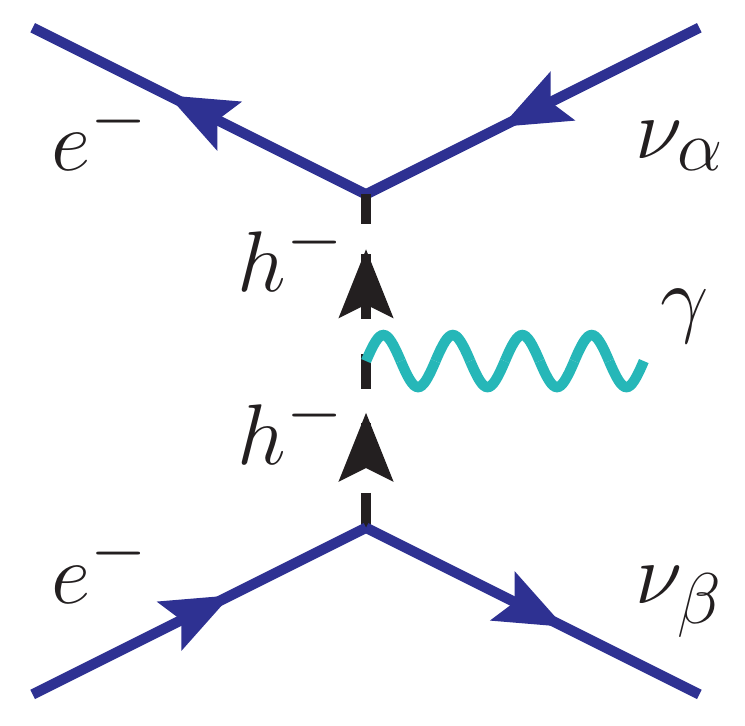} 
    \caption{Feynman diagrams for charged scalar contributions to monophoton signal at LEP.}
    \label{fig:monophoton}
\end{figure}
%%%%%%%%%%%%%%%%%%%%%%%%%%%%%%%%%%%%%%%%%%%%%%%%%%%%%%
Large neutrino NSI with electrons inevitably leads to a new contribution to the monophoton process $e^+ e^- \to \nu\bar{\nu}\gamma$ that can be constrained using  LEP data~\cite{Berezhiani:2001rs}. In the SM, this process occurs via $s$-channel $Z$-boson exchange and $t$- channel $W$-boson exchange, with the photon being emitted from either the initial state electron or positron or the intermediate state $W$ boson. In the Zee model, we get additional contributions from $t$-channel charged scalar exchange (see Fig.~\ref{fig:monophoton}). Both light and heavy charged scalars will contribute, but given the mass bound on the heavy states from LEP contact interaction, the dominant contribution will come from the light charged scalar. 

The total cross section for the process $e^+e^-\to \nu_\alpha \bar{\nu}_\beta \gamma$ can be expressed 
as 
$\sigma = \sigma^{\rm SM}+\sigma^{\rm NS}$, where $\sigma^{\rm SM}$ is the
SM cross section (for $\alpha=\beta$) and $\sigma^{\rm NS}$ represents the sum of the pure non-standard  contribution due to the charged scalar and its interference with the SM contribution. Note that since the charged scalar only couples to right-handed fermions, there is no interference  with the $W$-mediated process (for $\alpha=\beta=e$). Moreover, for either $\alpha$ or $\beta$ not equal to $e$, the $W$ contribution is absent. For $\alpha\neq \beta$, the $Z$ contribution is also absent. 

The monophoton process has been investigated carefully by all four LEP experiments~\cite{Tanabashi:2018oca}, but the most stringent limits on the cross section come from the L3 experiment, both on~\cite{Acciarri:1998vf} and off~\cite{Achard:2003tx} $Z$-pole. We use these results to derive constraints on the charged scalar mass and Yukawa coupling. The constraint $|\sigma - \sigma^{\rm exp}| \leq \delta\sigma^{\rm exp}$, where $\sigma^{\rm exp} \pm \delta\sigma^{\rm exp}$ is the experimental result, can be expressed in the following form:  
\be
\left| 1 + \frac{\sigma^{\rm NS}}{\sigma^{\rm SM}} 
- \frac{\sigma^{\rm exp}}{\sigma^{\rm SM}} \right|
\ \leq \  \left( \frac{\sigma^{\rm exp}}{\sigma^{\rm SM}} \right)
\left( \frac{\delta\sigma^{\rm exp}}{\sigma^{\rm exp}} \right) \, .
\ee
We evaluate the ratio $\sigma^{\rm exp}/\sigma^{\rm SM}$ by combining the L3 results~\cite{Acciarri:1998vf,Achard:2003tx} with an accurate computation of the SM cross section, both at $Z$-pole and off $Z$-pole. Similarly, we compute the 
ratio $\sigma^{\rm NS}/ \sigma^{\rm SM}$ 
numerically as a function of the charged scalar mass $m_{h^+}$ and the Yukawa coupling $Y_{\alpha\beta}\sin\varphi$. For comparison of cross sections at $Z$-pole, we adopt the same event acceptance criteria as in Ref.~\cite{Acciarri:1998vf}, i.e., we allow photon energy within the range 1 GeV $< E_{\gamma}<$ 10 GeV and the angular acceptance 45$\degree< \theta_{\gamma}<135 \degree$. Similarly, for the off $Z$-pole analysis, we adopt the same event topology as described in Ref.~\cite{Achard:2003tx}: i.e.,  14$\degree< \theta_{\gamma}<166 \degree$, 1 GeV $< E_{\gamma}$, and $p_T^{\gamma}>0.02\sqrt{s}$. We find that the off $Z$-pole measurement imposes more stringent bound than the $Z$-pole measurement bound. As we will see in the next section (see Fig.~\ref{fig:dnsi}), the monophoton constraints are important especially for the NSI involving tau-neutrinos. We also note that our monophoton constraints are somewhat weaker than those derived in Ref.~\cite{Fox:2011fx} using an effective four-fermion approximation.

%%%%%%%%%%%%%%%%%%%%%%%%%%%%%%%%%%%%%%%%%%%%%%%
\subsection{NSI predictions} \label{sec:NSIZee}

\begin{figure}[t!]
    \centering
    \subfigure[]{
        \includegraphics[scale=0.55]{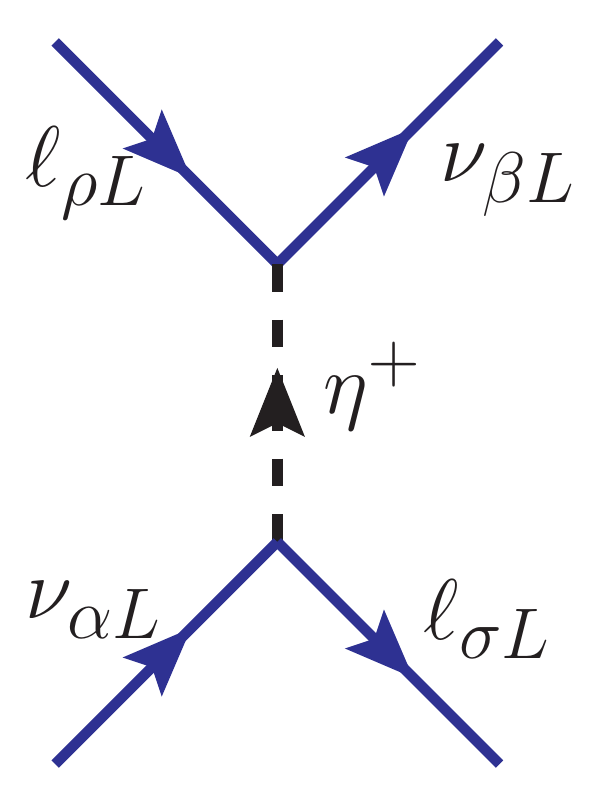}}
        \hspace{2mm}
    \subfigure[]{
         \includegraphics[scale=0.5]{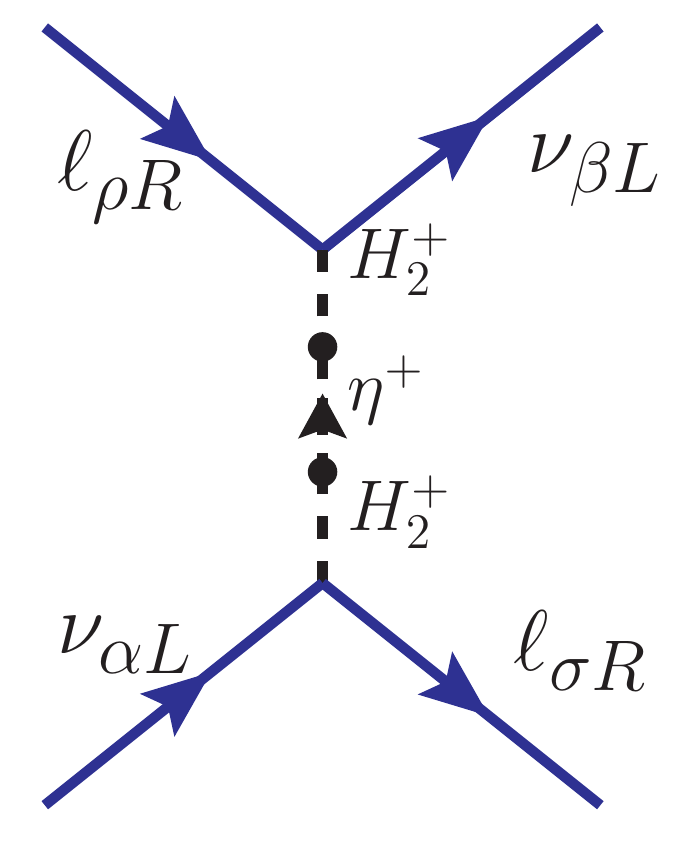}}
           \hspace{2mm}
   \subfigure[]{
         \includegraphics[scale=0.5]{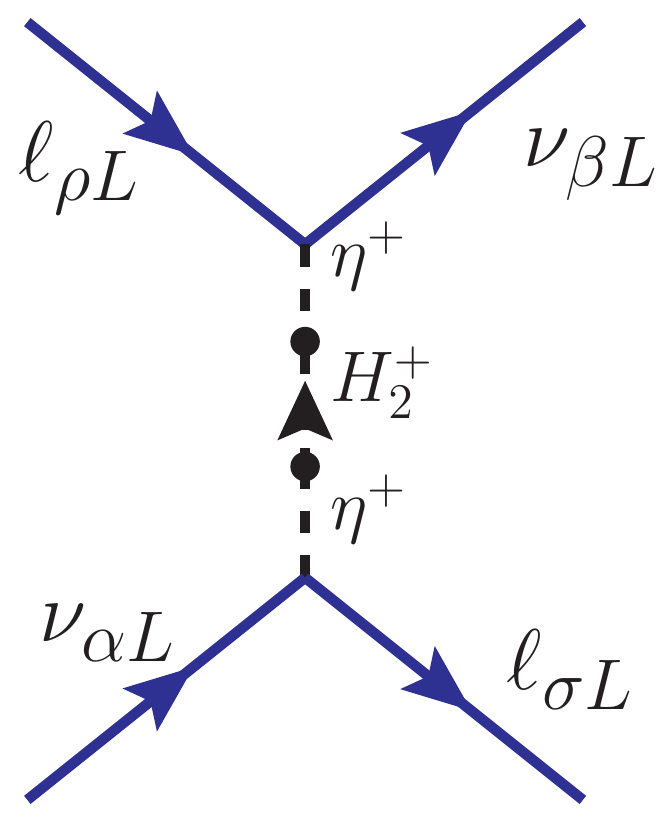}}
           \hspace{2mm}
        \subfigure[]{
         \includegraphics[scale=0.52]{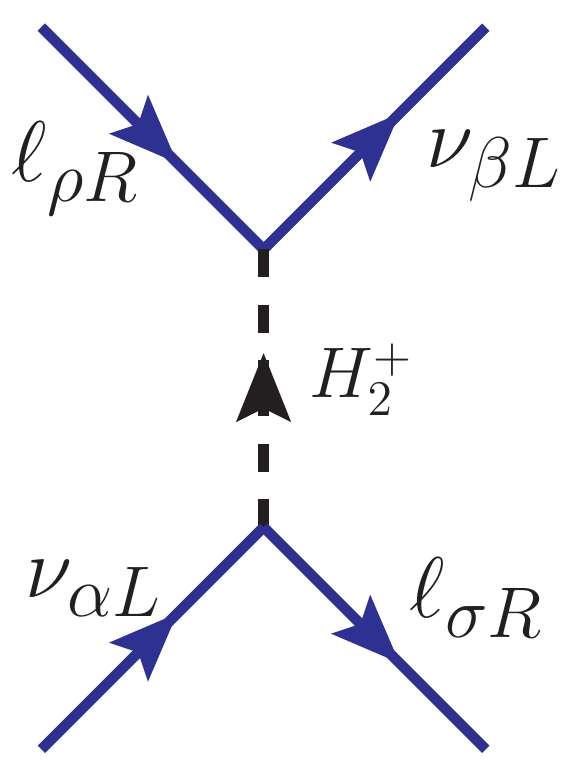}}
    \caption{Tree-level NSI induced by the exchange of charged scalars in the Zee model. Diagrams (a) and (d) are due to the pure singlet and doublet charged scalar components, while (b) and (c) are due to the mixing between them. }
    \label{nsi1}
 \end{figure}

The new singly-charged scalars $\eta^+$ and $H_2^+$ in the Zee Model induce NSI at tree level as shown in  Fig.~\ref{nsi1}. Diagrams (a) and (d) are induced by the pure singlet and doublet components of the charged scalar fields and depend on the Yukawa couplings $f$ and $Y$ respectively (cf.~Eqs.~\eqref{Lfab} and \eqref{eq:Yuk1}). On the other hand, diagrams (b) and (c) are induced by the mixing between the singlet and doublet fields, and depend on the combination of Yukawa couplings and the mixing angle $\varphi$ (cf.~Eq.~\eqref{mixphi}). As mentioned in Sec.~\ref{sec:neutrino}, satisfying the neutrino mass requires the product $f\cdot Y$ to be small. For $Y\sim {\cal O}(1)$, we must have $f\sim 10^{-8}$ to get $m_\nu\sim 0.1$ eV (cf.~Eq.~\eqref{nuMass}). In this case, the NSI from Figs.~\ref{nsi1} (a) and (c) are heavily suppressed. So we will only consider diagrams (b) and (d) for the following discussion and work in the mass basis for the charged scalars, where $\eta^+$ and $H_2^+$ are replaced by $h^+$ and $H^+$ respectively (cf.~Eq.~\eqref{eq:charged}).   

 %%%%%%%%%
 
The effective NSI Lagrangian for the contribution from Fig.~\ref{nsi1} (b) is given by   
\begin{eqnarray}
 \label{4foperator}
    \mathcal{L}_{\rm eff} &\ = \ & \sin^2{\varphi} \frac{Y_{\alpha \rho } Y_{ \beta \sigma}^\star }{m_{h^+}^2} (\Bar{\nu}_{\alpha L} \, \ell_{\rho R}) (\Bar{\ell}_{\sigma R} \, \nu_{\beta L})
      \nonumber\\
    &=& -\frac{1}{2} \sin^2{\varphi} \frac{Y_{\alpha \rho } Y_{ \beta \sigma}^\star  }{m_{h^+}^2}  (\Bar{\nu}_\alpha \gamma^\mu P_L \nu_\beta) (\Bar{\ell}_\sigma \gamma_\mu P_R \ell_\rho)\, ,
    \label{eq:Leff1}
    \end{eqnarray}
where in the second step, we have used the Fierz transformation. Comparing Eq.~\eqref{eq:Leff1} with Eq.~\eqref{NSI-NC}, we obtain the $h^+$-induced matter NSI parameters (setting $\rho=\sigma=e$)
\begin{align}
   \varepsilon_{\alpha \beta}^{(h^+)}   \ = \ & \frac{1}{4 \sqrt{2}G_F}  \frac{Y_{\alpha e} Y_{ \beta e}^{\star} }{m_{h^+}^2} \sin^2{\varphi}\, .
   \label{eq:nsi1}
   \end{align}
%   \\
 %   &=& - 2 \sqrt{2} G_F \varepsilon_{\alpha \beta}^{\rho \sigma} ( \Bar{\nu}_\alpha \gamma^\mu P_L \nu_\beta) ( \Bar{l}_\rho \gamma_\mu P_R l_\sigma )\\
%\text{where} \nonumber\\
%\end{eqnarray}
%where $h^+$ is the lightest charged scalar. For neutrinos propagating in the ordinary matter we have following canonical NSI parameter:
%    \begin{equation}
%         \varepsilon_{\alpha \beta}^m = \varepsilon_{\alpha \beta}^{ee} = \frac{1}{4 \sqrt{2}} \sin^2{\varphi} \frac{Y_{\alpha e }^{\star} Y_{ \beta e} }{G_F \, m_{h^+}^2}
%         \label{nsieqn}
%    \end{equation}
Thus, the diagonal NSI parameters $\varepsilon_{\alpha\alpha}$ depend on the Yukawa couplings $|Y_{\alpha e}|^2$, and are always positive in this model, whereas the off-diagonal ones  $\varepsilon_{\alpha\beta}$ (with $\alpha\neq \beta$) involve the product $Y_{\alpha e} Y_{ \beta e}^{\star}$ and can be of either sign, or even complex. Also, we have a correlation between the diagonal and off-diagonal NSI:
\begin{align}
     |\varepsilon_{\alpha \beta}| \ = \ \sqrt{\varepsilon_{\alpha \alpha} \varepsilon_{\beta \beta}} \, ,
     \label{eq:sum-rule}
\end{align}
which is a distinguishing feature of the model. 

Similarly, Fig.~\ref{nsi1} (d) gives the $H^+$-induced matter NSI contribution:
\begin{equation}
    \varepsilon_{\alpha \beta}^{(H^+)} \ = \  \frac{1}{4 \sqrt{2}G_F}  \frac{Y_{\alpha e } Y_{ \beta e}^{\star} }{ m_{H^+}^2} \cos^2{\varphi}\, .
    \label{eq:nsi2}
\end{equation}
%%%%%%%%%
Hence, the total matter NSI induced by the charged scalars in the Zee model can be expressed as
\begin{tcolorbox}[enhanced,ams align,
  colback=gray!30!white,colframe=white]
%\begin{equation}
 %  \boxed{ 
   \varepsilon_{\alpha \beta} \ \equiv  \ \varepsilon_{\alpha \beta}^{(h^+)} + \varepsilon_{\alpha \beta}^{(H^+)} \ = \ \frac{1}{4 \sqrt{2}G_F} Y_{\alpha e } Y_{ \beta e}^{\star}\left(\frac{ \sin^2{\varphi}}{ m_{h^+}^2}  +   \frac{\cos^2{\varphi} }{m_{H^+}^2}\right) \, .
   %}
         \label{nsieqntot}
%\end{equation}
\end{tcolorbox}

To get an idea of the size of NSI induced by Eq.~\eqref{nsieqntot}, let us take the diagonal NSI parameters from the light charged scalar contribution in Eq.~\eqref{eq:nsi1}: 
\begin{align}
   \varepsilon_{\alpha \alpha}^{(h^+)}   \ = \ & \frac{1}{4 \sqrt{2}G_F}  \frac{|Y_{\alpha e}|^2  }{m_{h^+}^2} \sin^2{\varphi}\, .
   \label{eq:nsi3}
   \end{align}
Thus, for a given value of $m_{h^+}$, the NSI are maximized for maximum allowed values of $|Y_{\alpha e}|$ and  $\sin{\varphi}$. Following Eq.~\eqref{eq:effmu}, we set the trilinear coupling $\lambda_{hh^+h^-}\to 0$, thus minimizing the constraints from Higgs signal strength. We also assume $\lambda_3=\lambda_8$ to get 
\begin{align}
    \mu \ = \ \frac{\sqrt 2 \lambda_8 v}{\sin 2\varphi} \, .
\end{align}
Now substituting this into Eq.~\eqref{mixphi}, we obtain 
\begin{align}
    \sin^2\varphi \ \simeq \ \frac{\lambda_8 v^2}{2(m_{H^+}^2-m_{h^+}^2)} \, .
    \label{eq:sinphi}
\end{align}
Furthermore, assuming the heavy charged and neutral scalars to be mass-degenerate, the LEP contact interaction constraints (cf.~Sec.~\ref{sec:contact}) require 
\begin{align}
    \frac{m_{H^+}^2}{|Y_{\alpha e}|^2} \ \gtrsim \  \frac{\Lambda_\alpha^2}{8\pi} \, ,
    \label{eq:lep2}
\end{align}
where $\Lambda_\alpha=10$ TeV, 7.9 TeV and 2.2 TeV for $\alpha=e,\mu,\tau$, respectively~\cite{LEP:2003aa}. 
Combining Eqs.~\eqref{eq:nsi3}, \eqref{eq:sinphi} and \eqref{eq:lep2}, we obtain 
\begin{align}
    \varepsilon_{\alpha\alpha}^{\rm max} \ \simeq \ \frac{\lambda_8 v^2}{m_{h^+}^2} \frac{\pi}{\sqrt{2} G_F \Lambda_{\alpha}^2}
\end{align}
%%%%%%%%%%%%%%%%%%%%%%%%%%%%%%%%%%%%%%%%%%%%%%%%%%%%%%%%%%%%%%%%%%%%
Using benchmark values of $m_{h^+} = 100 \, \text{GeV}$ and  $\lambda_8 = 3$, we obtain:
%%%%%%%%%%%%%%%%%%%%%%%%%%%%%%%%%%%%%%%%%%%%%%%%%%%%%%%%%%%%%%%%%%%%
\begin{equation}
    \varepsilon_{ee}^{\text{max}} \ \approx \  3.5 \% \, , \hspace{5mm}  
    \varepsilon_{\mu \mu}^{\text{max}} \ \approx \  5.6 \% \, , \hspace{5mm} \varepsilon_{\tau \tau}^{\text{max}} \approx  71.6 \% \, . 
    \label{eq:NSI-est}
\end{equation}
%%%%%%%%%%%%%%%%%%%%%%%%%%%%%%%%%%%%%%%%%%%%%%%%%%%%%%%%%%%%%%%%%%%%
Although a rough estimate, this tells us that observable NSI can be obtained in the Zee model, especially in the $\tau$ sector. To get a more accurate prediction of the NSI in the Zee model and to reconcile large NSI with all relevant theoretical and experimental constraints, we use Eq.~\eqref{nsieqntot} to numerically calculate the NSI predictions, as discussed below. 
%%%%%%%%%%%%%%%%%%%%%%%%
\subsubsection{Heavy neutral scalar case} \label{sec:Zee_heavy}
\begin{figure}[!t]
\vspace{-1.0cm}
\centering
  \subfigure[]{
    \includegraphics[height=8cm,width=0.45\textwidth]{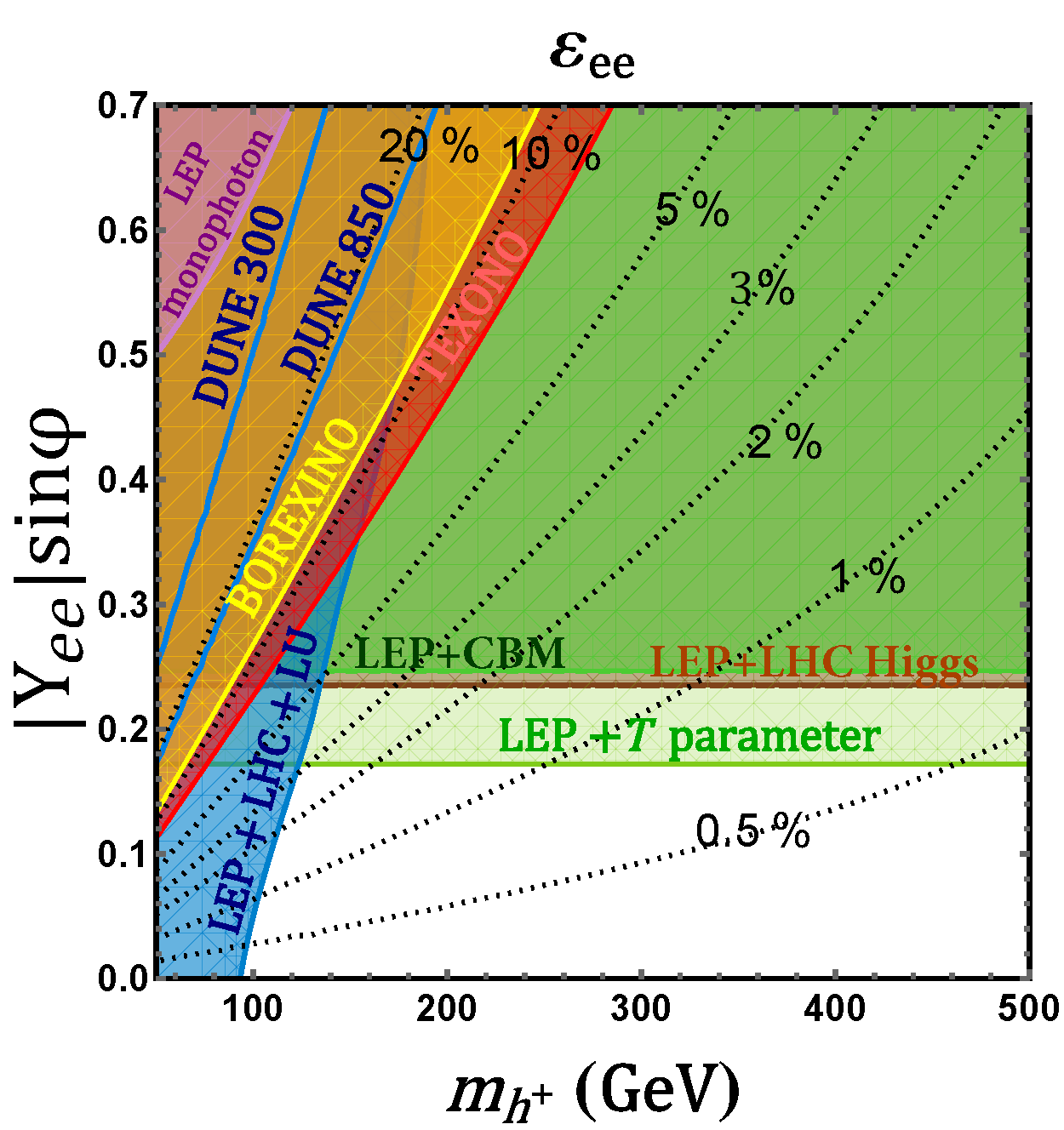}}
% \hspace{0.1in}
 \subfigure[]{
     \includegraphics[height=8cm,width=0.45\textwidth]{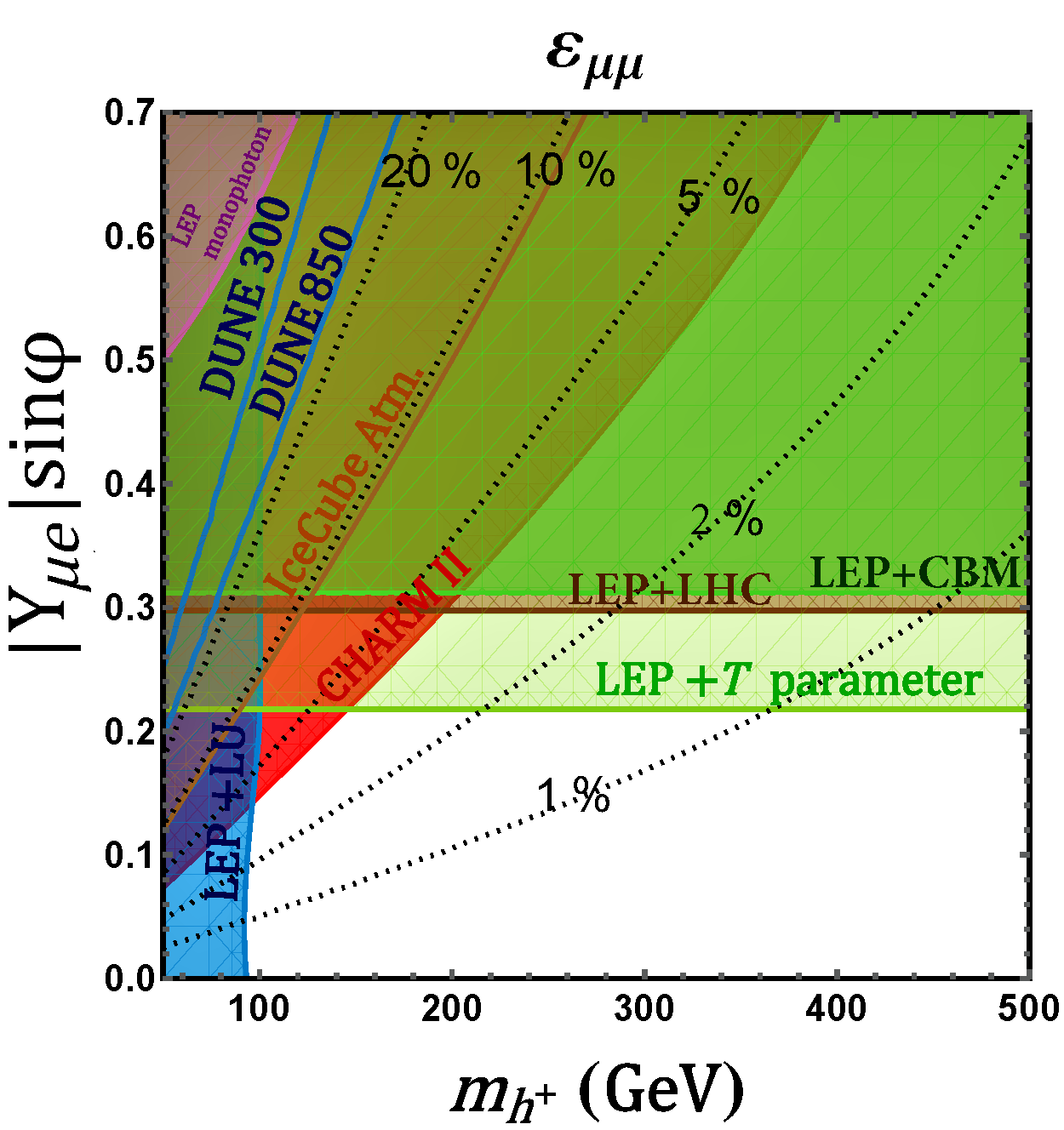}} \\
     \subfigure[]{
      \includegraphics[height=8cm,width=0.45\textwidth]{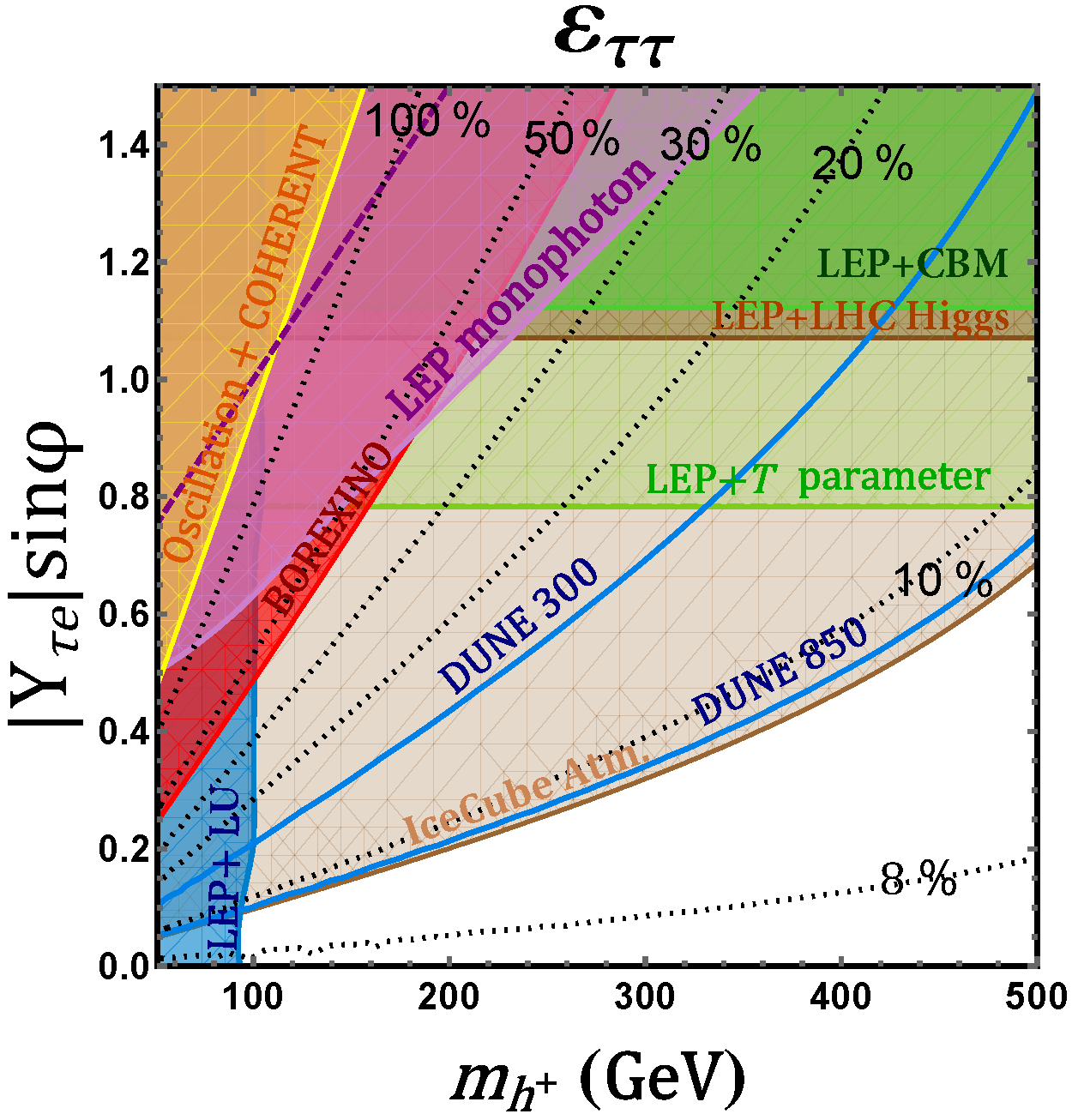}}
      \vspace{-0.5cm}
    \caption{Zee model predictions for diagonal NSI  $(\varepsilon_{ee},\, \varepsilon_{\mu\mu},\,\varepsilon_{\tau\tau})$ are shown by the black dotted contours. Color-shaded regions are excluded by various theoretical and experimental constraints: Blue-shaded region excluded by direct searches from LEP and LHC (Sec.~\ref{sec:colliderZee}) and/or lepton universality (LU) tests in $W$ decays (Sec.~\ref{sec:Wuniv}); purple-shaded region by off $Z$-pole LEP monophoton search (cf.~Sec.~\ref{sec:monop}), with the purple dashed line in (c) indicating a weaker limit from on $Z$-pole LEP search; light green, brown and deep green-shaded regions respectively by $T$ parameter (Sec.~\ref{sec:ewpt}), precision Higgs data (Sec.~\ref{sec:HiggsOb}), and charge-breaking minima (Sec.~\ref{sec:CBM}), each combined with LEP contact interaction constraint (Sec.~\ref{sec:contact}). In addition, we show the direct constraints on NSI from neutrino-electron scattering experiments (red/yellow-shaded), like  CHARM~\cite{Barranco:2007ej}, TEXONO~\cite{Deniz:2010mp}  and BOREXINO~\cite{Agarwalla:2019smc}, from IceCube atmospheric neutrino data~\cite{Esmaili:2013fva} (light brown), as well as the global-fit constraints from neutrino oscillation+COHERENT data~\cite{Esteban:2018ppq} (orange-shaded). We also show the future DUNE sensitivity (blue solid lines), for both 300 kt.MW.yr and 850 kt.MW.yr exposure~\cite{dev_pondd}.}
    \label{fig:dnsi}
  \vspace{-1.7cm}
\end{figure}

%%%%%%%%%%%%%%%%%%%%%%%%%%%%
First, we consider the case with heavy neutral and charged scalars, so that the LEP contact interaction constraints (cf.~Sec.~\ref{sec:contact}) are valid. To be concrete, we have fixed the heavy charged scalar mass $m_{H^+}=700$ GeV and the quartic couplings $\lambda_3=\lambda_8=3$. In this case, the heavy charged scalar contribution to NSI in Eq.~\eqref{nsieqntot} can be ignored. The NSI predictions in the light charged scalar mass versus Yukawa coupling plane are shown by black dotted contours in Fig.~\ref{fig:dnsi} for diagonal NSI and Fig.~\ref{fig:offdnsi} for off-diagonal NSI. The theoretical constraints on $\sin\varphi$ from charge-breaking minima (cf.~Sec.~\ref{sec:CBM}) and $T$-parameter (cf.~Sec.~\ref{sec:ewpt}) constraints are shown by the light and dark green-shaded regions, respectively. Similarly, the Higgs precision data constraint (cf.~Sec.~\ref{sec:HiggsOb}) on $\sin\varphi$ is shown by the brown-shaded region. To cast these constraints into limits on $Y_{\alpha e}\sin\varphi$, we have used the LEP contact interaction limits on $Y_{\alpha e}$ (cf.~Sec.~\ref{sec:contact}) for diagonal NSI, and similarly, the cLFV constraints (cf.~Sec.~\ref{sec:lfv}) for off-diagonal NSI, and combined these with the CBM, $T$-parameter and Higgs constraints,  which are all independent of the light charged scalar mass.  Also shown in Figs.~\ref{fig:dnsi} and~\ref{fig:offdnsi} are the LEP and/or LHC constraints on light charged scalar (cf.~Sec.~\ref{sec:colliderZee}) combined with the lepton universality constraints from $W$ and $\tau$ decays (cf.~Secs.~\ref{sec:Wuniv} and \ref{sec:taudecay}), which exclude the blue-shaded region below $m_{h^+}\sim 100$ GeV. In addition, the LEP monophoton constraints from off $Z$-pole search  (cf.~Sec.~\ref{sec:monop}) are shown in Fig.~\ref{fig:dnsi} by the light purple-shaded region. The corresponding limit from LEP on $Z$-pole search (shown by the purple dashed line in Fig.~\ref{fig:dnsi} (c) turns out to be weaker. %With these constraints imposed, the maximum allowed values of the NSI parameters are given by
%\begin{align}
%&   \varepsilon_{ee}^{\rm max} \ = \ 0.032 \, , \qquad 
 %   \varepsilon_{\mu\mu}^{\rm max} \ = \ 0.07 \, ,\qquad 
%      \varepsilon_{\tau\tau}^{\rm max} \ = \ 0.62 \, , \nonumber \\
%  & \varepsilon_{e\mu}^{\rm max} \ = \ 1.5\times 10^{-5} \, , \qquad  
 %  \varepsilon_{e\tau}^{\rm max} \ = \ 0.0056 \, , \qquad 
%   \varepsilon_{\mu\tau}^{\rm max} \ = \ 0.0034 \, .
%   \label{eq:Zeemax}
%\end{align}

The model predictions for NSI are then compared with the current direct experimental constraints from neutrino-electron scattering experiments (red/yellow-shaded), and the global-fit constraints (orange-shaded)~\cite{Esteban:2018ppq} which include the neutrino oscillation data~\cite{Tanabashi:2018oca}, as well as the recent results from COHERENT experiment~\cite{Akimov:2017ade};\footnote{For related NSI studies using the COHERENT data, see e.g.~Refs.~\cite{Coloma:2017egw, Coloma:2017ncl, Liao:2017uzy, AristizabalSierra:2018eqm,  Altmannshofer:2018xyo}.} see Table~\ref{tab:Zee} for more details. For neutrino-electron scattering constraints, we only considered the constraints on  $\varepsilon_{\alpha\beta}^{eR}$~\cite{Davidson:2003ha, Barranco:2007ej, Deniz:2010mp, Agarwalla:2019smc}, since the dominant NSI in the Zee model always involves right-handed electrons (cf.~Eq.~\eqref{eq:Leff1}).  For $\varepsilon_{\mu\mu}$, we have rederived the CHARM II limit following Ref.~\cite{Davidson:2003ha}, but using the latest PDG value for $s_w^2=0.22343$ (on-shell)~\cite{Tanabashi:2018oca}. Specifically, we used the CHARM II measurement of the $Z$-coupling to right-handed electrons $g_R^e=0.234\pm 0.017$ obtained from their $\nu_\mu e\to \nu e$ data~\cite{Vilain:1994qy} and compared with the SM value of $(g_R^e)_{\rm SM}=s_w^2$ to obtain a $90\%$ CL limit on $\varepsilon_{\mu\mu}<0.038$, which is slightly weaker than the limit of 0.03 quoted in Ref.~\cite{Barranco:2007ej}. Nevertheless, the CHARM limit turns out to be the strongest in realizing maximum $\varepsilon_{\mu\mu}$ in the Zee model, as shown in Fig.~\ref{fig:dnsi} (b).

There is a stronger constraint on $|\varepsilon_{\tau \tau}-\varepsilon_{\mu\mu}|< 9.3\%$ from the IceCube atmospheric neutrino oscillation data~\cite{Fornengo:2001pm, Esmaili:2013fva, Day:2016shw}. In general, this bound can be evaded even for large NSI, if e.g. both  $\varepsilon_{\mu \mu}$ and $\varepsilon_{\tau \tau}$ are large and there is a cancellation between them. However, in the Zee model, such cancellation cannot be realized, because we can only allow for one large diagonal NSI at a time, otherwise there will be stringent constraints from cLFV (cf.~Sec.~\ref{sec:lfv}). For instance, making both $\varepsilon_{\mu \mu}$ and $\varepsilon_{\tau \tau}$ large necessarily implies a large $\varepsilon_{\mu \tau}$ (due to the relation given by  Eq.~\eqref{eq:sum-rule}), which is severely constrained by $\tau^-\to \mu^- e^-e^+$ (cf.~Table~\ref{3ldecay} and Fig.~\ref{fig:offdnsi} (a)) and also by IceCube itself~\cite{Esmaili:2013fva, Salvado:2016uqu, Aartsen:2017xtt}.  Therefore, the bound on $\varepsilon_{\tau \tau}-\varepsilon_{\mu\mu}$ is equally applicable to both $\varepsilon_{\mu \mu}$ and $\varepsilon_{\tau \tau}$. This is shown by the brown-shaded regions in Fig.~\ref{fig:dnsi} (b) and (c), respectively. This turns out to be the most stringent constraint for $\varepsilon_{\tau \tau}$, although the model allows for much larger NSI, as shown by the black dotted contours in Fig.~\ref{fig:dnsi} (c).

For completeness, we also include in Fig.~\ref{fig:dnsi} global-fit constraints from neutrino oscillation plus scattering experiments~\cite{Esteban:2018ppq}.\footnote{We use the constraints on $\varepsilon_{\alpha\beta}^{p}$ from  Ref.~\cite{Esteban:2018ppq}, assuming that these will be similar for $\varepsilon_{\alpha\beta}^{e}$ due to charge-neutrality in matter.} The global-fit analysis assumes the simultaneous presence of all $\varepsilon_{\alpha\beta}$'s, and therefore, the corresponding limits on each $\varepsilon_{\alpha\beta}$ are much weaker than the ones derived from oscillation or scattering data alone, due to parameter degeneracies. For instance, the global-fit constraint on $\varepsilon_{\tau \tau}\in [-35\%, 140\%]$ (cf.~Table~\ref{tab:Zee}) is significantly affected by the presence of nonzero $\varepsilon_{ee}$ and $\varepsilon_{e\tau}$~\cite{Friedland:2004ah}, which were set to zero in the IceCube analysis of Ref.~\cite{Esmaili:2013fva}.

Also shown in Fig.~\ref{fig:dnsi} (blue solid lines) are the future sensitivity at long-baseline neutrino oscillation experiments, such as DUNE with 300 kt.MW.yr and 850 kt.MW.yr of exposure, derived at 90\% CL using {\tt  GloBES3.0}~\cite{Huber:2007ji} with the DUNE CDR simulation configurations~\cite{Alion:2016uaj}. Here we have used $\delta~({\rm true})=-\pi/2$ for the true value of the Dirac {\cal CP} phase and marginalized over all other oscillation parameters~\cite{dev_pondd}. We find that even the most futuristic DUNE sensitivity will not be able to surpass the current constraints on the Zee model. On the other hand, the current neutrino scattering experiments like COHERENT and atmospheric neutrino experiments such as IceCube should be able to probe a portion of the allowed parameter space for $\varepsilon_{\mu\mu}$ and $\varepsilon_{\tau\tau}$, respectively.

%%%%%%%%%%%%%%%%%%%%%%%%%%%%%%%%%%%%%%%%%%%%%%%%%
\subsubsection{Light neutral scalar case}\label{sec:Zee_light}
\begin{figure}[t!]
\vspace{-1.0cm}
\centering
\subfigure[]{
    \includegraphics[height=8cm,width=0.45\textwidth]{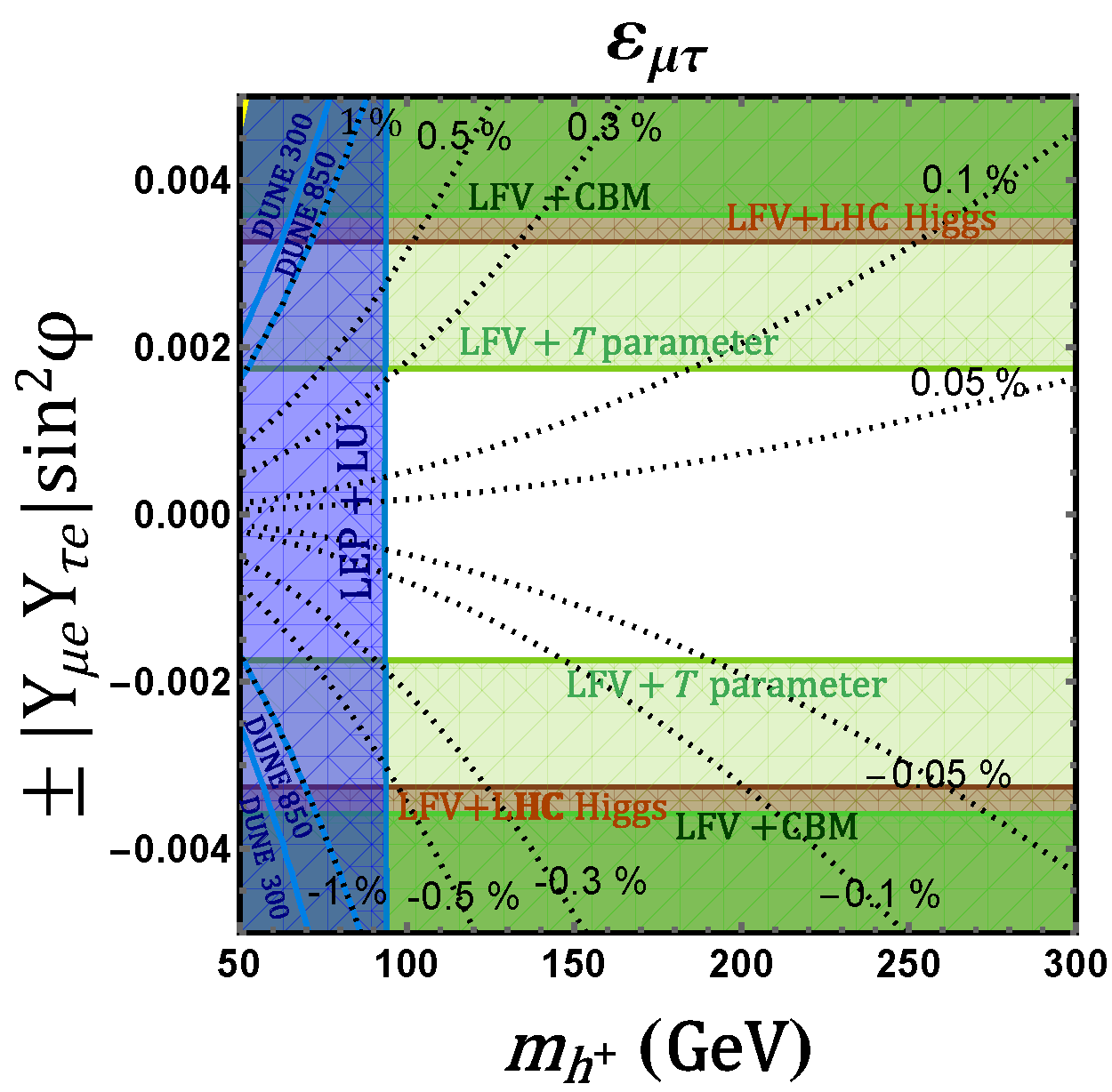}}
% \hspace{0.1in}
     \subfigure[]{\includegraphics[height=8cm,width=0.45\textwidth]{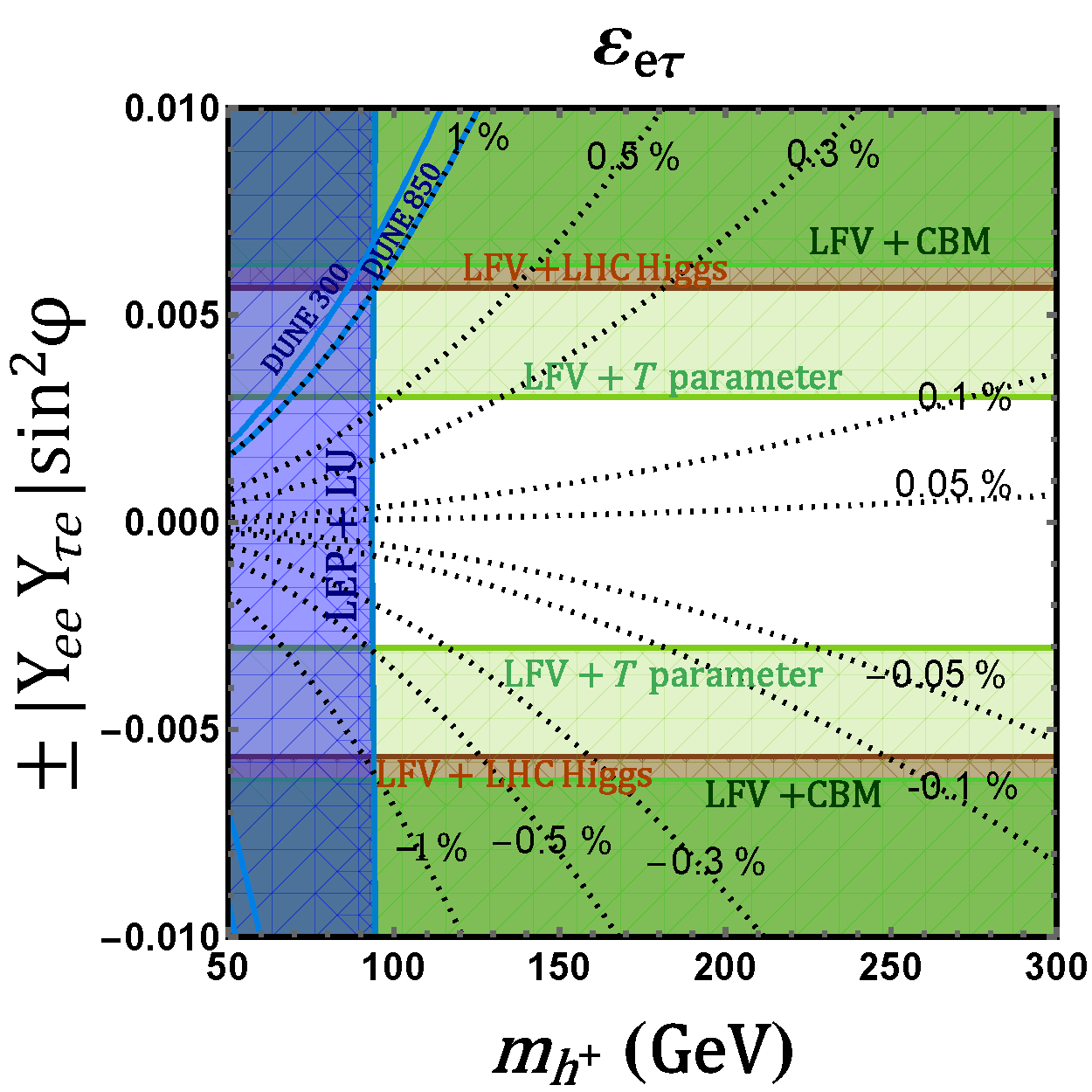}}\\
    \subfigure[]{
      \includegraphics[height=8cm,width=0.45\textwidth]{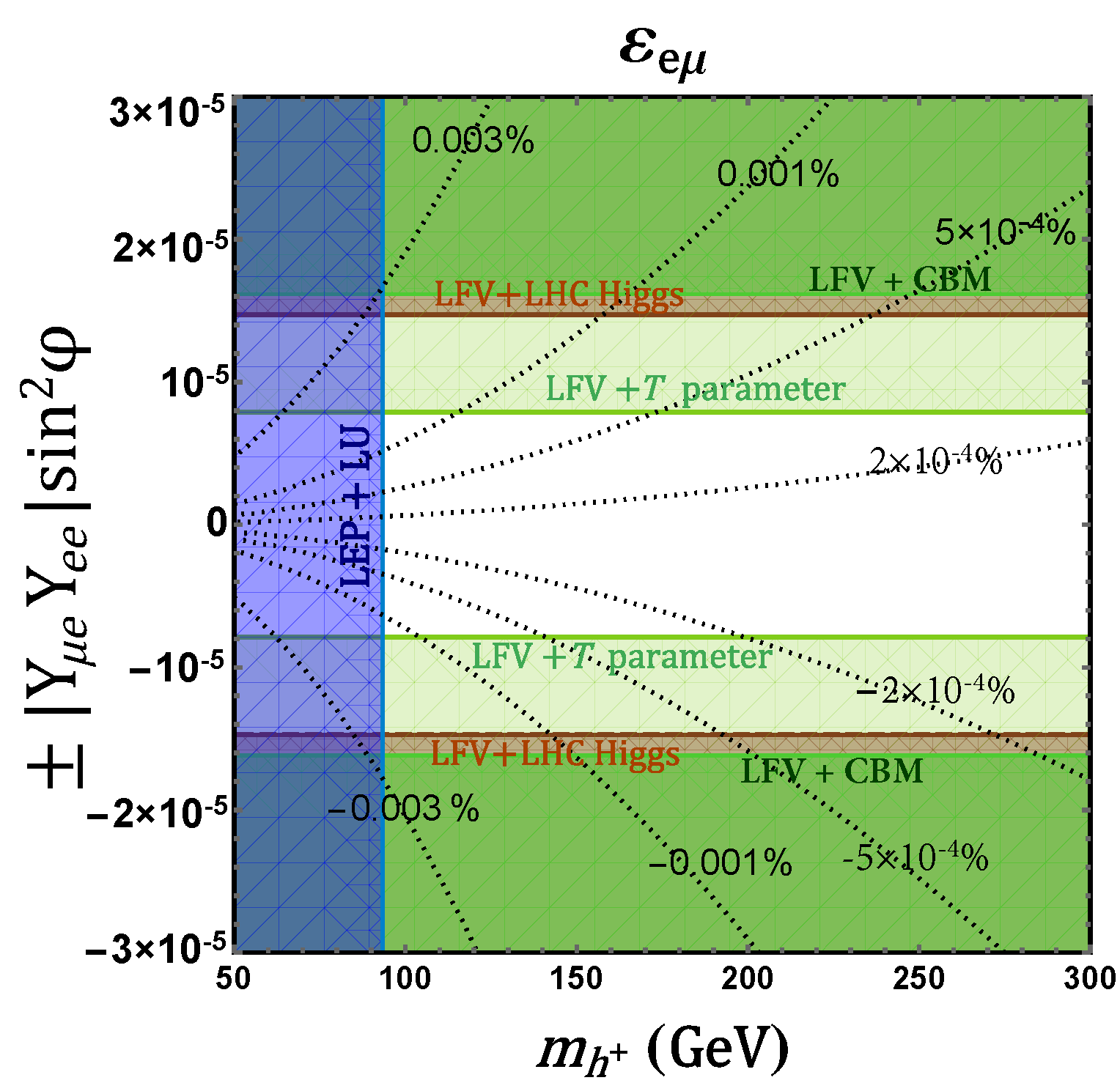}}
    \caption{Zee model predictions for off-diagonal NSI $(\varepsilon_{e\mu},\, \varepsilon_{\mu\tau},\,\varepsilon_{e\tau})$ are shown by black dotted contours. Color-shaded regions are excluded by various theoretical and experimental constraints. Blue-shaded region is excluded by direct searches from LEP and LHC (Sec.~\ref{sec:colliderZee}) and/or lepton universality (LU) tests in $W$ decays (Sec.~\ref{sec:Wuniv}). Light green, brown and deep green-shaded regions are excluded respectively by  $T$-parameter (Sec.~\ref{sec:ewpt}), precision Higgs data (Sec.~\ref{sec:HiggsOb}), and charge-breaking minima (Sec.~\ref{sec:CBM}), each combined with cLFV constraints (Sec.~\ref{sec:lfv}). The current NSI constraints from neutrino oscillation and scattering experiments are weaker than the cLFV constraints, and do not appear in the shown parameter space. The future DUNE sensitivity is shown by blue solid lines, for both 300 kt.MW.yr and 850 kt.MW.yr exposure~\cite{dev_pondd}. }
    \label{fig:offdnsi}
    \vspace{-0.5cm}
\end{figure}

\begin{figure}[!t]
%\vspace{-1.0cm}
\centering
  \subfigure[]{
    \includegraphics[height=8cm,width=0.45\textwidth]{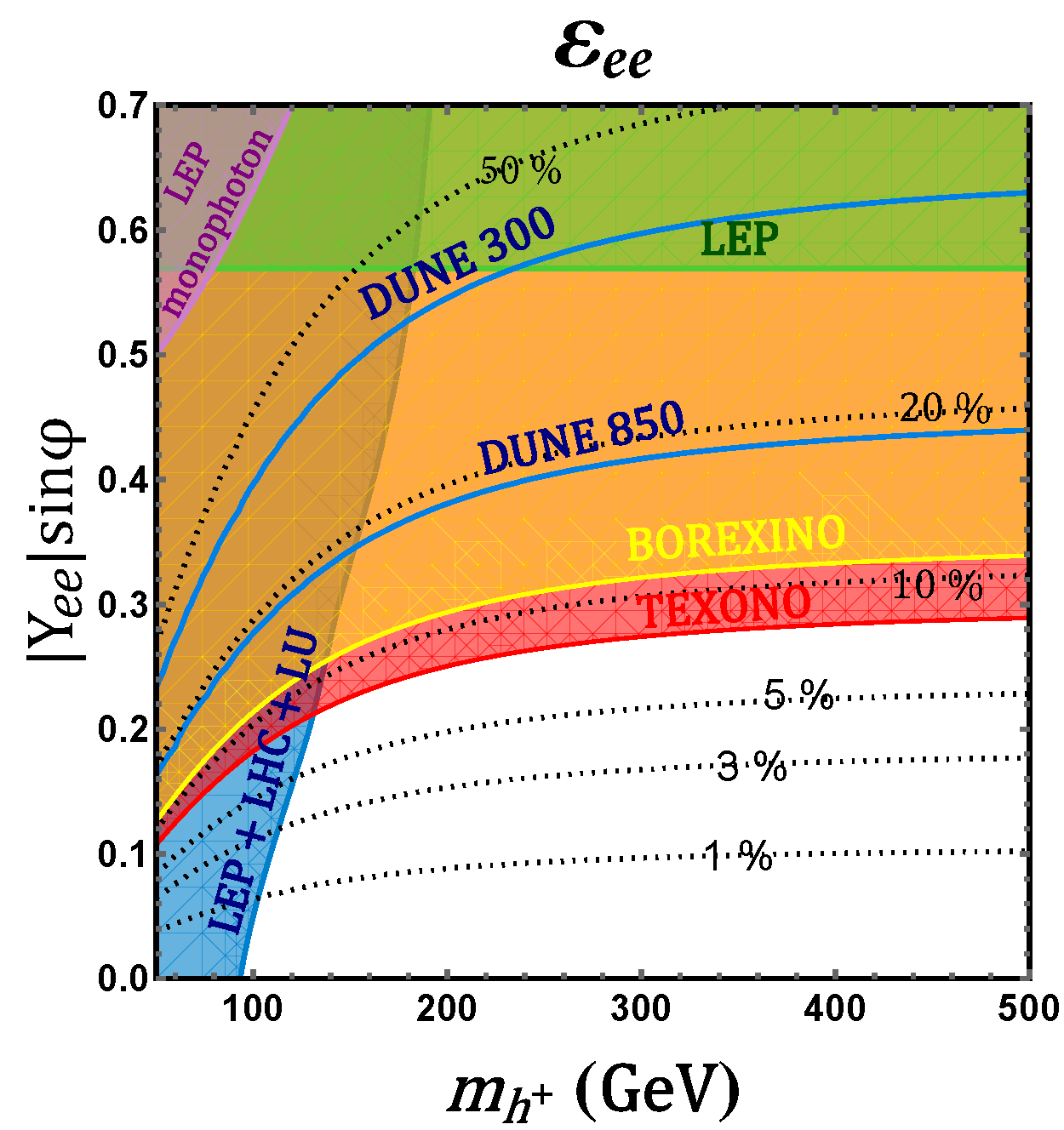}}
% \hspace{0.1in}
 \subfigure[]{
     \includegraphics[height=8cm,width=0.45\textwidth]{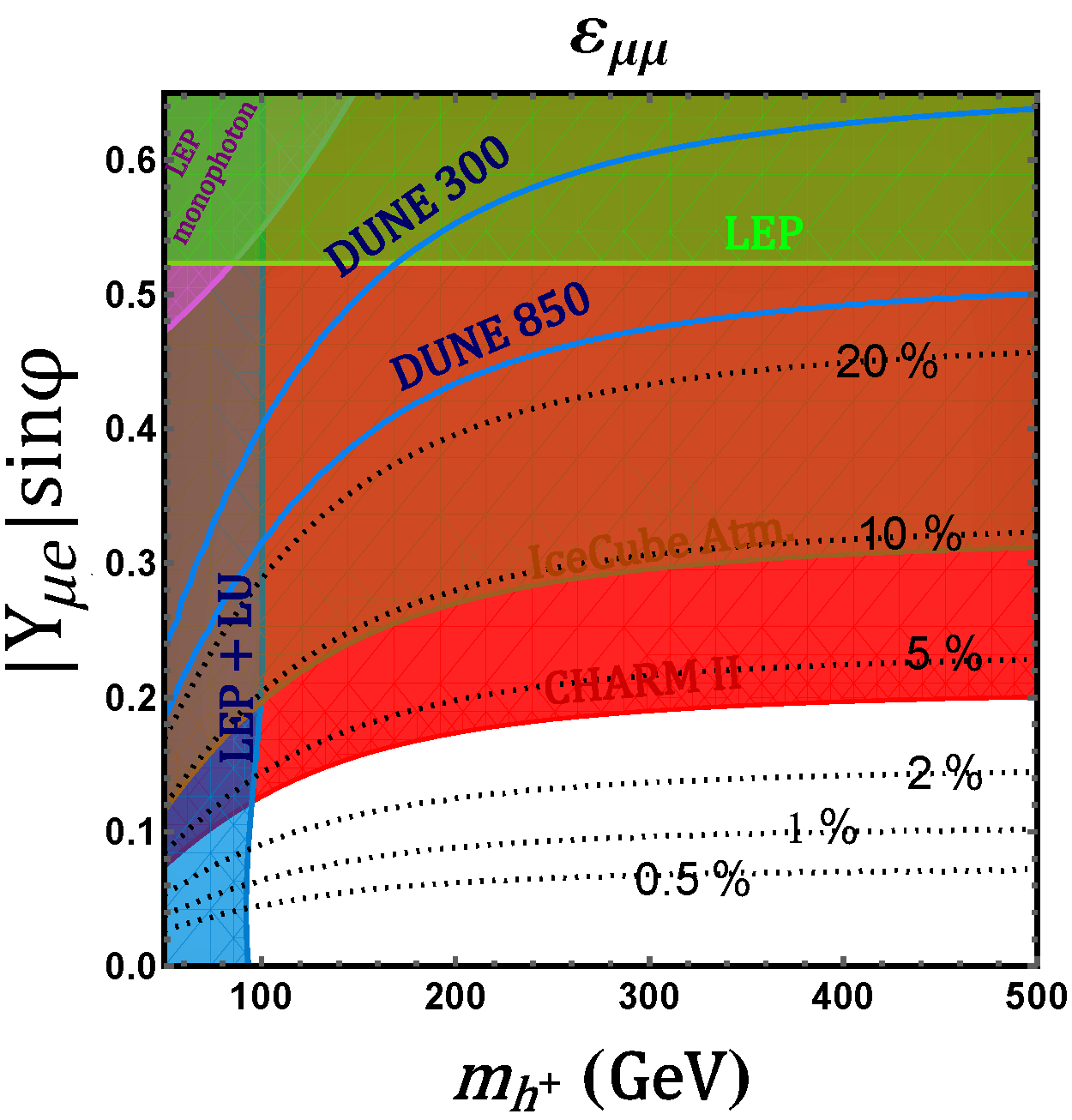}} \\
     \subfigure[]{
      \includegraphics[height=8cm,width=0.45\textwidth]{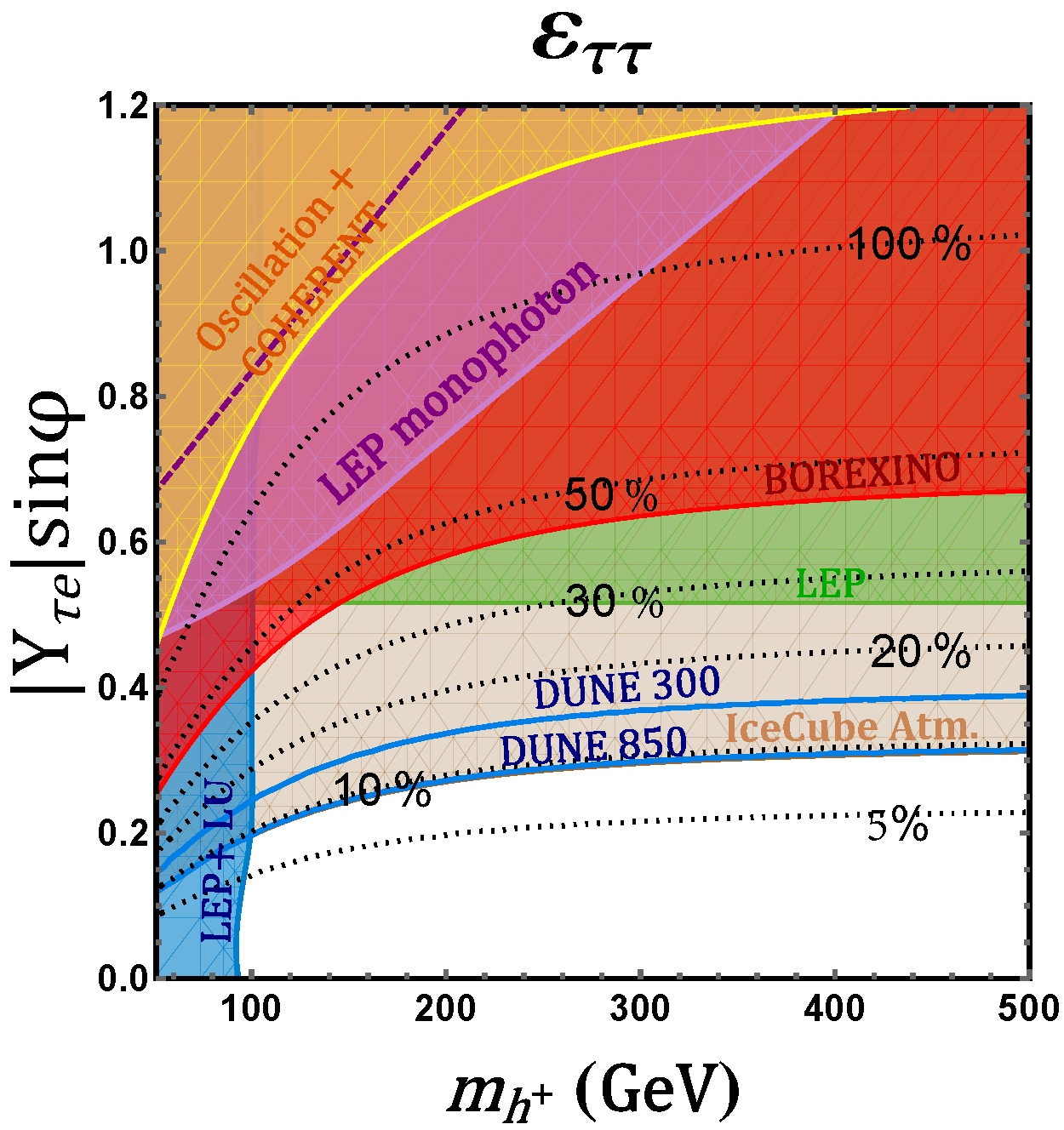}}
      \vspace{-0.5cm}
    \caption{Zee model predictions for diagonal NSI for light neutral scalar case. Here we have chosen $m_{H^+}=130$ GeV. Labeling of the color-shaded regions is the same as in Fig.~\ref{fig:dnsi}, except for the LEP dilepton constraint (green-shaded region) which replaces the $T$-parameter, CBM and 
    LHC Higgs constraints.}
    \label{fig:dnsispcl}
%  \vspace{-1.7cm}
\end{figure}

%%%%%%%%%%%%%%%%%%%%%%%%%%%%%%
Now we consider the case where the neutral scalars $H$ and $A$ are light, so that the LEP contact interaction constraints (cf.~\ref{sec:contact}) are not applicable. In this case, both $h^+$ and $H^+$ contributions to the NSI in Eq.~\eqref{nsieqntot} should be kept. For concreteness, we fix $m_{H^+}=130$ GeV to allow for the maximum $H^+$ contribution to NSI while avoiding the lepton universality constraints on $H^+$ (cf.~Sec.~\ref{sec:Wuniv}). We also choose the neutral scalars $H$ and $A$ to be nearly mass-degenerate with the charged scalar $H^+$, so that the $T$-parameter and CBM constraints are easily satisfied. The Higgs decay constraints can also be significantly relaxed in this case by making $\lambda_{hh^+h^-}\to 0$ in Eq.~\eqref{eq:effmu}. The NSI predictions for this special choice of parameters are shown in Fig.~\ref{fig:dnsispcl}. Note that for higher $m_{h^+}$, the NSI numbers are almost constant, because of the $m_{H^+}$ contribution which starts dominating. We do not show the off-diagonal NSI plots for this scenario, because the cLFV constraints still cannot be overcome (cf.~Fig.~\ref{fig:offdnsi}). 

Taking into account all existing constraints and this possibility of light $h^+$ and $H^+$, the maximum possible allowed values of the NSI parameters in the Zee model are shown in the second column of Table~\ref{tab:Zee}, along with the combination of the relevant constraints limiting each NSI parameter (shown in parentheses). Thus, we find that for the diagonal NSI, one can get maximum $\varepsilon_{ee}$ of 8\%, $\varepsilon_{\mu \mu}$ of 3.8\%, and  $\varepsilon_{\tau \tau}$ of 9.3\%, only limited by direct experimental searches (TEXONO, CHARM and IceCube, respectively). Thus, the future neutrino experiments could probe diagonal NSI in the Zee model. As for the off-diagonal NSI, they require the presence of at least two non-zero Yukawa couplings $Y_{\alpha e}$, and their products are all heavily constrained from cLFV; therefore, one cannot get sizable off-diagonal NSI in the Zee model that can be probed by any neutrino scattering or oscillation experiment in the foreseeable future.

%%%%%%%%%%%%%%%%%%%%%%%%%%%%%%%%%%%%%%%%%%%%%%%%%%%%%%%%%%%%%%%%%%%%%%%%%%%%%%%% 
%\clearpage
%%%%%%%%%%%%%%%%%%%%%%%%%%%%%%%%%%%%%%%%%%%%%%%%%%%%%%%%%%%%%%%%%%%%%%%%%%%%%%%%%%%%%%%%%
\clearpage
%%%%%%%%%%%%%%%%%%%%%%%%%%%%%%%%%%%%%

%%%%%%%%%%%%%%%%%%%%%%%%%%%%%%%%%%%%%%%%%%%%%%%%%%%%%%%%%%%%%%%%%%%%%%%%%%%%%%%% 
%\clearpage

%%%%%%%%%%%%%%%%%%%%%%%%%%%%%%%
\begin{table}[!t]
    \centering
    \scriptsize
    \begin{tabular}{|c|c|c|c|c|}
    \hline \hline
        \textbf{NSI} & \textbf{Zee Model} & \textbf{Individual} & \textbf{Global-fit} & \textbf{DUNE} \\
        & \textbf{Prediction (Max.)} & \textbf{constraints} & \textbf{ constraints}~\cite{Esteban:2018ppq} & {\bf sensitivity}~\cite{dev_pondd}  \\ \hline \hline
        
          $\varepsilon_{ee}$       & 0.08 & $[-0.07, 0.08]$ \cite{Deniz:2010mp} & $[-0.010, 2.039]$ & $[-0.185, 0.380]$   \\ 
          & (TEXONO)& & &  ($[-0.130, 0.185]$) \\ \hline
          $\varepsilon_{\mu \mu}$  & 0.038 & $[-0.03, 0.03]$ \cite{Barranco:2007ej} & $[-0.364, 1.387]$ & $[-0.290, 0.390]$   \\ 
          & (CHARM)& $[-0.017, 0.038]$ (ours) & & ($[-0.192, 0.240]$) \\ \hline
          $\varepsilon_{\tau \tau}$& 0.093  & $[-0.093,0.093]$ \cite{Esmaili:2013fva} & $[-0.350, 1.400]$ & $[-0.360, 0.145]$    \\ 
          & (IceCube)& & & ($[-0.120, 0.095]$) \\ \hline
          $\varepsilon_{e \mu}$    & $1.5\times 10^{-5}$  & $[-0.13, 0.13]$ \cite{Barranco:2007ej} & $[-0.179, 0.146]$ & $[-0.025, 0.052]$  \\ 
          & (LEP  + LU + cLFV + $T$-param.)& & & ( $[-0.017, 0.040]$) \\ \hline
          $\varepsilon_{e \tau}$   & 0.0056 & $[-0.19, 0.19]$ \cite{Deniz:2010mp}  & $[-0.860, 0.350]$ & $[-0.055, 0.023]$    \\ 
          & (LEP  + LU + cLFV + $T$-param.)& & & ($[-0.042, 0.012]$) \\ \hline
          $\varepsilon_{\mu \tau}$ & 0.0034 & $[-0.10, 0.10]$ \cite{Barranco:2007ej} & $[-0.035, 0.028]$ & $[-0.0.015, 0.013]$   \\ 
          & (LEP  + LU + cLFV + $T$-param)& & & ($[-0.010, 0.010]$) \\ \hline \hline
    \end{tabular}
    \caption{Maximum allowed NSI (with electrons) in the Zee model, after imposing constraints from CBM (Sec.~\ref{sec:CBM}), $T$-parameter (Sec.~\ref{sec:ewpt}), cLFV searches (Sec.~\ref{sec:lfv}), LEP contact interaction (Sec.~\ref{sec:contact}), direct collider searches   (Sec.~\ref{sec:colliderZee}), lepton universality (LU) in $W$ decays (Sec.~\ref{sec:Wuniv}), LHC Higgs data (Sec.~\ref{sec:HiggsOb}), and  LEP monophoton searches (Sec.~\ref{sec:monop}). We also impose the individual constraints, taking one NSI parameter at a time, from either neutrino-electron scattering or neutrino oscillation experiments (as shown in the third column), like CHARM-II~\cite{Barranco:2007ej},  TEXONO~\cite{Deniz:2010mp} and  BOREXINO~\cite{Agarwalla:2019smc} (only $\epsilon_{\alpha\beta}^{eR}$ are considered, cf.~Eq.~\eqref{eq:Leff1}) or IceCube~\cite{Esmaili:2013fva} as well as the global-fit constraints (as shown in the fourth column), taking all NSI parameters simultaneously,  from neutrino oscillation+COHERENT data~\cite{Esteban:2018ppq} (only $\varepsilon_{\alpha\beta}^p$ are considered), whichever is stronger. The maximum allowed value for each NSI parameter is obtained after scanning over the light charged Higgs mass (see Figs.~\ref{fig:dnsi} and \ref{fig:offdnsi}) and the combination of all relevant constraints limiting the NSI are shown in parentheses in the second column. In the last column, we also show the future DUNE sensitivity for 300 kt.MW.yr exposure (and 850 kt.MW.yr in parentheses)~\cite{dev_pondd}.}
    \label{tab:Zee}
\end{table}

%%%%%%%%%%%%%%%%%%%%%%%%%%%%%%%%%%%%%%%%%%%%%%%
\subsection{Consistency with neutrino oscillation data}\label{sec:neu}
%%%%%%%%%%%%%%%%%%%%%%%%%%%%%%%%%%%%%%%%%%%%%%%
In this section, we show that the choice of the Yukawa coupling matrix used to maximize our NSI parameter values is consistent with the neutrino oscillation data. The neutrino mass matrix in the Zee model is given by Eq.~\eqref{nuMass} which is diagonalized by the unitary transformation 
\begin{equation}
    U_{\rm PMNS}^T M_\nu \, U_{\rm PMNS} \ = \ \widehat{M}_\nu \, ,
\end{equation}
where $\widehat{M}_\nu={\rm diag}(m_1,m_2,m_3)$ is the diagonal mass matrix with the eigenvalues $m_{1,2,3}$ and $U_{\rm PMNS}$ is the 3 $\times$ 3 lepton mixing matrix. In the standard parametrization~\cite{Tanabashi:2018oca}, 
\begin{equation}
    U_{\rm PMNS} \ = \ \left(
\begin{array}{ccc}
 c_{12} c_{13} & c_{13} s_{12} &
   e^{-i \delta } s_{13} \\
 -c_{23} s_{12}- c_{12}
   s_{13} s_{23}e^{i \delta } & c_{12}
   c_{23}- s_{12}
   s_{13} s_{23}e^{i \delta } & c_{13} s_{23} \\
 s_{12} s_{23}- c_{12}
   c_{23} s_{13} e^{i \delta } & - c_{12}
   s_{23}-c_{23} s_{12} s_{13} e^{i \delta }  & c_{13} c_{23} \\
\end{array}
\right) \, ,
\label{eq:PMNS}
\end{equation}
where $c_{ij} \equiv \cos{\theta_{ij}}$,  $s_{ij} \equiv \sin{\theta_{ij}}$, $\theta_{ij}$ being the mixing angle between different flavor eigenstates $i$ and $j$, and $\delta$ is the Dirac $\mathcal{CP}$ phase.  We diagonalize the neutrino mass matrix \eqref{nuMass} numerically, assuming certain forms of the Yukawa coupling matrices given below.  The unitary matrix thus obtained is converted to the mixing angles $\theta_{ij}$  using the following relations from Eq.~\eqref{eq:PMNS}:
\begin{equation}
    s_{12}^2 \ = \ \frac{|U_{e2}|^2}{1-|U_{e3}|^2}, \, \qquad  s_{13}^2 \ = \ |U_{e3}|^2, \, \qquad  s_{23}^2  \ = \  \frac{|U_{\mu3}|^2}{1-|U_{e3}|^2} \, .
\end{equation}
%%%%%%%%%%%%%%%%%%%
Since the NSI expressions in Eq.~\eqref{nsieqntot} depend on $Y_{\alpha e}$ (the first column of the Yukawa matrix), we choose the following three sets of benchmark points (BPs) for Yukawa textures to satisfy all the cLFV constraints, see Tables~\ref{lgdecay} and \ref{3ldecay}. For simplicity, we also take all the elements of Yukawa matrix to be real.  
\begin{eqnarray}
 {\rm BP~I}:~ Y &\ = \ &   \left(
         \begin{array}{ccc}
             Y_{ee}  & 0  & Y_{e \tau}  \\
            0  & Y_{\mu \mu}  & Y_{\mu \tau}  \\
             0  &  Y_{\tau \mu}  & Y_{\tau \tau}  \\
              \end{array}
        \right), \label{Ymat1} \\
     {\rm BP~II}:~ Y &\ = \ & \left(
         \begin{array}{ccc}
             0  & Y_{e \mu}  & Y_{e\tau}   \\
              Y_{\mu e}  & 0  & Y_{\mu \tau}  \\
              0  &  Y_{\tau \mu}  & Y_{\tau \tau}  \\
              \end{array}
        \right), \label{Ymat2}\\
     {\rm BP~III}:~ Y& \ = \ &   \left(
         \begin{array}{ccc}
             Y_{ee}  & 0  &  Y_{e\tau}  \\
              0  & Y_{\mu \mu}  & Y_{\mu \tau}  \\
              Y_{\tau e}  &  0  & Y_{\tau \tau}  \\
              \end{array}
        \right)\label{Ymat3}
\end{eqnarray} 
For BP I, substituting $Y$ from Eq.~\eqref{Ymat1} in Eq.~\eqref{nuMass}, we get a symmetric neutrino mass matrix as follows:
%%%%%%%%%%%%%%
\begin{eqnarray}
M_\nu \ & = & \ a_0 \left(
        \begin{array}{ccr}
        m_{11} & m_{12} & m_{13} \\
         m_{12} & m_{22} & m_{23} \\
        m_{13} &  m_{23} & m_{33}
\end{array}
\right) \, ,
\end{eqnarray}
where $a_0 = \kappa f_{\mu \tau} Y_{e e} $ fixes the overall scale, and the entries in $M_\nu$ are given by 
\begin{eqnarray}
m_{11} \ & = & \ 2 m_\tau  x_2 \, y_{13}\, , \nonumber\\ 
m_{12} \ & = & \ -m_e x_1 y_{11} + m_\tau  y_{13} + m_\mu \, x_1 \, y_{22} + m_\tau \, x_2 \, y_{23} \, , \nonumber \\
m_{13} \ & = & \ -m_e x_2 y_{11} + m_\mu x_1 y_{32} + m_\tau \, x_2\, y_{33} \, , \nonumber \\
m_{22} \ & = & \ 2 \, m_\tau y_{23} \, , \nonumber \\
m_{23} \ & = & \  - m_\mu \,y_{22} + m_\tau  y_{33} \, , \nonumber \\
m_{33} \ & = & \ -2 m_\mu \, y_{32}  \, , \nonumber
\label{MatY1}
\end{eqnarray}
%%%%%%%%%%%%%%%
%%%%%%%%%%%%%%%%%%%%%%%%%
and we have defined the ratios $x_1 = \frac{f_{e \mu}}{f_{\mu \tau}}$ , $x_2 = \frac{f_{e \tau}}{f_{\mu \tau}}$, $y_{13} = \frac{Y_{e \tau}}{Y_{e e}}$, $y_{22} = \frac{Y_{\mu \mu}}{Y_{e e}}$, $y_{23} = \frac{Y_{\mu \tau}}{Y_{e e}}$, $y_{32} = \frac{Y_{ \tau \mu}}{Y_{e e}}$, and $y_{33} = \frac{Y_{\tau \tau}}{Y_{e e}}$. 
Similarly, for BPs II and III, one can absorb $Y_{\mu \mu}$ and $Y_{\tau \tau}$ respectively in the overall factor $a_0$ to get the mass matrix parameters in terms of the ratios $x_i$ and $y_{ij}$. 

For each set of Yukawa structure, we show in Table~\ref{nuTab} the best-fit values of the parameters $x_i$, $y_{ij}$ and $a_0$. For BP I and II, we obtain inverted hierarchy (IH) and for BP III, we get normal hierarchy (NH) of neutrino masses. The model predictions for the neutrino oscillation parameters in each case are shown in Table~\ref{nuTab2}, along with the 3$\sigma$ allowed range from a recent {\tt NuFit4} global analysis~ \cite{Esteban:2018azc}. It is clear that the fits for all the three sets are in very good agreement with the observed experimental values. We note here that the {\tt NuFit4} analysis does not include any NSI effects, which might affect the fit results; however, it is sufficient for the consistency check of our benchmark points. A full global analysis of the oscillation data in presence of NSI to compare with our benchmark points is beyond the scope of this work.

\begin{table}[!t]
 \centering
\scriptsize
\begin{tabular}{|c|c|c|c|c|c|c|c|c|c|c|c|c|}
\hline \hline
BP & $x_1$  & $x_2$ & $y_{11}$ & $y_{12}$ & $y_{13}$ & $y_{21}$ & $y_{22}$ & $y_{23}$ & $y_{31}$ & $y_{32}$ & $y_{33}$ & $a_0(10^{-9})$   \\ \hline \hline
BP I (IH) & $-7950$ & 34 & $-1.0$ & 0 & $-0.01$ & 0 & 0.001 & 0.08 & 0 & 0.05 & 0.70 & 0.017 \\ \hline
BP II (IH) & 14 & 4.7 & 0 & 0.05 & 0.01 & 1.0 & 0 & 0.02 & 0 & 0.06 & 0.03 & 0.19 \\ \hline
BP III (NH)& $-9.9$ & 0.27 & 0.01 & 0 & 0.07 & 0 & 0.13 & $-0.007$ & $-1.0$ & 0 & $-0.036$ &0.6 \\
     \hline \hline
\end{tabular}
\caption{Values of parameters chosen for different sets of Yukawa structure given in Eqs.~\eqref{Ymat1}-\eqref{Ymat3} to fit the neutrino oscillation data. }
\label{nuTab}
\end{table}

\color{black}
\begin{table}[t!]
\centering
\begin{tabular}{|c|c|c|c|c|}
\hline \hline
 {\textbf{Oscillation}}   &   {\textbf{$3\sigma$ allowed range}} &  \multicolumn{3}{c|}{\textbf{Model prediction}}\\
 \cline{3-5}
 {\bf parameters} & \bf{from {\tt NuFit4}}~\cite{Esteban:2018azc} & BP I (IH) &  BP II (IH) &  BP III (NH) \\ \hline \hline
 $\Delta m_{21}^2 (10^{-5}~{\rm eV}^2$)  &   6.79 - 8.01   &  7.388 & 7.392  & 7.390 \\ \hline
 $\Delta m_{23}^2 (10^{-3}~{\rm eV}^2) $(IH) &   2.412 - 2.611  & 2.541 &  2.488 & - \\ 
 $\Delta m_{31}^2 (10^{-3}~{\rm eV}^2) $(NH) &   2.427 - 2.625   & - & -  & 2.505 \\ \hline
  $\sin^2{\theta_{12}}$   &   0.275 - 0.350 &   0.295 &  0.334 & 0.316\\ \hline
 $\sin^2{\theta_{23}}$  (IH) &   0.423 - 0.629  &   0.614 &  0.467 & - \\
 $\sin^2{\theta_{23}}$ (NH)  &   0.418 - 0.627  &  -  &  - & 0.577 \\ \hline 
  $\sin^2{\theta_{13}} $  (IH) &   0.02068 - 0.02463  &   0.0219 &  0.0232 & - \\
  $\sin^2{\theta_{13}}  $(NH)  &   0.02045 - 0.02439  &  -  &-   & 0.0229  \\\hline \hline
\end{tabular}
\caption{$3\sigma$ allowed ranges of the neutrino oscillation parameters from a recent global-fit~\cite{Esteban:2018azc} (without NSI), along with the model predictions for each BP.}
\label{nuTab2}
\end{table}
%%%%%%%%%%
\begin{figure}[t!]
$$
\includegraphics[height=5cm,width=0.49\textwidth]{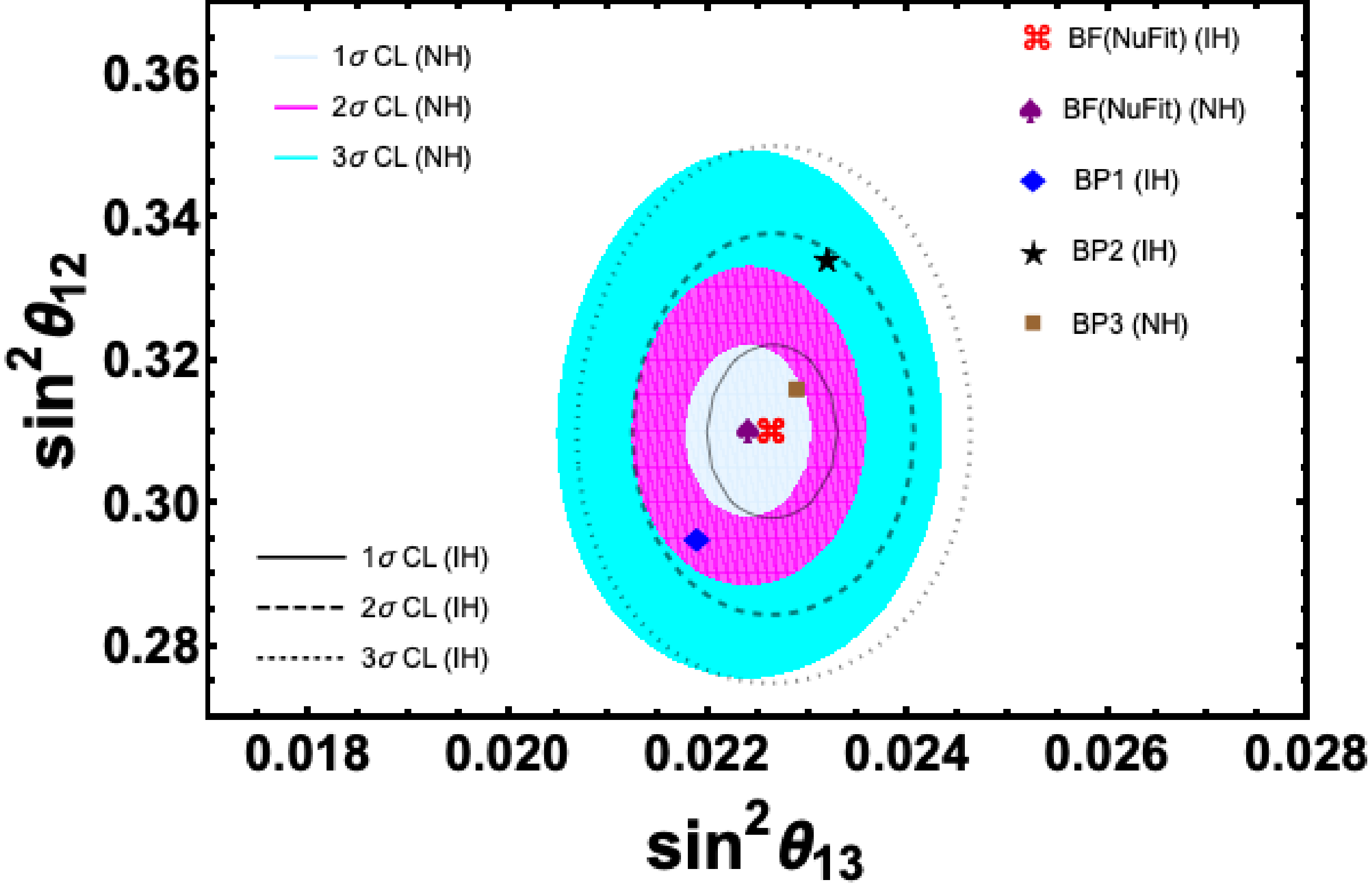}\hspace{2mm}
\includegraphics[height=5cm,width=0.49\textwidth]{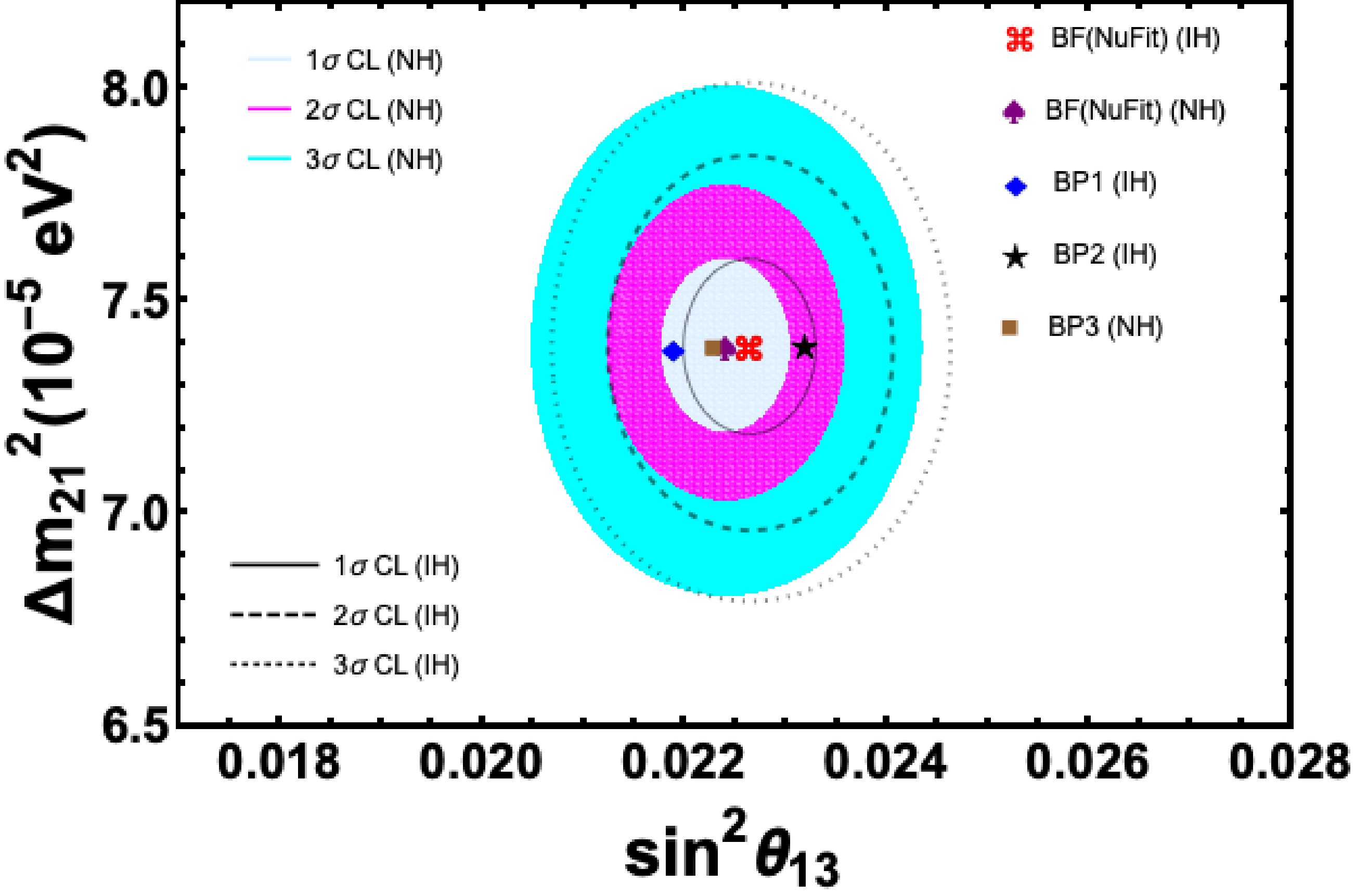}
$$
$$
\includegraphics[height=5cm,width=0.49\textwidth]{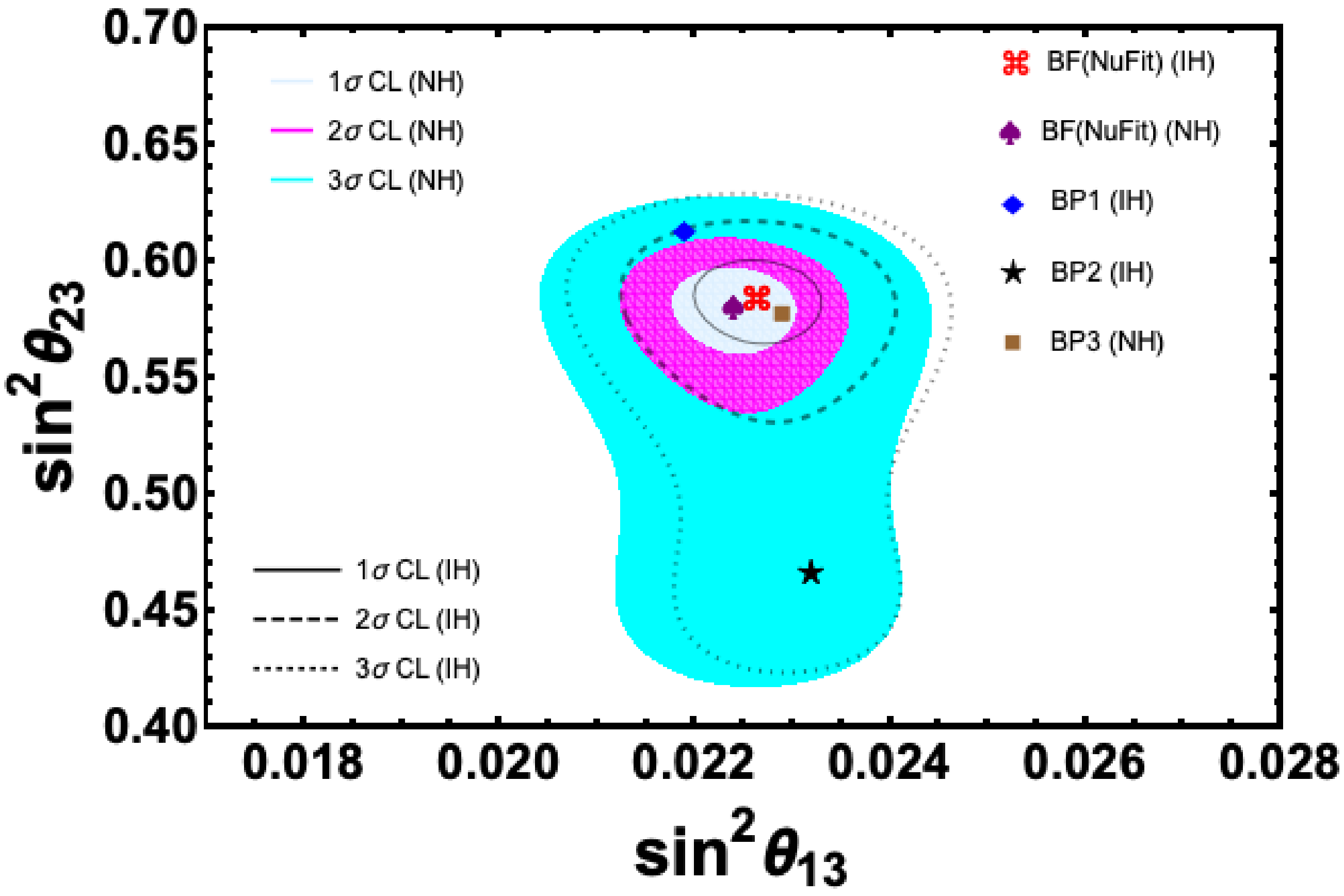}\hspace{2mm}
\includegraphics[height=5cm,width=0.49\textwidth]{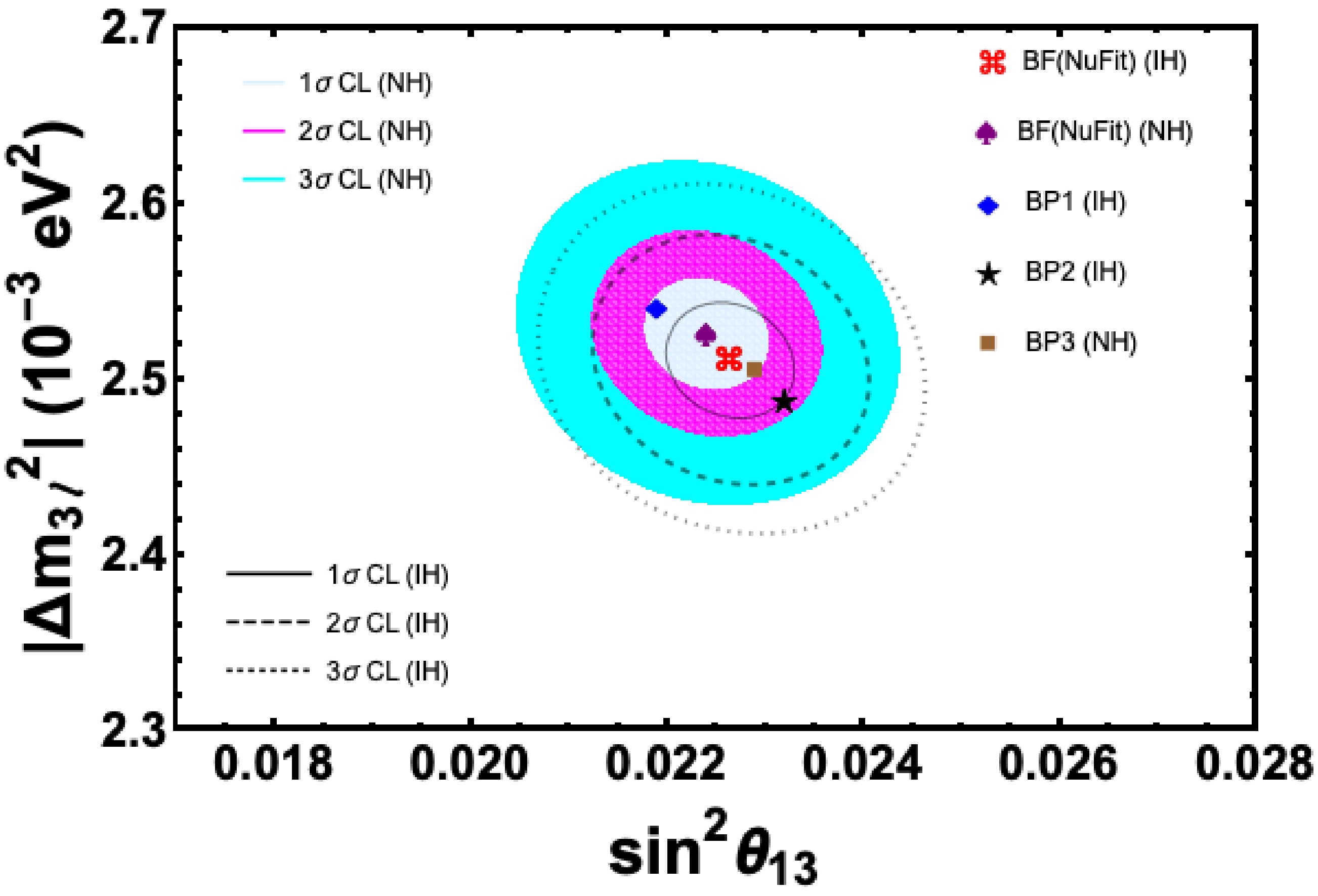}
$$
\caption{Global oscillation analysis obtained from {\tt NuFit4}~ \cite{Esteban:2018azc} for both Normal hierarchy (NH) and Inverted hierarchy (IH) compared with our model benchmark points (BP1, BP2, BP3). Gray, Magenta, and Cyan colored contours represent $1\sigma, 2\sigma, \text{and} \, 3\sigma $ CL contours for NH, whereas solid, dashed, and dotted lines respectively correspond to $1\sigma, 2\sigma, \text{and} \, 3\sigma $ CL contours for IH. Red, purple, and (blue, black, brown) markers are respectively best-fit from NuFit for IH and NH, and benchmark points I, II and III for Yukawa structures given in Eqs.~\eqref{Ymat1}-\eqref{Ymat3}. }
    \label{nufit}
\end{figure}
%%%%%%%%%%%%%%%%%%%%%%%%%%%%%%%
%%%%%%%%%%%%%%%%%%%%%%%%%%%%%%%%

%%%%%%%%%%%
In addition to the best fit results in the tabulated format, we also display them in Fig.~\ref{nufit} in the two-dimensional projections of $1\sigma$, $2\sigma$  and $3\sigma$  confidence regions of the global-fit results~\cite{Esteban:2018azc} (without inclusion of the Super-K atmospheric $\Delta \chi^2$-data). Colored regions (grey, magenta, cyan) are for normal hierarchy,  whereas regions enclosed by solid, dashed, dotted lines are for inverted hierarchy. The global-fit best-fit points, along with the model predictions for each benchmark point, are shown for comparison. It is clear that the theoretical predictions are within the observed 3$\sigma$ range in each case. 

%

%%%%%%%%%%%%%%%%%%%%%%%%%%%%%%%%%%%%%%%%%%%%%%%
\section{NSI in one-loop leptoquark model} \label{sec:LQ}

There are only four kinds of scalar LQs that can interact with the neutrinos at the renormalizable level in the SM  (see Table~\ref{tab:glossory}): $Ld^c\Omega$, $LQ\chi^\star$, $LQ\bar{\rho}$ and $Lu^c\delta$.\footnote{The LQ fields $\Omega, \, \chi^\star, \, \bar{\rho},\, \delta$ are often denoted as $S_1, \, S_3, \, R_2, \, \widetilde{R}_2$ respectively~\cite{Buchmuller:1986zs}.}   
%\begin{align}
%    \overline{Q_L^c}\chi^\star L \, , \quad 
%    \overline{Q_L^c} \bar{\rho}L \, , \quad 
%    \bar{u}_R \eta L \, , \quad {\rm and}~~  
%    \bar{d}_R \Omega L \, ,
%\end{align}
%where the quantum numbers of the relevant LQs under the SM gauge group are given in .
%\begin{align}
%    \chi^\star\left(\overline{\bf 3},{\bf 1},+\frac{1}{3}\right) \, ,\quad 
%    \bar{\rho}\left(\overline{\bf 3},{\bf 3},\frac{1}{3}\right) \, , \quad
 %   \eta\left({\bf 3},{\bf 2},\frac{7}{6}\right)\, , \quad 
 %   \Omega\left({\bf 3},{\bf 2},\frac{1}{6}\right) \, .
%\end{align}
In this section and next, we discuss neutrino mass models with various combinations of these LQs. Our focus is again the range of neutrino NSI that is possible in these models. We note in passing that all these scalar LQ scenarios have gained recent interest in the context of semileptonic $B$-decay anomalies, viz., $R_{D^{( \star)}}$ and $R_{K^{( \star)}}$ (see e.g.,~\cite{Aebischer:2019mlg}). But it turns out that none of these scalar LQ models can simultaneously explain both $R_{D^{( \star)}}$ and $R_{K^{( \star)}}$~\cite{Angelescu:2018tyl}.

We start with a LQ variant of the Zee model that generates small neutrino masses at one-loop level, via the operator is ${\cal O}_{3b}$ (cf.~Eq.~(\ref{eq:O3})). It turns out that ${\cal O}_{3b}$ will induce neutrino masses at one-loop, while $\mathcal{O}_{3a}$, owing to the $SU(2)_L$ index structure, will induce $m_\nu$ at the two-loop level.  A UV complete model of $\mathcal{O}_{3a}$ will be presented in Sec. \ref{subsec:color2}. More precisely, the model of this section corresponds to $\mathcal{O}_3^8$ of Table~\ref{tab:O3}, which involves two LQ fields and no new fermions.  All other realizations of $\mathcal{O}_3$ will be analyzed in subsequent sections.  

The phenomenology of the basic LQ model generating $\mathcal{O}_3^8$ will be analyzed in detail in this section, and the resulting maximum neutrino NSI will be obtained.  The constraints that we derive here on the model parameters can also be applied, with some modifications, to the other $\mathcal{O}_3$ models, as well as other one-loop, two-loop and three-loop LQ models discussed in subsequent sections. 

To realize operator $\mathcal{O}_{3b}$  the $SU(2)_L$ doublet and singlet scalars of the Zee model \cite{Zee:1980ai} are replaced by $SU(2)_L$ doublet and singlet LQ fields.  
This model has been widely studied in the context of $R$-parity breaking supersymmetry, where the LQ fields are identified as the $\widetilde{Q}$ and $\widetilde{d^c}$ fields of the MSSM \cite{Hall:1983id,Barbier:2004ez,Dreiner:1997uz}.  For a non-supersymmetric description and analysis of the model, see Ref. \cite{AristizabalSierra:2007nf}.  

The gauge symmetry of the model denoted as $\mathcal{O}_3^8$ is the same as the SM: $SU(3)_c \times SU(2)_L \times U(1)_Y$. In addition to the SM Higgs doublet $H\left({\bf 1},{\bf 2},\frac{1}{2}\right)$, two $SU(3)_c$ triplet LQ fields 
    $ \Omega \left({\bf 3},{\bf 2}, \frac{1}{6}\right)  =  \left(
           \omega^{2/3}, 
           \omega^{-1/3} \right)$ and 
        $\chi^{-1/3} \left({\bf 3},{\bf 1},-\frac{1}{3}\right)$
        %\,, 
%\end{equation}
are introduced. 
%where the numbers in superscript denote the electric charge. 
%%%%%%%%%%%%%%%%%%%%%%%%%%%%       
The Yukawa Lagrangian relevant for neutrino mass generation in the model is given by
\begin{eqnarray}
    \mathcal{L}_{Y} & \ \supset \ & \lambda_{\alpha\beta} L_\alpha^i d_{\beta}^c \Omega^j \epsilon_{ij}+ \lambda_{\alpha\beta}' L_\alpha^i Q_\beta^j \chi^{\star} \epsilon_{ij} + {\rm H.c.} \nonumber \\
    & \ \equiv \ & \lambda_{\alpha\beta}\left(\nu_{\alpha } d_{\beta }^c  \omega^{-1/3} - \ell_{\alpha } d_{\beta }^c  \omega^{2/3} \right) + \lambda_{\alpha\beta}' \left(\nu_{\alpha } d_{\beta }  - \ell_{\alpha } u_{\beta } \right) \chi^{\star} + {\rm H.c.} \, 
    \label{lagLQ}
\end{eqnarray}
Here $\{\alpha,\beta\}$ are family indices and \{$i,j$\} are $SU(2)_L$ indices as before. 
As in the Zee model, a cubic scalar coupling is permitted, given by
\begin{equation}
     V \ \supset \ \mu H^\dagger \Omega \chi^{\star}+ {\rm H.c.} \ \equiv \ \mu \left(\omega^{2/3} H^-  + \omega^{-1/3} \overline{H}^0\right) \chi^{\star}  + {\rm H.c.} \, 
     \label{pot_lepto}
\end{equation}
which ensures lepton number violation.

Once the neutral component of the SM Higgs doublet acquires a VEV, the cubic term in the scalar potential~\eqref{pot_lepto} will generate mixing between the  $\omega^{-1/3}$ and $\chi^{-1/3}$ fields, with the mass matrix given by:
\begin{equation}
       M_{\rm LQ}^2 \ = \ \begin{pmatrix}
          m_\omega^2 & \mu v/\sqrt{2} \\
           \mu^\star v/ \sqrt{2} & m_\chi^2 
        \end{pmatrix},
\end{equation}
where $m_\omega^2$ and $m_\chi^2$ include the bare mass terms plus a piece of the type $\lambda v^2$ arising from the SM Higgs VEV.  The physical states are denoted as $\{ X_1^{-1/3},X_2^{-1/3} \}$, defined as
\begin{eqnarray}
    X_1 &\ = \ & \cos\alpha\, \omega + \sin\alpha\,  \chi \, , \nonumber \\
    X_2 & \ = \ & -\sin\alpha \,\omega + \cos\alpha\, \chi \, ,
    \label{eq:X12}
\end{eqnarray}
with the mixing angle given by 
\begin{equation}
    \tan 2\alpha \ = \ \frac{-\sqrt{2} \, \mu v}{m_\chi^2 - m_\omega^2}\,.
    \label{eq:tan2a}
\end{equation}
The squared mass eigenvalues of these states are:
\begin{equation}
    m_{1,2}^2 \ = \ \frac{1}{2} \left[ m_\omega^2 + m_\chi^2 \mp \sqrt{(m_\omega^2 - m_\chi^2)^2 + 4 \mu^2 v^2} \right] \,.
    \label{eq:mX12}
\end{equation}
%%%%%%%%%%%%%%%%%%%%%%%%%
\begin{figure}[t!]
    \centering
    \includegraphics[scale=0.5]{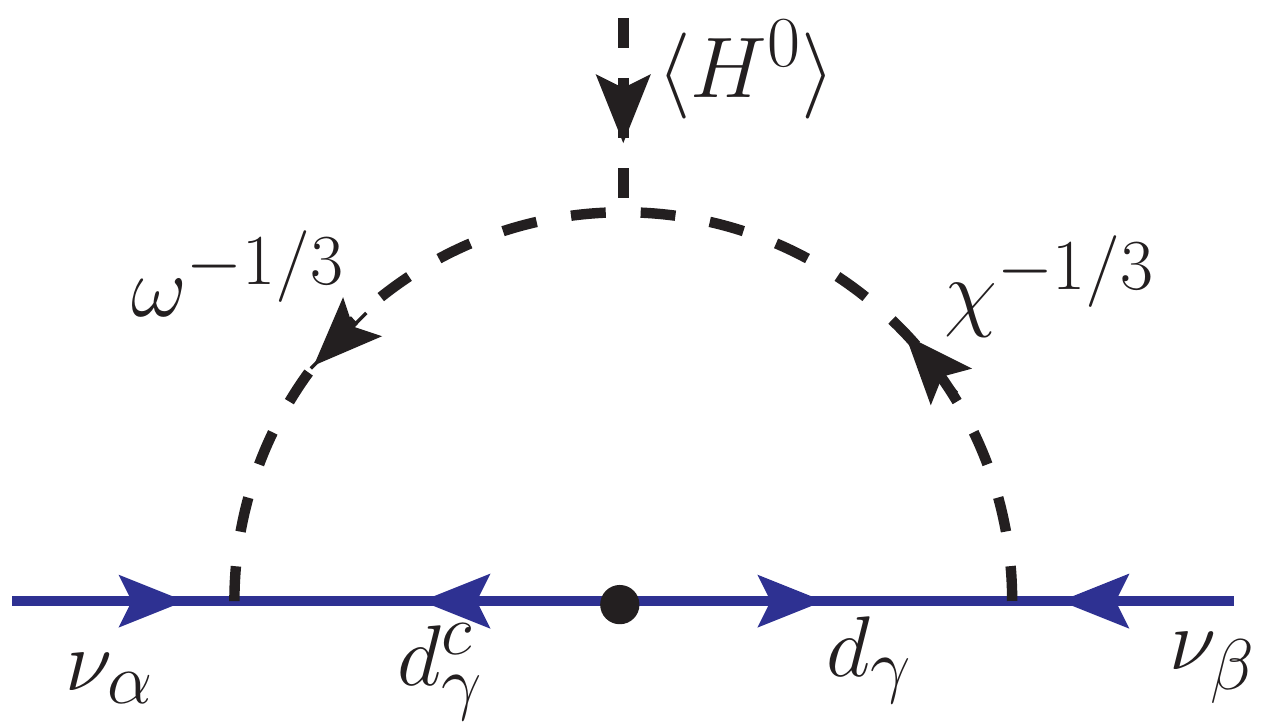}
    \caption{One-loop diagram inducing neutrino mass in the LQ 
    model. This is the model ${\cal O}_3^8$ of Table~\ref{tab:O3}. In SUSY models with $R$-parity violation, $\omega^{-1/3}$ is identified as $\tilde{d}$ and $\chi^{\star 1/3}$ as $\tilde{d^c}$.}
    \label{coloredzee}
\end{figure}

Neutrino masses are induced via the one-loop diagram shown in Fig.~\ref{coloredzee}. The mass matrix is given by:
\begin{equation}
    \textcolor{black}{M_\nu \ = \ \frac{3 \sin 2\alpha}{32 \pi^2} \log \left(\frac{m_1^2}{m_2^2}\right) (\lambda M_d \lambda'^T + \lambda' M_d \lambda^T)}\,.
    \label{eq:Mnucz}
\end{equation}
Here $M_d$ is the diagonal down-type quark mass matrix.  
Acceptable neutrino masses and mixing can arise in the model for a variety of parameters.  Note that the induced $M_\nu$ is proportional to the down-quark masses, the largest being $m_b$. In the spirit of maximizing neutrino NSI, which are induced by either the $\omega^{-1/3}$ or the $\chi^{-1/3}$ field, without relying on their mixing, we shall adopt a scenario where the couplings $\lambda_{\alpha\beta}$ are of order one, while $\lambda'_{\alpha\beta} \ll 1$.  Such a choice would realize small neutrino masses.  One could also consider $\lambda' \sim {\cal O}(1)$, with $\lambda \ll 1$ as well.  However, in the former case, there is a GIM-like suppression in the decay rate for $\ell_\alpha\rightarrow \ell_\beta +\gamma$ \cite{Babu:2010vp}, which makes the model with $\lambda \sim {\cal O}(1), \lambda' \ll 1$ somewhat less constrained from cLFV, and therefore we focus on this scenario.  The reason for this suppression will be elaborated in Sec.~\ref{sec:llgLQ}.

 %%%%%%%%%%%%%%%%%%%%%%%%%%
 \subsection{Low-energy constraints} \label{sec:lowconstraints}
 
 One interesting feature of the LQ model presented in this section is that the radiative decay $\ell_\alpha \rightarrow \ell_\beta + \gamma$ is suppressed in the model due to a GIM-like cancellation. On the other hand, $\mu-e$ conversion in nuclei gives a stringent constraint on the Yukawa couplings of the model, as do the trilepton decays of the lepton to some extent.  Since the product $|\lambda \lambda' |\ll 1$ in order to generate the correct magnitude of the neutrino masses (cf.~Eq.~(\ref{eq:Mnucz})), we shall primarily consider the case where $|\lambda'| \ll 1$ with $|\lambda|$ being of order one.  This is the case where the constraints from radiative decays are nonexistent. If on the other hand, $|\lambda| \ll 1$ and $|\lambda'|$ is of order unity, then these radiative decays do provide significant constraints.  This situation will be realized in other LQ models as well; so we present constraints on the model of this section in this limit as well.  The processes that are considered are: $\ell_\alpha \to \ell_\beta + \gamma$, $\mu-e$ conversion in nuclei, $\ell_\alpha \to \bar{\ell}_\beta \ell_\gamma \ell_\delta$ (with at least two of the final state leptons being of same flavor), $\tau \to \ell \pi$, $\tau \to \ell\eta$, $\tau \to \ell \eta'$ (where $\ell = e$ or $\mu)$, and APV.
%%%%%%%%%%%%%%%%%%%%%%%%%
\subsubsection{Atomic parity violation}  \label{sec:APV}
\begin{figure}[t!]
    \centering
    \subfigure[]{
        \includegraphics[scale=0.45]{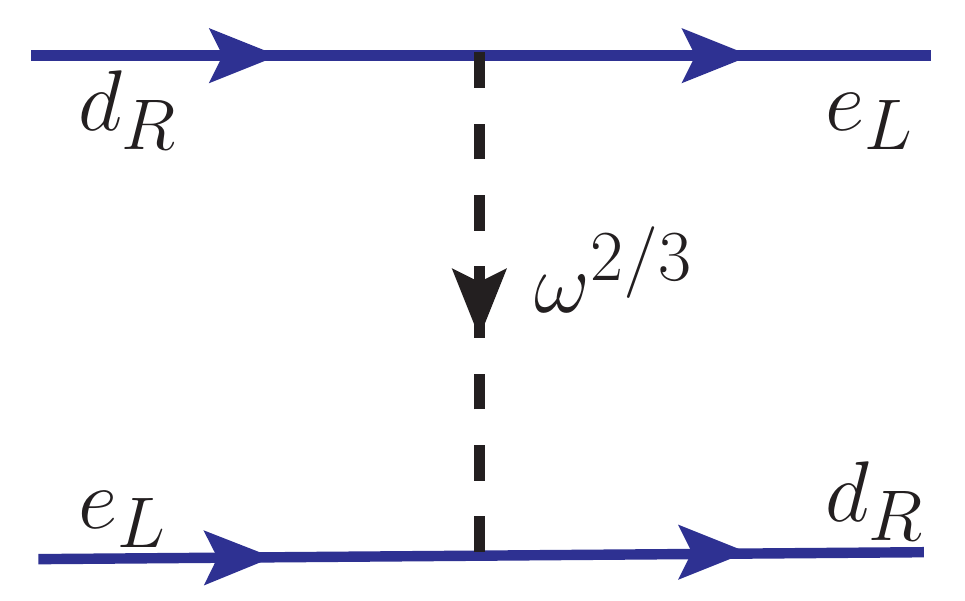}}
        \hspace{5mm}
        \subfigure[]{
        \includegraphics[scale=0.45]{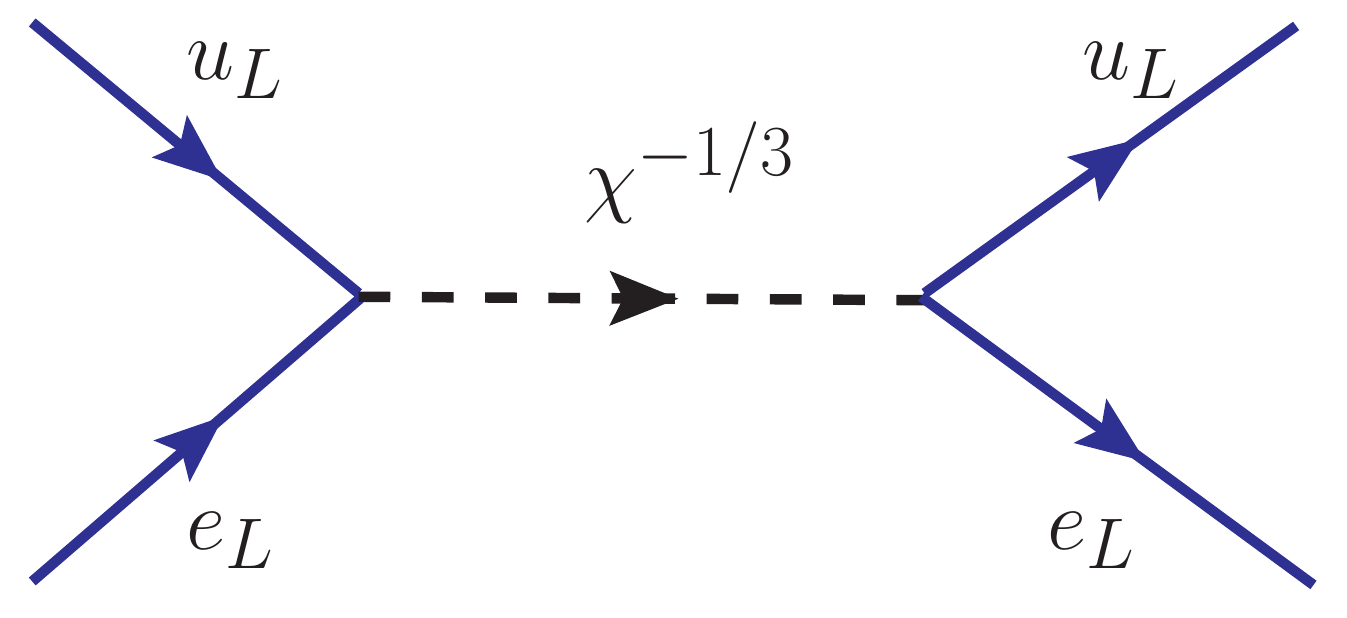}}
    \caption{Doublet and singlet LQ contribution to APV at tree-level.}
    \label{atomic_parity}
 \end{figure}

 The strongest constraints on the $\lambda_{ed}$ and $\lambda'_{ed}$ couplings come from atomic parity violation (APV)~\cite{Roberts:2014bka}, analogous to the $R$-parity violating supersymmetric case~\cite{Kao:2009fg}. The diagrams shown in Fig.~\ref{atomic_parity} lead to the following effective couplings between up/down quarks and electrons: 
 \begin{align}
 {\cal L}_{\rm eff} \ = \  & \frac{|\lambda_{ed}|^2}{m_\omega^2}\left(\bar{e}_L d_R \right)\left(\bar{d}_R e_L\right)+ \frac{|\lambda'_{ed}|^2}{m_\chi^2}\left(\overline{e_L^c} u_L\right)\left(\bar{u}_L e_L^c\right) \nonumber \\
 \ = \ & -\frac{1}{2}\frac{|\lambda_{ed}|^2}{m_\omega^2}\left(\bar{e}_L \gamma^\mu e_L \right)\left(\bar{d}_R \gamma_\mu d_R \right) 
 +\frac{1}{2}\frac{|\lambda'_{ed}|^2}{m_\chi^2}\left(\bar{e}_L \gamma^\mu e_L\right)\left(\bar{u}_L \gamma_\mu u_L\right) \, ,
 \label{eq:LeffLQ}
 \end{align}
 where we have used the Fierz transformation in the second step. The parity-violating parts of these interactions are given by 
 \begin{align}
 {\cal L}_{\rm eff}^{\rm PV} \ = \ & \frac{1}{8}\frac{|\lambda_{ed}|^2}{m_\omega^2}\left[\left(\bar{e} \gamma^\mu \gamma^5 e \right)\left(\bar{d} \gamma_\mu d \right) -\left(\bar{e}\gamma^\mu e\right)\left(\bar{d}\gamma_\mu \gamma^5 d\right)\right] \nonumber \\
& \quad  -\frac{1}{8}\frac{|\lambda'_{ed}|^2}{m_\chi^2}\left[\left(\bar{e} \gamma^\mu \gamma^5 e\right)\left(\bar{u} \gamma_\mu u\right) +\left(\bar{e}\gamma_\mu e\right)\left(\bar{u}\gamma_\mu \gamma^5 u\right) \right]\, .
\label{eq:PV}
 \end{align}
 On the other hand, the parity-violating SM interactions at tree-level are given by 
 \begin{align}
 {\cal L}_{\rm SM}^{\rm PV} \ = \ \frac{G_F}{\sqrt 2}\sum_{q=u,d} \left[C_{1q}\left(\bar{e}\gamma^\mu \gamma^5 e\right)\left(\bar{q}\gamma_\mu q\right)+C_{2q}\left(\bar{e}\gamma^\mu e\right)\left(\bar{q}\gamma_\mu \gamma^5 q\right)\right] \, ,
 \end{align}
 with 
 \begin{align}
 & C_{1u} \ = \ -\frac{1}{2}+\frac{4}{3}s^2_w \, , \qquad C_{2u} \ = \ -\frac{1}{2}+2s^2_w \, ,\nonumber \\
& C_{1d} \ = \ \frac{1}{2}-\frac{2}{3}s^2_w \, , \qquad C_{2d} \ = \ \frac{1}{2}-2s^2_w \, .
 \end{align}
Correspondingly, the weak charge of an atomic nucleus with $Z$ protons and $N$ neutrons is given by 
\begin{align}
Q_w(Z,N) \ = \ -2\left[C_{1u}(2Z+N)+C_{1d}(Z+2N)\right] \ = \ (1-4s^2_w)Z-N \, ,
\end{align}
where $(2Z+N)$ and $(Z+2N)$ are respectively the number of up and down quarks in the nucleus. The presence of the new PV couplings in Eq.~\eqref{eq:PV} will shift the weak charge to 
\begin{align}
\delta Q_w(Z,N) \ = \ \frac{1}{2\sqrt 2 G_F} \left[(2Z+N)\frac{|\lambda'_{ed}|^2}{m_\chi^2}-(Z+2N)\frac{|\lambda_{ed}|^2}{m_\omega^2}\right] \, .
\label{eq:Qw}
\end{align}

There are precise experiments measuring APV in cesium, thallium, lead and bismuth~\cite{Safronova:2017xyt}. The most precise measurement comes from cesium (at the 0.4\% level~\cite{Porsev:2009pr}), so we will use this to derive constraints on LQ. 
For $_{55}^{133}$Cs, Eq.~\eqref{eq:Qw} becomes 
\begin{align}
\delta Q_w\left(_{55}^{133}{\rm Cs}\right) \ = \ \frac{1}{2\sqrt 2 G_F} \left(188\frac{|\lambda'_{ed}|^2}{m_\chi^2}-211\frac{|\lambda_{ed}|^2}{m_\omega^2}\right) \, .
\label{eq:Qw1}
\end{align}
Taking into account the recent atomic structure calculation~\cite{Roberts:2014bka}, the experimental value of the weak charge of $_{55}^{133}$Cs is given by~\cite{Tanabashi:2018oca}
\begin{align}
Q^{\rm exp}_w\left(_{55}^{133}{\rm Cs}\right) \ = \ -72.62\pm 0.43 \, ,
\label{eq:Qwexp}
\end{align}
whereas the SM prediction is ~\cite{Roberts:2014bka, Tanabashi:2018oca}
\begin{align}
Q^{\rm SM}_w\left(_{55}^{133}{\rm Cs}\right) \ = \ -73.23\pm 0.02 \, ,
\label{eq:Qwsm}
\end{align}
based on a global-fit to all electroweak observables with radiative corrections. Assuming new radiative corrections from LQ are small and saturating the difference between Eqs.~\eqref{eq:Qwexp} and \eqref{eq:Qwsm}, we obtain a $2\sigma$ allowed range of $\delta Q_w$:
\begin{align}
    -0.29 \ < \ \delta Q_w \ < \ 1.51 \, .
    \label{eq:delqw}
\end{align} 
Comparing this with Eq.~\eqref{eq:Qw1}, we obtain the corresponding $2\sigma$ bounds on $\lambda_{ed}$ and $\lambda'_{ed}$ as a function of the LQ mass as follows:
\begin{align}
|\lambda_{ed}| \ < \ 0.21\left(\frac{m_\omega}{{\rm TeV}}\right) \, , \qquad 
|\lambda'_{ed}| \ < \ 0.51\left(\frac{m_\chi}{{\rm TeV}}\right) \, .
\label{eq:APV}
\end{align}
The APV constraint on down-quark coupling of the LQ is stronger than the up-quark coupling constraint due to the fact that the experimental value of $Q_w$ (cf.~Eq.~\eqref{eq:Qwexp}) is $1.5\sigma$ larger than the SM prediction (cf.~Eq.~\eqref{eq:Qwsm}), while the doublet LQ contribution to $Q_w$ goes in the opposite direction (cf.~Eq.~\eqref{eq:Qw1}).  
%This is also summarized in Table~\ref{tab:constraint_LQ}. 

%\begin{table}[]
%    \centering
 %    \begin{tabular}{|c|c|c|}
%\hline \hline
%    \textbf{Process } & \textbf{Exp. bound} &  \textbf{Constraint}  \\ \hline
%     APV & $\left|\delta Q_w\left(_{55}^{133}{\rm Cs}\right) \right| < 0.76$~\cite{Porsev:2009pr} & $|\lambda_{ed}| < 0.3 \left(\frac{m_\omega}{1~\text{TeV}}\right)$  \\
%      & & $|\lambda'_{ed}| < 0.5 \left(\frac{m_\chi}{1~ \text{TeV}}\right)$ \\ \hline
   %  $\mu \to e^+ e^- e^-$   &  $|\lambda_{13} \lambda_{23}| < 7.6 \times 10^{-3} $ \\
    % & \\\hline
%  $\mu - e$ conversion & \cite{Bertl:2006up}   &  $|\lambda_{e d} \lambda_{\mu d}^\star| < 3.3 \times 10^{-7} \left(\frac{m_\omega}{1000 \text{GeV}}\right)^2$   
%\\ \hline
% \end{tabular}
%    \caption{Constraints on Yukawa couplings and LQ masses from APV (Sec.~\ref{sec:APV}), $\mu-e$ conversion (Sec.~\ref{sec:mueconv}), $\ell_\alpha\to \bar{\ell}_\beta\ell_\gamma \ell_\delta$ (Sec.~\ref{sec:muto3e}) and $\ell_\alpha\to \ell_\beta\gamma$ (Sec.~\ref{sec:llgLQ}).}
%    \label{tab:constraint_LQ}
%\end{table}
%%%%%%%%%%%%%%%%%%%%%%
 \subsubsection{\texorpdfstring{$\mu-e$}{mue} conversion} \label{sec:mueconv}
 %%%%%%%%%%%%%%%%%%%%%%%%%%%
 \begin{figure}[!t]
     \centering
     \includegraphics[scale=0.45]{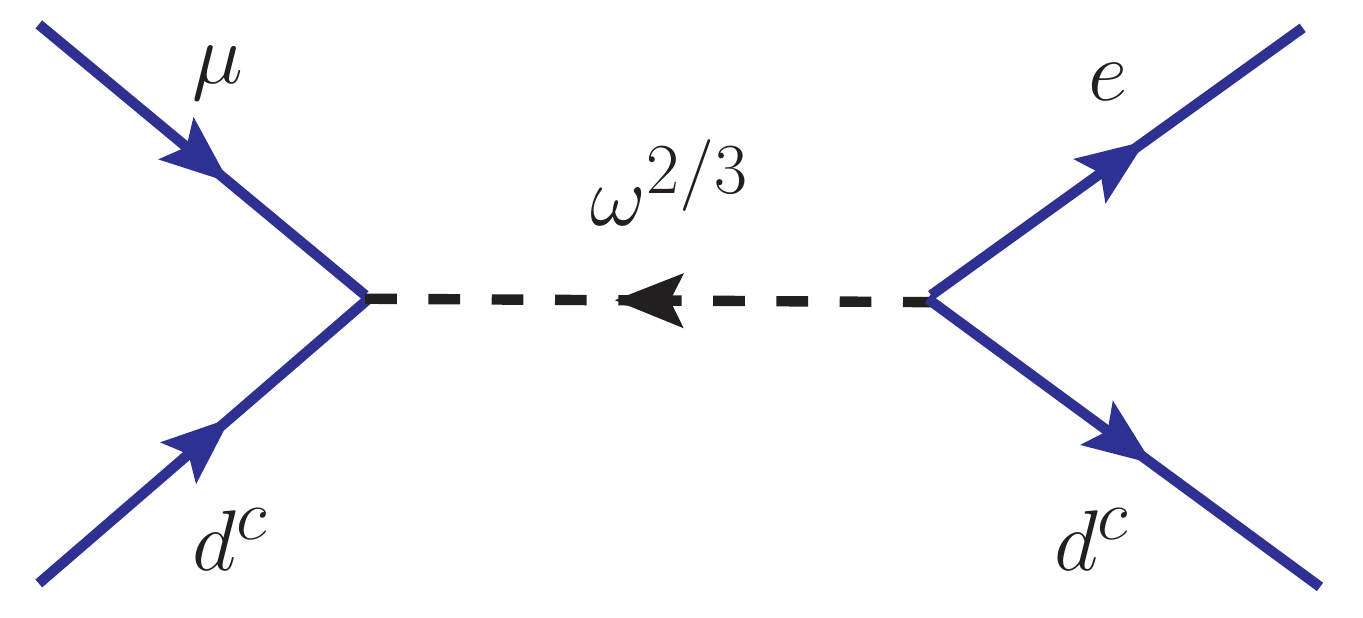}\hspace{2mm}
     \includegraphics[scale=0.45]{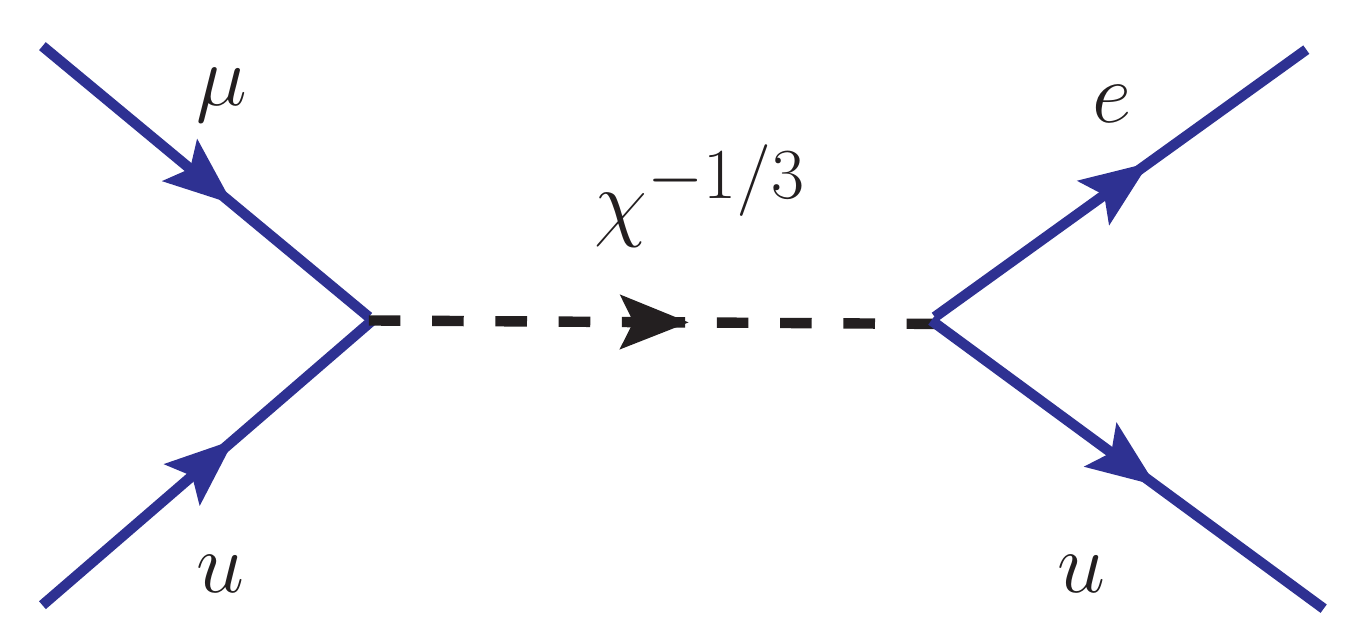}
     \caption{Feynman diagrams leading to $\mu-e$ conversion at tree-level in the doublet-singlet LQ model.}
     \label{fig:feynmutoeLQ}
 \end{figure}
 %%%%%%%%%%%%%%%%%%%%%%%%%%%%
 Another constraint on the LQ model being discussed comes from the cLFV process of coherent $\mu - e $ conversion in nuclei ($\mu N \to e N)$. We will only consider the tree-level contribution as shown in Fig.~\ref{fig:feynmutoeLQ}, since the loop-level contributions are sub-dominant.  Following the general procedure described in Ref.~\cite{Kuno:1999jp}, we can write down the branching ratio for this process as~\cite{Babu:2010vp}
 \begin{align}
 {\rm BR}(\mu N\to eN) \ \simeq \  \frac{|\vec{p}_e|E_e m_\mu^3 \alpha^3Z^4_{\rm eff}F_p^2}{64\pi^2 Z\Gamma_N}(2A-Z)^2\left(\frac{|\lambda^\star_{ed}\lambda_{\mu d}|}{m^2_\omega}+\frac{|\lambda^{\prime\star}_{ed}\lambda'_{\mu d}|}{m^2_\chi}\right)^2 \, , 
 \label{eq:mueconv}
 \end{align}
 where $\vec{p}_e$ and $E_e$ are the momentum and energy of the outgoing electron respectively, $Z$ and $A$ are the atomic number and mass number of the nucleus respectively, $Z_{\rm eff}$ is the effective atomic number, $F_p$  is the nuclear matrix element, and $\Gamma_N$ is the muon capture rate of the nucleus. Here we take $|\vec{p}_e|\simeq E_e\simeq m_\mu$ and use the values of $Z_{\rm eff}$ and $F_p$ from Ref.~\cite{Kitano:2002mt}, and the value of $\Gamma_N$ from Ref.~\cite{Suzuki:1987jf}. Comparing the model predictions from Eq.~\eqref{eq:mueconv} with the experimental limits for different nuclei~\cite{Kaulard:1998rb, Bertl:2006up, Honecker:1996zf}, we obtain the constraints on the Yukawa couplings (either $\lambda$ or $\lambda'$) and LQ mass as shown in Table~\ref{tab:mueconvLQ}. 
 \begin{table}[t!]
 \small
    \centering
    \begin{tabular}{|c|c|c|c|c|c|}
    \hline\hline
        {\bf Nucleus} & {\bf Experimental } & $Z_{\rm eff}$ & $F_p$ & $\Gamma_N$ \cite{Suzuki:1987jf}& {\bf Constraint} \\ 
        & {\bf Limit} &~\cite{Kitano:2002mt}&\cite{Kitano:2002mt} & ~$(10^{6}~{\rm s}^{-1})$ & {\bf on} $|\lambda_{ed}^\star \lambda_{\mu d}|$ \\ \hline \hline
       $_{22}^{48}$Ti  & BR $< \ 6.1 \times 10^{-13}$~\cite{Kaulard:1998rb} & 17.6 & 0.54 & 2.59 & $ < 4.30 \times 10^{-6} \left(\frac{m_{\omega}}{\text{TeV}}\right)^2$  \\ \hline
     $_{79}^{197}$Au  & BR $< \ 7.0 \times 10^{-13}$~\cite{Bertl:2006up} & 33.5 & 0.16 & 13.07 & $  < 4.29 \times 10^{-6}\left(\frac{m_{\omega}}{\text{TeV}}\right)^2$\\ \hline
      $_{82}^{208}$Pb  & BR $< \ 4.6 \times 10^{-11}$~\cite{Honecker:1996zf} & 34.0 & 0.15 & 13.45 & $ < 3.56 \times 10^{-5}\left(\frac{m_{\omega}}{\text{TeV}}\right)^2$  \\ \hline\hline
    \end{tabular}
    \caption{Constraints on Yukawa couplings and LQ masses from $\mu-e$ conversion in different nuclei. For $|\lambda_{ed}^{\prime\star} \lambda'_{\mu d}|$, the same constraints apply, with $m_\omega$ replaced by $m_\chi$. }
    \label{tab:mueconvLQ}
\end{table}

\subsubsection{\texorpdfstring{$\ell_\alpha\to \bar{\ell}_\beta\ell_\gamma \ell_\delta$}{ell} decay}  \label{sec:muto3e}

LQs do not induce trilepton decays of the type $\mu \to 3e$ at the tree-level.  However, they do induce such processes at the loop level.  There are LQ mediated $Z$ and photon penguin diagrams, as well as box diagrams.  These contributions have been evaluated for the LQ model of this section in Ref.~\cite{Babu:2010vp}.  With the Yukawa couplings $\lambda$ being of order one, but with $|\lambda'| \ll 1$, the branching ratio for $\mu^- \to e^+e^-e^-$ decay is given by \cite{Babu:2010vp}
\begin{equation}
{\rm BR}(\mu \to 3e) \ = \ \left(\frac{3\sqrt{2}}{32 \pi^2 G_F}\right)^2C_{dd}^L\frac{|\lambda_{ed}\lambda_{\mu d}^\star|^2}{m_\omega^4} \, ,
\label{muto3eLQ}
\end{equation}
where 
\begin{align}
C_{dd}^L  \ = \ & \frac{1}{7776}\Bigg[72 e^4 \left(\log\frac{m_\mu^2}{m_\omega^2}\right)^2-108(3e^4+2e^2|\lambda_{ed}|^2)\log\left(\frac{m_\mu^2}{m_\omega^2}\right)  \nonumber \\
& \qquad \qquad 
+(449+68 \pi^2) e^4+ 486 e^2|\lambda_{ed}|^2+243|\lambda_{ed}|^4
\Bigg]~.
\end{align}
Here we have kept only those couplings that are relevant for neutrino NSI, and we have assumed that there are no accidental cancellations among various contributions. Using BR($\mu \rightarrow 3e)< 1.0 \times 10^{-12}$ \cite{Bertl:1985mw}, we obtain
\begin{equation}
|\lambda_{ed} \lambda_{\mu d}^\star| \ < \ 4.4 \times 10^{-3} \left(\frac{m_\omega}{\rm TeV}\right)^2 \left(1+1.45 |\lambda_{ed}|^2 + 0.81 |\lambda_{ed}|^4\right)^{-1/2}~.
\label{eq:mu3eLQ}
\end{equation}
Analogous constraints from $\tau \to 3e$ and $\tau \to 3 \mu$ are less stringent.  For example, from BR($\tau \to 3e)< 1.4 \times 10^{-8}$~\cite{Amhis:2016xyh}, and using Eq.~(\ref{muto3eLQ}) with a multiplicative factor of BR($\tau \to \bar{\nu}_\ell \ell \nu_\tau) = 0.174$, we obtain
\begin{equation}
|\lambda_{ed} \lambda_{\tau d}^\star| \ < \ 1.2 \left(\frac{m_\omega}{\rm TeV}\right)^2 (1+1.96 |Y_{ed}|^2 + 1.50 |Y_{ed}|^4)^{-1/2}~.
\label{eq:tau3lLQ1}
\end{equation}
Similarly, from BR($\tau \to 3\mu) < 1.2 \times 10^{-8}$~\cite{Amhis:2016xyh} we obtain
\begin{equation}
|\lambda_{\mu d} \lambda_{\tau d}^\star| \ < \ 1.1 \left(\frac{m_\omega}{\rm TeV}\right)^2 (1+1.96 |Y_{\mu d}|^2 + 1.50 |Y_{\mu d}|^4)^{-1/2}~.
\label{eq:tau3lLQ2}
\end{equation}

The constraint on $|\lambda_{ed} \lambda_{\mu d}^\star|$ from the trilepton decay (cf.~Eq.~\eqref{eq:mu3eLQ}) turns out to be weaker than those from $\mu-e$ conversion (cf.~Table~\ref{tab:mueconvLQ}). Similarly, the constraints on $|\lambda_{ed} \lambda_{\tau d}^\star|$ and $|\lambda_{\mu d} \lambda_{\tau d}^\star|$ from the trilepton decay (cf.~Eqs.~\eqref{eq:tau3lLQ1} and \eqref{eq:tau3lLQ2}) turn out to be weaker than those from semileptonic tau decays (cf.~Table~\ref{tab:semilep}).

\subsubsection{\texorpdfstring{$ \ell_\alpha \to \ell_\beta \gamma$}{llj} constraint} \label{sec:llgLQ}
%%%%%%%%%%%%%%%%%%%%%%%%%%%%%%%%%%%%%%%%%%%%%%%
\begin{figure}[!t]
    \centering
    \includegraphics[scale=0.6]{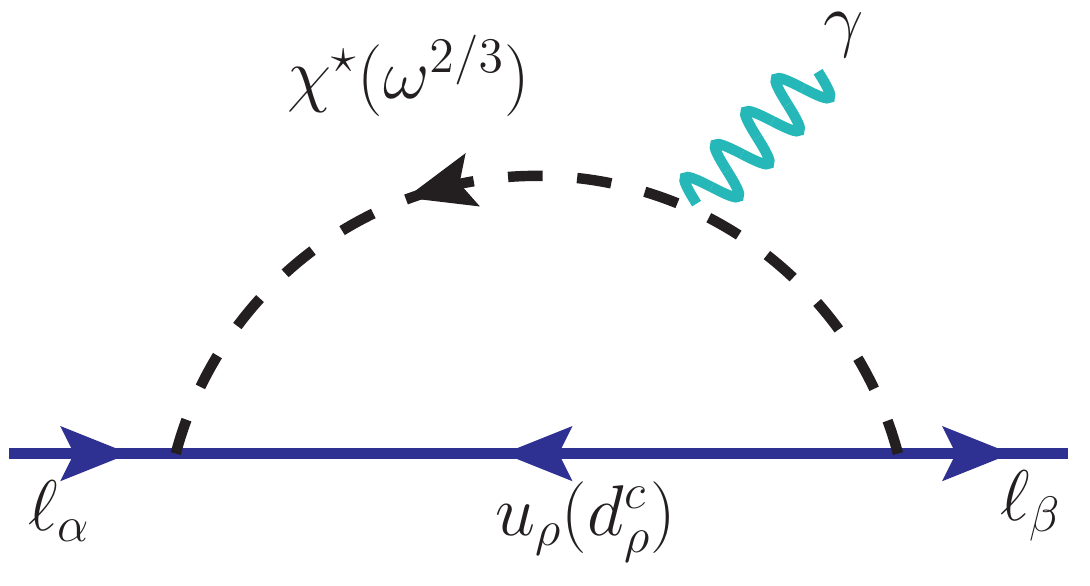}
    \includegraphics[scale=0.6]{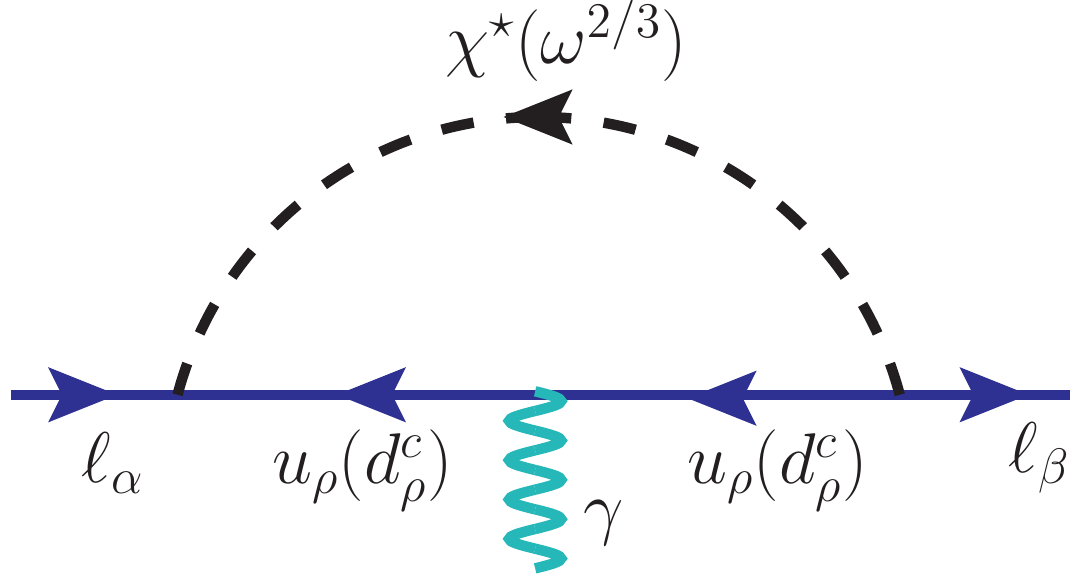}
    \caption{One-loop Feynman diagrams for $\ell_\alpha \to \ell_\beta \gamma$ processes mediated by LQ. }
    \label{fig:llg_LQ}
\end{figure}
The lepton flavor violating radiative decay $\ell_\alpha \to \ell_\beta + \gamma$ arises via one-loop diagrams with the exchange of LQ fields (see Fig.~\ref{fig:llg_LQ}).  These diagrams are analogous to Fig.~\ref{mutoeg}, but with the charged and neutral scalars replaced by LQ scalars. Note that the photon can be emitted from either the LQ line, or the internal fermion line.  It turns out that the LQ Yukawa coupling matrix $\lambda$ leads to suppressed decay rates for $\ell_\alpha \to \ell_\beta + \gamma$, owing to a GIM-like cancellation.  The coupling of the $\omega^{2/3}$ LQ has the form $\ell_{\alpha L}\overline{d_{\beta R}^c} \omega^{2/3}$, which implies that $Q_B = 2/3$ and $Q_F = -1/3$ in Eq.~(\ref{rate_rad}).  Consequently, the rate becomes proportional to a factor which is at most of order $(m_b^2/m_\omega^2)^2$.   Thus,  the off-diagonal couplings of $\lambda$ are unconstrained by these decays.

On the other hand, the $\chi^{-1/3}$ LQ field does mediate $\ell_\alpha \to \ell_\beta + \gamma$  decays, proportional to the Yukawa coupling matrix $\lambda'$.  The relevant couplings have the form $\bar{u}_L \ell_L \chi^\star$, which implies that $Q_F = -2/3$ and $Q_B = 1/3$ in Eq.~(\ref{rate_rad}).  We find the decay rate to be
\begin{equation}
\Gamma(\ell_\alpha \to \ell_\beta + \gamma) \ = \ \frac{9 \alpha}{576}\frac{|\lambda'_{\beta d}\lambda^{\prime\star}_{\alpha d}|^2}{(16\pi^2)^2} \frac{m_\alpha^5}{m_\chi^4}~,
\label{eq:llgLQ}
\end{equation}
where $9=3^2$ is a color factor. Here we have assumed $t=m_F^2/m_B^2\to 0$, since the LQ is expected to be much heavier than the SM charged leptons to satisfy the  experimental constraints. The limits on the products of Yukawa couplings from these decays are listed in Table~\ref{lqlfv1}.

%%%%%%%%%%%%%%%%%%%%%%%%%%%%%%%%%%%%%%%%%%%%%%%
\begin{table}[!htb]
    \centering
    \begin{tabular}{|c|c|c|}
    \hline \hline
        {\bf Process}             & {\bf Exp. limit}                     & {\bf Constraint} \\ \hline \hline
        \rule{0pt}{10pt}  $\mu \to e \gamma$    & BR < 4.2 $\times 10^{-13}$~\cite{TheMEG:2016wtm} & $|\lambda'_{e d} \lambda^{\prime\star}_{\mu d}| $ \textless \, $2.4 \times 10^{-3} \left(\frac{m_\chi}{\text{TeV}}\right)^2$     \\
        \rule{0pt}{15pt}  $\tau \to e \gamma$   & BR < 3.3 $\times 10^{-8}$~\cite{Aubert:2009ag} & $|\lambda'_{e d} \lambda^{\prime\star}_{\tau d}| $ \textless \, $ 1.6 \left(\frac{m_\chi}{\text{TeV}}\right)^2 $      \\
        \rule{0pt}{15pt}  $\tau \to \mu \gamma$ & BR < 4.4 $\times 10^{-8}$~\cite{Aubert:2009ag} &  $|\lambda^{\prime\star}_{\mu d} \lambda'_{\tau d}| $ \textless \, $1.9 \left(\frac{m_\chi}{\text{TeV}}\right)^2    $     \\
   \hline \hline
    \end{tabular}
    \caption{Constraints on the Yukawa couplings $\lambda'$ as a function of the singlet LQ mass from $ \ell_\alpha \to \ell_\beta \gamma$ processes.}
    \label{lqlfv1}
\end{table}
%%%%%%%%%%%%
%%%%%%%%%%%%%%%%%%%%%%%%%%%%%%
\subsubsection{Semileptonic tau decays} \label{sec:tauLQ}
\begin{figure}[!t]
    \centering
    \includegraphics[scale=0.5]{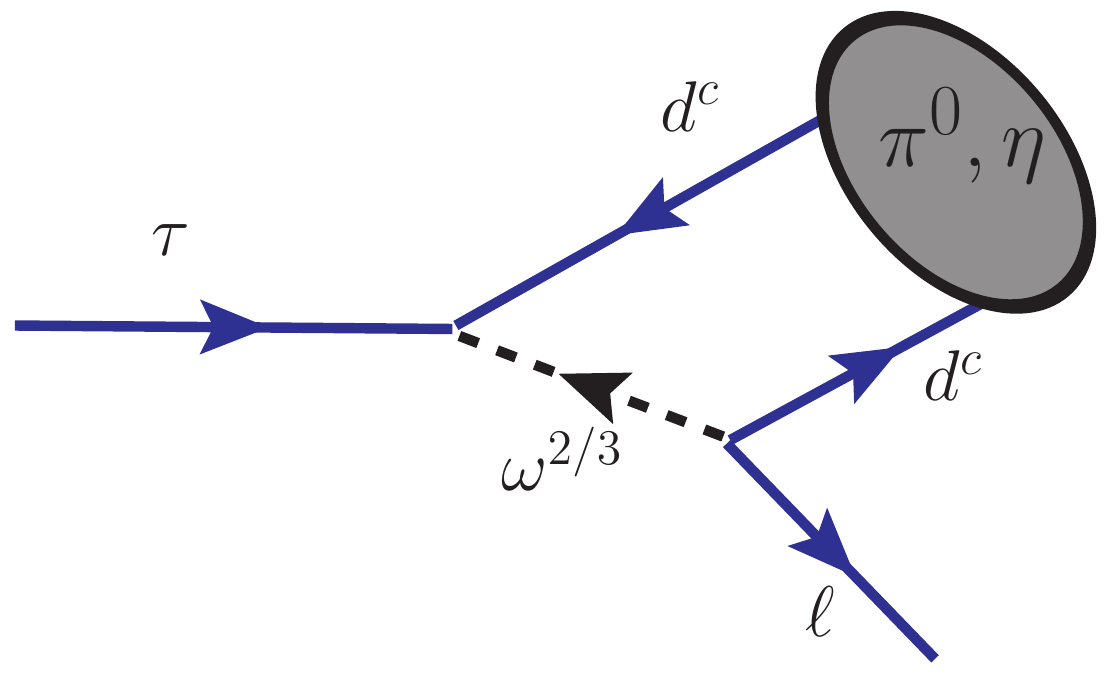} \hspace{7mm}
    \includegraphics[scale=0.5]{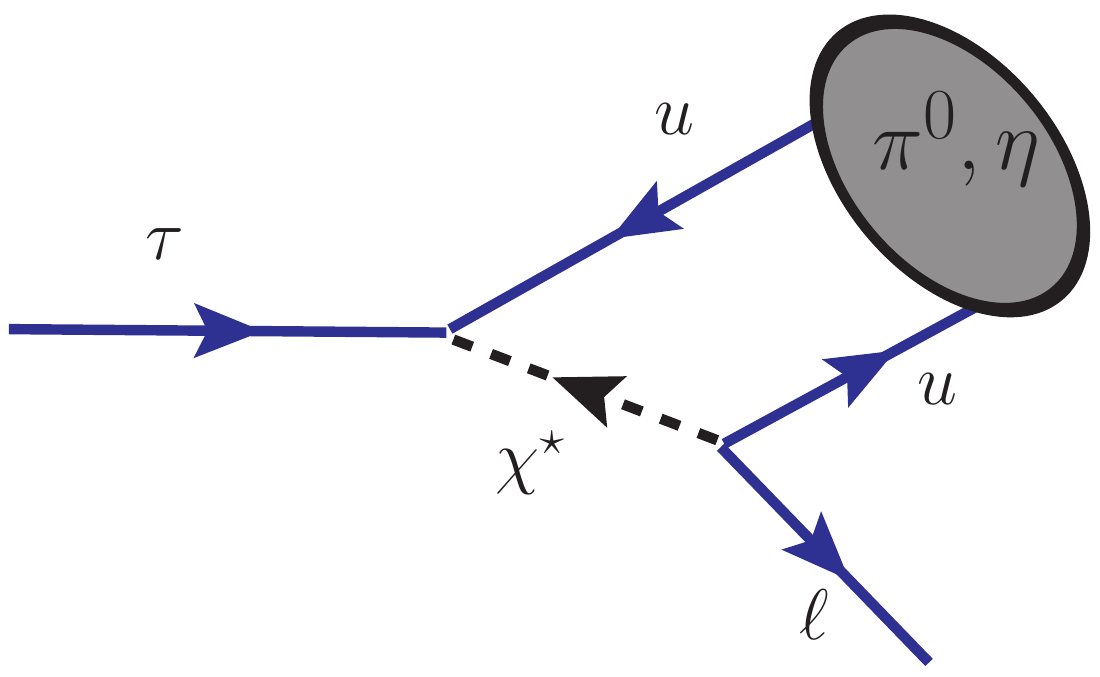}
    \caption{Feynman diagram for $\tau \to \mu \pi^0$ ($\mu \eta$, $\mu \eta'$) and $\tau \to e \pi^0$ ($e \eta$, $e \eta'$) decays.}
    \label{taupidecay}
\end{figure}

The decays $\tau^- \to \ell^- \pi^0,\, \ell^- \eta,\, \ell^- \eta'$, with $\ell = e$ or $\mu$ will occur at tree level mediated by the doublet LQ $\omega^{2/3}$ or the singlet LQ $\chi^{-1/3}$. The relevant Feynman diagrams are shown in Fig.~\ref{taupidecay}.  The decay rate for $\tau^- \to \ell^- \pi^0$ mediated by $\omega$ LQ is given by
\begin{eqnarray}
\Gamma_{\tau \rightarrow \ell \pi^{0}} & \ = \ & \frac{\left|\lambda_{\ell d} \lambda_{\tau d}^{\star}\right|^{2}}{1024 \pi} \frac{f_{\pi}^{2} m_{\tau}^{3}}{m_{\omega}^{4}}{\cal F}_\tau(m_\ell,m_\pi) \, ,
\label{tautoeta}
\end{eqnarray}
where 
\begin{align}
    {\cal F}_\tau(m_\ell,m_\pi) \ = \ &  \left[
\left(1-\frac{m_\ell^2}{m_\tau^2}\right)^2-\left(1+\frac{m_\ell^2}{m_\tau^2}\right)\frac{m_\pi^2}{m_\tau^2} \right]  \left[1-\left(\frac{m_\ell}{m_\tau}+ \frac{m_\pi}{m_\tau}  \right)^2 \right]^{1/2} \nonumber \\ & \qquad \qquad \times \left[1-\left(\frac{m_\ell}{m_\tau}-\frac{m_\pi}{m_\tau}\right)^2  \right]^{1/2} \, .
\label{eq:Ftau}
\end{align}
If this decay is mediated by the $\chi$ LQ, the same relation will hold, up to a factor of $|V_{ud}|^2$, with the replacement $(\lambda,\,m_\omega) \rightarrow (\lambda',\,m_\chi)$.  
%\begin{equation}
%\Gamma_{\tau \rightarrow \ell_\alpha \eta}=\frac{\left|\lambda_{\alpha d} \lambda_{\tau d}%^{\star}\right|^{2}}{1024 \pi} \frac{f_{\eta}^{2} m_{\tau}^{3}}{m_{\omega}^{4}}\left(1- %%\frac{3m_\ell^2}{m_{\tau}^2}- \frac{2m_{\eta}^{2}}{m_{\tau}^2}\right)
%\end{equation}
The rates for $\tau^- \to \ell^- \eta$ and $\tau^- \to \ell^- \eta'$ can be obtained from Eq.~(\ref{tautoeta}) by the replacement $(f_\pi,\,m_\pi) \rightarrow (m_\eta,\, f_\eta^q)$ and 
$(m_{\eta'},\, f_{\eta'}^q)$ respectively.  Here we have defined the matrix elements to be
\begin{eqnarray}
\langle \pi^0(p)|\bar{u} \gamma^\mu \gamma^5 u  |0 \rangle & \ = \ & -\langle \pi^0(p)|\bar{d} \gamma^\mu \gamma^5 d  |0 \rangle  \ = \ - i \frac{f_\pi}{\sqrt{2}} p^\mu \, ,\label{eq:pimatrix} \\
\langle \eta(p)|\bar{u} \gamma^\mu \gamma^5 u  |0 \rangle & \ = \ & \langle \eta(p)|\bar{d} \gamma^\mu \gamma^5d  |0 \rangle  \ = \ - i \frac{f_\eta^q}{\sqrt{2}} p^\mu \, ,\label{eq:etamatrix} \\
\langle \eta'(p)|\bar{u} \gamma^\mu \gamma^5u  |0 \rangle & \ = \ & \langle \eta'(p)|\bar{d} \gamma^\mu \gamma^5d  |0 \rangle  \ = \ - i \frac{f_{\eta'}^q}{\sqrt{2}} p^\mu~. \label{eq:etapmatrix}
\end{eqnarray}
The sign difference in Eq.~\eqref{eq:pimatrix} is due to the fact that the state $|\pi^0\rangle=(u\bar{u}-d\bar{d})/\sqrt{2}$. As for $|\eta\rangle$ and $|\eta'\rangle$ states, these are obtained from the mixing of the  flavor states $|\eta_q \rangle = (\bar{u} u + \bar{d}d)/\sqrt{2}$ and $|\eta_s\rangle = \bar{s}s$:
\begin{eqnarray}
|\eta\rangle &\ = \ & \cos\phi\, |\eta_q\rangle - \sin\phi\, |\eta_s \rangle, \nonumber\\
|\eta'\rangle & \ = \ & \sin\phi\, |\eta_q \rangle + \cos\phi \,|\eta_s \rangle~.
\end{eqnarray}
The matrix elements entering semileptonic $\tau$ decays are then related as
\begin{equation}
f_\eta^q \ = \ \cos\phi f_q \, ,\qquad f_{\eta'}^q \ = \ \sin\phi f_q
\end{equation}
where $f_q$ is defined through
\begin{equation}
\langle \eta_q(p)|\bar{q} \gamma^\mu \gamma^5q  |0 \rangle \ = \ - i \frac{f_q}{\sqrt{2}} p^\mu ~.
\end{equation}
The mixing angle $\phi$ and the decay parameter $f_q$ have been determined to be~\cite{Beneke:2002jn}
\begin{equation}
\phi \ = \ (39.3 \pm 1)^0 \, , \qquad  f_q \ = \ (1.07 \pm 0.02) f_\pi~.
\end{equation}
Using these relations, and with $f_\pi \simeq 130$ MeV, we have $f_\eta^q \simeq 108$ MeV and $f_{\eta'}^q \simeq 89$ MeV \cite{Li:2005rr}.  Using these values and the experimental limits on the semileptonic branching ratios \cite{Tanabashi:2018oca}, we obtain limits on products of Yukawa couplings as functions of the LQ mass, which are listed in Table \ref{tab:semilep}. It turns out that these limits are the most constraining for off-diagonal NSI mediated by LQs.

We should mention here that similar diagrams as in Fig.~\ref{taupidecay} will also induce alternative pion and $\eta$-meson decays: $\pi^0\to e^+e^-$ and $\eta\to \ell^+\ell^-$ (with $\ell=e$ or $\mu$). In the SM, ${\rm BR}(\pi^0\to e^+e^-)=6.46\times 10^{-8}$~\cite{Tanabashi:2018oca}, compared to ${\rm BR}(\pi^0\to \gamma\gamma)\simeq 0.99$. Specifically, the absorptive part of $\pi^0\to e^+e^-$ decay rate\footnote{The dispersive part of $\pi^0\to e^+e^-$ decay rate is found to be 32\% smaller than the absorptive part in the vector meson dominance~\cite{Babu:1982yz}.}  is given by~\cite{Berman:1960zz, Babu:1982yz} 
\begin{align}
   \frac{\Gamma_{\rm absp}(\pi^0\to e^+e^-)}{\Gamma(\pi^0\to \gamma\gamma)} \ = \ \frac{1}{2}\alpha^2 \left(\frac{m_e}{m_\pi}\right)^2\frac{1}{\beta}\left(\log\frac{1+\beta}{1-\beta}\right)^2 \, ,
\end{align}
where $\beta=\sqrt{1-4m_e^2/m_\pi^2}$. For LQ mediation, the suppression factor $(m_e/m_\pi)^2\sim 1.4\times 10^{-5}$ is replaced by the factor $(m_\pi/m_\omega)^4\sim 3.3\times 10^{-16}$ for a TeV-scale LQ.  Similar suppression occurs for the $\eta$ decay processes $\eta\to \ell^+\ell^-$ (with $\ell=e$ or $\mu$)~\cite{Geffen:1965zz, Babu:1982yz}. Therefore, both pion and $\eta$ decay constraints turn out to be much weaker than those from $\tau$ decay given in Table~\ref{tab:semilep}.

\begin{table}[!t]
    \centering
    \begin{tabular}{|c|c|c|}
    \hline \hline
        {\bf Process} & {\bf Exp. limit}~\cite{Tanabashi:2018oca} & {\bf Constraint}  \\
        \hline \hline
        \rule{0pt}{15pt} $\tau \rightarrow \mu \pi^{0}$ & BR < $1.1 \times 10^{-7} $ & $|\lambda_{\mu d} \lambda_{\tau d}^{\star}| < 9.3 \times 10^{-2} \left(\frac{m_\omega}{ \text{TeV}}\right)^2$ \\
       \rule{0pt}{15pt}  $\tau \rightarrow e \pi^{0}$ & BR < $8 \times 10^{-8}$& $|\lambda_{e d} \lambda_{\tau d}^\star| < 7.9 \times 10^{-2} \left(\frac{m_\omega}{\text{TeV}}\right)^2$  \\
       \rule{0pt}{15pt}  $\tau \rightarrow \mu \eta$  & BR < $6.5 \times 10^{-8}$ & $|\lambda_{\mu d} \lambda_{\tau d}^\star| <9.5 \times 10^{-2} \left(\frac{m_\omega}{ \text{TeV}}\right)^2$ \\
      \rule{0pt}{15pt}  $\tau \rightarrow e \eta$ & BR < $9.2 \times 10^{-8}$ & $|\lambda_{e d} \lambda_{\tau d}^\star| <1.1 \times 10^{-1} \left(\frac{m_\omega}{\text{TeV}}\right)^2$ \\
       \rule{0pt}{15pt}  $\tau \rightarrow \mu \eta'$  & BR < $1.3 \times 10^{-7}$ & $|\lambda_{\mu d} \lambda_{\tau d}^\star|< 2.3 \times 10^{-1} \left(\frac{m_\omega}{\text{TeV}}\right)^2$ \\
      \rule{0pt}{15pt}  $\tau \rightarrow e \eta'$ & BR < $1.6 \times 10^{-7}$ & $|\lambda_{e d} \lambda_{\tau d}^\star|<2.5 \times 10^{-1} \left(\frac{m_\omega}{\text{TeV}}\right)^2$ \\
       \hline \hline
    \end{tabular}
    \caption{Constraints on couplings and the LQ mass from semileptonic tau decays. Exactly the same constraints apply to $\lambda'$ couplings, with $m_\omega$ replaced by $m_\chi$.}
    \label{tab:semilep}
\end{table}

%%%%%%%%%%%%%%%%%%%%%%%%%%%%%%%%%%
%%%%%%%%%%%%%%%%%%%%%%%%%%%%%%%%
\subsubsection{Rare \texorpdfstring{$D$}{D}-meson decays} \label{sec:Dmeson}
%%%%%%%%%%%%%%%%%%%%%%%%%%%%%%%%%%%%%%
\begin{figure}[!t]
    \centering
    \includegraphics[scale=0.6]{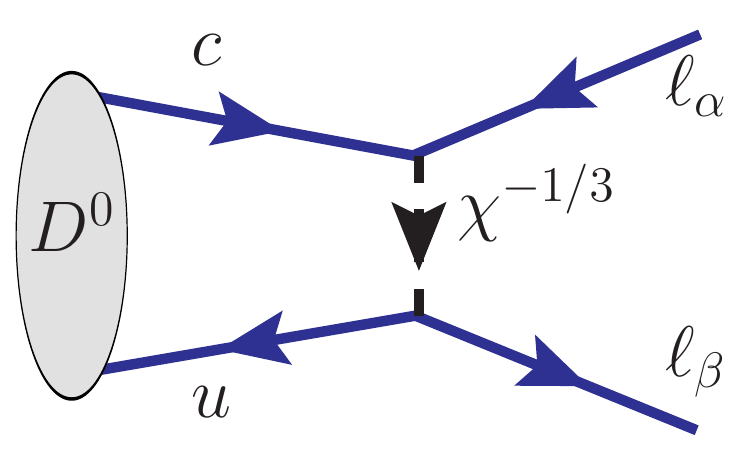}\hspace{10mm}
    \includegraphics[scale=0.5]{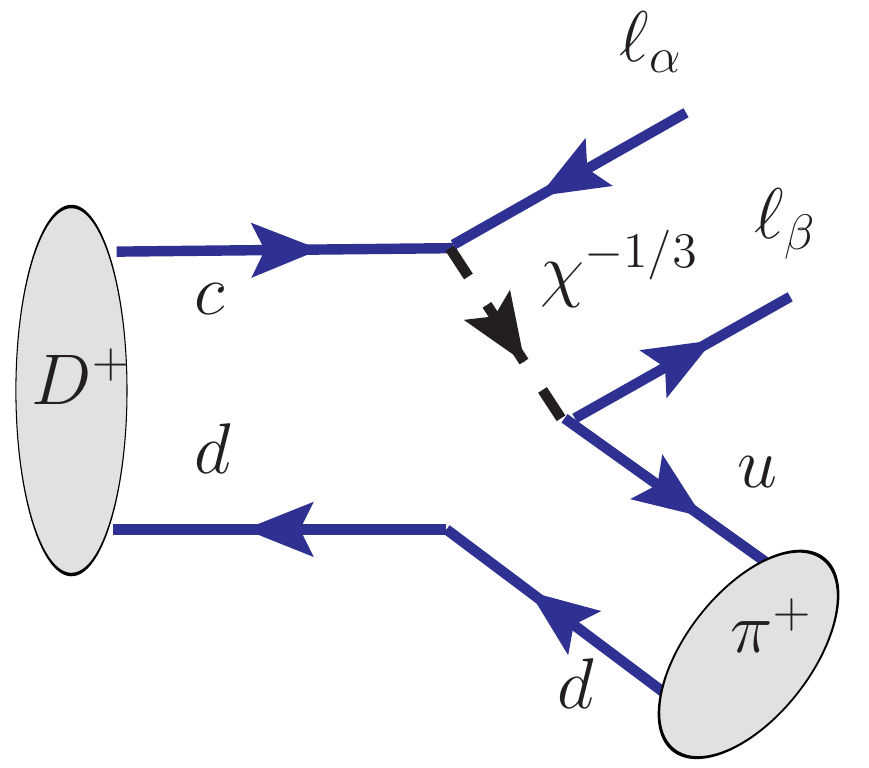}
    \caption{Feynman diagram for rare leptonic and semileptonic $D$-meson decays mediate by the $\chi$ LQ.}
    \label{fig:Dmeson}
\end{figure}

%%%%%%%%%%%%%%%%%%%%%%%%%%%%%%%%%%%%%%

The coupling matrix $\lambda'$ of Eq. (\ref{lagLQ}) contains, even with only diagonal entries,  flavor violating couplings in the quark sector.  To see this, we write the interaction terms in a basis where the down quark mass matrix is diagonal.  Such a choice of basis is always available and conveniently takes care of the stringent constraints in the down-quark sector, such as from rare kaon decays.  The $\chi$ LQ interactions with the physical quarks, in this basis, read as
\begin{equation}
-{\cal L}_Y \ \supset \ \lambda'_{\alpha d} \left(\nu_\alpha d \chi^\star - \ell_\alpha V_{id}^\star u_i \chi^\star\right) + {\rm H.c.}
\end{equation}
Here $V$ is the CKM mixing matrix.  In particular, the Lagrangian contains the following terms:
\begin{equation}
-{\cal L}_Y \ \supset \  -\lambda'_{\alpha d} \left(V_{ud}^\star \ell_\alpha u \chi^\star + V_{cd}^\star \ell_\alpha c  \chi^\star\right) + {\rm H.c.}
\end{equation}
The presence of these terms will result in the rare decays $D^0 \rightarrow \ell^+ \ell^-$ as well as $D \rightarrow \pi \ell^+ \ell^-$ where $\ell = e,\,\mu$.  The partial width for the decay $D^0 \rightarrow \ell^+ \ell^-$ is given by
\begin{align}
\Gamma_{D^{0} \rightarrow \ell_\alpha^{-} \ell_\alpha^{+}} \ & = \ \frac{|\lambda'_{\alpha d} \lambda_{\alpha d}^{\prime \star}|^{2}|V_{ud} V_{cd}^\star|^{2}}{128 \pi} \frac{m_{\ell}^2 f_D^2 m_D}{m_{\chi}^{4}} 
\left(1-\frac{4m_{\ell}^2}{m_{D}^2}\right)^{1/2}~.
\end{align}
Here we have used the effective Lagrangian arising from integrating out the $\chi$ field to be
\begin{equation}
{\cal L}_{\rm eff} \ = \ \frac{\lambda'_{\alpha d} \lambda^{\prime \star}_{\beta d}}{2 m_\chi^2}(\bar{u}_L \gamma^\mu c_L) (\bar{\ell}_{\beta L} \gamma^\mu \ell_{\alpha L})
\label{eff_D0}
\end{equation}
and the hadronic matrix element
\begin{equation}
\langle D^0| \bar{u} \gamma_\mu \gamma^5 c|0\rangle \ = \ - i f_D p_\mu~.
\end{equation}
Using $f_D = 200$ MeV, we list the constraint arising from this decay in Table \ref{tab:D0decay}. It will turn out that the NSI parameter $\varepsilon_{\mu\mu}$ will be most constrained by the limit $D^0 \to \mu^+\mu^-$, in cases where $\chi$ LQ is the mediator.  Note that this limit only applies to $SU(2)_L$ singlet and triplet LQ fields, and not to the doublet LQ field $\Omega$.  The doublet LQ field always has couplings to a $SU(2)_L$ singlet quark field, which does not involve the CKM matrix, and thus has not quark flavor violation arising from $V$.

The semileptonic decay $D^+ \rightarrow \pi^+ \ell^+ \ell^-$ is mediated by the same effective Lagrangian as in Eq.~(\ref{eff_D0}). The hadronic matrix element is now given by
\begin{equation}
\langle \pi^+(p_2)|\bar{u} \gamma_\mu c| D^+(p_1)\rangle \ = \ F_+(q^2) (p_1+p_2)_\mu + F_-(q^2)(p_1-p_2)_\mu
\end{equation}
with $q^2 = (p_1-p_2)^2$.  Since the $F_-(q^2)$ term is proportional to the final state lepton mass, it can be ignored.  For the form factor $F_+(q^2)$ we use
\begin{equation}
F_+(q^2) \ = \ \frac{f_D}{f_\pi} \frac{g_{D^\star D\pi}}{1-q^2/m_{D^\star}^2}~.
\end{equation}
For the $D^\star \to D \pi$ decay constant we use $g_{D^\star D\pi} = 0.59$ \cite{Burdman:2003rs}. Vector meson dominance hypothesis gives very similar results \cite{Babu:1987xe}.  With these matrix elements, the decay rate is given by
\begin{align}
\Gamma_{D^+ \to \pi^+\ell_\alpha^+\ell_\beta^-} \ & = \ \left[\frac{|\lambda'_{\alpha d} \lambda^{\prime\star}_{\beta d}|}{4m_\chi^2}\frac{f_D}{f_\pi}g_{D^\star D\pi}|V_{ud}V_{cd}^{\star}|\right]^2\frac{1}{64\pi^3 m_D}{\cal F} \, .
\end{align}
The function ${\cal F}$ is defined as
\begin{align}
    {\cal F} \ &= \ \frac{m_{D^\star}^2}{12m_D^2}\left[-2m_D^6+9m_D^4m_{D^\star}^2-6m_D^2m_{D^\star}^4-6(m_{D^\star}^2-m_D^2)^2m_{D^\star}^2\log\left(\frac{m_{D^\star}^2-m_D^2}{m_{D^\star}^2}\right)\right] \nonumber ~.
\end{align}
Note that in the limit of infinite $D^\star$ mass, this function ${\cal F}$ reduces to $m_D^6/24$.  The numerical value of the function is ${\cal F} \simeq 2.98$ GeV$^6$.  
Using $f_D=200$ MeV, $f_\pi=130$ MeV, $g_{D^\star D\pi}=0.59$ and the experimental upper limits on the corresponding branching ratios ~\cite{Tanabashi:2018oca}, we obtain bounds on the $\lambda'$ couplings as shown in Table~\ref{tab:D0decay}. These semileptonic $D$ decays have a mild effect on the maximal allowed NSI.  Note that the experimental limits on $D^0 \rightarrow \pi^0 \ell^+ \ell^-$ are somewhat weaker than the $D^+$ decay limits and are automatically satisfied when the $D^+$ semileptonic rates are satisfied.  

\begin{table}[!t]
    \centering
    \begin{tabular}{|c|c|c|}
    \hline \hline
        {\bf Process} & {\bf Exp. limit}~\cite{Tanabashi:2018oca} & {\bf Constraint} \\
        \hline \hline
        $D^{0} \rightarrow e^{+} e^{-}$ & BR < $ 7.9 \times 10^{-8} $ & $|\lambda'_{e d}|<16.7 \left(\frac{m_\chi}{ \text{TeV}}\right)$ \\
        $D^{0} \rightarrow \mu^{+} \mu^{-}$  & BR < $6.2 \times 10^{-9} $ & $|\lambda'_{\mu d}|<0.614 \left(\frac{m_\chi}{\text{TeV}}\right)$ \\
        $D^{+} \rightarrow \pi^{+} e^+ e^-$  & BR < $1.1 \times 10^{-6} $ & $|\lambda'_{e d}|<0.834 \left(\frac{m_\chi}{\text{TeV}}\right)$ \\
        $D^{+} \rightarrow \pi^{+} \mu^+ \mu^-$  & BR < $7.3 \times 10^{-8} $ & $|\lambda'_{\mu d}|<0.426 \left(\frac{m_\chi}{\text{TeV}}\right)$ \\
        $D^{+} \rightarrow \pi^{+} e^+ \mu^-$  & BR < $3.6 \times 10^{-6} $ & $|\lambda'_{\mu d} \lambda^{\prime \star}_{e d}|<1.28  \left(\frac{m_\chi}{\text{TeV}}\right)^{2}$ \\
       
       \hline \hline
    \end{tabular}
    \caption{Constraints on the $\chi$ LQ Yukawa couplings from $D^0 \to \ell^+ \ell^-$ and $D^+ \to \pi^+ \ell^+ \ell^-$ decays.}
    \label{tab:D0decay}
\end{table}
%%%%%%%%%%%%%%%%%%%%%%%%%%%%%%%%%%%
\subsection{Contact interaction constraints} \label{sec:contactlq}
%%%%%%%%%%%%%%%%%%%%%%%%%%%%%%%%%%%%%%%}
High-precision measurements of inclusive $e^\pm p\to e^\pm p$ scattering cross sections at HERA with maximum $\sqrt s=320$ GeV~\cite{Abramowicz:2019uti} and $e^+e^-\to q\bar{q}$ scattering cross sections at LEP II with maximum $\sqrt s=209$ GeV~\cite{LEP:2003aa} can be used in an effective four-fermion interaction theory to set limits on the new physics scale $\Lambda>\sqrt{s}$ that can be translated into a bound in the LQ mass-coupling plane. This is analogous to the LEP contact interaction bounds derived in the Zee model~\ref{sec:contact}.  Comparing the effective LQ Lagrangian~\eqref{eq:LeffLQ} with Eq.~\eqref{eq:eff} (for $f=u,d$), we see that for the doublet LQ, the only relevant chirality structure is $LR$, whereas for the singlet LQ, it is $LL$, with $\eta^d_{LR}=\eta^u_{LL}=-1$. The corresponding experimental bounds on $\Lambda^-$ and the resulting constraints on LQ mass and Yukawa coupling are given in Table~\ref{tab:LEPcontactlq}. 
\begin{table}[!t]
    \centering
    \begin{tabular}{||c|c|c|c|c||}
    \hline \hline
     \textbf{LQ} &   \multicolumn{2}{c||}{\bf LEP} & \multicolumn{2}{c||}{\bf HERA} \\ \cline{2-5}
     {\bf type} & {\bf Exp. bound}~\cite{LEP:2003aa} & {\bf Constraint} & {\bf Exp. bound}~\cite{Abramowicz:2019uti} & {\bf Constraint} \\
     \hline\hline
     $\omega^{2/3}$ & $\Lambda_{LR}^- >5.1$ TeV & $\frac{m_{\omega}}{|\lambda_{ed}|}>1.017$ TeV &   
      $\Lambda_{LR}^->4.7$ TeV   &   $\frac{m_{\omega}}{|\lambda_{ed}|}>0.937$ TeV \\ \hline
     $\chi^{-1/3}$ & $\Lambda_{LL}^- >3.7$ TeV & 
     $\frac{m_{\chi}}{|\lambda_{ed}|}>0.738$ TeV &  
      $\Lambda_{LL}^->12.8$ TeV &    $\frac{m_{\chi}}{|\lambda_{ed}|}>2.553$ TeV \\ \hline \hline
    \end{tabular}
    \caption{Constraints on the ratio of LQ mass and the Yukawa coupling from LEP~\cite{LEP:2003aa} and HERA~\cite{Abramowicz:2019uti} contact interaction bounds.}
    \label{tab:LEPcontactlq}
\end{table}

\begin{figure}[htb!]
$$
\includegraphics[width=0.90\textwidth]{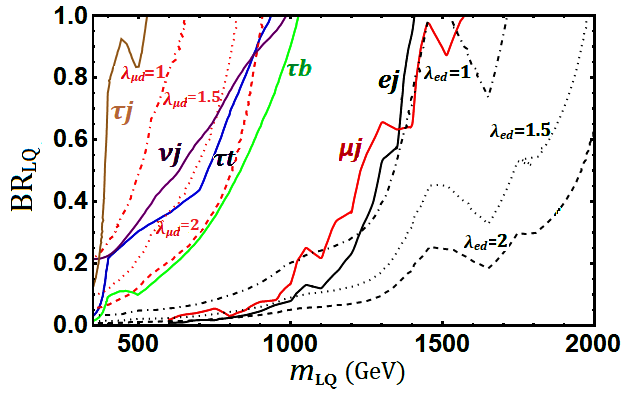}
$$
\caption{LHC constraints on scalar LQ in the LQ mass and branching ratio plane.  For a given channel, the branching ratio is varied from 0 to 1, without specifying the other decay modes which compensate for the missing branching ratios to add up to one. Black, red, green, blue, brown and purple solid lines represent present bounds from the pair production process at the LHC, i.e., looking for $e^+ e^- jj, \mu^+ \mu^- j j, \tau^+ \tau^- b\bar{b}, \tau^+ \tau^- t\bar{t} $, $\tau^+\tau^- jj$ and $\nu \bar{\nu} jj$ signatures respectively. These limits are independent of the LQ Yukawa coupling. On the other hand, black (red) dashed, dotted and dot-dashed lines indicate the bounds on LQ mass from the single production in association with one charged lepton for LQ couplings $\lambda_{ed~(\mu d)}=2,1.5$ and 1 respectively for first (second) generation LQ. }
\label{fig:collq1}
\end{figure}

In principle, one could also derive an indirect bound on LQs from the inclusive dilepton measurements at the LHC, because the LQ will give an additional $t$-channel contribution to the process $pp\to \ell^+\ell^-$. However, for a TeV-scale LQ as in our case, the LHC contact interaction bounds~\cite{Aaboud:2017buh, Sirunyan:2018ipj} with $\sqrt s=13$ TeV are not applicable. Recasting the LHC dilepton searches in the fully inclusive category following Ref.~\cite{Faroughy:2016osc} yields constraints weaker than those coming from direct LQ searches shown in Fig.~\ref{fig:collq1}. 
%One has to perform a dedicated analysis, comparing the dilepton kinematic distributions in presence of LQ interactions with the data to put bounds on the LQ mass and couplings. Given the $\sim 10\%$ uncertainties in the double differential cross sections at high $m_{\ell\ell}$ regime~\cite{Aad:2016zzw}, we expect these constraints to be weaker than the ones.
%%%%%%%%%%%%%%%%%%%%%%%%%%%%%%
\subsection{LHC constraints} \label{sec:highconstraints}
%%%%%%%%%%%%%%%%%%%%%%%%%%%%%%
 \begin{figure}[t!]
         \centering
         \subfigure[]{
         \includegraphics[scale=0.45]{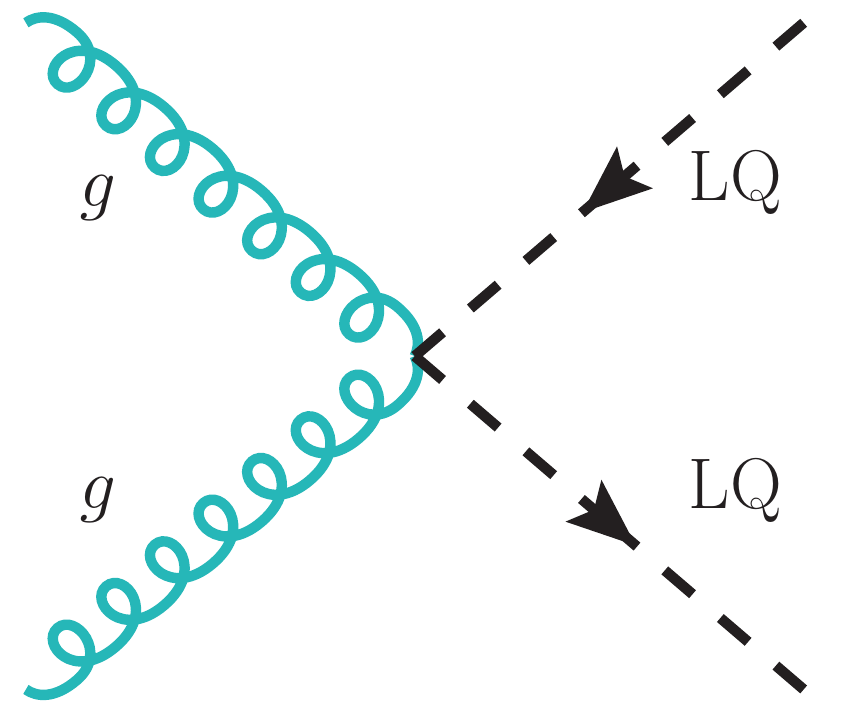}}
          \hspace{0.15in}
          \subfigure[]{
         \includegraphics[scale=0.45]{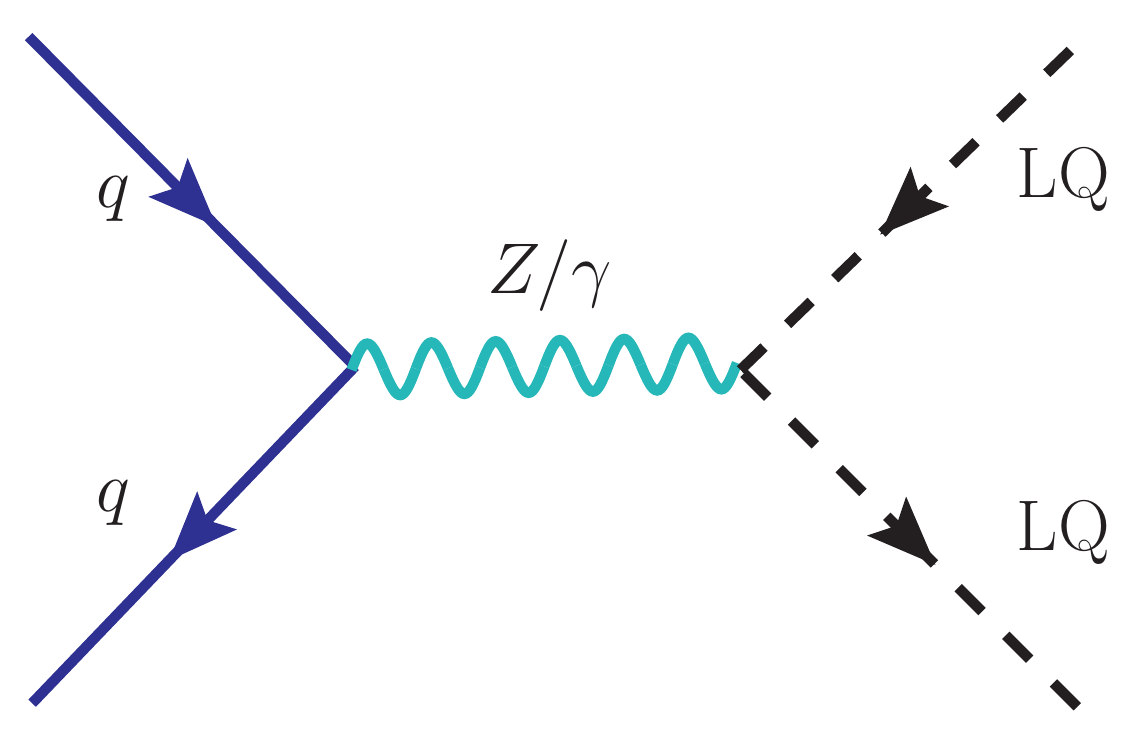}}
         \hspace{0.15in}
         \subfigure[]{
         \includegraphics[scale=0.45]{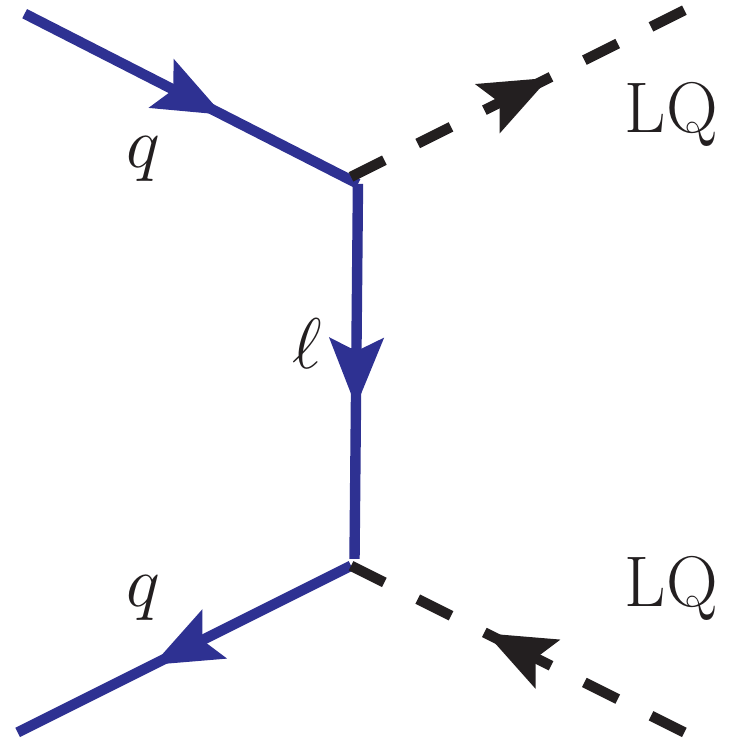}} \\
         \subfigure[]{
         \includegraphics[scale= 0.45]{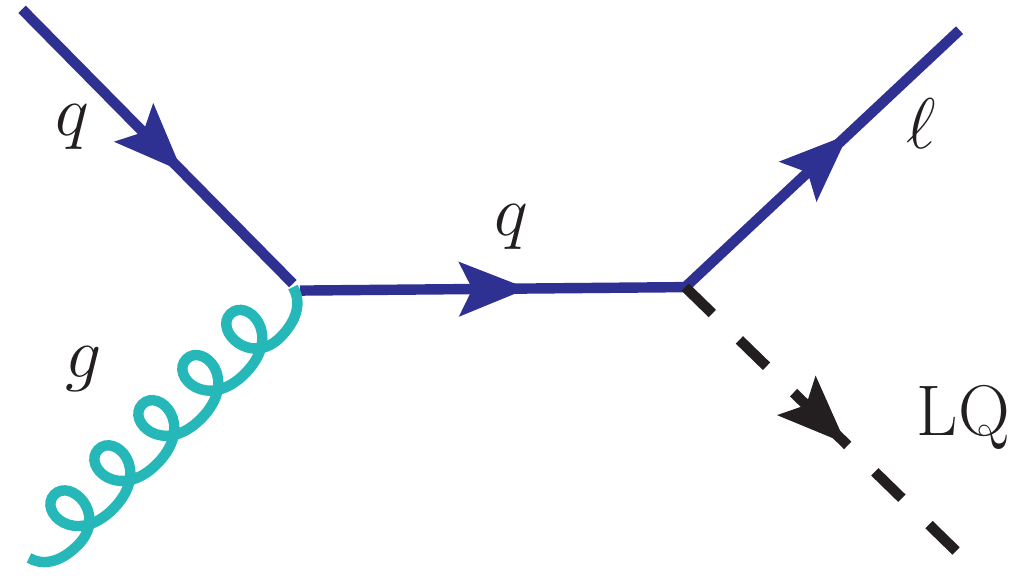}}
         \hspace{0.15in}
         \subfigure[]{
        \includegraphics[scale=0.45]{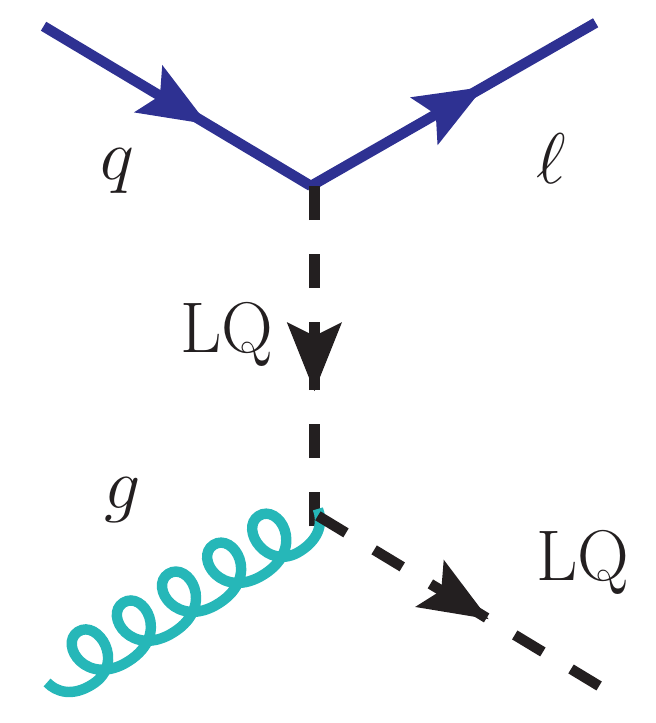} }
         \caption{Feynman diagrams for pair- and single-production of  LQ at the LHC.}
         \label{fig:feyn3}
    \end{figure}
%%%%%%%%%%%%%%%%%%%%%%%%%%%%%%%%%%%%%%%%%%%%%%%%%%%%%%%%%

In this section, we derive the LHC constraints on the LQ mass and Yukawa couplings which will be used in the next section for NSI studies. 
%%%%%%%%%%%%%%%%%%%%%%%%%%%%%%
\subsubsection{Pair production} \label{sec:pairLQ}
%%%%%%%%%%%%%%%%%%%%%%%%%%%%%%
At hadron colliders, LQs can be pair-produced through either $gg$ or $q\bar{q}$ fusion, as shown in Fig.~\ref{fig:feyn3} (a), (b) and (c). Since LQs are charged under $SU(3)_c$, LQ pair production at LHC is a QCD-driven process, solely determined by the LQ mass and strong coupling constant, irrespective of their Yukawa couplings. Although there is a $t$-channel diagram [cf.~Fig.~\eqref{fig:feyn3} (c)] via charged lepton exchange through which LQ can be pair-produced via quark fusion process, this cross-section is highly suppressed compared to the $s$-channel pair production cross-section. 
%Considering the pair-production channel, the current limits from $\sqrt s=13$ TeV LHC data exclude at 95\% CL LQ masses up to 1.43  TeV for first generation~\cite{CMS:2018sxp}, 1.25 TeV for second generation~\cite{Aaboud:2019jcc} and 1.02 TeV for third generation~\cite{Sirunyan:2018vhk, Aaboud:2019bye} assuming $50 \%$ branching ratio scenario.  

There are dedicated searches for pair production of first~\cite{CMS:2018sxp,Khachatryan:2015vaa}, second~\cite{Aaboud:2019jcc,Khachatryan:2015vaa,Sirunyan:2018kzh} and third generation~\cite{Sirunyan:2018vhk, Aaboud:2019bye,Sirunyan:2018kzh} LQs at the LHC. Given the model Lagrangian~\ref{lagLQ}, we are interested in the final states containing either two charged leptons and two jets ($\ell\ell jj$), or two neutrinos and two jets ($\nu \nu jj$). Note that for the doublet LQ $\Omega = (\omega^{2/3},\omega^{-1/3})$, the jets will consist of down-type quarks, while for the singlet LQ $\chi^{-1/3}$, the jets will be of up-type quarks. For the light quarks $u,d,c,s$, there is no distinction made in the LHC LQ searches; therefore, the same limits on the corresponding LQ masses will apply to both doublet and singlet LQs. The only difference is for the third-generation LQs, where the limit from $\tau^+\tau^-b\bar{b}$ final state is somewhat stronger than that from  $\tau^+\tau^-t\bar{t}$ final state~\cite{Aaboud:2019bye,Sirunyan:2018kzh}.

In Fig.~\ref{fig:collq1}, we have shown the LHC limits on LQ mass as a function of the corresponding branching ratios for each channel. For a given channel, the branching ratio is varied from 0 to 1, without specifying the other decay modes which compensate for the missing branching ratios to add up to one. For matter NSI, the relevant LQ couplings must involve either up or down quark. Thus, for first and second generation LQs giving rise to NSI, we can use $e^+ e^- jj$ and $\mu^+ \mu^- j j$ final states from LQ pair-production at LHC to impose stringent bounds on the $\lambda_{\alpha d}$ and $\lambda'_{\alpha d}$ couplings (with $\alpha=e,\mu$) which are relevant for NSI involving electron and muon flavors. There is no dedicated search for LQs in the $\tau^+ \tau^- j j$ channel to impose similar constraints on $\lambda_{\tau d}$ and $\lambda'_{\tau d}$ relevant for tau-flavor NSI. There are searches for third generation LQ~\cite{Sirunyan:2018vhk, Aaboud:2019bye} looking at $\tau^+ \tau^- b\bar{b}$ and $\tau^+ \tau^- t\bar{t}$ signatures which are not relevant for NSI, since we do not require $\lambda'_{\tau t}$ (for $\chi^{-1/3}$) or $\lambda_{\tau b}$ (for $\omega^{2/3}$) couplings. For constraints on $\lambda_{\tau d}$, we recast the $\tau^+ \tau^- b\bar{b}$ search limits~\cite{Sirunyan:2018vhk, Aaboud:2019bye,Sirunyan:2018kzh} taking into account the $b$-jet misidentification as light jets, with an average rate of $1.5\%$ (for a $b$-tagging efficiency of 70\%)~\cite{Chatrchyan:2012jua}. As expected, this bound is much weaker, as shown in Fig.~\ref{fig:collq1}. 

However, a stronger bound on NSI involving the tau-sector comes from $\nu\bar{\nu}jj$ final state. From the Lagrangian~\eqref{lagLQ}, we see that the same $\lambda_{\tau d}$ coupling that leads to $\tau^+\tau^-dd$ final state from the pair-production of  $\omega^{2/3}$ also leads to  $\nu_\tau \bar{\nu}_\tau dd$ final state from the pair-production of the $SU(2)_L$ partner LQ $\omega^{-1/3}$, whose mass cannot be very different from that of $\omega^{2/3}$ due to electroweak precision data constraints (similar to the Zee model case, cf.~Sec.~\ref{sec:ewpt}). Since the final state neutrino flavors are indistinguishable at the LHC, the $\nu\bar{\nu}jj$ constraint will equally apply to all $\lambda_{\alpha d}$ (with $\alpha=e,\mu,\tau$) couplings which ultimately restrict the strength of tau-sector NSI, as we will see in the next subsection. The same applies to the $\lambda'_{\tau d}$ couplings of the singlet LQ $\chi^{-1/3}$, which are also restricted by the $\nu\bar{\nu}jj$ constraint. 

%%%%%%%%%%%%%%%%%%%%%%%%%%%%%%
\subsubsection{Single production} \label{sec:singleLQ}
%%%%%%%%%%%%%%%%%%%%%%%%%%%%%%

LQs can also be singly produced at the collider in association with charged leptons via $s$- and $t$- channel quark-gluon fusion processes, as shown in Fig.~\ref{fig:feyn3} (d) and (e). The single production limits, like the indirect low-energy constraints, are necessarily in the mass-coupling plane.  This signature is applicable to LQs of all generations. In Fig.~\ref{fig:collq1}, we have shown the collider constraints in the single-production channel for some benchmark values of the first and second generation LQ couplings $\lambda_{ed}$  and $\lambda_{\mu d}$ (since $d$ jets cannot be distinguished from $s$ jets) equal to 1, 1.5 and 2 by dot-dashed, dotted and dashed curves respectively. The single-production limits are more stringent than the pair-production limits only for large $\lambda_{ed}$, but not for $\lambda_{\mu d}$. There is no constraint in the $\tau j$ channel, and the derived constraint from $\tau b$ channel is too weak to appear in this plot. 
%%%%%%%%%%%%%%%%%%%%%%%%%%%%%%
\subsubsection{How light can the leptoquark be?} \label{sec:lightLQ}
%%%%%%%%%%%%%%%%%%%%%%%%%%%%%%
There is a way to relax the $\nu\bar{\nu}jj$ constraint and allow for smaller LQ masses for the doublet components. This is due to a new decay channel $\omega^{-1/3}\to \omega^{2/3}+W^-$ which, if kinematically allowed, can be used to suppress the branching ratio of $\omega^{-1/3}\to \nu d$ decay for relatively smaller values of $\lambda_{\alpha d}$ couplings, thereby reducing the impact of the $\nu\bar{\nu}jj$ constraint. The partial decay widths for  $\omega^{-1/3}\to \omega^{2/3}+W^-$ and $\omega^{-1/3}\to \nu_\alpha d_\beta$ are respectively given by 
\begin{align}
\Gamma (\omega^{-1/3} \to \omega^{2/3} W^-) \  = \ & \frac{1}{32\pi} \frac{\mwo^3}{v^2}  \left( 1- \frac{\mwt^2}{\mwo^2} \right)^2 \nonumber \\
& \times 
\left[\left\lbrace 1-\left(\frac{\mwt+\mw}{\mwo}\right)^2 \right\rbrace\left\lbrace 1-\left(\frac{\mwt-\mw}{\mwo}\right)^2 \right\rbrace \right]^{1/2} \, , \label{eq:omegaW}\\
\Gamma (\omega^{-1/3} \to \nu_\alpha d_\beta) \  = \ & \frac{|\lambda_{\alpha \beta}|^2}{16\pi}\mwo \, . 
\label{eq:omeganu}
\end{align}
In deriving Eq.~\eqref{eq:omegaW}, we have used the Goldstone boson equivalence theorem, and in Eq.~\eqref{eq:omeganu}, the factor in the denominator is not $8\pi$ (unlike the SM $h\to b\bar{b}$ case, for instance), because only one helicity state contributes. 

The lighter LQ $\omega^{2/3}$ in this case can only decay to $\ell_\alpha d_\beta$ with 100\% branching ratio. Using the fact that constraints from $\tau^+\tau^- jj$ channel are weaker, one can allow for $\omega^{2/3}$ as low as 522 GeV, as shown in Fig.~\ref{fig:collq1} by the solid brown curve, when considering the $\lambda_{\tau d}$ coupling alone. %This puts an absolute lower limit of ?? on the mass of $\omega^{-1/3}$, %as shown by the vertical dotted line in Fig.~\ref{} (c).   
This is, however, not applicable to the scenario when either $\lambda_{e d}$ or $\lambda_{\mu d}$ coupling is present, because of the severe constraints from $e^+e^-jj$ and $\mu^+\mu^-jj$ final states.

%%%%%%%%%%%%%%%%%%%%%%%%%%%%%%%%%%%%%%%%%%%%
\subsection{NSI prediction} \label{sec:NSILQ}
%%%%%%%%%%%%%%%%%%%%%%%%%%%%%%%%%%%%%%%%%%%%
\begin{figure}[t!]
    \centering
    \subfigure[]{
        \includegraphics[scale=0.5]{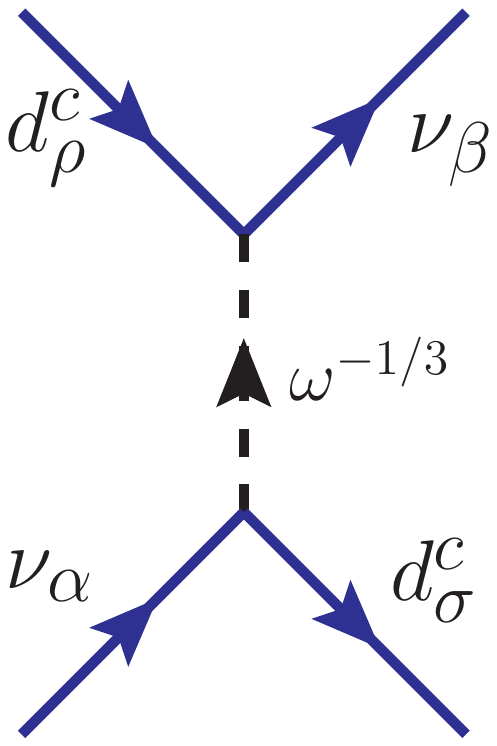}}
    \hspace{5mm}
    \subfigure[]{
        \includegraphics[scale=0.5]{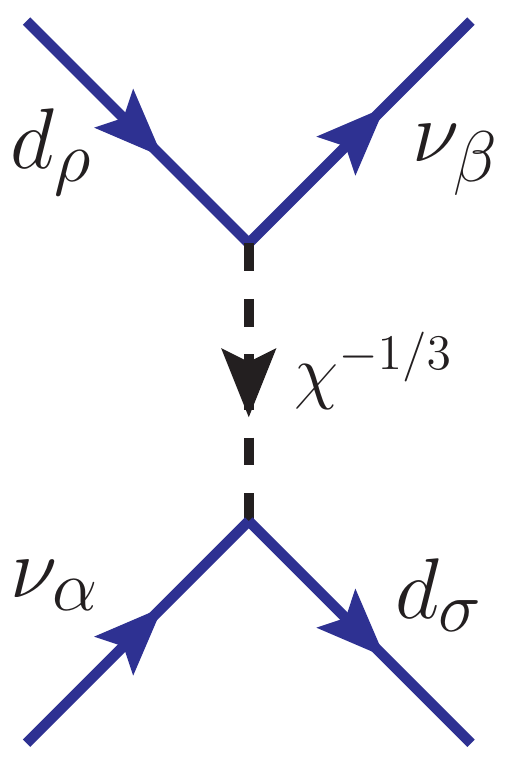}}
    \caption{Tree-level NSI diagrams with the exchange of heavy LQs: (a) for doublet LQ with Yukawa $\lambda \sim \mathcal{O}$(1), and (b) for singlet LQ with Yukawa $\lambda' \sim \mathcal{O}$(1).}
    \label{coloredzeeNSI}
 \end{figure}
The LQs $\omega^{-1/3}$ and $\chi^{-1/3}$ in the model have couplings with neutrinos and down-quark (cf.~Eq.~\eqref{lagLQ}), and therefore, induce NSI at tree level as shown in Fig. \ref{coloredzeeNSI} via either $\lambda$ or $\lambda'$ couplings. From Fig.~\ref{coloredzeeNSI}, we can write down the effective four-fermion Lagrangian as 
\begin{eqnarray}
    \mathcal{L}  & \ = \ & \frac{\lambda_{\alpha d}^\star \lambda_{\beta d} }{m_{\omega}^2} (\Bar{d}_R\nu_{\beta L}) (\Bar{\nu}_{\alpha L} d_R)+ \frac{\lambda_{\alpha d}^{\prime\star} \lambda'_{\beta d} }{m_{\chi}^2} (\Bar{d}_L\nu_{\beta L}) (\Bar{\nu}_{\alpha L} d_L)\nonumber \\ 
    & \ = \ & 
    -\frac{1}{2} \left[ \frac{\lambda_{\alpha d}^\star \lambda_{\beta d} }{m_{\omega}^2} (\Bar{d}_R \gamma^\mu d_R) (\Bar{\nu}_{\alpha L} \gamma_\mu \nu_{\beta L})+ \frac{\lambda_{\alpha d}^{\prime\star} \lambda'_{\beta d} }{m_{\chi}^2} (\Bar{d}_L \gamma^\mu d_L) (\Bar{\nu}_{\alpha L} \gamma_\mu \nu_{\beta L})\right]\, ,
    \label{eq:NSILQ}
\end{eqnarray}
  where we have used Fierz transformation in the second step. Comparing Eq.~\eqref{eq:NSILQ} with Eq.~\eqref{NSI-NC}, we obtain the NSI parameters 
\begin{eqnarray}
    \varepsilon_{\alpha \beta}^d & \ = \ & \frac{1}{4 \sqrt{2} \ G_F} \left(\frac{\lambda_{\alpha d}^{\star} \lambda_{\beta d} }{ m_\omega^2} + \frac{\lambda_{\alpha d}^{\prime\star} \lambda'_{\beta d} }{ m_\chi^2}\right)\, .
    \label{color_LQ}
\end{eqnarray}
For $Y_n(x)\equiv \frac{N_n(x)}{N_p(x)}= 1$, one can obtain the effective NSI parameters from Eq.~\eqref{eq:eps2} as
%\begin{equation}
%\boxed{
\begin{tcolorbox}[enhanced,ams align,
  colback=gray!30!white,colframe=white]
\varepsilon_{\alpha \beta} \ \equiv \ 3\varepsilon_{\alpha \beta}^d \ = \ \frac{3}{4 \sqrt{2} \ G_F}\left( \frac{\lambda_{\alpha d}^{\star} \lambda_{\beta d} }{ m_\omega^2}+ \frac{\lambda_{\alpha d}^{\prime\star} \lambda'_{\beta d} }{ m_\chi^2}\right) \, .
%}
\label{nsi_coloredzee}
%\end{equation}
\end{tcolorbox}
To satisfy the neutrino mass constraint [cf.~Eq.~\eqref{eq:Mnucz}], we can have either $\lambda_{\alpha d}^{\star} \lambda_{\beta d}$ or $\lambda_{\alpha d}^{\prime\star} \lambda'_{\beta d}$ of ${\cal O}(1)$, but not both simultaneously, for a given flavor combination $(\alpha,\beta)$. But we can allow for $\lambda_{\alpha d}^{\star} \lambda_{\beta d}$ and $\lambda_{\alpha' d}^{\prime\star} \lambda'_{\beta' d}$ simultaneously to be of ${\cal O}(1)$ for either $\alpha\neq\alpha'$ or $\beta\neq \beta'$, which will be used below to avoid some experimental constraints for the maximum NSI predictions. 
%As mentioned in Sec.~\ref{sec:lowconstraints}, the choice $\lambda'\ll 1$ and $\lambda\sim {\cal O}(1)$ is less constrained from cLFV. 
%%%%%%%%%%%%%%%%%%%%%%%%%%%%%%%%
\subsubsection{Doublet leptoquark} \label{subsec:doublet_NSI}
%%%%%%%%%%%%%%%%%%%%%%%%%%%%%%%%%%%
First, let us consider the doublet LQ contribution by focusing on the $\lambda$-couplings only. We show in Figs.~\ref{dnsiLQ} and \ref{offdnsiLQ}  the predictions for diagonal  $(\varepsilon_{ee},\, \varepsilon_{\mu\mu},\,\varepsilon_{\tau\tau})$ and off-diagonal  $(\varepsilon_{e\mu},\, \varepsilon_{\mu\tau},\,\varepsilon_{e\tau})$ NSI parameters respectively from Eq.~\eqref{nsi_coloredzee} by black dotted contours. Color-shaded regions in each plot are excluded by various theoretical and experimental constraints. In Figs.~\ref{dnsiLQ} (b) and (c), the yellow colored regions are excluded by perturbativity constraint, which requires the LQ coupling $\lambda_{\alpha d}<\sqrt{\frac{4\pi}{\sqrt 3}}$~\cite{DiLuzio:2017chi}. Red-shaded region in Fig.~\ref{dnsiLQ} (a) is excluded by the APV bound (cf.~Sec.~\ref{sec:APV}), while the brown and cyan regions are excluded by HERA and LEP contact interaction bounds, respectively (cf.~Table~\ref{tab:LEPcontactlq}).  Red-shaded region in Fig.~\ref{dnsiLQ} (c) is excluded by the global-fit constraint from neutrino oscillation+COHERENT data~\cite{Esteban:2018ppq}.   Blue-shaded regions in Figs.~\ref{dnsiLQ} (a) and (b) are excluded by LHC LQ searches (cf.~Fig.~\ref{fig:collq1}) in the pair-production mode for small $\lambda_{\alpha d}$ (which is independent of $\lambda_{\alpha d}$) and single-production mode for large $\lambda_{\alpha d}$) with $\alpha=e,\mu$. Here we have assumed 50\% branching ratio to $ej$ or $\mu j$, and the other 50\% to $\tau d$ in order to relax the LHC constraints and allow for larger NSI. Blue-shaded region in Fig.~\ref{dnsiLQ} (c) is excluded by the LHC constraint from the $\nu\bar{\nu}jj$ channel, where the vertical dashed line indicates the limit assuming ${\rm BR}(\omega^{-1/3}\to \nu d)=100\%$, and the unshaded region to the left of this line for small $\lambda_{\tau d}$ is allowed by opening up the $\omega^{-1/3}\to \omega^{2/3}W^-$ channel (cf.~Sec.~\ref{sec:lightLQ}). Note that we cannot completely switch off the $\omega^{-1/3}\to \nu d$ channel, because that would require $\lambda_{\tau d}\to 0$ and in this limit, the NSI will also vanish. 

The red line in Fig.~\ref{dnsiLQ} (b) is the suggestive limit on $\varepsilon_{\alpha\beta}^{dR}$ from NuTeV data~\cite{Davidson:2003ha} (cf.~Table~\ref{tab:LQ}). This is not shaded because there is a $2.7\sigma$ discrepancy of their $s_w^2$ measurement with the PDG average~\cite{Tanabashi:2018oca} and a possible resolution of this might affect the NSI constraint obtained from the same data. Here we have rederived the NuTeV limit following Ref.~\cite{Davidson:2003ha}, but using the latest value of $s_w^2$ (on-shell)~\cite{Tanabashi:2018oca} (without including NuTeV). Specifically, we have used the NuTeV measurement of the effective coupling $\left(\tilde{g}^\mu_{R}\right)^2=0.0310\pm 0.0011$ from $\nu_\mu q\to \nu q$ scatterings~\cite{Zeller:2001hh} which is consistent with the SM prediction of  $\left(\tilde{g}^\mu_{R}\right)_{\rm SM}^2=0.0297$. Here $\left(\tilde{g}^\mu_{R}\right)^2$ is defined as 
\begin{align}
    \left(\tilde{g}^\mu_{R}\right)^2 \ = \ \left(g_R^u+\varepsilon_{\mu\mu}^{uR}\right)^2+\left(g_R^d+\varepsilon_{\mu\mu}^{dR}\right)^2 \, ,
    \label{eq:gRmu}
\end{align}
where $g_R^u =  -\frac{2}{3}s_w^2$ and $g_R^d  =  \frac{1}{3}s_w^2$ 
are the $Z$ couplings to right-handed up and down quarks respectively. Only the right-handed couplings are relevant here, since the effective NSI Lagrangian~\eqref{eq:NSILQ} involves right-handed down-quarks for the doublet LQ component  $\omega^{2/3}$. In Eq.~\eqref{eq:gRmu}, setting $\varepsilon_{\mu\mu}^{uR}=0$ for this LQ model and comparing $\left(\tilde{g}^\mu_{R}\right)^2$ with the measured value, we obtain a 90\% CL on  $\varepsilon_{\mu\mu}^{dR}<0.029$, which should be multiplied by 3 (since $\varepsilon_{\alpha\beta}\equiv 3\varepsilon_{\alpha\beta}^{dR}$) to get the desired constraint on  $\varepsilon_{\alpha\beta}$ shown in Fig.~\ref{dnsiLQ} (b). 

Also note that unlike in the Zee model case discussed earlier, the IceCube limit on $|\varepsilon_{\tau\tau}-\varepsilon_{\mu\mu}|$~\cite{Esmaili:2013fva} is not shown in Figs.~\ref{dnsiLQ} (b) and (c). This is because the NSI parameters in the LQ model under consideration receive two contributions as shown in Eq.~\eqref{nsi_coloredzee}. Although we cannot have both $\lambda$ and $\lambda'$ contributions large for the {\it same} $\varepsilon_{\alpha\beta}$, it is possible to have a large $\lambda$ contribution to $\varepsilon_{\alpha\beta}$ and a large $\lambda'$ contribution to $\varepsilon_{\alpha'\beta'}$ (with either $\alpha\neq \beta$ or $\beta\neq \beta'$), thus evading the cLFV constraints (which are only applicable to either $\lambda$ or $\lambda'$ sectors), as well as the IceCube constraint on $|\varepsilon_{\tau\tau}-\varepsilon_{\mu\mu}|$, which is strictly applicable only in the limit of all $\varepsilon_{e\alpha}\to 0$. This argument can be applied to all the LQ models discussed in subsequent sections, with a few exceptions, when the NSI arises from only one type of couplings; see e.g. Eq.~\eqref{eq:714} and \eqref{eq:719}). So we will not consider the IceCube limit on $\varepsilon_{\mu\mu}$ and  $\varepsilon_{\tau\tau}|$ for our LQ NSI analysis, unless otherwise specified.

For $\varepsilon_{ee}$, the most stringent constraint comes from APV  (Sec.~\ref{sec:APV}), as shown by the red-shaded region in Fig.~\ref{dnsiLQ} (a) which, when combined with the LHC constraints on the mass of LQ, rules out the possibility of any observable NSI in this sector. Similarly, for $\varepsilon_{\mu\mu}$, the most stringent limit of 8.6\% comes from NuTeV. However, if this constraint is not considered, $\varepsilon_{\mu\mu}$ can be as large as 21.6\%. Similarly, $\varepsilon_{\tau\tau}$ can be as large as 34.3\%, constrained only by the LHC constraint on the LQ mass and perturbative unitarity constraint on the Yukawa coupling (cf.~Fig.~\ref{dnsiLQ} (c)). This is within the future DUNE sensitivity reach, at least for the 850 kt.MW.yr (if not 300 kt.MW.yr) exposure~\cite{dev_pondd}, as shown in Fig.~\ref{dnsiLQ} (c).
%Taking this into account and given that $\varepsilon_{\mu\mu}^{\rm max}=4.5\%$, the maximum allowed $\varepsilon_{\tau\tau}$ can only go up to 14\%.  
%Nonetheless, 

As for the off-diagonal NSI in Fig.~\ref{fig:offdnsi}, the LHC constraints (cf.~Sec.~\ref{sec:highconstraints}) are again shown by blue-shaded regions. The yellow-shaded region in Fig.~\ref{fig:offdnsi} (b) is from the combination of APV and perturbative unitarity constraints. However, the most stringent limits for all the off-diagonal NSI come from cLFV processes. In particular, $\tau\to \ell \pi^0$ and $\tau\to \ell \eta$ (with $\ell=e,\mu$) impose strong constraints (cf.~Sec.~\ref{sec:tauLQ}) on  $\varepsilon_{\mu\tau}$ and $\varepsilon_{e\tau}$, as shown in Figs.~\ref{offdnsiLQ} (a) and (b). For $\varepsilon_{e\mu}$, the most stringent limit comes from  $\mu-e$ conversion (cf.~Sec.~\ref{sec:mueconv}), as shown in Fig.~\ref{offdnsiLQ} (c). The maximum allowed NSI in each case is tabulated in Table~\ref{tab:LQ}, along with the current constraints from  neutrino-nucleon scattering experiments, like CHARM~\cite{Davidson:2003ha}, COHERENT~\cite{Coloma:2017ncl} and IceCube~\cite{Salvado:2016uqu}, as well as the global-fit constraints from neutrino oscillation+COHERENT data~\cite{Esteban:2018ppq} and future DUNE sensitivity~\cite{dev_pondd}. It turns out that the cLFV constraints have essentially ruled out the prospects of observing any off-diagonal NSI in this LQ model in future neutrino experiments. This is consistent with general arguments based on $SU(2)_L$ gauge-invariance~\cite{Gavela:2008ra}.

%%%%%%%%%%%%%%%%%%%%%%%%%%%%%%%%%%%
\subsubsection{Singlet leptoquark} \label{subsec:singlet_NSI}
\begin{figure}[t!]
 \vspace{-2.2cm}
 \centering
 \subfigure[]{
    \includegraphics[height=8cm,width=0.45\textwidth]{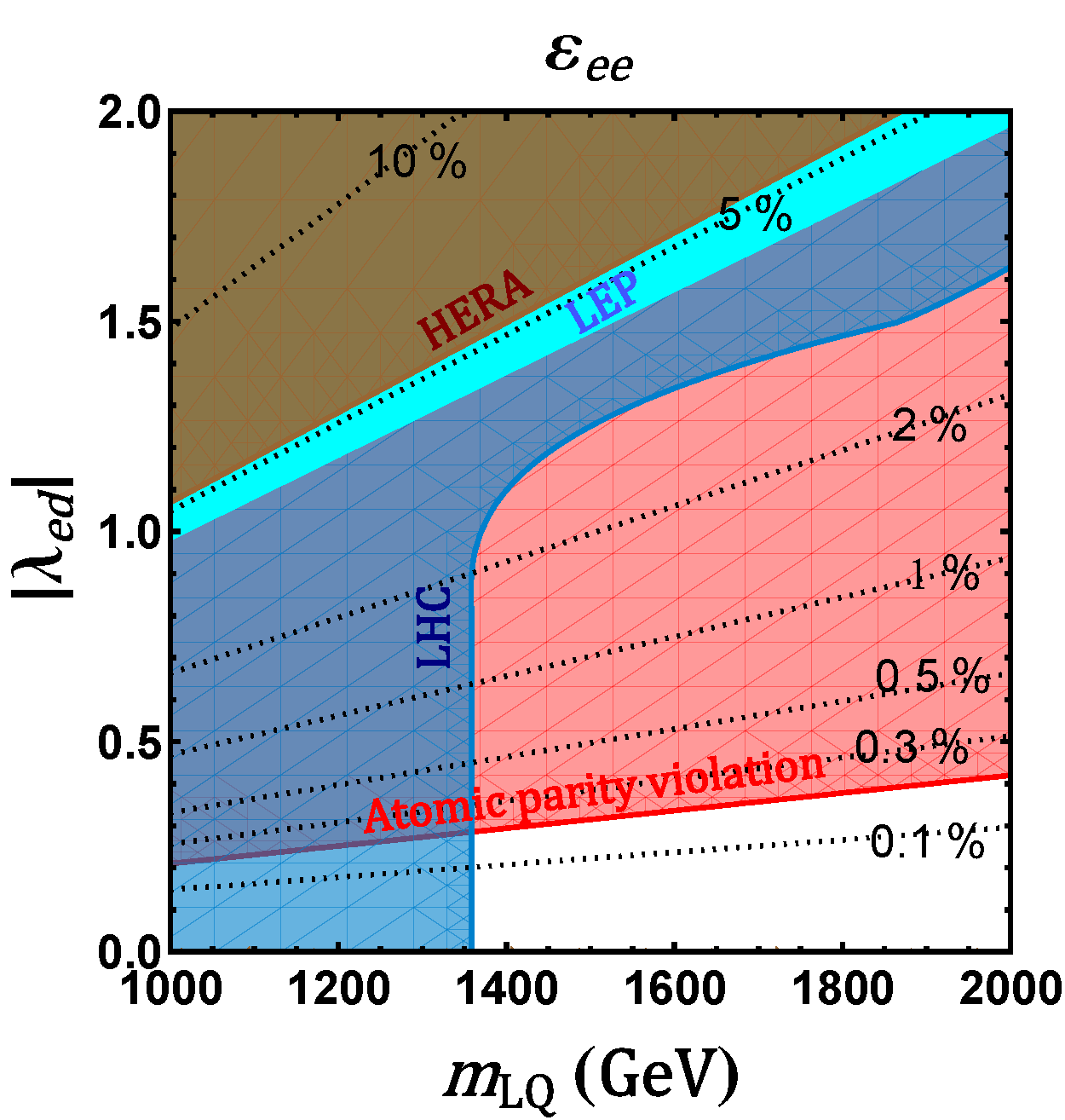} }
    \subfigure[]{
     \includegraphics[height=8cm,width=0.45\textwidth]{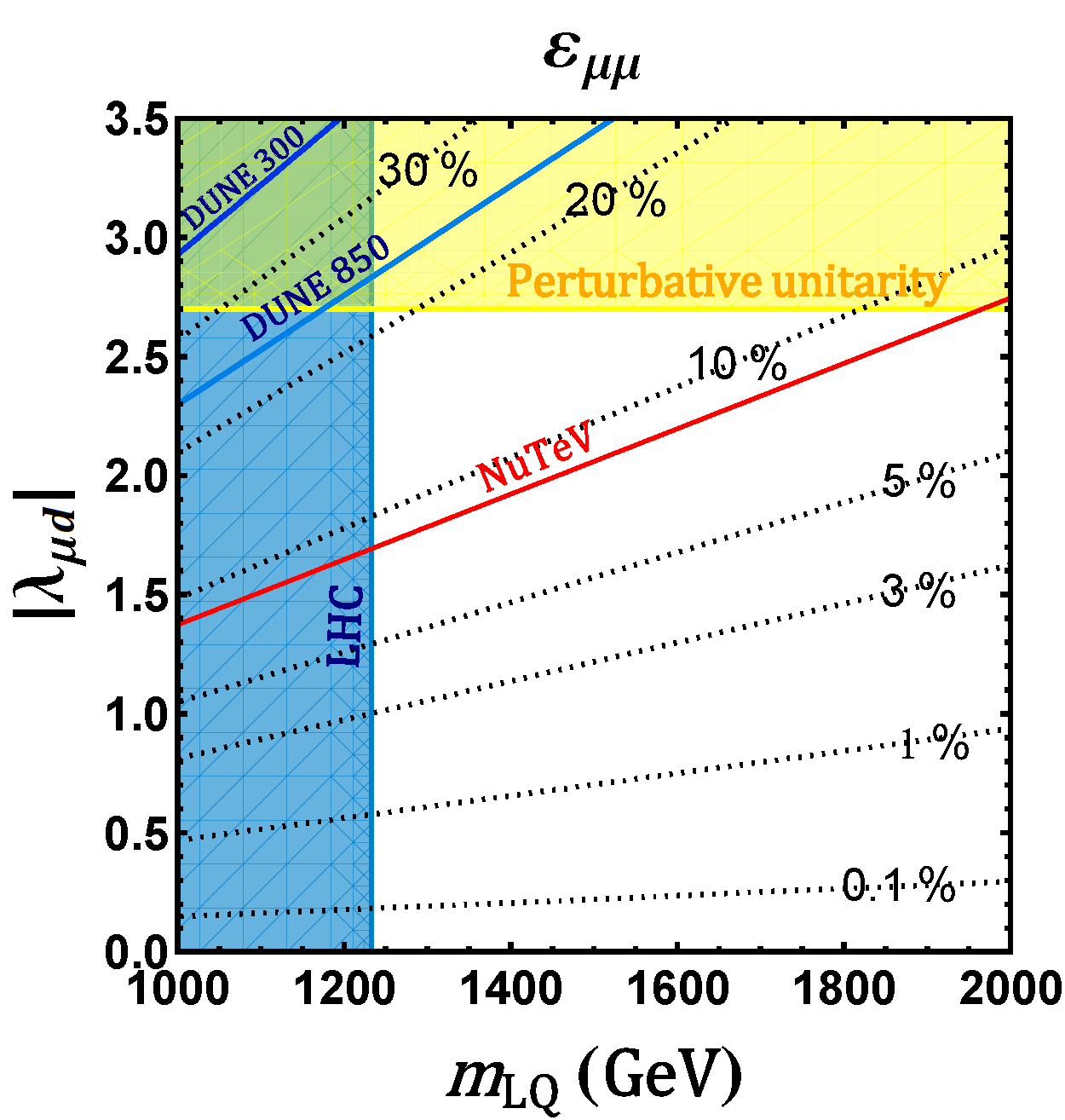}} \\
     \subfigure[]{
      \includegraphics[height=8cm,width=0.45\textwidth]{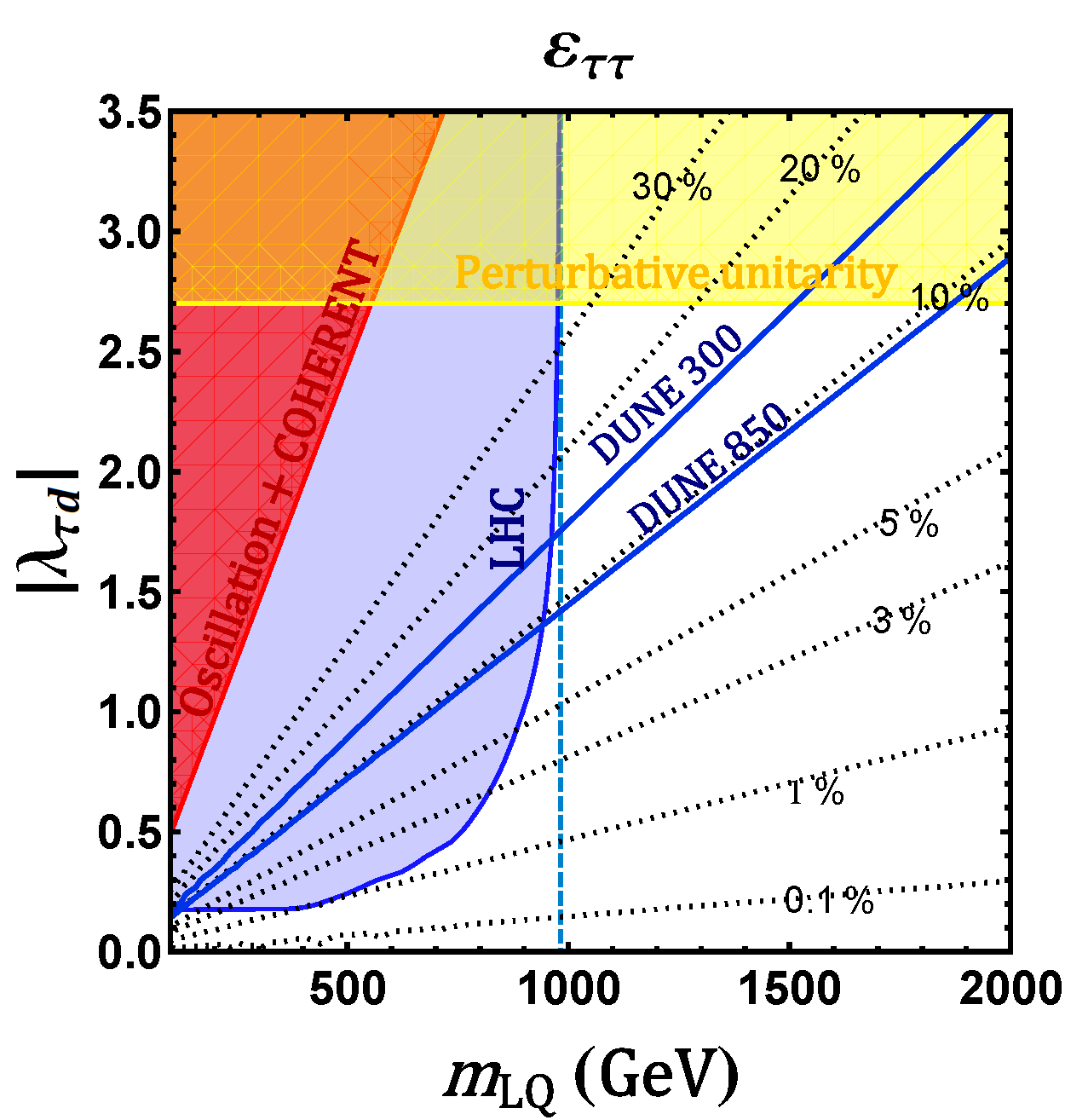}}
    \caption{Predictions for diagonal NSI $(\varepsilon_{ee},\, \varepsilon_{\mu\mu},\,\varepsilon_{\tau\tau})$ induced by doublet LQ in the one-loop LQ model are shown by black dotted contours. Color-shaded regions are excluded by various theoretical and experimental constraints.  Yellow colored region is excluded by perturbativity constraint on LQ coupling $\lambda_{\alpha d}$~\cite{DiLuzio:2017chi}.  Blue-shaded region is excluded by LHC LQ searches (Fig.~\ref{fig:collq1}) in subfigure (a) by $e+$jets channel (pair production for small $\lambda_{ed}$ and single-production for large $\lambda_{ed}$), in subfigure (b) by $\mu+$jets channel, and in subfigure (c) by $\nu$+jet channel. In (a), the red, brown and cyan-shaded regions are excluded by the  APV bound (cf.~Eq.~\ref{eq:APV}), HERA and LEP contact interaction bounds (cf.~Table~\ref{tab:LEPcontactlq}) respectively. In (b), the red line is the suggestive limit from  NuTeV~\cite{Davidson:2003ha}. In (c), the red-shaded region is excluded by  the global-fit constraint from neutrino oscillation+COHERENT data~\cite{Esteban:2018ppq}. 
    We also show the future DUNE sensitivity in blue solid lines for both 300 kt.MW.yr and 850 kt.MW.yr~\cite{dev_pondd}.}
    \label{dnsiLQ}
\end{figure}

\begin{figure}[t!]
  \vspace{-1.7cm}
\centering
\subfigure[]{
      \includegraphics[height=8cm,width=0.45\textwidth]{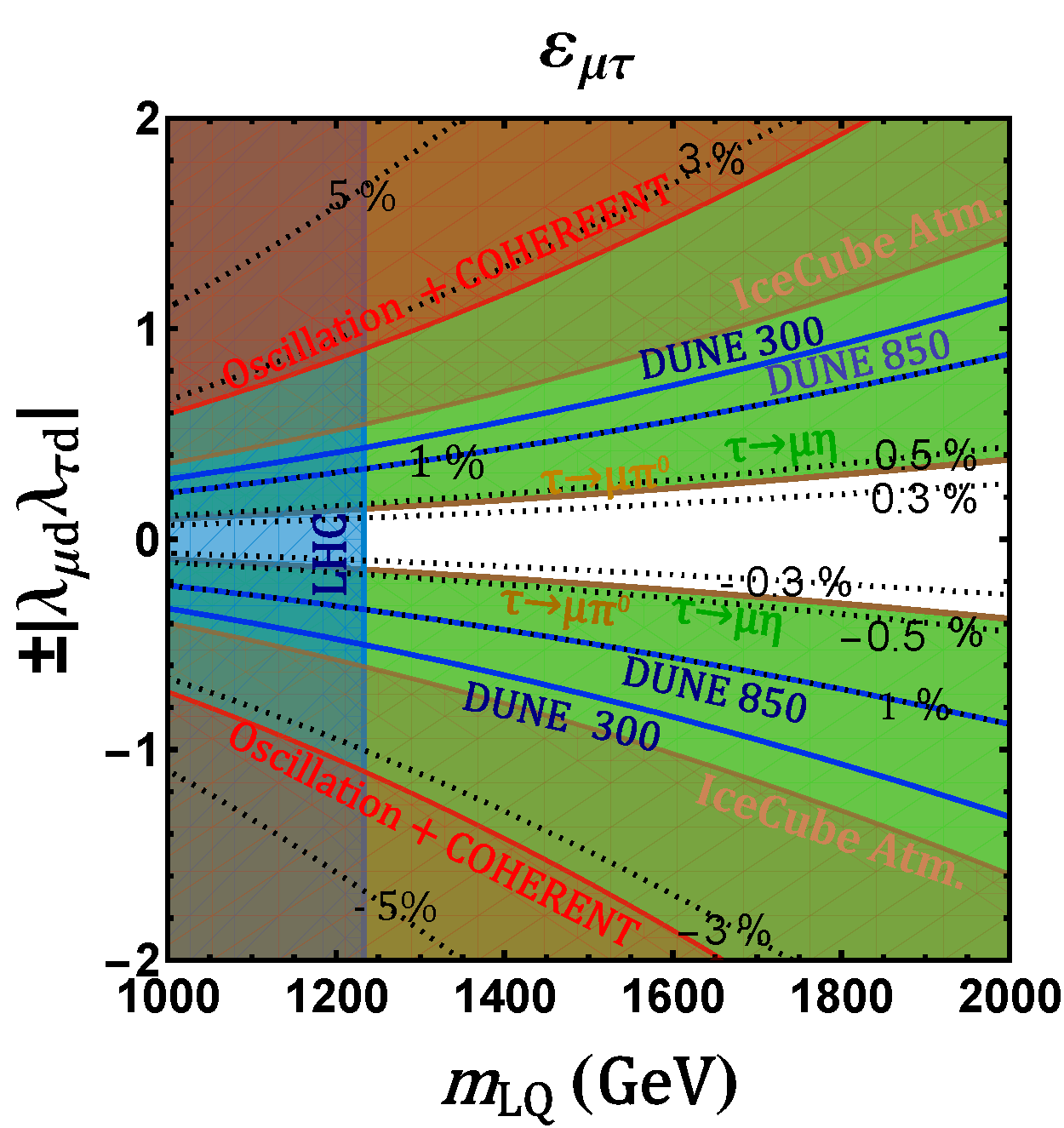}}
      \subfigure[]{
      \includegraphics[height=8cm,width=0.45\textwidth]{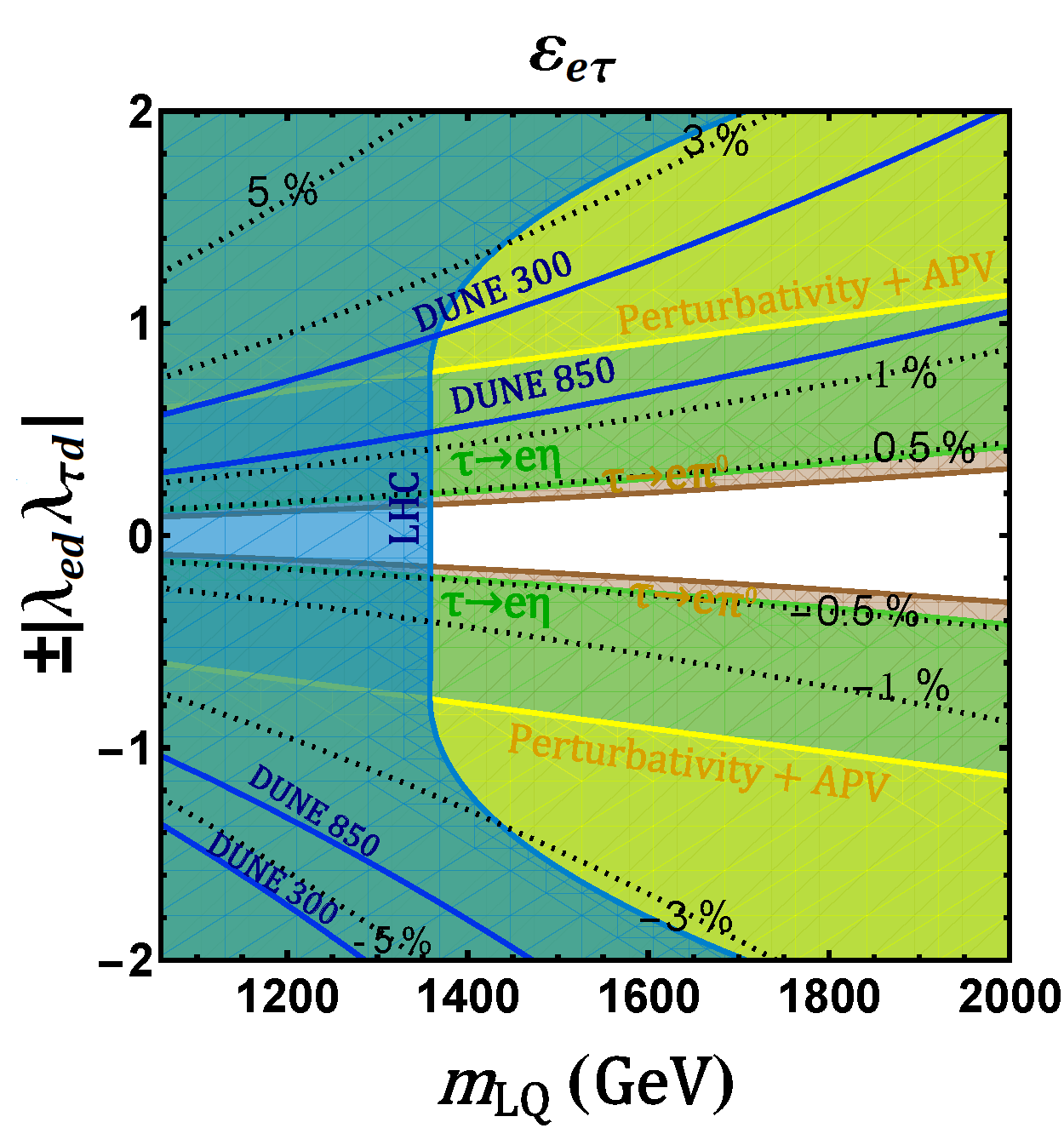}}\\
   \subfigure[]{
     \includegraphics[height=8cm,width=0.5\textwidth]{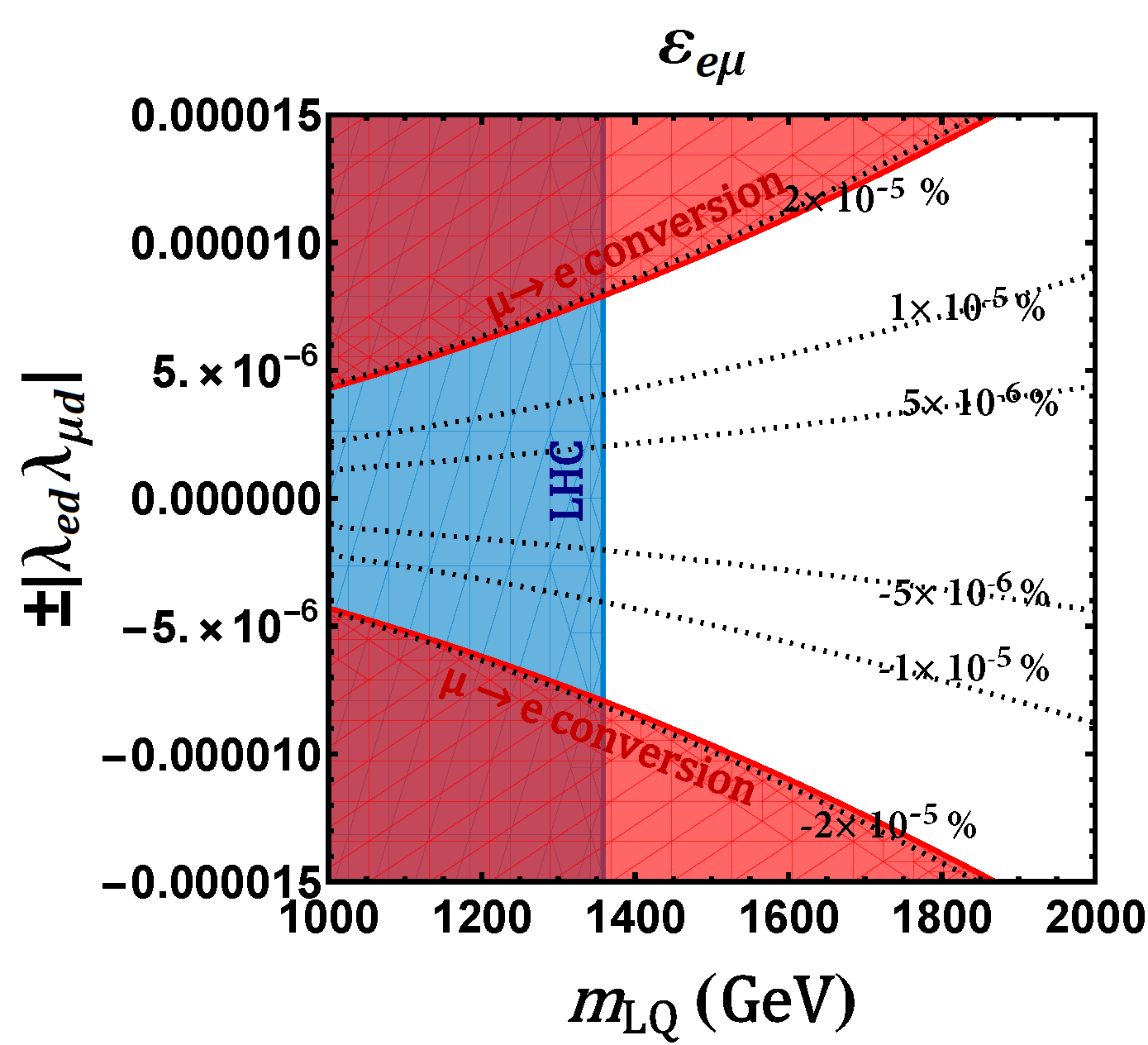}}
    \caption{Predictions for off-diagonal NSI $(\varepsilon_{e\mu},\, \varepsilon_{\mu\tau},\,\varepsilon_{e\tau})$ induced by the doublet LQ in the one-loop LQ model are shown by black dotted contours. Color-shaded regions are excluded by various theoretical and experimental constraints. Blue-shaded area is excluded by LHC LQ searches (cf.~Fig.~\ref{fig:collq1}). In (a) and (b), the brown and green-shaded regions are excluded by $\tau \to \ell \pi^0$ and $\tau\to \ell \eta$ (with $\ell=e,\mu$) constraints (cf.~Table~\ref{tab:semilep}). In (a), the red-shaded region is excluded by the global-fit constraint on NSI from neutrino oscillation+COHERENT data~\cite{Esteban:2018ppq}, and the light brown-shaded region is excluded by IceCube constraint~\cite{Salvado:2016uqu}. In (b), the yellow-shaded region is excluded by perturbativity constraint on LQ coupling $\lambda_{\alpha d}$~\cite{DiLuzio:2017chi} combined with APV  constraint (cf.~Eq.~\eqref{eq:APV}).  In (c), the red-shaded region is excluded by $\mu \to e$ conversion constraint.  Also shown in (b) are the future DUNE sensitivity in blue solid lines for both 300 kt.MW.yr and 850 kt.MW.yr~\cite{dev_pondd}. }
    \label{offdnsiLQ}
\end{figure}

\begin{figure}[t!]
\vspace{-1.0cm}
\centering
\subfigure[]{
      \includegraphics[height=8cm,width=0.45\textwidth]{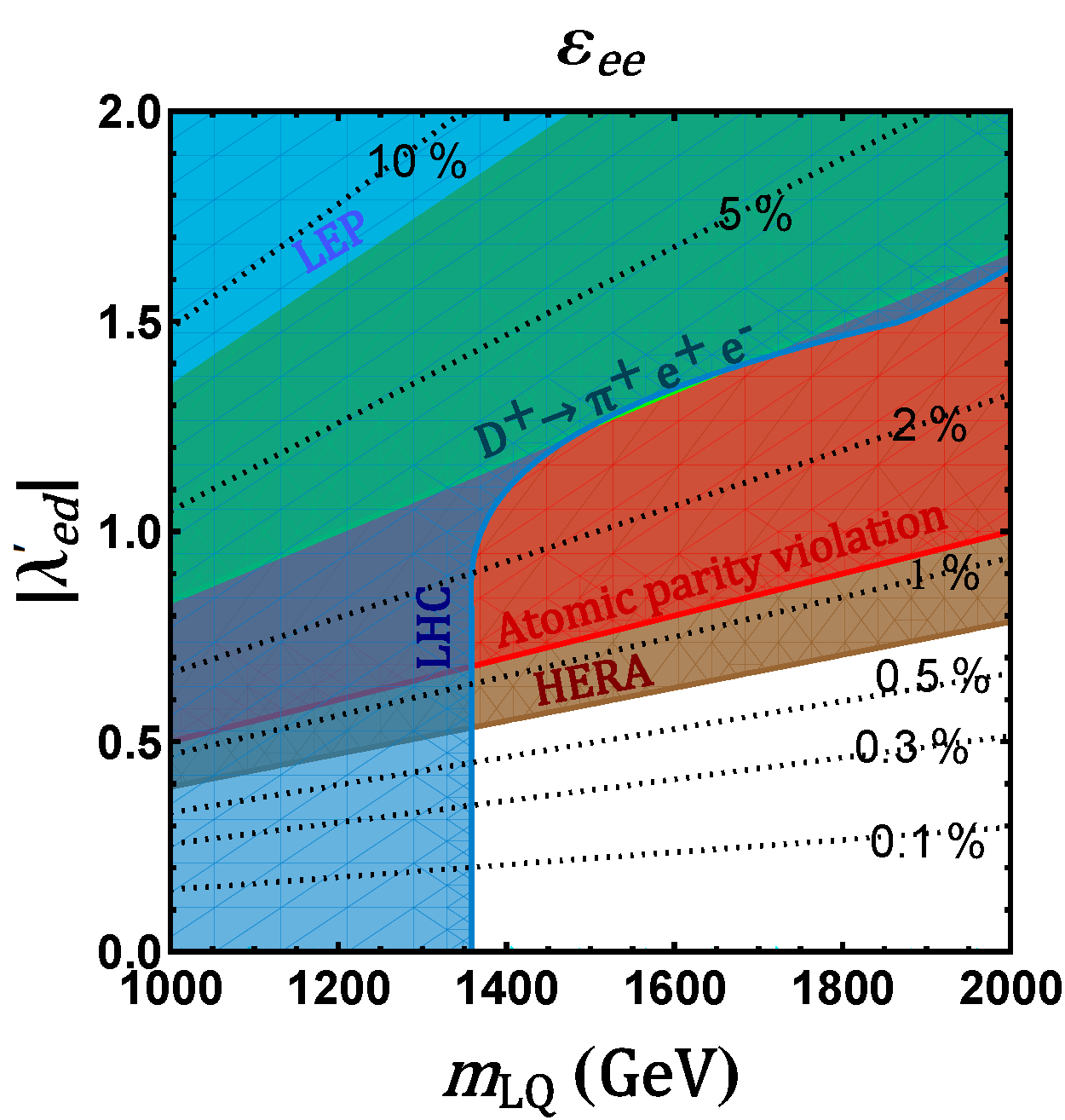}}
      \subfigure[]{
      \includegraphics[height=8cm,width=0.45\textwidth]{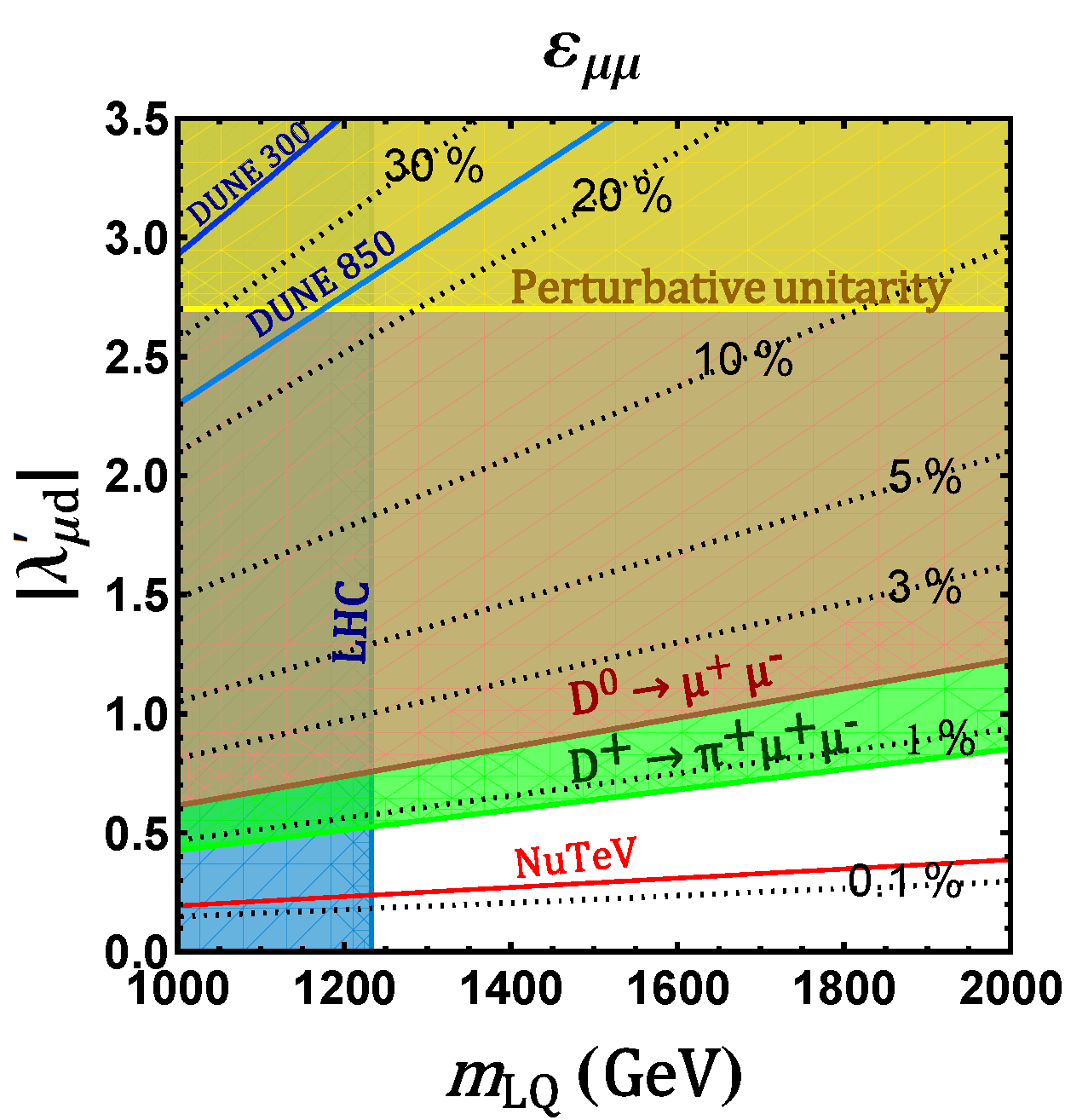}}\\
      \subfigure[]{
    \includegraphics[height=8cm,width=0.45\textwidth]{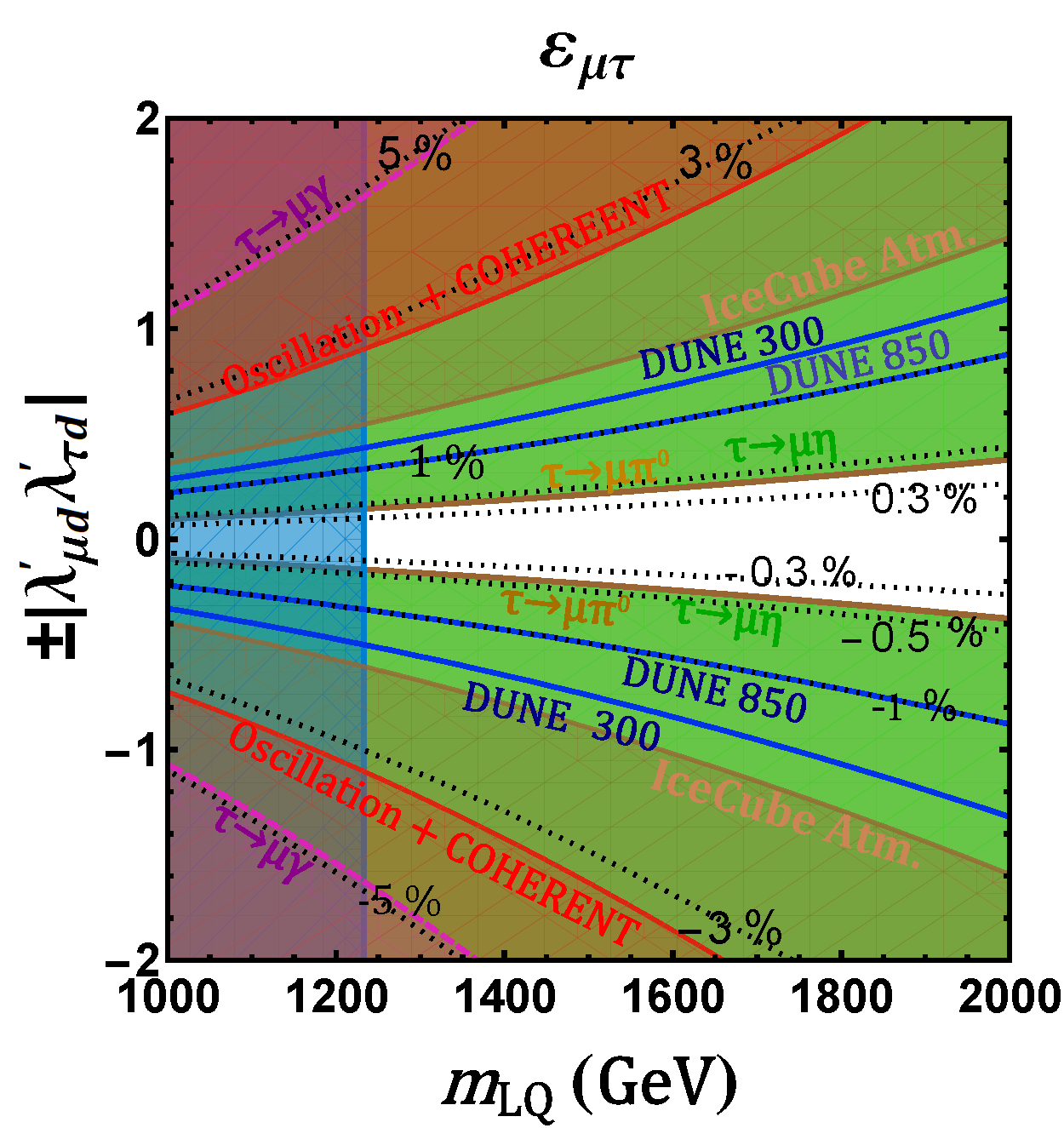}}
      \subfigure[]{
      \includegraphics[height=8cm,width=0.45\textwidth]{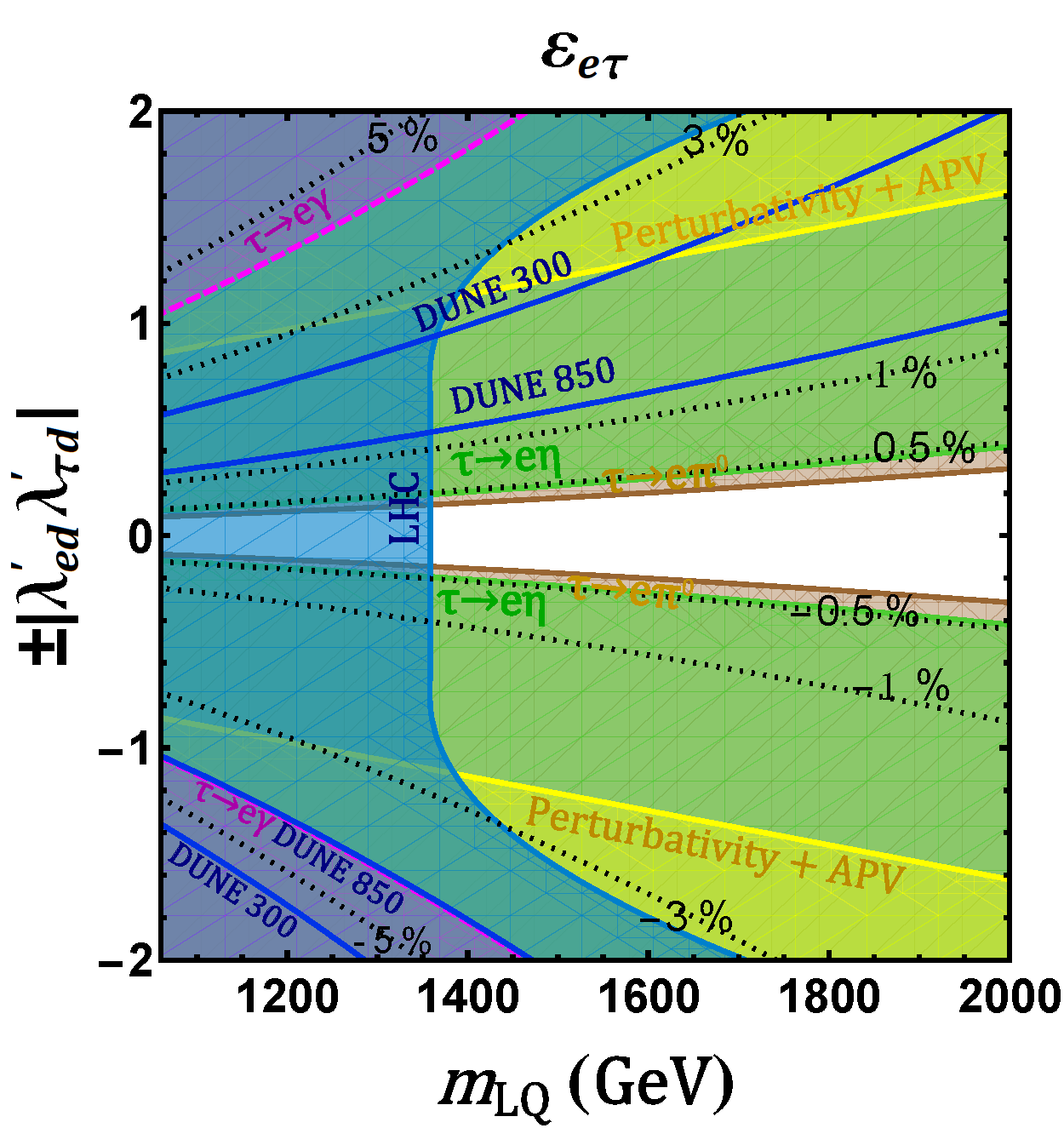}}
    \caption{Additional low-energy constraints on NSI induced by singlet LQ. Subfigure (a) has the same APV and LHC constraints as in  Fig.~\ref{fig:dnsi} (a), the modified HERA and LEP contact interaction bounds (cf.~Table~\ref{tab:LEPcontactlq}), plus the $D^+\to \pi^+ e^+e^-$ constraint, shown by green-shaded region (cf.~Sec.~\ref{sec:Dmeson}). Subfigure (b) has the same constraints as in  Fig.~\ref{fig:dnsi} (b), plus the $D^+\to \pi^+ \mu^+\mu^-$ constraint, shown by light-green-shaded region, and $D^0\to \mu^+\mu^-$ constraint shown by brown-shaded region (cf.~Sec.~\ref{sec:Dmeson}). Subfigure (c) has the same constraints as in  Fig.~\ref{fig:offdnsi} (a), plus the $\tau\to \mu\gamma$ constraint, shown by purple-shaded region. Subfigure (d) has the same constraints as in  Fig.~\ref{fig:offdnsi} (b), plus the $\tau\to e\gamma$ constraint, shown by purple-shaded region. 
    }
    \label{fig:nsiLQlp}
\end{figure}

%%%%%%%%%%%%%%%%%%%%%%%%%%%%%%%%%%%%%%
Now if we take the $\lambda'$ couplings instead of $\lambda$ in Eq.~\eqref{nsi_coloredzee}, the NSI predictions, as well as the constraints, can be analyzed in a similar way as in Figs.~\ref{dnsiLQ} and \ref{offdnsiLQ}. Here the APV (cf.~Eq.~\eqref{eq:APV}), as well as the LEP and HERA contact interaction constraints on $\varepsilon_{ee}$ (cf.~Table~\ref{tab:LEPcontactlq}) are somewhat modified. In addition, there are new constraints from $D^+\to \pi^+\ell^+\ell^-$ and $D^0\to \ell^+\ell^-$ (cf.~Sec.~\ref{sec:Dmeson}) for $\varepsilon_{ee}$ and $\varepsilon_{\mu\mu}$, as shown in Fig.~\ref{fig:nsiLQlp} (a) and (b). For $\varepsilon_{ee}$, the $D^+\to \pi^+e^+e^-$ constraint turns out to be much weaker than the APV constraint. The $D^0\to e^+e^-$ constraint is even weaker and does not appear in Fig.~\ref{fig:nsiLQlp} (a). However, for $\varepsilon_{\mu\mu}$, the $D^+\to \pi^+\mu^+\mu^-$ constraint turns out to be the strongest, limiting the maximum allowed value of $\varepsilon_{\mu\mu}$ to a mere 0.8\%, as shown in Fig.~\ref{fig:nsiLQlp} (b) and in Table~\ref{tab:LQ}. 
 
The NuTeV constraint also becomes more stringent here due to the fact that the singlet LQ $\chi$ couples to left-handed quarks (cf.~Eq.~\eqref{eq:NSILQ}). So it will affect the  effective coupling $\left(\tilde{g}^\ell_{L}\right)$. For $\varepsilon_{\mu\mu}$, we use the NuTeV measurement of  $\left(\tilde{g}^\mu_{L}\right)^2=0.3005\pm 0.0014$ from $\nu_\mu q\to \nu q$ scatterings~\cite{Zeller:2001hh} which is $2.7\sigma$ smaller than the SM prediction of  $\left(\tilde{g}^\mu_{L}\right)_{\rm SM}^2=0.3043$. Here $\left(\tilde{g}^\mu_{L}\right)^2$ is defined as 
\begin{align}
    \left(\tilde{g}^\mu_{L}\right)^2 \ = \ \left(g_L^u+\varepsilon^{uL}_{\mu\mu}\right)^2+\left(g_L^d+\varepsilon^{dL}_{\mu\mu}\right)^2 \, ,
    \label{eq:gLmu}
\end{align}
where $g_L^u  =  \frac{1}{2}-\frac{2}{3}s_w^2$ and $g_L^d =  -\frac{1}{2}+\frac{1}{3}s_w^2$.  For the SM prediction, we have used the latest PDG value for on-shell $s_w^2=0.22343$ from a global-fit to electroweak data (without NuTeV)~\cite{Tanabashi:2018oca}. Comparing $\left(\tilde{g}^\mu_{L}\right)^2$ with the measured value, we derive a 90\% CL constraint
 %
%%%%%%%%%%%%%%%%%%%%%%%%%%%%%
%%%%%%%%%%%%%%%%%%%%%%%%%%%%%%
 %%%%%%%%%%%%%%%%%%%%%%%%%%%%%%
%%%%%%%%%%%%%%%%%%%%%%%%
\clearpage
\noindent on $0.0018< \varepsilon_{\mu\mu}<0.8493$. Note that this prefers a non-zero $\varepsilon_{\mu\mu}$ at 90\% CL ($1.64\sigma$) because the SM  with $\varepsilon_{\mu\mu}=0$ is $2.7\sigma$ away and also because there is a cancellation between $g_L^d$ (which is negative) and $\varepsilon_{\mu\mu}$ (which is positive) in Eq.~\eqref{eq:gLmu} to lower the value of $\left(\tilde{g}^\mu_{L}\right)^2$ to within $1.64\sigma$ of the measured value.
%%%%%%%%%%%%%%%%%%%%%%%%%%%%%%%%%%%%%%%%%%%%%
\begin{table}[!t]
    \centering
    \scriptsize
    \begin{tabular}{|c|c|c|c|c|c|}
    \hline \hline
         & \multicolumn{2}{c|}{\textbf{LQ model prediction (Max.)}} & \textbf{Individual} & \textbf{Global-fit} & \textbf{DUNE} \\ \cline{2-3}
     \textbf{NSI}   & {\bf Doublet} & {\bf Singlet} & \textbf{constraints}  & \textbf{constraints} \cite{Esteban:2018ppq} & {\bf sensitivity}~\cite{dev_pondd} \\ 
        \hline \hline
          $\varepsilon_{ee}$  & 0.004 & 0.0069 & $[-1.8, 1.5]$ \cite{Davidson:2003ha}  & $[-0.036, 1.695]$ & $[-0.185, 0.380]$    \\ 
          & (LHC + APV) & (LHC+HERA) & & & ($[-0.130,0.185]$) \\ \hline
          $\varepsilon_{\mu \mu}$  & 0.216 & 0.0086   & $[-0.024, 0.045]$ \cite{Davidson:2003ha} & $[-0.309, 1.083]$ & $[-0.290,0.390]$    \\ 
          & (LHC+PU) & ($D \to \pi \mu \mu$) & $[0.0277, 0.0857]$ (ours) & & ($[-0.192, 0.240]$) \\ \hline
          $\varepsilon_{\tau \tau}$&  \multicolumn{2}{c|}{0.343}  & $[-0.225, 0.99]$ \cite{Coloma:2017ncl} & $[-0.306, 1.083]$ & $[-0.360,0.145]$  \\ 
           & \multicolumn{2}{c|}{(LHC + Unitarity)} & & & ($[-0.120,0.095]$) \\ \hline
          $\varepsilon_{e \mu}$    & \multicolumn{2}{c|}{$1.5 \times 10^{-7}$}  & $[-0.21, 0.12]$ \cite{Coloma:2017ncl}  & $[-0.174, 0.147]$ & $[-0.025,0.052]$  \\ 
           & \multicolumn{2}{c|}{(LHC + $\mu-e$ conv.)} & & & ($[-0.017, 0.040]$) \\ \hline
          $\varepsilon_{e \tau}$   & \multicolumn{2}{c|}{0.0036} & $[-0.39, 0.36]$ \cite{Coloma:2017ncl}  & $[-0.618, 0.330]$ & $[-0.055,0.023]$\\ 
           & \multicolumn{2}{c|}{(LHC + $\tau \to e \pi^0) $}& & & ($[-0.042,0.012]$) \\ \hline
          $\varepsilon_{\mu \tau}$ & \multicolumn{2}{c|}{0.0043} & $[-0.018, 0.0162]$ \cite{Salvado:2016uqu}  & $[-0.033, 0.027]$ & $[-0.015, 0.013]$  \\ 
           & \multicolumn{2}{c|}{(LHC + $\tau \to \mu \pi^0 $)} & & & ($[-0.010,0.010]$) \\ 
    \hline \hline
    \end{tabular}
    \caption{Maximum allowed NSI (with $d$-quarks) in the one-loop LQ model, after imposing the constraints from APV (Sec.~\ref{sec:APV}), cLFV (Secs.~\ref{sec:mueconv}, \ref{sec:tauLQ}, \ref{sec:Dmeson}), LEP and HERA contact interaction (Sec.~\ref{sec:contactlq}), perturbative unitarity and collider (Sec.~\ref{sec:highconstraints}) constraints. We also impose the constraints from neutrino-nucleon scattering experiments, like CHARM II~\cite{Davidson:2003ha}, NuTeV~\cite{Davidson:2003ha},  COHERENT~\cite{Coloma:2017ncl} and IceCube~\cite{Salvado:2016uqu}, as well as the global-fit constraints from neutrino oscillation+COHERENT data~\cite{Esteban:2018ppq}, whichever is stronger. The scattering and global-fit constraints are on $\varepsilon_{\alpha\beta}^d$, so it has been scaled by a factor of 3 for the constraint on $\varepsilon_{\alpha\beta}$ in the Table. The maximum allowed value for each NSI parameter is obtained after scanning over the LQ mass (see Figs.~\ref{dnsiLQ} and \ref{offdnsiLQ}) and the combination of the relevant constraints limiting the NSI are shown in parentheses in the second column. The same numbers are applicable for the doublet and singlet LQ exchange, except for $\varepsilon_{ee}$ where the APV constraint is weaker than HERA (Fig.~\ref{fig:nsiLQlp} (a))) and for $\varepsilon_{\mu\mu}$ which has an additional constraint from $D^+\to \pi^+\mu^+\mu^-$ decay (see Fig.~\ref{fig:nsiLQlp} (b)). In the last column, we also show the future DUNE sensitivity~\cite{dev_pondd} for 300 kt.MW.yr exposure (and 850 kt.MW.yr in parentheses).}
    \label{tab:LQ}
\end{table}

For the off-diagonal sector, there are new constraints from $\tau\to \ell\gamma$ relevant for $\varepsilon_{\mu\tau}$ and $\varepsilon_{e\tau}$, as shown in Figs.~\ref{fig:nsiLQlp} (c) and (d). However, these are less stringent than the $\tau\to \ell \pi^0$ and $\tau\to \ell \eta$ constraints discussed before. 
There are no new constraints for $\varepsilon_{\tau\tau}$ and $\varepsilon_{e\mu}$ that are stronger than those shown in Figs.~\ref{dnsiLQ} (c) and \ref{offdnsiLQ} (c) respectively, so we do not repeat these plots again in Fig.~\ref{fig:nsiLQlp}. 

%%%%%%%%%%%%%%%%%%%%%%%%%%%%%%%%%%
\section{NSI in a triplet leptoquark model}\label{sec:CCSVO39}
%%%%%%%%%%%%%%%%%%%%%%%%%%%%%%%%%%%%%%%%%%%%%%%
\begin{figure}[t!]
    \centering
    \includegraphics[scale=0.5]{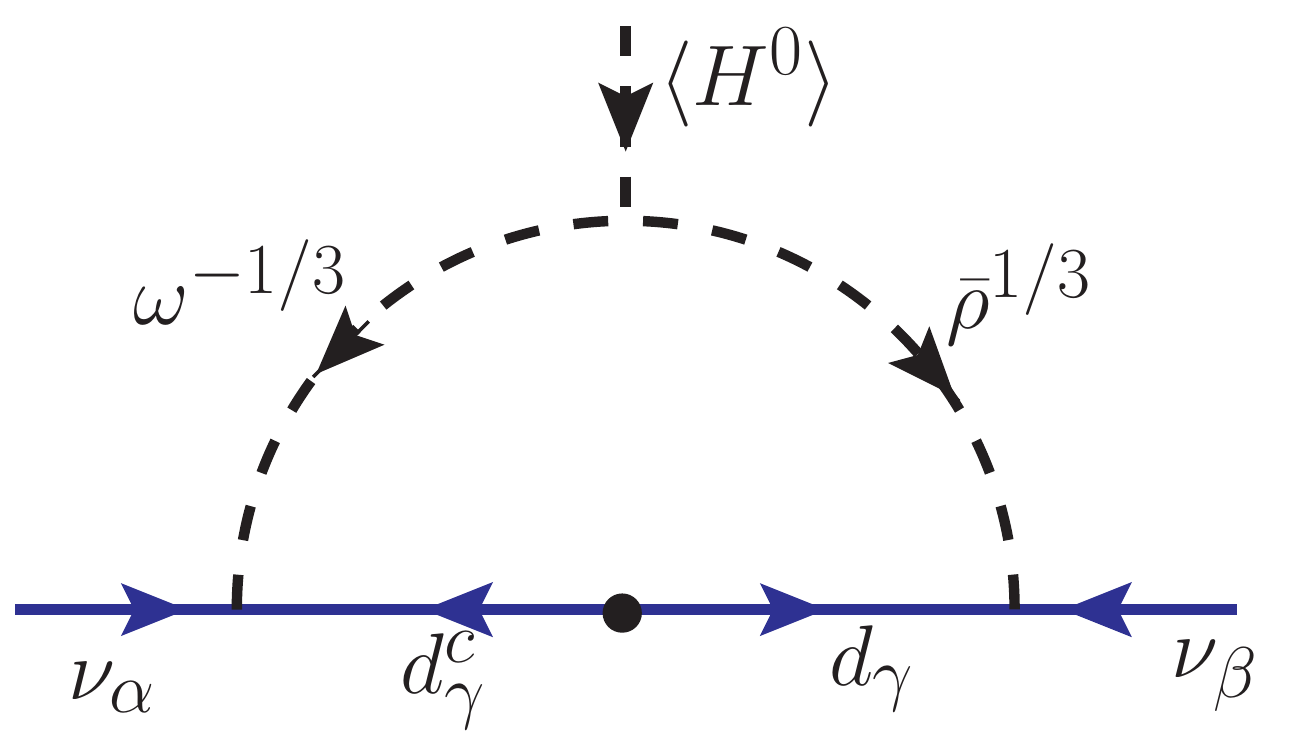}
    \caption{Neutrino mass generation in the one-loop model with both doublet and triplet LQs. This is the $\mathcal{O}_3^9$ model of Table \ref{tab:O3}~\cite{Cai:2014kra}.}
    \label{CCSVO39}
\end{figure}
%%%%%%%%%%%%%%%%%%%%%%%%%%%%%%%%%%%%%%%%%%%%%%%
This is the $\mathcal{O}_3^9$ model of Table~\ref{tab:O3} \cite{Cai:2014kra}.  
In this model, two new fields are introduced -- an $SU(2)_L$-triplet scalar LQ $\bar{\rho}\left({\bf \bar{3}},{\bf 3},\frac{1}{3}\right) = \left(\bar{\rho}^{4/3}, \,\bar{\rho}^{1/3}, \, \bar{\rho}^{-2/3}\right)$ and an $SU(2)_L$-doublet LQ $\Omega\left({\bf 3},{\bf 2},\frac{1}{6}\right)=\left(\omega^{2/3}, \, \omega^{-1/3}\right)$. The relevant Lagrangian for the neutrino mass generation can be written as
\begin{eqnarray}
   -\mathcal{L}_Y & \supset & \lambda_{\alpha \beta} L_\alpha d_\alpha^c \Omega + \lambda_{\alpha \beta}^\prime L_\alpha Q_\beta \Bar{\rho} + {\rm H.c.} \ = \ \lambda_{\alpha \beta} \left(\nu_\alpha d_\beta^c \omega^{-1/3} - \ell_\alpha d_\beta^c \omega^{2/3}\right)  \nonumber \\
   &&  \qquad + \lambda_{\alpha \beta}^\prime \left[\ell_\alpha d_\beta \Bar{\rho}^{4/3} - \frac{1}{\sqrt 2}\left(\nu_\alpha d_\beta + \ell_\alpha u_\beta\right)\Bar{\rho}^{1/3}  + \nu_\alpha u_\beta \Bar{\rho}^{-2/3}\right]+ {\rm H.c.} \, 
   \label{eq:lagO39ex}
\end{eqnarray}
These interactions, along with the potential term 
\begin{align}
    V \ \supset \ \mu \widetilde{\Omega} \rho H + {\rm H.c.} \ = \ & \mu \bigg[\omega^{\star 1/3} \rho^{-4/3} H^+ + \frac{1}{\sqrt 2}\left(\omega^{\star 1/3} H^0 - \omega^{\star -2/3} H^+\right)\rho^{-1/3} \nonumber \\
    & \qquad  - \omega^{\star -2/3} \rho^{2/3} H^0\bigg] + {\rm H.c.} \, ,
\end{align}
 where $\bar{\rho}$ is related to $\rho$ by charge conjugation as $\rho\left({\bf {3}},{\bf 3},-\frac{1}{3}\right) = \left(\rho^{2/3},\, -\rho^{-1/3},\, \rho^{-4/3}\right)$, 
induce neutrino mass at one-loop level via the ${\cal O}_3^9$ operator in the notation of Ref.~\cite{Cai:2014kra}, as shown in Fig.~\ref{CCSVO39}.  
The neutrino mass matrix can be estimated as
\begin{equation}
    M_\nu \ \sim \ \frac{1}{16\pi^2}\frac{\mu v}{M^2}\left(\lambda M_d \lambda^{\prime T}   + \lambda^{\prime}M_d\lambda^T \right) \, ,
    \label{eq:MnuO39}
\end{equation}
where $M_d$ is the diagonal down-type quark mass matrix and $M \equiv {\rm max}(m_\omega, m_\rho)$.  The NSI parameters read as
%\begin{equation}
%   \boxed{  
\begin{tcolorbox}[enhanced,ams align,
  colback=gray!30!white,colframe=white]
   \varepsilon_{\alpha \beta} \ = \ \frac{3}{4 \sqrt{2}G_F} \left(\frac{\lambda_{\alpha d}^{\star} \lambda_{\beta d} }{ m_\omega^2} + \frac{\lambda_{\alpha u}^{\prime\star} \lambda'_{\beta u}}{ m_{\rho^{-2/3}}^2} + \frac{\lambda_{\alpha d}^{\prime\star} \lambda'_{\beta d}}{ 2m_{\rho^{1/3}}^2}\right)  \, .
   %}
      \label{eq:NSI-O39}
%\end{equation}
\end{tcolorbox}
Note that both $\lambda$ and $\lambda'$ cannot be large at the same time due to neutrino mass constraints (cf.~Eq.~\eqref{eq:MnuO39}). 
For $\lambda\gg \lambda'$, this expression is exactly the same as the doublet LQ contribution derived in Eq.~\eqref{nsi_coloredzee} and the corresponding maximum NSI can be read off from Table~\ref{tab:LQ} for the doublet component. 

On the other hand, for $\lambda'\gg \lambda$, the third 
term in Eq.~\eqref{eq:NSI-O39} is analogous to the down-quark induced singlet LQ NSI given in Eq.~\eqref{nsi_coloredzee} (except for the Clebsch-Gordan factor of $(1/\sqrt 2)^2$), whereas the second term is a new contribution from the up-quark sector. Note that both terms  depend on the same Yukawa coupling $\lambda'_{\alpha u}=\lambda'_{\alpha d}$ in the Lagrangian~\eqref{eq:lagO39ex}. This is unique to the triplet LQ model, where neutrinos can  have sizable couplings to both up and down quarks simultaneously, without being in conflict with the neutrino mass constraint. As a result, some of the experimental constraints quoted in Sec.~\ref{sec:LQ} which assumed the presence of only down-quark couplings of LQ will be modified in the triplet case, as discussed below:  
%%%%%%%%%%%%%%%%%%%%%%%%%%%%
\subsection{Atomic parity violation} 
The shift in the weak charge given by Eq.~\eqref{eq:Qw} is modified to 
\begin{align}
\delta Q_w(Z,N) \ = \ \frac{1}{2\sqrt 2 G_F} \left[(2Z+N)\frac{|\lambda'_{eu}|^2}{2m_{\rho^{1/3}}^2}-(Z+2N)\frac{|\lambda'_{ed}|^2}{m_{\rho^{4/3}}^2}\right] \, .
\label{eq:Qwt}
\end{align}
Assuming $m_{\rho^{1/3}}=m_{\rho^{4/3}}\equiv m_\rho$ and noting that $\lambda'_{\alpha u}=\lambda'_{\alpha d}$ in Eq.~\eqref{eq:lagO39ex}, we obtain 
\begin{align}
    \delta Q_w\left(_{55}^{133}{\rm Cs}\right) \ = \ -\frac{117}{2\sqrt 2 G_F}\frac{|\lambda'_{ed}|^2}{m_\rho^2} \, .
\end{align}
Comparing this with the $2\sigma$ allowed range~\eqref{eq:delqw}, we obtain the modified constraint
\begin{align}
    |\lambda'_{ed}| \ < \ 0.29 \left(\frac{m_\rho}{\rm TeV}\right) \, ,
\end{align}
which is weaker (stronger) than that given by Eq.~\eqref{eq:APV} for the $SU(2)_L$-doublet (singlet) LQ alone. 
%This is due to the partial cancellation between the $\rho^{-2/3}$ and $\rho^{1/3}$ contributions in Eq.~\eqref{eq:Qwt}. 

\subsection{\texorpdfstring{$\mu-e$}{mue} conversion} 
From Eq.~\eqref{eq:mueconv}, we see that for the triplet case, the rate of $\mu-e$ conversion will be given by 
\begin{align}
 {\rm BR}(\mu N\to eN) \ \simeq \  \frac{|\vec{p}_e|E_e m_\mu^3 \alpha^3Z^4_{\rm eff}F_p^2}{64\pi^2 Z\Gamma_N}(2A-Z)^2\left(\frac{|\lambda^{\prime\star}_{ed}\lambda'_{\mu d}|}{m^2_{\rho^{4/3}}}+\frac{|\lambda^{\prime\star}_{eu}\lambda'_{\mu u}|}{2m^2_{\rho^{1/3}}}\right)^2 \, , 
 \label{eq:mueconv1}
 \end{align}
 For degenerate $\rho$-mass and $\lambda'_{\ell d}=\lambda'_{\ell u}$, we obtain the rate to be $(3/2)^2$ times larger than that given in Eq.~\eqref{eq:mueconv}. Therefore, the constraints on $|\lambda^{\prime\star}_{ed}\lambda'_{\mu d}|$ given in Table~\ref{tab:mueconvLQ} will be a factor of 3/2 stronger. 

\subsection{Semileptonic tau decays} The semileptonic tau decays such as $\tau^-\to \ell^-\pi^0,~\ell^-\eta,~\ell^-\eta'$ will have two contributions from $\bar{\rho}^{1/3}$ and $\bar{\rho}^{4/3}$.  The relevant terms in the Lagrangian~\eqref{eq:lago35ex} are 
\begin{align}
    -{\cal L}_Y & \ \supset \ \lambda'_{\alpha\beta}\left(-\frac{1}{\sqrt 2}\ell_\alpha u_\beta \bar{\rho}^{1/3}+\ell_\alpha d_\beta \bar{\rho}^{4/3}\right)+{\rm H.c.} \nonumber \\
    & \ \supset \ \lambda'_{\tau d}\left(-\frac{1}{\sqrt 2}\tau V^\star_{ud} u \bar{\rho}^{1/3}+\tau d \bar{\rho}^{4/3}\right)+\lambda_{\ell d}\left(-\frac{1}{\sqrt 2}\ell V^\star_{ud} u \bar{\rho}^{1/3}+\ell d \bar{\rho}^{4/3}\right)+{\rm H.c.} \, ,
\end{align}
where we have assumed a basis with diagonal down-type quark sector. Using the matrix element~\eqref{eq:pimatrix}, we find the modified decay rate for $\tau^-\to \ell^-\pi^0$ from Eq.~\eqref{tautoeta}:
\begin{align}
    \Gamma_{\tau\to \ell \pi^0} \ = \  \frac{\left|\lambda'_{\ell d} \lambda_{\tau d}^{\prime\star}\right|^{2}}{1024 \pi} f_{\pi}^{2} m_{\tau}^{3}{\cal F}_\tau(m_\ell,m_\pi)\left( \frac{1}{m^2_{\rho^{4/3}}}-\frac{1}{2m^2_{\rho^{-1/3}}}\right)^2 \, .
\end{align}
Thus, for $m_{\rho^{-1/3}}=m_{\rho^{4/3}}$, the $\tau^-\to \ell^-\pi^0$ decay rate is suppressed by a factor of 1/4, compared to the doublet or singlet LQ case (cf.~Eq.~\eqref{tautoeta}). So the constraints on $\lambda'_{\ell d} \lambda_{\tau d}^{\star}$ from $\tau\to \ell \pi^0$ shown in Table~\ref{tab:semilep} will be a factor of 2 weaker in the triplet LQ case.

On the other hand, using the matrix element~\eqref{eq:etamatrix}, we find that the modified decay rate for $\tau^-\to \ell^-\eta$ becomes
\begin{align}
    \Gamma_{\tau\to \ell \eta} \ = \  \frac{\left|\lambda'_{\ell d} \lambda_{\tau d}^{\prime \star}\right|^{2}}{1024 \pi} f_{\eta}^{2} m_{\tau}^{3}{\cal F}_\tau(m_\ell,m_\eta)\left( \frac{1}{m^2_{\rho^{4/3}}}+\frac{1}{2m^2_{\rho^{-1/3}}}\right)^2 \, ,
\end{align}
which is enhanced by a factor of 9/4 for $m_{\rho^{-1/3}}=m_{\rho^{4/3}}$, compared to the doublet or singlet LQ case. So the constraints on $\lambda_{\ell d} \lambda_{\tau d}^{\star}$ from $\tau\to \ell \eta$ shown in Table~\ref{tab:semilep} will be a factor of 3/2 stronger in the triplet LQ case. The same scaling behavior applies to $\tau\to \ell \eta'$ constraints. These modified constraints are summarized in Table~\ref{tab:taudecay_triplet}. 

\begin{table}[!t]
    \centering
    \begin{tabular}{|c|c|c|}
    \hline \hline
        {\bf Process} & {\bf Exp. limit}~\cite{Tanabashi:2018oca} & {\bf Constraint}  \\
        \hline \hline
        \rule{0pt}{15pt} $\tau \rightarrow \mu \pi^{0}$ & BR < $1.1 \times 10^{-7} $ & $|\lambda'_{\mu d} \lambda_{\tau d}^{\prime\star}| < 1.9 \times 10^{-1} \left(\frac{m_\rho}{ \text{TeV}}\right)^2$ \\
       \rule{0pt}{15pt}  $\tau \rightarrow e \pi^{0}$ & BR < $8 \times 10^{-8}$& $|\lambda'_{e d} \lambda_{\tau d}^{\prime\star}| < 1.6 \times 10^{-1} \left(\frac{m_\rho}{\text{TeV}}\right)^2$  \\
       \rule{0pt}{15pt}  $\tau \rightarrow \mu \eta$  & BR < $6.5 \times 10^{-8}$ & $|\lambda'_{\mu d} \lambda_{\tau d}^{\prime\star}| <6.3 \times 10^{-2} \left(\frac{m_\rho}{ \text{TeV}}\right)^2$ \\
      \rule{0pt}{15pt}  $\tau \rightarrow e \eta$ & BR < $9.2 \times 10^{-8}$ & $|\lambda'_{e d} \lambda_{\tau d}^{\prime\star}| <7.3 \times 10^{-2} \left(\frac{m_\rho}{\text{TeV}}\right)^2$ \\
       \rule{0pt}{15pt}  $\tau \rightarrow \mu \eta'$  & BR < $1.3 \times 10^{-7}$ & $|\lambda'_{\mu d} \lambda_{\tau d}^{\prime\star}|< 1.5 \times 10^{-1} \left(\frac{m_\rho}{\text{TeV}}\right)^2$ \\
      \rule{0pt}{15pt}  $\tau \rightarrow e \eta'$ & BR < $1.6 \times 10^{-7}$ & $|\lambda'_{e d} \lambda_{\tau d}^{\prime\star}|<1.7 \times 10^{-1} \left(\frac{m_\rho}{\text{TeV}}\right)^2$ \\
       \hline \hline
    \end{tabular}
    \caption{Constraints on couplings and the LQ mass from semileptonic tau decays in the triplet LQ case. Here we have assumed all the triplet fields ($\bar{\rho}^{4/3}$, $\bar{rho}^{1/3}$, $\bar{\rho}^{-2/3}$) to have the same mass $m_\rho$.}
    \label{tab:taudecay_triplet}
\end{table}

\subsection{\texorpdfstring{$\ell_\alpha\to \ell_\beta+\gamma$}{ellalpha}}
The cLFV decay $\ell_\alpha\to \ell_\beta+\gamma$ arises via one-loop diagrams with the exchange of $\bar{\rho}$ LQ fields, analogous to Fig.~\ref{fig:llg_LQ}. The relevant couplings in Eq.~\eqref{eq:lagO39ex} have the form $\ell u\bar{\rho}^{1/3}=\overline{u^c}P_L\ell \bar{\rho}^{1/3}$ for which $Q_F=-2/3$ and $Q_B=1/3$ in the general formula~\eqref{rate_rad}, whereas for the couplings $\ell d\bar{\rho}^{4/3}=\overline{d^c}P_L\ell \bar{\rho}^{4/3}$, we have $Q_F=1/3$ and $Q_B=4/3$. Substituting these charges in Eq.~\eqref{rate_rad} and taking the limit $t=m_F^2/m_B^2\to 0$ (since the LQs are expected to be much heavier than the SM charged leptons), we obtain 
\begin{align}
    \Gamma(\ell_\alpha\to \ell_\beta+\gamma) \ = \ \frac{9\alpha}{256}\frac{|\lambda'_{\beta d}\lambda_{\alpha d}^{\prime\star}|}{(16\pi^2)^2}\frac{m_\alpha^5}{m_\rho^4} \, ,
    \label{eq:llgO39}
\end{align}
where $9=3^2$ is a color factor and we have assumed $m_{\rho^{-1/3}}=m_{\rho^{4/3}}$. The rate in Eq.~\eqref{eq:llgO39} is 9/4 times larger than that given in Eq.~\eqref{eq:llgLQ} for the singlet LQ case. Therefore, the constraints on $|\lambda'_{\beta d}\lambda_{\alpha d}^{\prime\star}|$ derived in Table~\ref{lqlfv1} will be weakened by a factor of 3/2. 

\subsection{\texorpdfstring{$D$}{D}-meson decays} 
The $\ell_\alpha u_\beta\bar{\rho}^{1/3}$ and $\ell_\alpha d_\beta \bar{\rho}^{4/3}$ terms in Eq.~\eqref{eq:lago35ex} induce  flavor violating quark decays. Following the discussion in Sec.~\ref{sec:Dmeson}, we work in a basis where the down quark mass matrix is diagonal, so there are no constraints from rare kaon decays. However, the $\ell_\alpha u_\beta \bar{\rho}^{1/3}$ term in Eq.~\eqref{eq:lago35ex} now becomes $\ell_\alpha V^\star_{id}u_i\bar{\rho}^{1/3}$ which induces  $D^0\to \ell^+\ell^-$ and $D^+\to \pi^+\ell^+\ell^-$ decays. The analysis will be the same as in Sec.~\ref{sec:Dmeson}, except that the $\lambda'_{\alpha d}$ couplings will now be replaced by $\lambda'_{\alpha d} /\sqrt{2}$. Correspondingly, the constraints on $|\lambda'_{\alpha d}|$ given in Table~\ref{tab:D0decay} will be $\sqrt 2$ times weaker. For instance, 
\begin{align}
   | \lambda'_{\mu d}| \ < \ \left\{\begin{array}{ll} 0.868 \left(\frac{m_\rho}{\rm TeV}\right) & \quad {\rm from}~D^0\to \mu^+\mu^-\\
    0.602\left(\frac{m_\rho}{\rm TeV}\right) & \quad  {\rm from}~D^+\to \pi^+\mu^+\mu^- 
    \end{array}\right. \, .
\end{align}

\subsection{Contact interaction constraints} 
The LEP and HERA contact interaction bounds discussed in Sec.~\ref{sec:contactlq} will also be modified in the triplet LQ case. Here, the interactions are only of $LL$ type, but the effective Yukawa coupling is $\sqrt{3/2}$ times that of the singlet case in Table~\ref{tab:LEPcontactlq}. The modified constraint is given by 
\begin{align}
    \frac{m_\rho}{|\lambda'_{ed}|} \ = \ \sqrt{\frac{3}{16\pi}}\:\Lambda_-^{LL} \ > \  \left\{\begin{array}{ll} 0.904~{\rm TeV} & \quad {\rm from~LEP} \\
    3.127~{\rm TeV} & \quad {\rm from~HERA}
    \end{array}\right. 
    \, .
    \end{align}
 
 \subsection{LHC constraints} 
 The LHC constraints on the $\bar{\rho}$ fields will be similar to the discussion in Sec.~\ref{sec:highconstraints}. Comparing the Lagrangians~\eqref{lagLQ} and \eqref{eq:lago35ex}, we see that  $\bar{\rho}^{1/3}$ will have the same decay modes to $\nu j$ and $\ell j$, and therefore, the same constraints as the singlet $\chi^{-1/3}$ discussed in Sec.~\ref{subsec:singlet_NSI}. In our analysis, we have assumed degenerate mass spectrum for all the triplet LQ fields. But we note here that the $\bar{\rho}^{-2/3}$ component can in principle be lighter, since it can only decay to $\nu j$ for which the constraints are weaker (cf.~Fig.~\ref{fig:collq1}). However, the mass splitting between $\bar{\rho}^{-2/3}$ and $\bar{\rho}^{1/3}$ cannot be more than $\sim 100$ GeV from $T$-parameter constraints, analogous to the charged scalar case discussed in Sec.~\ref{sec:ewpt} (cf.~Fig.~\ref{fig:ewp2}). In that case, the limit on $m_{\rho^{1/3}}$ for 50\% branching ratio to $\nu j$ and $\ell j$ channels (since they are governed by the same $\lambda'_{\alpha d}$ coupling), one can allow for $m_{\rho^{-2/3}}$ as low as 800 GeV or so. 
 
\subsection{NSI prediction}  
\begin{figure}[t!]
 \vspace{-1.7cm}
 \centering
 \subfigure[]{
    \includegraphics[height=8cm,width=0.45\textwidth]{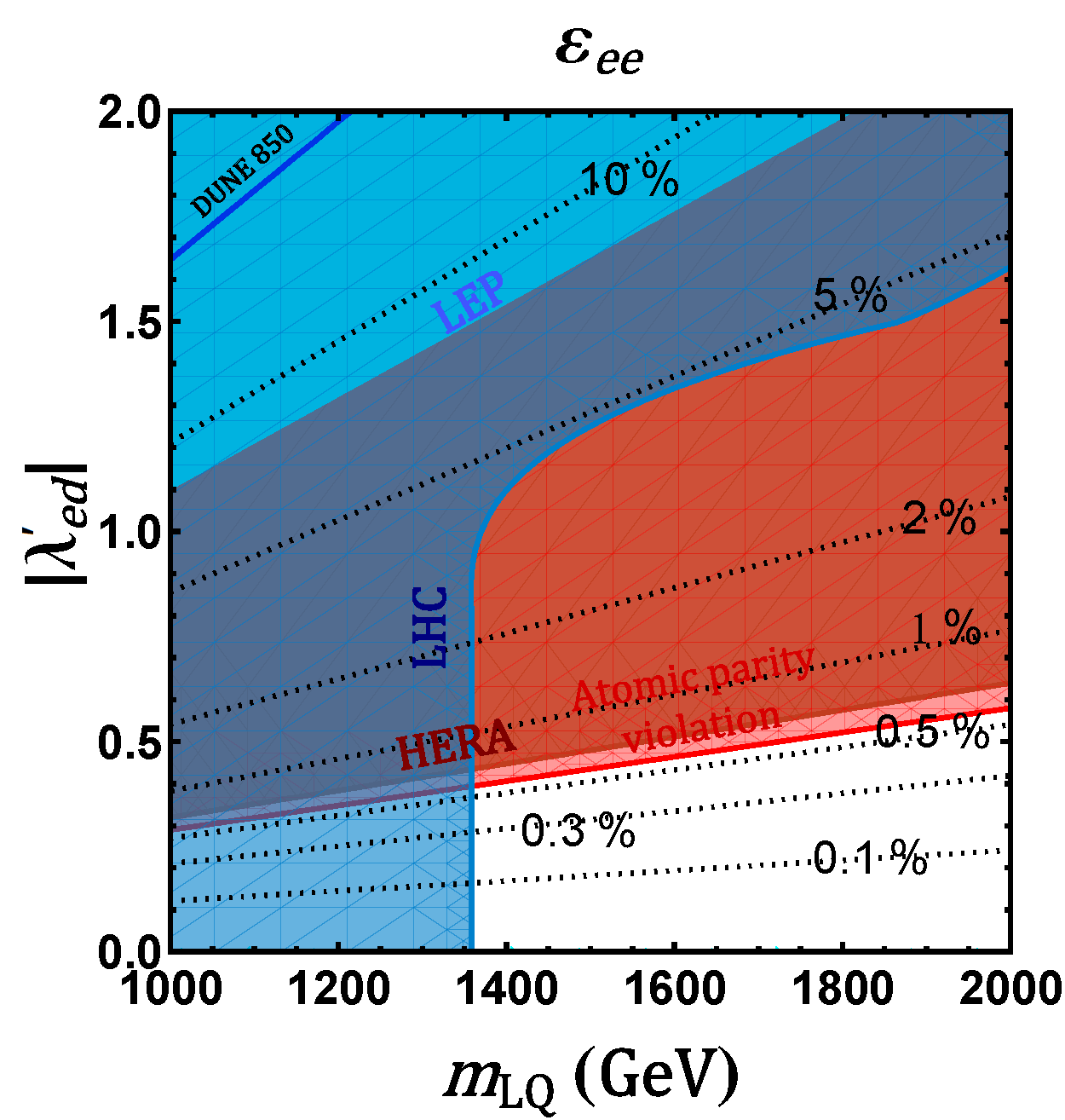} }
    \subfigure[]{
     \includegraphics[height=8cm,width=0.45\textwidth]{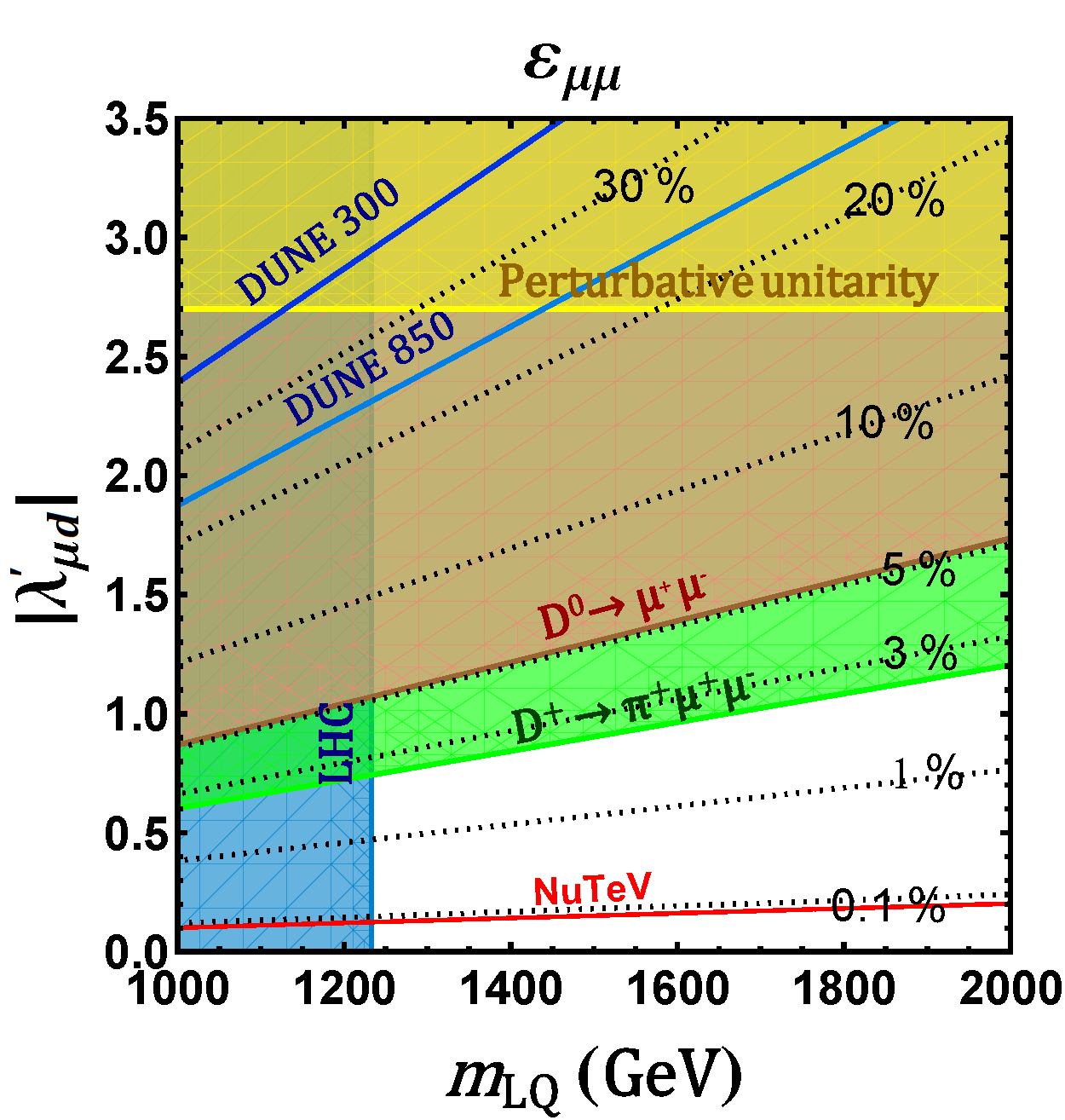}} \\
     \subfigure[]{
      \includegraphics[height=8cm,width=0.45\textwidth]{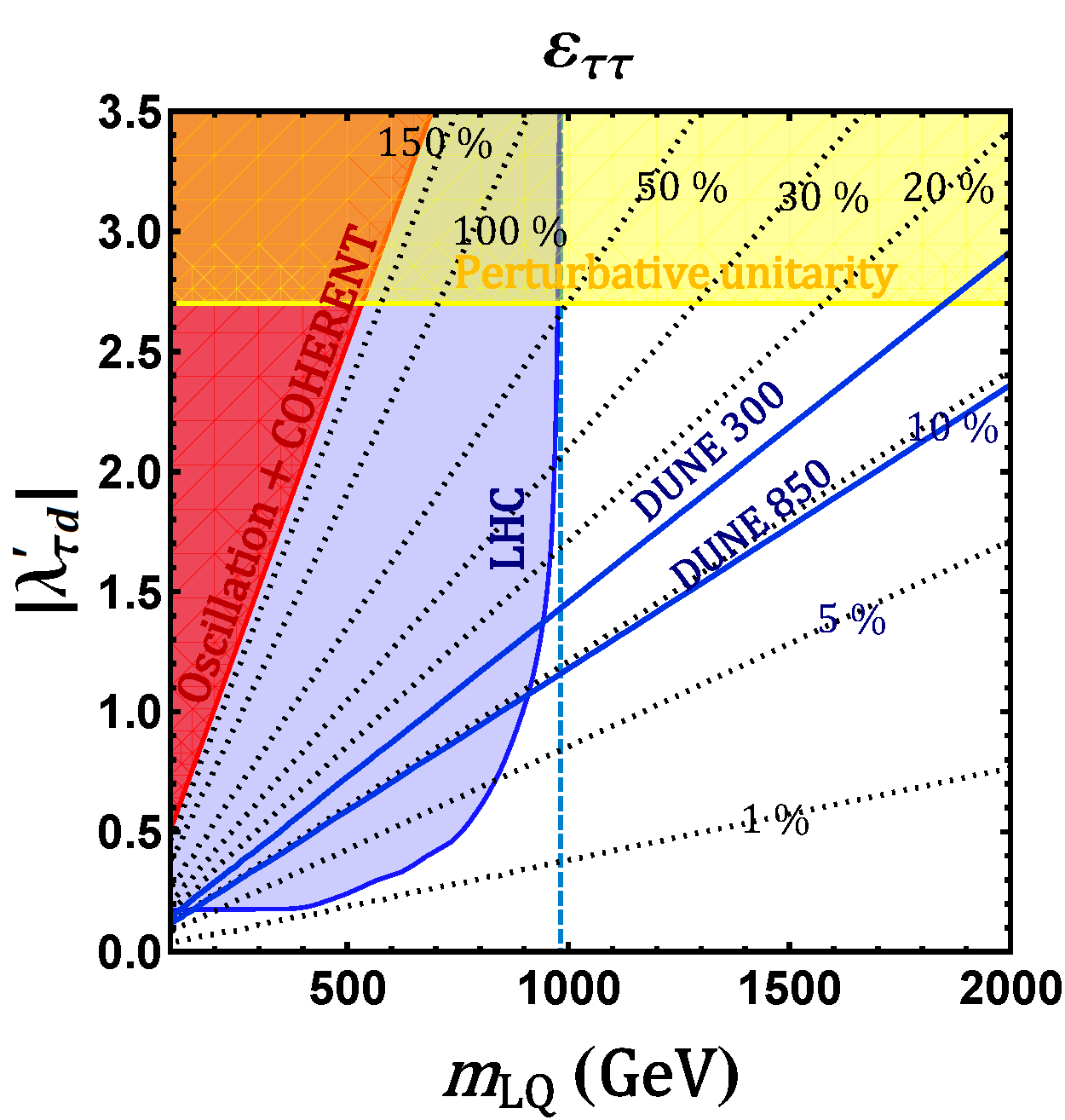}}
    \caption{Predictions for diagonal NSI $(\varepsilon_{ee},\, \varepsilon_{\mu\mu},\,\varepsilon_{\tau\tau})$ induced by the triplet LQ are shown by black dotted contours. Color-shaded regions are excluded by various theoretical and experimental constraints. The labels are same as in Fig.~\ref{dnsiLQ}.} 
    %Yellow colored region is excluded by perturbativity constraint on LQ coupling $\lambda'_{\alpha d}$~\cite{DiLuzio:2017chi}.  Blue shaded region is excluded by LHC LQ searches (Fig.~\ref{fig:collq1}) in subfigure (a) by $e+$jets channel (pair production for small $\lambda'_{ed}$ and single-production for large $\lambda'_{ed}$), in subfigure (b) by $\mu+$jets channel, and in subfigure (c) by $\nu$+jet channel. In (a), the red, brown and cyan shaded regions are excluded by the  APV bound (cf.~Eq.~\ref{eq:APV}), HERA and LEP contact interaction bounds (cf.~Table~\ref{tab:LEPcontactlq}) respectively. In (b), the orange shaded region is excluded by NuTeV~\cite{Davidson:2003ha}. In (c), the red shaded region is excluded by  the global-fit constraint from neutrino oscillation+COHERENT data~\cite{Esteban:2018ppq}. 
   % We also show the future DUNE sensitivity in blue solid lines for both 300 kt.MW.yr and 850 kt.MW.yr~\cite{dev_pondd}.}
    \label{dnsiLQ3}
\end{figure}

\begin{figure}[t!]
  \vspace{-1.7cm}
\centering
\subfigure[]{
      \includegraphics[height=8cm,width=0.45\textwidth]{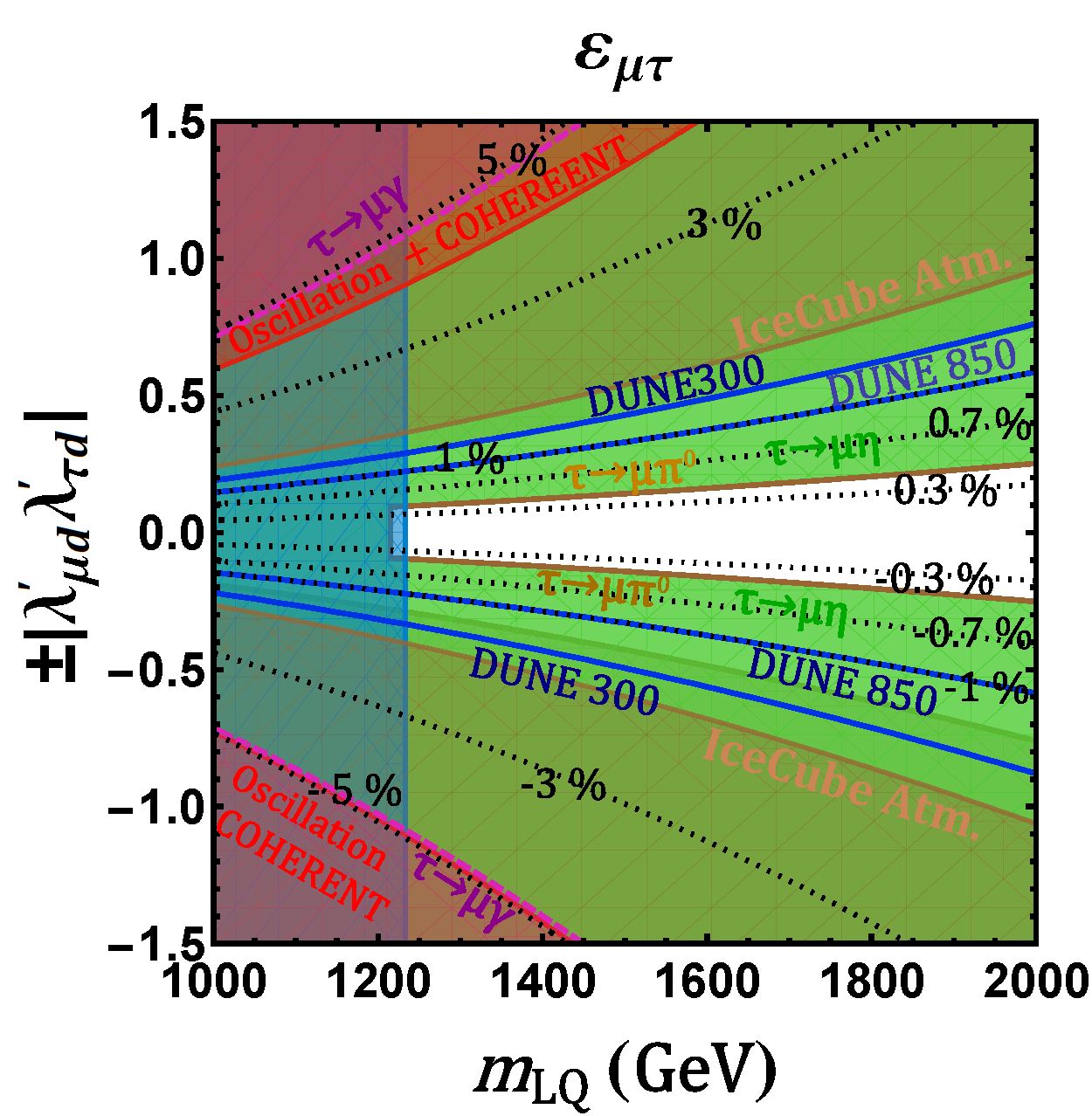}}
      \subfigure[]{
      \includegraphics[height=8cm,width=0.45\textwidth]{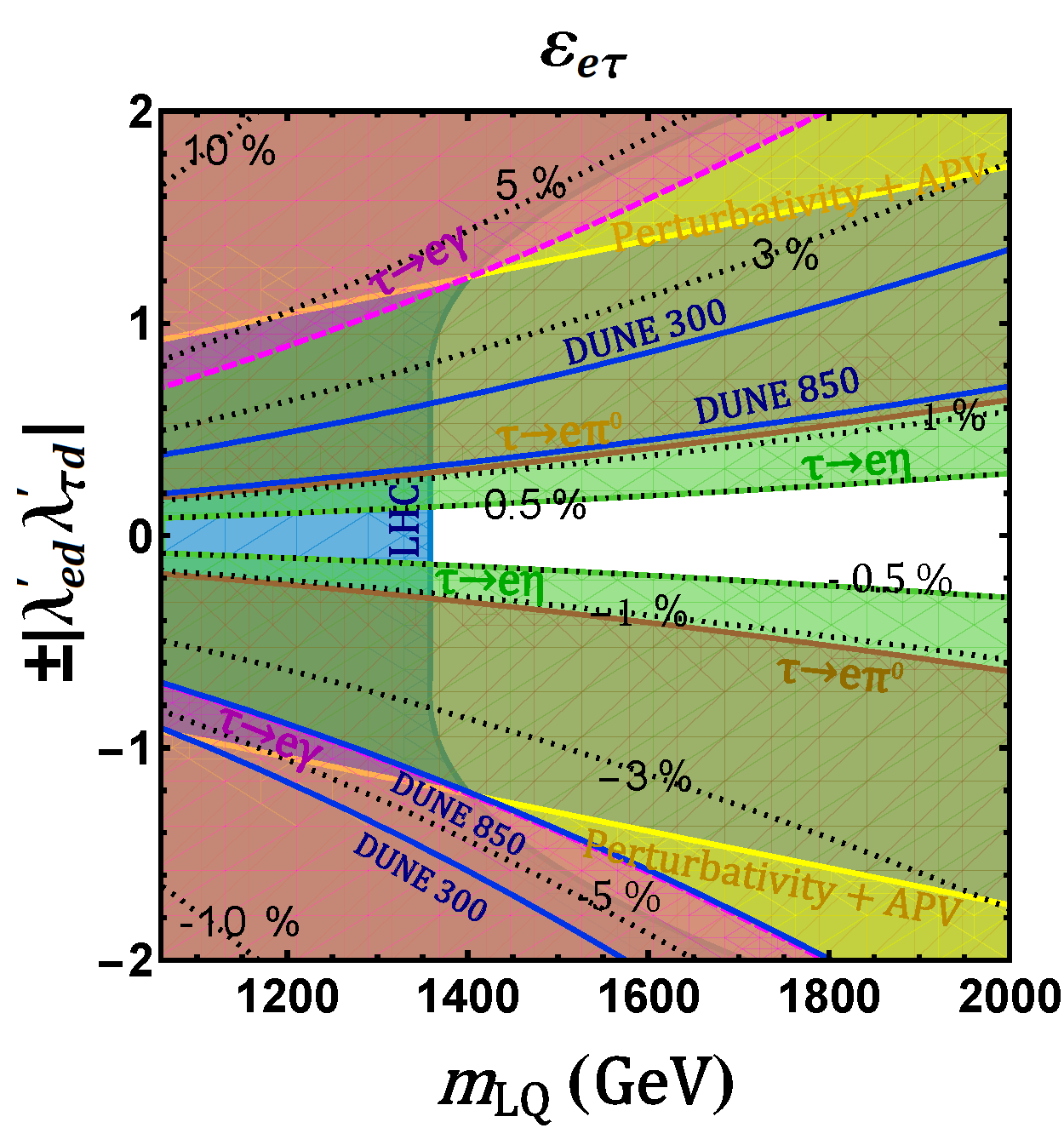}}\\
   \subfigure[]{
     \includegraphics[height=8cm,width=0.5\textwidth]{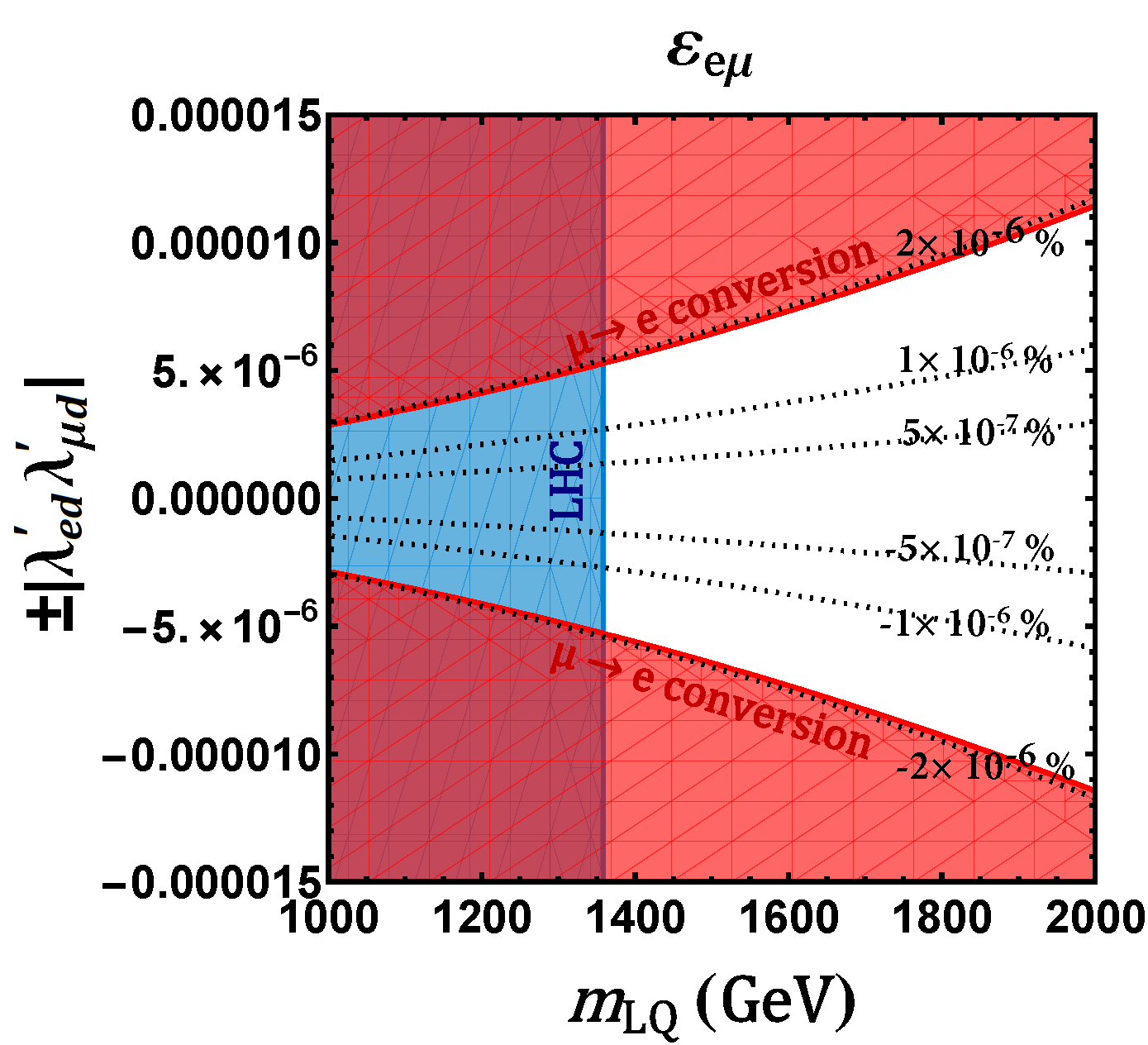}}
    \caption{Predictions for off-diagonal NSI $(\varepsilon_{e\mu},\, \varepsilon_{\mu\tau},\,\varepsilon_{e\tau})$ induced by the triplet LQ are shown by black dotted contours. Color-shaded regions are excluded by various theoretical and experimental constraints. The labels are same as in Fig.~\ref{offdnsiLQ}.} 
    %Blue shaded area is excluded by LHC LQ searches (cf.~Fig.~\ref{fig:collq1}). In (a) and (b), the brown and green shaded regions are excluded by $\tau \to \ell \pi^0$ and $\tau\to \ell \eta$ (with $\ell=e,\mu$) constraints (cf.~Table~\ref{tab:semilep}). In (a), the red shaded region is excluded by the global-fit constraint on NSI from neutrino oscillation+COHERENT data~\cite{Esteban:2018ppq}. In (b), the yellow shaded region is excluded by perturbativity constraint on LQ coupling $\lambda_{\alpha d}$~\cite{DiLuzio:2017chi} combined with APV  constraint (cf.~Eq.~\eqref{eq:APV}).  In (c), the red shaded region is excluded by $\mu \to e$ conversion constraint.  Also shown in (b) are the future DUNE sensitivity in blue solid lines for both 300 kt.MW.yr and 850 kt.MW.yr~\cite{dev_pondd}. }
    \label{offdnsiLQ3}
\end{figure}

Taking into account all the constraints listed above, we show in Figs.~\ref{dnsiLQ3} and \ref{offdnsiLQ3}  the predictions for diagonal  $(\varepsilon_{ee},\, \varepsilon_{\mu\mu},\,\varepsilon_{\tau\tau})$ and off-diagonal  $(\varepsilon_{e\mu},\, \varepsilon_{\mu\tau},\,\varepsilon_{e\tau})$ NSI parameters respectively from Eq.~\eqref{eq:NSI-O39} by black dotted contours. Color-shaded regions in each plot are excluded by various theoretical and experimental constraints, as in Figs.~\ref{dnsiLQ} and \ref{offdnsiLQ}. The main difference is in the NuTeV constraint shown in Fig.~\ref{dnsiLQ3} (b), which is more stringent than those shown in Figs.~\ref{dnsiLQ} (b) and \ref{fig:nsiLQlp} (b). The reason is that in presence 
of both $\varepsilon^{uL}_{\mu\mu}$ and $\varepsilon^{dL}_{\mu\mu}$ as in this LQ model (cf.~\eqref{eq:lagO39ex}), the total contribution to $\left(\tilde{g}^\mu_{L}\right)^2$ in Eq.~\eqref{eq:gLmu} is always positive, and therefore, any nonzero $\varepsilon_{\mu\mu}$ will make the discrepancy worse than the SM case of $2.7\sigma$. Therefore, we cannot impose a 90\% CL ($1.64\sigma$) constraint from NuTeV in this scenario. The line shown in Fig.~\ref{dnsiLQ3} (b) corresponds to the $3\sigma$ constraint on  $\varepsilon_{\mu\mu}<0.0007$, which is subject to the same criticism as the discrepancy with the 
SM, and therefore, we have not shaded the NuTeV exclusion region and do not consider it while quoting the maximum allowed NSI. 

%Similarly, in Fig.~\ref{dnsiLQ3} (c), the global-fit constraint from oscillation+COHERENT data~\cite{Esteban:2018ppq} is slightly stronger from that given in Fig.~\ref{dnsiLQ} (c). This is because the constraint is now on the combination of $3\varepsilon^u_{\alpha\beta}+3\varepsilon^d_{\alpha \beta}/2$.

%%%%%%%%%%%%%%%%%%%%%%%%%%%%%%

%\clearpage

 %%%%%%%%%%%%%%%%%%%%%%%%%%%%%%
%%%%%%%%%%%%%%%%%%%%%%%%
%\clearpage
From Figs.~\ref{dnsiLQ3} and \ref{offdnsiLQ3}, we find the maximum allowed values of the NSI parameters in the triplet LQ model to be  
%whose maximum values are the same as down-quark induced doublet LQ NSI in Table~\ref{tab:LQ}, except for  $\varepsilon_{\mu \mu}$, where the constraint from NuTeV is a factor of five stronger~\cite{Davidson:2003ha}. Similarly, , whose maximum values are also given in Table~\ref{tab:LQ}. The extra factor of 1/2 for this term in Eq.~\eqref{eq:epsrho} is due to the Clebsch-Gordan coefficient of $1/\sqrt 2$ in Eq.~\eqref{eq:lago35ex}. Combining both doublet and singlet LQ-induced NSI constraints, we find that the maximum NSI allowed in this model are as follows:
\begin{align}
&   \varepsilon_{ee}^{\rm max} \ = \ 0.0059 \, , \qquad 
    \varepsilon_{\mu\mu}^{\rm max} \ = \ 0.0007 \, ,\qquad 
      \varepsilon_{\tau\tau}^{\rm max} \ = \ 0.517 \, , \nonumber \\
  & \varepsilon_{e\mu}^{\rm max} \ = \ 1.9\times 10^{-8} \, , \qquad  
   \varepsilon_{e\tau}^{\rm max} \ = \ 0.0050 \, , \qquad 
   \varepsilon_{\mu\tau}^{\rm max} \ = \ 0.0038 \, .
   \label{eq:NSI-rho}
\end{align}
This is also summarized in Fig.~\ref{fig:summaryplot} and in  Table~\ref{Table_Models}.

%%%%%%%%%%%%%%%%%%%%%%%%%%%%%%%%%%%%%%%%%%%%%%%%%%%%%%%%%%%%%%%%%%%%%%%%%%%%%%%%%%%%%%%%%%%%%%
%%%%%%%%%%%%%%%%%%%%%%%%%%%%%
\section{Other type-I radiative models} \label{sec:other_type1}
%%%%%%%%%%%%%%%%%%%%%%%%%%%%%%
In this section, we briefly discuss the NSI predictions in other type-I radiative models at one-, two- and three-loops. In each case, we present the new particle content, model Lagrangian, Feynman diagrams for neutrino mass generation and expressions for neutrino mass, followed by the expression for NSI parameters. The maximum NSI allowed in each model is summarized in Table~\ref{Table_Models}.  
%%%%%%%%%%%%%%%%%%%%%%%%%%%%%%%%
\subsection{One-loop models} \label{sec:other_1loop}
\subsubsection{Minimal radiative inverse seesaw model} \label{subsec:mrism}
\begin{figure}[!t]
\centering
     \includegraphics[scale=0.5]{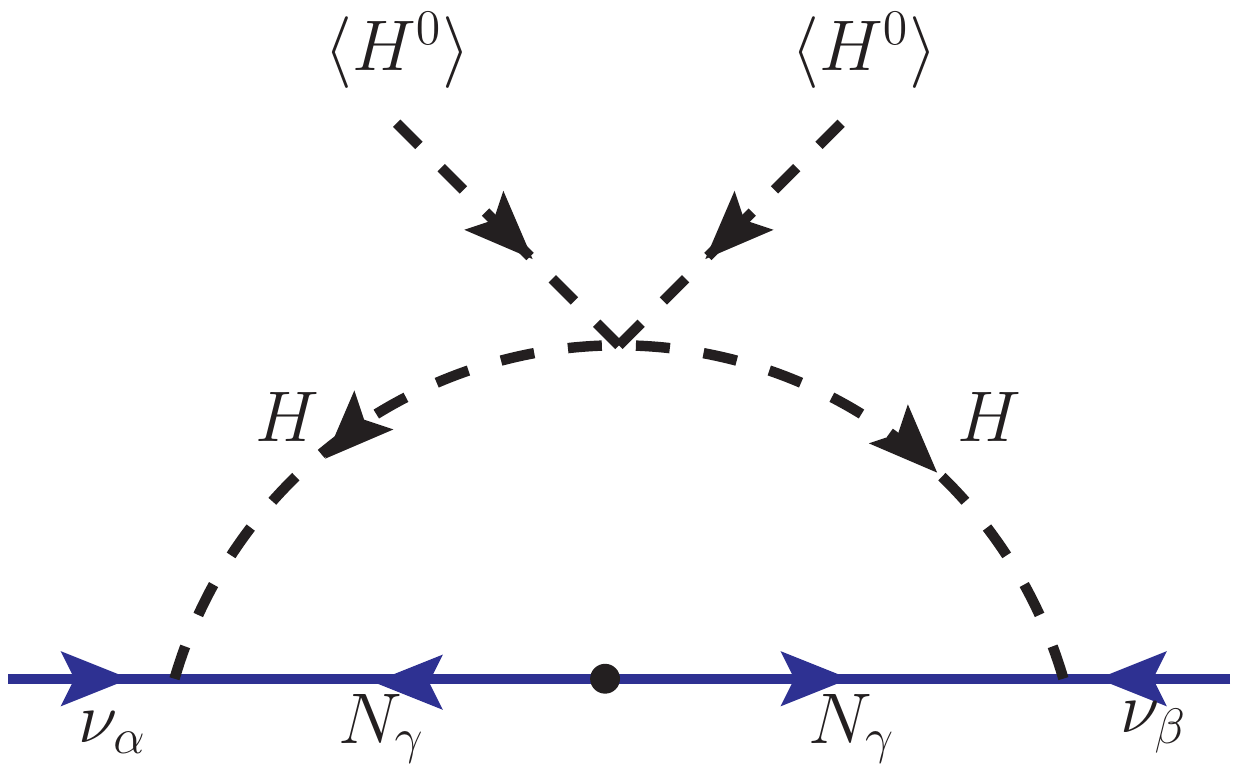}
     \includegraphics[scale=0.5]{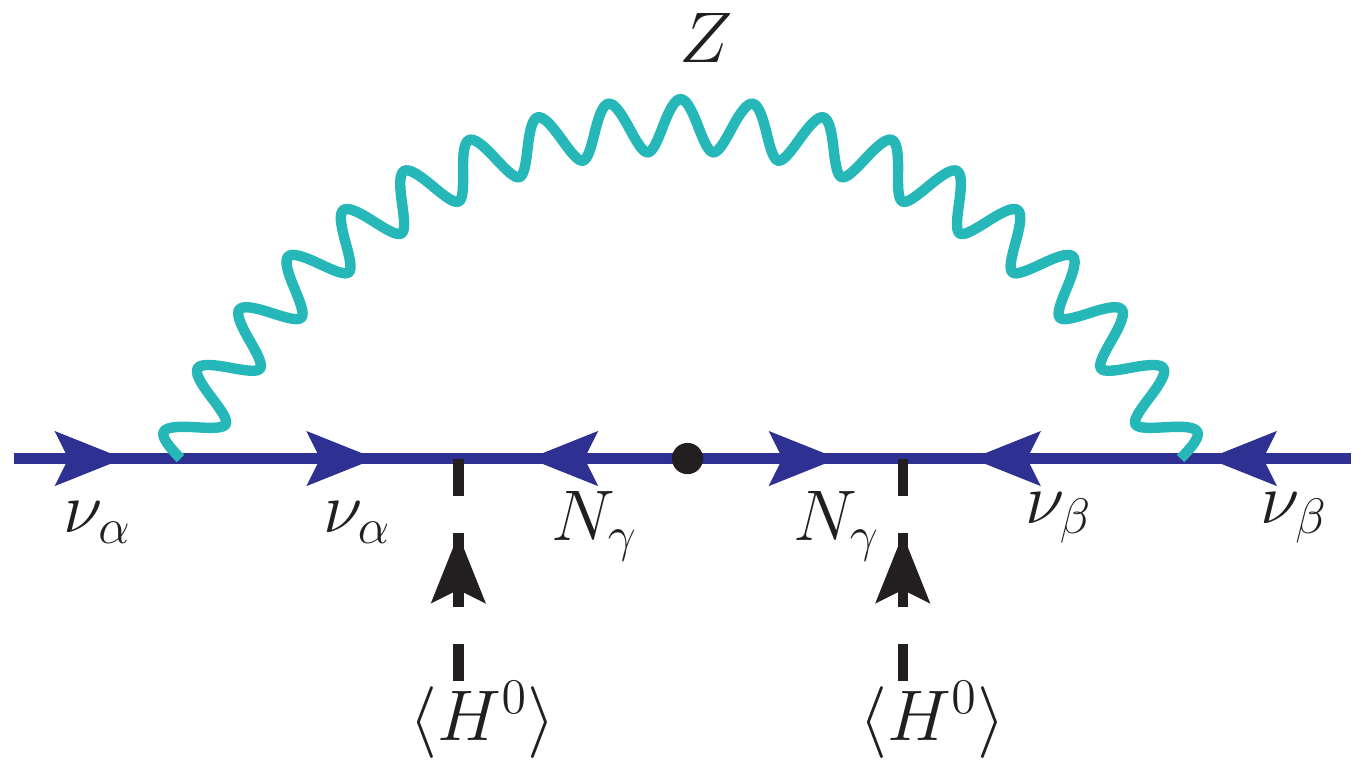}
    \caption{One-loop neutrino mass in the minimal radiative inverse seesaw model~\cite{Dev:2012sg}. This model induces the operator ${\cal O}'_2$ of Eq.~\eqref{eq:opmrism}. }
    \label{MRIS}
\end{figure}
This is an exception to the general class of type-I radiative models, where the new particles running in the loop will always involve a  scalar boson. In this model, the SM Higgs and $Z$ bosons are the mediators, with the new particles being SM-singlet fermions.\footnote{There is yet another possibility where the mediators could be new vector bosons; however, this necessarily requires some new gauge symmetry and other associated Goldstone bosons to cancel the UV divergences.} The low-energy effective operator that leads to neutrino mass in this model is the dimension-7 operator 
\begin{equation}
{\cal O}'_2 \ = \ L^i L^j H^k H^l \epsilon_{ik} \epsilon_{jl} (H^\dagger H)~.
\label{eq:opmrism}
\end{equation}
However, this mechanism is only relevant when the dimension-5 operator given by Eq.~\eqref{O1} that leads to the tree-level neutrino mass through the seesaw mechanism is forbidden due to some symmetry. This happens in the minimal radiative inverse seesaw model~\cite{Dev:2012sg}. In the usual inverse seesaw model~\cite{Mohapatra:1986bd}, one adds two sets of SM-singlet fermions, $N$ and $S$, with opposite lepton numbers. The presence of a Majorana mass term for the $S$-field, i.e.,  $\mu_S S S$ leads to a tree-level neutrino mass via the standard inverse seesaw mechanism~\cite{Mohapatra:1986bd}. However, if one imposes a global $U(1)$ symmetry under which the $S$-field is charged, then the $\mu_S S S$ term can be explicitly forbidden at tree-level.\footnote{This can be done, for instance, by adding a singlet scalar field $\sigma$ with a global $U(1)$ charge of +2, and by making $N$ and $S$ oppositely charged under this $U(1)$, viz., $N(-1)$ and $S(+1)$, so that the $S  \sigma S$ term is forbidden, but $N  \sigma N$ and $\overline{S}\sigma N$ are allowed. Furthermore, this global $U(1)$ symmetry can be gauged, e.g., in an $E_6$ GUT embedding, where the fundamental representation {\bf 27} breaks into ${\bf 16}_1+{\bf10}_{-2}+{\bf 1}_{4}$ under $SO(10)\times U(1)$. The $\nu$ and $N$ belong to the ${\bf 16}_1$ subgroup, while the $S$ belongs to ${\bf 1}_4$. Adding two scalars $\sigma,\sigma'$ with $U(1)$ charges $-2$ and $-5$ respectively allows the Dirac mass term $\overline N\sigma S$  and Majorana mass term $N  \sigma'N$  in Eq.~\eqref{eq:invsw}, but not the Majorana mass terms $S  \sigma^{(')}S$.} In this case, the only lepton number breaking term that is allowed is the Majorana mass term for the $N$-field, i.e., $\mu_R N  N$. It can be shown that this term by itself does not give rise to neutrino mass at tree-level, but a non-zero neutrino mass is inevitably induced at one-loop through the diagram shown in Fig.~\ref{MRIS} involving the SM Higgs doublet (which gives rise to two diagrams involving the SM Higgs and $Z$-boson after electroweak symmetry breaking~\cite{Dev:2012sg}).  One can see that the  low-energy effective operator that leads to neutrino mass in this model is the $d=7$ operator ${\cal O}'_1$ of Eq. (\ref{Op})  by cutting Fig.~\ref{MRIS} at one of the $H$-legs in the loop.

The relevant part of the Yukawa Lagrangian of this model is given by   
\begin{equation}
  -  \mathcal{L}_Y \ \supset \ Y_{\alpha\beta} \overline{L}_\alpha H N_{\beta} + \overline{S}_{\rho \alpha}(M_N)_{\rho \alpha} N_{\alpha } + \frac{1}{2} N^T_{\alpha }C (\mu_{R})_{\alpha \beta} N_{\beta } + {\rm H.c.} \, 
    \label{eq:invsw}
\end{equation}
After electroweak symmetry breaking, evaluating the self-energy diagrams that involve the $Z$-boson and Higgs boson (cf.~Fig.~\ref{MRIS}), the neutrino mass reads as (in the limit $\mu_R\ll M_N$)~\cite{Dev:2012sg, Dev:2012bd}: 
\begin{equation}
    M_{\nu} \ \simeq \ \frac{\alpha_w}{16\pi m_W^2} (M_D \mu_R M_D^T) \left[\frac{x_h}{x_N-x_H}\log\left(\frac{x_N}{x_H}\right)+\frac{3x_Z}{x_N-x_Z}\log\left(\frac{x_N}{x_Z}\right)\right] \, ,
\end{equation}
where $M_D\equiv Yv/\sqrt{2}$, $\alpha_w\equiv g^2/4\pi$, $x_N=m_N^2/m_W^2$, $x_H=m_H^2/m_W^2$ and $x_Z=m_Z^2/m_W^2$, and we have assumed $M_N=m_N{\bf 1}$ for simplicity. 

The NSI in this model arise due to the fact that the light $SU(2)_L$-doublet neutrinos $\nu$ mix with the singlet fermions $N$ and $S$, due to which the $3\times 3$ lepton mixing matrix is no longer unitary. The neutrino-nucleon and neutrino-electron interactions proceed as in the SM via $t$-channel exchange of $W$ and $Z$ bosons, but now with modified strength because of the non-unitarity effect, that leads to NSI~\cite{Blennow:2016jkn}. If only one extra Dirac state with mass larger than $\sim$ GeV (such that it cannot be produced in accelerator neutrino oscillation experiments, such as DUNE) mixes with the three light states with mixing parameters $U_{\alpha 4}$ (with $\alpha=e,\mu,\tau$), we can write the NSI parameters as
%\begin{align}
%\boxed{
%\begin{empheq}[box=\fbox]{align} %[box=\widebox]{gather}
\begin{tcolorbox}[enhanced,ams align,
  colback=gray!30!white,colframe=white]
\varepsilon_{ee} \  = & \ \left(\frac{Y_{n}}{2}-1\right)|U_{e 4}|^2, \qquad 
\varepsilon_{\mu\mu} \  =  \ \frac{Y_{n}}{2} |U_{\mu 4}|^2,\qquad 
\varepsilon_{\tau\tau} \  =  \ \frac{Y_{n}}{2}|U_{\tau 4}|^2, \nonumber\\
\varepsilon_{e\mu} \  = & \ \frac{1}{2}\left(Y_{n}-1\right)U_{e 4} U^{\star}_{\mu 4}, \qquad 
\varepsilon_{e\tau} \  = \ \frac{1}{2}\left(Y_{n}-1\right)U_{e 4} U^{\star}_{\tau 4}, \qquad 
\varepsilon_{\mu\tau} \  = \ \frac{Y_n}{2}U_{\mu 4} U^{\star}_{\tau 4} \, .
%}
\label{eq:NSI_mrism}
%\end{align}
%\end{empheq}
\end{tcolorbox}
Here $Y_{n}=N_n/N_e$ is the ratio of the average number density of neutrons and electrons in matter. Note that for $Y_n\to 1$ which is approximately true for neutrino propagation in earth matter, we get vanishing $\varepsilon_{e\mu}$ and $\varepsilon_{e\tau}$ up to second order in $U_{\alpha 4}$.\footnote{This result is in disagreement with Ref.~\cite{Blennow:2016jkn}, where they have $\varepsilon_{\alpha\beta}=\frac{1}{2}U_{\alpha 4} U^{\star}_{\beta 4}$ for all the off-diagonal NSI parameters, which cannot be the case, because for $\alpha=e$, both CC and NC contributions are present, whereas for $\alpha\neq e$, only the NC contribution matters.}  Taking into account all the experimental constraints on $U_{\alpha 4}U_{\beta 4}^\star$ from neutrino oscillation data in the averaged-out regimes, beta decay, rare meson decay, beam dump experiments, cLFV searches, collider constraints from LEP and LHC, as well as electroweak precision constraints~\cite{Atre:2009rg, Deppisch:2015qwa, deGouvea:2015euy, Alonso:2012ji, Blennow:2016jkn, Antusch:2014woa}, the maximum NSI parameters allowed in this model are summarized in Table~\ref{Table_Models}.  We find that~\cite{Blennow:2016etl} 
\begin{align}
    |\varepsilon_{ee}^{\rm max}| \ = \ 1.3\times 10^{-3} \, , \qquad 
    \varepsilon_{\mu\mu}^{\rm max} \ = \ 2.2\times 10^{-4} \, , \qquad 
    \varepsilon_{\tau\tau}^{\rm max} \ = \ 2.8\times 10^{-3} \, , \nonumber \\
    \varepsilon_{e\mu}^{\rm max} \ = \ 3.5\times 10^{-5} \, , \qquad 
    \varepsilon_{e\tau}^{\rm max} \ = \ 1.4\times 10^{-4} \, , \qquad 
    \varepsilon_{\mu\tau}^{\rm max} \ = \ 1.2\times 10^{-3} \, . 
    \label{eq:maxNSI_mrsim}
\end{align}
For $\varepsilon_{e\mu}$ and $\varepsilon_{e\tau}$, we have used $Y_n=1.051$ (for average value all over the earth) in Eq.~\eqref{eq:NSI_mrism}, in addition to the cLFV constraints on $U_{e 4}U_{\mu 4}^\star$ and $U_{e 4}U_{\tau 4}^\star$. The maximum NSI values listed above (and also summarized in Table~\ref{Table_Models}) are obtained for a relatively heavy sterile neutrino (with mass larger than the electroweak-scale), so that the stringent low-energy constraints from beam dump and meson decays can be avoided, and the only relevant constraint comes from the electroweak precision data~\cite{Antusch:2014woa}.

The NSI expressions~\eqref{eq:NSI_mrism} also apply to two-loop radiative models with two $W$-boson exchange~\cite{Petcov:1984nz, Babu:1988ig, Davidson:2006tg}. However, the maximum NSI obtainable in these models will be much smaller than the estimate in Eq.~\eqref{eq:maxNSI_mrsim} because the sterile neutrino in this case is required to be heavier for successful neutrino mass generation at two-loop. 

\subsubsection{One-loop model with vectorlike leptons} \label{subsec:CCSVO21}
%%%%%%%%%%%%%%%%%%%%%%%%%%%%%%%%%%%%%%%%%%
\begin{figure}[t!]
    \centering
    \includegraphics[scale=0.6]{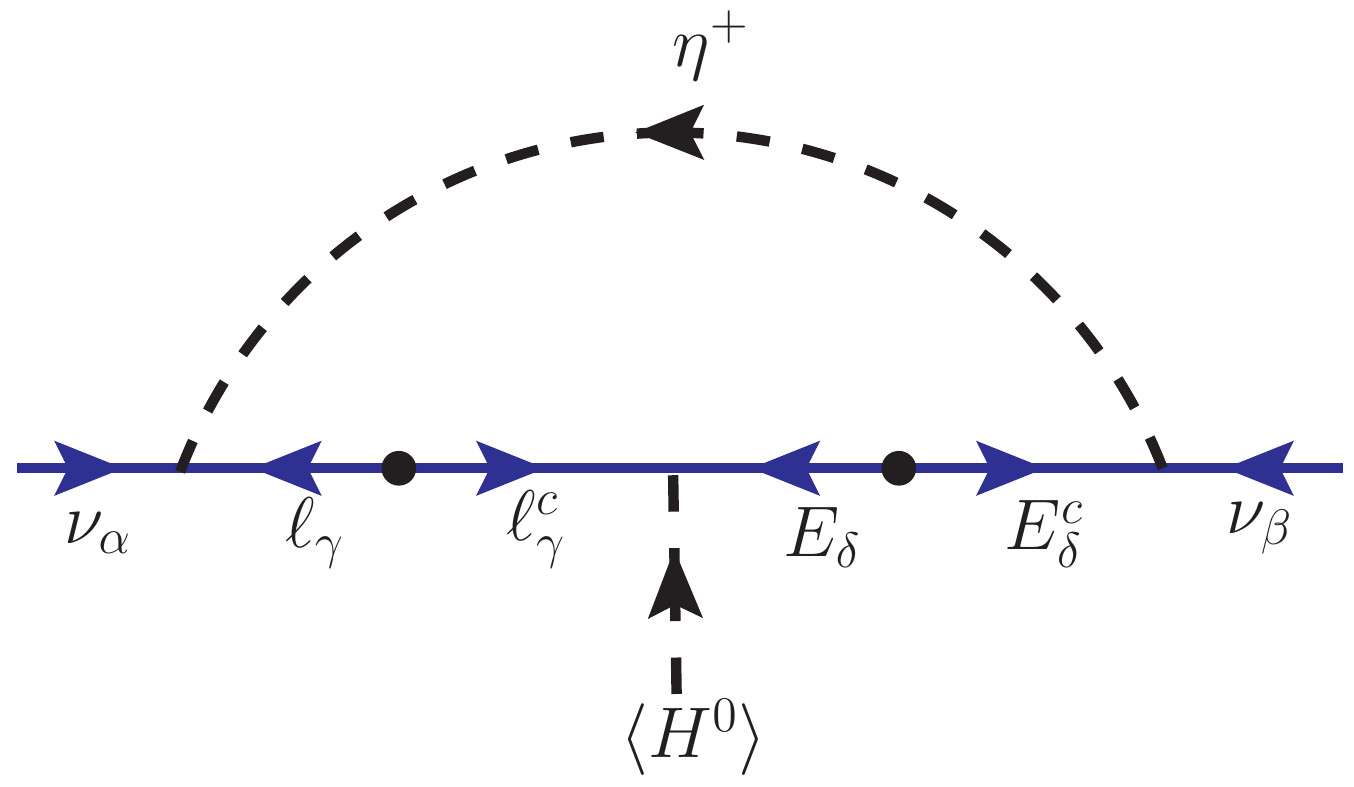}
    \caption{Neutrino mass generation in the one-loop model with vectorlike leptons.  This is the $\mathcal{O}_2^1$ model of Table \ref{tab:O2}~\cite{Cai:2014kra}.}
    \label{CCSVO21}
\end{figure}
%%%%%%%%%%%%%%%%%%%%%%%%%%%%%%%%%%%%%%%%%
This model~\cite{Cai:2014kra} utilizes the same $d=7$ operator ${\cal O}_2 = L^i L^j L^k e^c H^l \epsilon_{ij} \epsilon_{kl}$ (cf.~Eq.~\eqref{eq:O2}), as in the Zee model to generate a one-loop neutrino mass. The new particles added are a scalar singlet $\eta^+({\bf 1},{\bf 1},1)$ and a vectorlike lepton $\psi\left({\bf 1},{\bf 2},-\frac{3}{2}\right)=(E, \, F^{--})$, which give rise to the ${\cal O}_2^1$ operator $L(LL)(e^cH)$ (cf.Table~\ref{tab:O2}). 
Neutrino mass is generated via the one-loop diagram shown in Fig.~\ref{CCSVO21}. The relevant Lagrangian for the neutrino mass generation reads:
\begin{equation}
    -\mathcal{L} \ \supset \  f_{\alpha \beta} L_\alpha L_\beta \eta^+ + y_{\alpha \beta}^\prime L_\alpha \psi_\beta^c \eta^- + y_{\alpha \beta} \ell^c_\alpha \psi_\beta  H + m_\psi \psi \psi^c  + {\rm H.c.} \,
\end{equation}
where $\psi^c= (F^{++}, \, -E^c)$ and $H\left({\bf 1},{\bf 2},\frac{1}{2}\right)$ is the SM Higgs doublet. 
Expanding the first two terms, we get
\begin{equation}
     -\mathcal{L} \ \supset \ f_{\alpha \beta} (\nu_\alpha \ell_\beta \eta^+ - \ell_\alpha \nu_\beta \eta^+) - y_{\alpha \beta}^\prime (\nu_\alpha E^+_\beta \eta^- + \ell_\alpha E^{++}_\beta \eta^-) + {\rm H.c.} \,  
     \label{eq:lag21ex}
\end{equation}
The neutrino mass matrix can be estimated as
\begin{equation}
    M_\nu \ \sim \ \frac{1}{16 \pi^2}\frac{v}{M^2}\left( f \, M_\ell \, y   \, M_E \, y^{\prime T} + y^{\prime } M_E y^T M_\ell f^T\right) \, ,
    \label{eq:Mnu21}
\end{equation}
where $M_\ell$ is the diagonal mass matrix for the SM charged leptons, $M_E$ is the diagonal mass matrix for the vector-like leptons with eigenvalues $m_{E_i}$, and $M \equiv {\rm max}(m_\eta, m_{E_i})$. Note that just one flavor of $\psi$ is not sufficient, because in this case, the neutrino mass matrix~\eqref{eq:Mnu21} would have a flavor structure given by $(f M_\ell - M_\ell f)$, which has all the diagonal entries zero, similar to the Zee-Wolfenstein model~\cite{Wolfenstein:1980sy}.  Such a structure is ruled out by observed neutrino oscillation data. Thus, we require at least two flavors of $\psi$, in which case the diagonal entries of $M_\nu$ are nonzero, and the model is consistent with experiments.

NSI in this model are induced by the $f$-type couplings in Eq.~\eqref{eq:lag21ex}, similar to the $f$-couplings in the Zee model Lagrangian~\eqref{eq:Zeefex}. The NSI parameters read as
\begin{tcolorbox}[enhanced,ams align,
  colback=gray!30!white,colframe=white]
%\begin{equation}
 %\boxed{   
 \varepsilon_{\alpha \beta} \ \equiv \ \varepsilon_{\alpha \beta}^{ee} \ = \ \frac{1}{\sqrt{2}G_F}\frac{f_{ e \alpha}^{\star}f_{ e \beta }  }{m_{\eta^+}^2}  \, .
 %}
    \label{eq:nsio21}
%\end{equation}
\end{tcolorbox}
Due to the antisymmetric nature of the $f$ couplings, the only relevant NSI parameters in this case are $\varepsilon_{\mu \tau}$, $\varepsilon_{\mu \mu}$, and $\varepsilon_{\tau \tau}$. These are severely constrained by cLFV searches and universality of charged currents~\cite{Nebot:2007bc}, as shown in Table~\ref{tab:zee-babu}. This is similar to the case of Zee-Babu model discussed later in Sec.~\ref{sec:zeebabu}. Since the singly-charged scalar mass has to be above $\sim 100$ GeV to satisfy the LEP constraints (cf.~Sec.~\ref{sec:colliderZee}), we obtain from Eq.~\eqref{eq:nsio21} and Table~\ref{tab:zee-babu}  the following maximum values: 
\begin{align}
 &   \varepsilon_{ee}^{\rm max} \ = \ 0 \, , \qquad 
    \varepsilon_{\mu\mu}^{\rm max} \ = \ 9.1\times 10^{-4} \, ,\qquad 
      \varepsilon_{\tau\tau}^{\rm max} \ = \ 3.0\times 10^{-3} \, , \nonumber \\
  & \varepsilon_{e\mu}^{\rm max} \ = \ 0 \, , \qquad  
   \varepsilon_{e\tau}^{\rm max} \ = \ 0 \, , \qquad 
   \varepsilon_{\mu\tau}^{\rm max} \ = \ 3.0\times 10^{-3} \, .
   \label{eq:maxZB}
\end{align}
This is also summarized in Table~\ref{Table_Models}.
\begin{table}[!t]
    \centering
    \begin{tabular}{|c|c|c|}
    \hline\hline
    {\bf Observable} & {\bf Exp. limit} & {\bf Constraint} \\ \hline \hline
        $\mu \rightarrow e \gamma$ & BR < $4.2 \times 10^{-13}$~\cite{TheMEG:2016wtm} & $|f_{e \tau}^{\star} f_{\mu \tau}|$ <  $1.09 \times 10^{-3} \left(\frac{m_{h^+}}{ \mathrm{TeV}}\right)^{2}$ \\ \hline
      $\tau \rightarrow e \gamma$ & BR < $3.3 \times 10^{-8}$~\cite{Aubert:2009ag} & $|f_{e \mu}^{\star} f_{\mu \tau}|$ <  $0.71 \left(\frac{m_{h^+}}{\mathrm{TeV}}\right)^{2}$ \\ \hline
        $\tau \rightarrow \mu \gamma$ & BR < $4.4 \times 10^{-8}$~\cite{Aubert:2009ag} & $|f_{e \mu}^\star f_{e \tau}|$ <  $0.82 \left(\frac{m_{h^+}}{\mathrm{TeV}}\right)^{2}$ \\ \hline
       lep./had. univ. &  $ \displaystyle{\sum_{q=d, s, b}} |V_{u q}^{\rm exp}|^{2} = 0.9992 \pm 0.0011 $~\cite{Tanabashi:2018oca} & $|f_{e \mu}|^{2} < 0.015 \left(\frac{m_{h^+}}{\mathrm{TeV}}\right)^{2}$ \\ \hline
   $\mu/e$ univ.   & $ g_{\mu}^{\rm exp} / g_{e}^{\rm exp}=1.0001 \pm 0.0020$~\cite{Tanabashi:2018oca} & $ \big| |f_{\mu \tau}|^{2}-|f_{e \tau}|^{2} \big|<0.05 \left(\frac{m_{h^+}}{ \mathrm{TeV}}\right)^{2} $ \\ \hline
    $\tau/\mu$ univ.   & $ g_{\tau}^{\rm exp} / g_{\mu}^{\rm exp}=1.0004 \pm 0.0022$~\cite{Tanabashi:2018oca} & $ \big| |f_{e \tau}|^{2}-|f_{e \mu}|^{2} \big|<0.06 \left(\frac{m_{h^+}}{ \mathrm{TeV}}\right)^{2} $ \\ \hline
     $\tau/e$ univ.   & $ g_{\tau}^{\rm exp} / g_{e}^{\rm exp}=1.0004\pm 0.0023$ ~\cite{Tanabashi:2018oca} & $\big| |f_{\mu \tau}|^{2}-|f_{e \mu}|^{2} \big|<0.06 \left(\frac{m_{h^+}}{ \mathrm{TeV}}\right)^{2} $ \\ \hline\hline
    \end{tabular}
    \caption{Constraints on the singly-charged scalar Yukawa couplings~\cite{Nebot:2007bc}. Here $g^{\rm exp}_\alpha$ stands for the effective gauge coupling extracted from muon and tau decays in the different leptonic channels.}
    \label{tab:zee-babu}
\end{table}
%%%%%%%%%%%%%%%%%%%%%%%%%%%%%%%%%%%%%%
%%%%%%%%%%%%%%%%%%%%%%%%%%%%%%

%%%%%%%%%
%%%%%%%%%%%%%%%%%%%%%%%%%%%%%%%%%%%%%%%%%

\subsubsection{\texorpdfstring{$SU(2)_L$}{su2l}-singlet leptoquark model with vectorlike quark} \label{subsec:CCSVO34}
%%%%%%%%%%%%%%%%%%%%%%%%%%%%%%%%%%%%%%%%%%%%%%%%%%%%%%%%%%%%%%%%%%%%%%%%%
\begin{figure}[t!]
    \centering
    \includegraphics[scale=0.6]{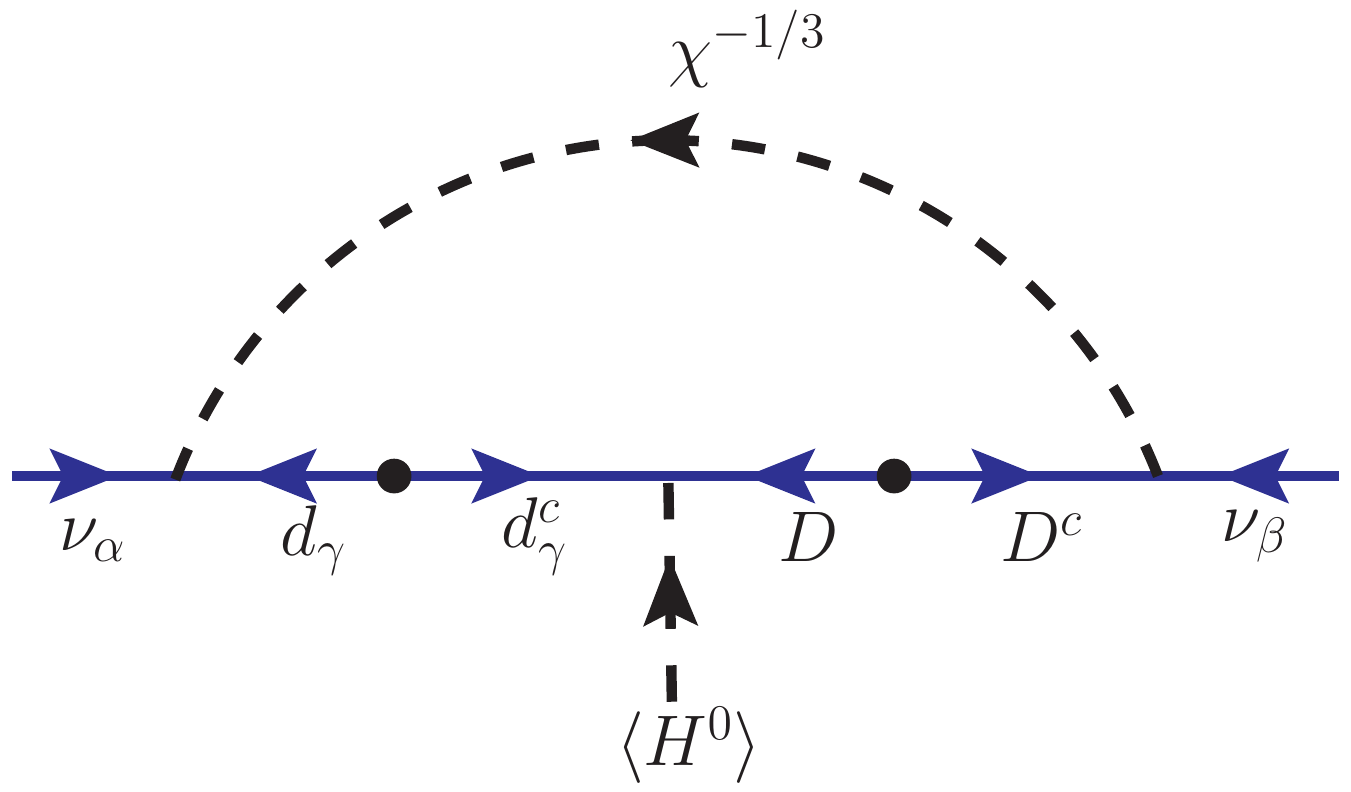}
    \caption{Neutrino mass generation in the one-loop singlet LQ model with vectorlike quarks.  This is the $\mathcal{O}_3^4$ model of Table \ref{tab:O3}~\cite{Cai:2014kra}.}
    \label{CCSVO34}
\end{figure}
%%%%%%%%%%%%%%%%%%%%%%%%%%%%%%%%%%%%%%%%%%%%%%%%%%%%%%%%%%%%%%%%%%%%%%%%%
This model~\cite{Cai:2014kra} is the $\mathcal{O}_3^4$ realization of the dimension-7 operator ${\cal O}_3$ (cf.~Table~\ref{tab:O3}). The new particles introduced are a scalar LQ singlet $\chi\left({\bf 3}, {\bf 1}, -\frac{1}{3}\right)$ and a vectorlike quark doublet ${\cal Q}\left({\bf 3},{\bf 2},-\frac{5}{6}\right)=\left(D^{-1/3}, \, X^{-4/3}\right)$. 
Neutrino mass is generated at one-loop level as shown in  Fig.~\ref{CCSVO34}. The  $Q Q \chi^\star$ and $d^c u^c \chi$ interaction terms, allowed by gauge invariance, are forbidden by demanding baryon-number conservation in order to avoid rapid proton decay. The relevant Lagrangian for the neutrino mass generation reads as 
\begin{equation}
   -\mathcal{L}_Y \ \supset \  \lambda_{\alpha \beta} L_\alpha Q_\beta \chi^\star + \lambda_\alpha^{\prime} L_\alpha \mathcal{Q}^c \chi + f_{\alpha} d_\alpha^c \mathcal{Q} H + f_{\alpha \beta}' \ell_\alpha^c u_\beta^c \chi + {\rm H.c.} \, 
\end{equation}
Expanding the first two terms, we get
\begin{equation}
     -\mathcal{L}_Y \ \supset \  \lambda_{\alpha \beta} (\nu_\alpha d_\beta \chi^ \star - \ell_\alpha u_\beta \chi^\star) - \lambda_\alpha^\prime (\nu_\alpha D^c \chi + \ell_\alpha X^c \chi) \, .
     \label{eq:lag34ex}
\end{equation}
The neutrino mass matrix can be estimated as
\begin{equation}
    M_\nu \ \sim \ \frac{1}{16 \pi^2}\frac{v}{M^2}\left(\lambda M_d f M_D \lambda^{\prime T}+\lambda^{\prime}M_Df^TM_d\lambda^T\right) \, ,
\end{equation}
where $M_d$ is the diagonal down-type quark mass matrix, $M_D$ is the mass matrix for the down-type VQ with eigenvalues $m_{D_i}$, and $M \equiv {\rm max}(m_\chi, m_{D_i})$. With a single copy of VQ quarks, the rank of $M_\nu$ is two, implying that the lightest neutrino has zero mass at the one-loop order. This model can lead to consistent neutrino oscillation phenomenology. 

NSI in this model are induced by the $\lambda$-type interactions in Eq.~\eqref{eq:lag34ex}:
\begin{tcolorbox}[enhanced,ams align,
  colback=gray!30!white,colframe=white]
%\begin{equation}
 %    \boxed{ 
     \varepsilon_{\alpha \beta} \ = \ \frac{3}{4 \sqrt{2}G_F} \frac{\lambda_{\alpha d}^{\star} \lambda_{\beta d} }{ m_\chi^2} \, .
     %}
      \label{eq:714}
%\end{equation}
\end{tcolorbox}
This is similar to the singlet LQ contribution in Eq.~\eqref{nsi_coloredzee}, with the important exception that the NSI get contribution only from the $\lambda$-couplings, and therefore, the IceCube limits on $|\varepsilon_{\mu\mu}-\varepsilon_{\tau\tau}|<9.3\%$ cannot be avoided, just like in the Zee model case. The corresponding maximum NSI can be read off from Table~\ref{tab:LQ}, except for $\varepsilon_{\tau\tau}$:  
\begin{align}
 &   \varepsilon_{ee}^{\rm max} \ = \ 0.0069 \, , \qquad 
    \varepsilon_{\mu\mu}^{\rm max} \ = \ 0.0086 \, ,\qquad 
      \varepsilon_{\tau\tau}^{\rm max} \ = \ 0.093 \, , \nonumber \\
  & \varepsilon_{e\mu}^{\rm max} \ = \ 1.5\times 10^{-7} \, , \qquad  
   \varepsilon_{e\tau}^{\rm max} \ = \ 0.0036 \, , \qquad 
   \varepsilon_{\mu\tau}^{\rm max} \ = \ 0.0043 \, .
   \label{eq:NSI-singlet}
\end{align}
This is also summarized in Table~\ref{Table_Models}.
%%%%%%%%%%%%%%%%%%%%%%%%%%%%%%%%%%%%%%%%%
%%%%%%%%%%%%%%%%%%%%%%%%%%%%%%%%%%%%%%%%%
%%%%%%%%%%%%%%%%%%%%%%%%%%%%%%%%%%%%%%%%%

%%%%%%%%%%%%%%%%%%%%%%%%%%%%%%%%%%%%%%%%%%%%%%%%%%%%%%%%%%%%%%%%%%%%%%%%%%%%%%%%%%%%%%%%%%%%%%
%%%%%%%%%%%%%%%%%%%%%%%%%%%%%%%%%%%%%%%%%%%%%%%%%%%%%%%%%%%%%%%%%%%%%%%%%%%%%%%%%%%%%%%%%%%%%%
\subsubsection{\texorpdfstring{$SU(2)_L$}{su2ll} -doublet leptoquark model with vectorlike quark }\label{sec:CCSVO36}
%%%%%%%%%%%%%%%%%%%%%%%%%%%%%%%%%%%%%%%%%%%%%%%
\begin{figure}
    \centering
    \includegraphics[scale=0.5]{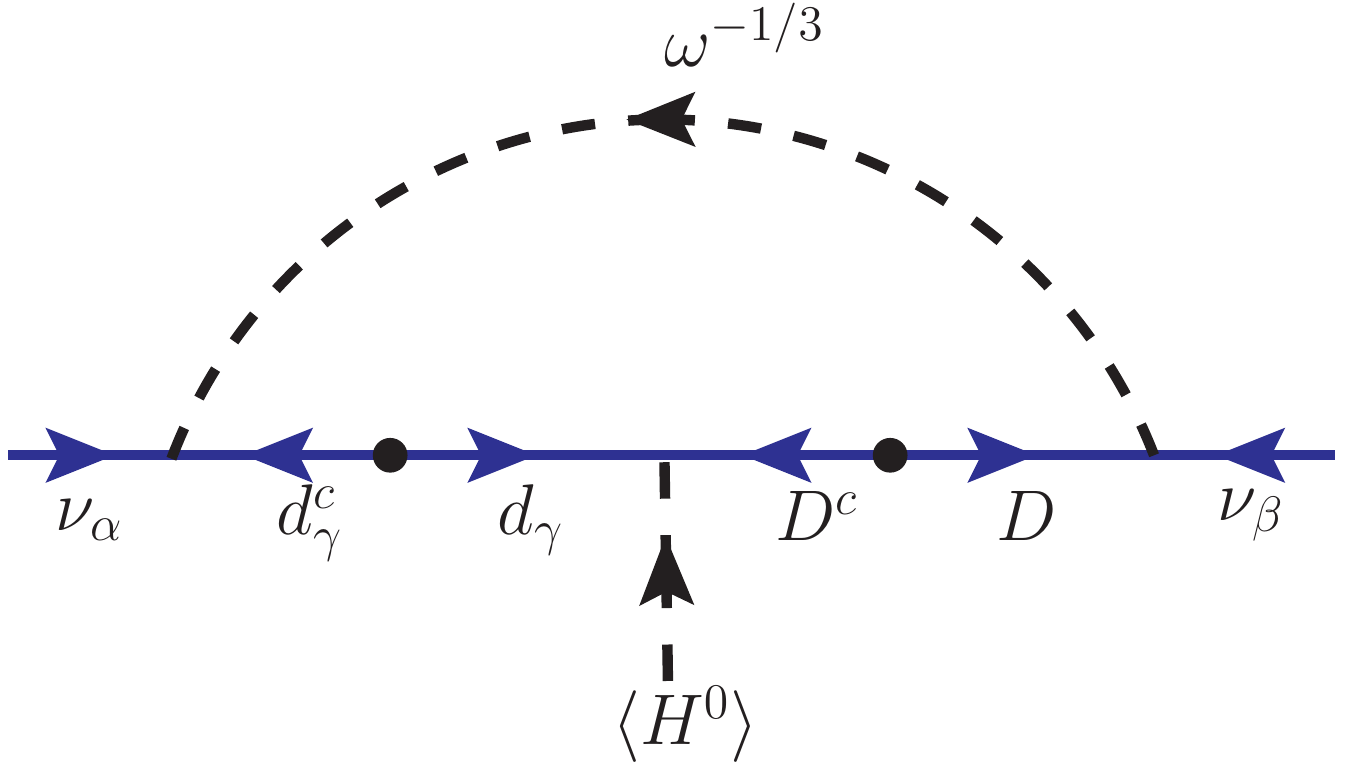}
    \caption{Neutrino mass generation in the one-loop doublet LQ model with vectorlike quarks. This is the model $\mathcal{O}_3^6$ of Table \ref{tab:O3}~\cite{Cai:2014kra}.}
    \label{CCSVO36}
\end{figure}
%%%%%%%%%%%%%%%%%%%%%%%%%%%%%%%%%%%%%%%%%%%%%%%
This is referred to as $\mathcal{O}_3^6$ in Table \ref{tab:O3}.  
The model has an $SU(2)_L$-doublet LQ $\Omega\left({\bf 3},{\bf 2},\frac{1}{6}\right)=\left(\omega^{2/3}, \, \omega^{-1/3}\right)$ and an $SU(2)_L$-triplet vectorlike quark $\Sigma\left({\bf 3},{\bf 3},\frac{2}{3}\right)= \left(Y^{5/3}, \, U^{2/3}, \, D^{-1/3} \right)$. 
Neutrino mass is generated at one-loop level via the Feynman diagram  shown in Fig.~\ref{CCSVO36}.  The relevant Lagrangian for the neutrino mass generation can be written as
\begin{eqnarray}
   -\mathcal{L}_Y &\ \supset \ & M_\Sigma \Sigma \Sigma^c + \left( \lambda_{\alpha \beta} L_\alpha d_\beta^c \Omega + \lambda_\alpha^\prime Q_\alpha \Sigma^c H + \lambda_\alpha^{\prime \prime} L_\alpha \Sigma \widetilde{\Omega} + {\rm H.c.} \right) \, ,
  \label{eq:lago36}
   \end{eqnarray}
   where $\widetilde{\Omega}=i\tau_2 \Omega^\star$ is the isospin conjugate field. Expanding the terms in Eq.~\eqref{eq:lago36}, we obtain
   \begin{align}
   -\mathcal{L}_Y \ \supset \  & M_\Sigma \left(YY^c + DD^c + UU^c\right) + \left[\lambda_{\alpha \beta} \left(\nu_\alpha \omega^{-1/3} - \ell_\alpha \omega^{2/3}\right) d_\beta^c \right. \nonumber \\
    & \left.+\lambda_\alpha^\prime \left\{u_\alpha Y^c H^+ + \frac{1}{\sqrt{2}}\left(u_\alpha H^0 + d_\alpha H^+\right) U^c + d_\alpha D^c H^0 \right\} \right. \nonumber \\
    & \left. +\lambda_\alpha^{\prime \prime} \left\{\nu_\alpha D \omega^{\star 1/3} - \frac{1}{\sqrt 2}\left(-\nu_\alpha \omega^{\star -2/3} + \ell_\alpha \omega^{\star 1/3}\right)U - \ell_\alpha Y \omega^{\star-2/3}\right\}+{\rm H.c.}\right] \, .
\end{align}
The neutrino mass can be estimated as 
\begin{equation}
    M_\nu \ \sim \ \frac{1}{16\pi^2}\frac{v}{M^2}\left(\lambda M_d \lambda^\prime M_D \lambda^{\prime \prime T} +\lambda^{\prime\prime}M_D\lambda^{\prime T}M_d\lambda^T\right) \, ,
\end{equation}
where $M_d$ and $M_D$ are the diagonal down quark mass matrix and vectorlike quark mass matrix respectively, and $M \equiv {\rm max}(m_\omega, m_{D_i})$, with $m_{D_i}$ being the  eigenvalues of $M_D$. As in previous models with one copy of vectorlike fermion, the rank of $M_\nu$ is two in this model, implying that the lightest neutrino is massless at the one-loop level.

NSI in this model are induced by the doublet LQ component $\omega^{-1/3}$. The NSI parameters read as
\begin{tcolorbox}[enhanced,ams align,
  colback=gray!30!white,colframe=white]
%\begin{equation}
%     \boxed{ 
     \varepsilon_{\alpha \beta} \ = \ \frac{3}{4 \sqrt{2}G_F} \frac{\lambda_{\alpha d}^{\star} \lambda_{\beta d} }{ m_\omega^2} \, .
     %}
      \label{eq:719}
%\end{equation}
\end{tcolorbox}
This expression is similar to the doublet LQ contribution in Eq.~\eqref{nsi_coloredzee}, with the exception that the IceCube limits on $|\varepsilon_{\mu\mu}-\varepsilon_{\tau\tau}|<9.3\%$ cannot be avoided. The corresponding maximum NSI can be read off from Table~\ref{tab:LQ}, except for $\varepsilon_{\mu\mu}$ and $\varepsilon_{\tau\tau}$:   
\begin{align}
 &   \varepsilon_{ee}^{\rm max} \ = \ 0.004 \, , \qquad 
    \varepsilon_{\mu\mu}^{\rm max} \ = \ 0.093 \, ,\qquad 
      \varepsilon_{\tau\tau}^{\rm max} \ = \ 0.093 \, , \nonumber \\
  & \varepsilon_{e\mu}^{\rm max} \ = \ 1.5\times 10^{-7} \, , \qquad  
   \varepsilon_{e\tau}^{\rm max} \ = \ 0.0036 \, , \qquad 
   \varepsilon_{\mu\tau}^{\rm max} \ = \ 0.0043 \, .
   \label{eq:maxNSI_BJ}
\end{align}
This is also summarized in Table~\ref{Table_Models}.
%

%%%%%%%%%%%%%%%%%%%%%%%%%%%%%%%%%%%%%%%%%%%%%%%%%%%%%%%%%%%%%%%%%%%%%%%%%%%%%%%%%%%%%%%%%%%%%%

%%%%%%%%%%%%%%%%%%%%%%%%%%
\subsubsection{Model with \texorpdfstring{$SU(2)_L$}{su2LL}-triplet leptoquark and vectorlike quark}\label{sec:CCSVO35}
%%%%%%%%%%%%%%%%%%%%%%%%%%%%%%%%%%%%%%%%%%%%%%%
\begin{figure}[t!]
    \centering
    \includegraphics[scale=0.5]{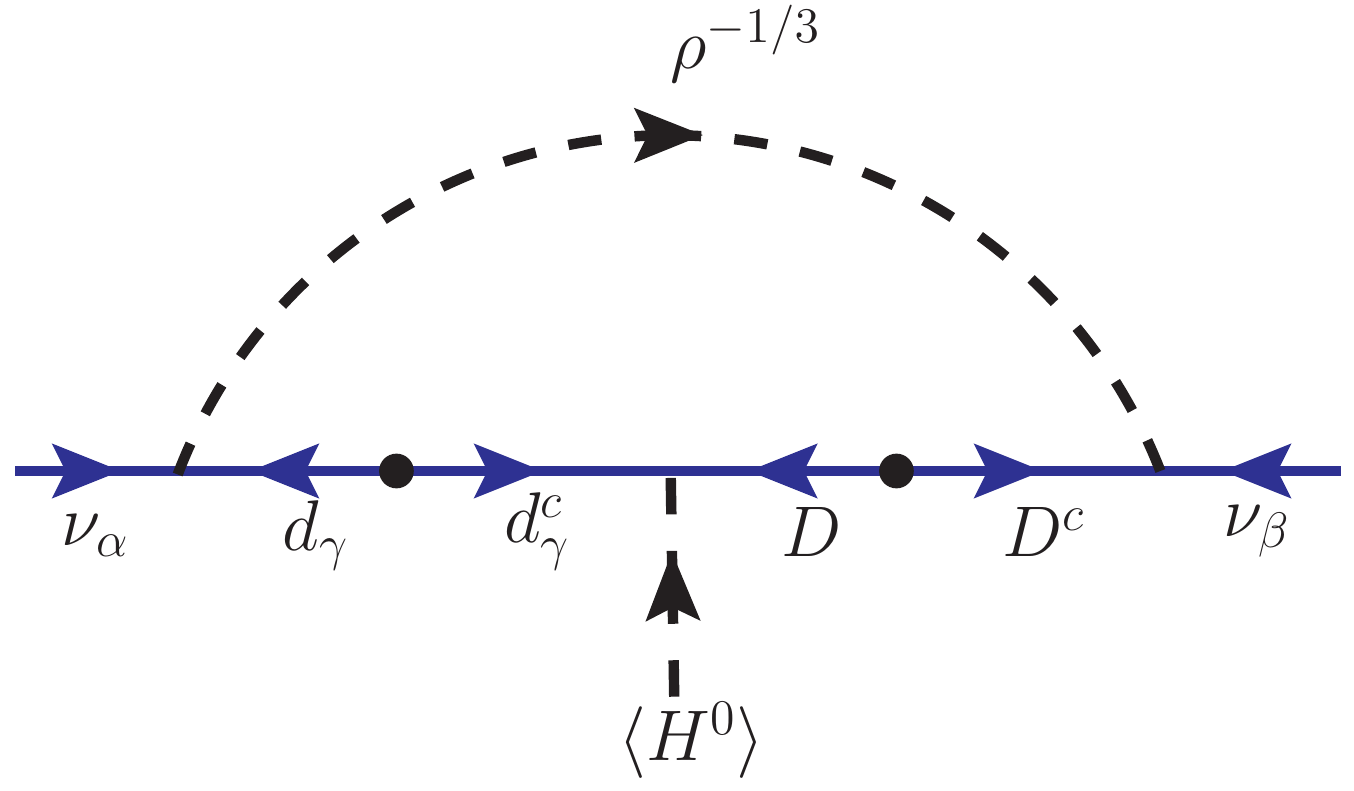}
    \caption{Neutrino mass generation in the one-loop triplet LQ model with vectorlike quarks.  This model corresponds to $\mathcal{O}_3^5$ of Table \ref{tab:O3} \cite{Cai:2014kra}.}
    \label{CCSVO35}
\end{figure}
%%%%%%%%%%%%%%%%%%%%%%%%%%%%%%%%%%%%%%%%%%%%%%%
This is based on the operator ${\cal O}_3^5$ (see Table~\ref{tab:O3}) which is realized by adding an $SU(2)_L$-triplet $\bar{\rho}\left({\bf \bar{3}},{\bf 3},\frac{1}{3}\right) = \left(\bar{\rho}^{4/3}, \, \bar{\rho}^{1/3}, \, \bar{\rho}^{-2/3}\right)$  and a vectorlike quark doublet  $\mathcal{Q}\left({\bf 3},{\bf 2},-\frac{5}{6}\right) = \left(D^{-1/3}, \, X^{-4/3}\right)$. Neutrino mass is generated at one-loop level, as shown as Fig.~\ref{CCSVO35}. There is also a two-loop diagram involving $\rho^{2/3}$, which is not considered here, as that would be sub-dominant to the one-loop diagram. The interaction term $Q Q \rho$ is forbidden by demanding baryon-number conservation to avoid proton decay. The relevant Lagrangian for the neutrino mass generation can be written as 
\begin{eqnarray}
  -\mathcal{L}_Y &\ \supset \ & M_\mathcal{Q} \mathcal{Q} \mathcal{Q}^c + (\lambda_{\alpha \beta} L_\alpha Q_\beta \Bar{\rho}+  \lambda_\alpha^{\prime} L_\alpha Q^c \rho + y_\alpha d_\alpha^c \mathcal{Q} H + {\rm H.c.} ) \, ,
  \label{eq:lag035}
  \end{eqnarray}
  where $\bar{\rho}$ is related to $\rho$ by charge conjugation as $\rho\left({\bf {3}},{\bf 3},-\frac{1}{3}\right) = \left(\rho^{2/3},\, -\rho^{-1/3},\, \rho^{-4/3}\right)$. Expanding the terms in Eq.~\eqref{eq:lag035}, we get 
 \begin{align}
 -\mathcal{L}_Y &\ \supset \  M_\mathcal{Q}\left(DD^c + X X^c\right) + \left[\lambda_{\alpha \beta} \left\{\nu_\alpha u_\beta \Bar{\rho}^{-2/3} - \frac{1}{\sqrt{2}}\left(\nu_\alpha d_\beta + \ell_\alpha u_\beta\right) \Bar{\rho}^{1/3} + \ell_\alpha d_\beta \Bar{\rho}^{4/3} \right\}\right. \nonumber \\
  &\left. \qquad \qquad + \lambda_\alpha^\prime \left \{\nu_\alpha X^c \rho^{-4/3} + \frac{1}{\sqrt 2}\left(\ell_\alpha X^c - \nu_\alpha D^c\right) \rho^{-1/3} - \ell_\alpha D^c \rho^{2/3}\right\} \right. \nonumber \\
  & \left. \qquad \qquad +  y_\alpha \left(D H^0 - H^+ X\right) d_\alpha^c+{\rm H.c.}\right] \, .
  \label{eq:lago35ex}
\end{align}
The neutrino mass can be estimated as
\begin{equation}
    M_\nu \ \sim \  \frac{1}{16\pi^2}\frac{v}{M^2}\left(\lambda\, M_d\, y\, M_D\, \lambda^{\prime T} + \lambda^{\prime}\,M_D\, y^T M_d\,\lambda^T\right) \, ,
    \label{eq:MnuO35}
\end{equation}
where $M_d$ and $M_D$ are the diagonal mass matrices for down-type quark and vectorlike quark fields, and $M = {\rm max}(m_{D_i}, m_\rho)$, with $m_{D_i}$ being the eigenvalues of $M_D$. With a single copy of the vectorlike quark, the matrices $y$ and $\lambda'$ are $3 \times 1$ dimensional.  Consequently the rank of $M_\nu$ is two, which would imply that the lightest neutrino mass $m_1 =0$ at the one-loop level. Realistic neutrino mixing can however be generated, analogous to the model of Ref. \cite{Zee:1985id,Babu:1988ki}.  

NSI in this model are induced by both $\bar{\rho}^{-2/3}$ and $\bar{\rho}^{1/3}$ fields, which couple to up and down quarks respectively (cf.~Eq.~\eqref{eq:lago35ex}). The NSI parameters read as
\begin{tcolorbox}[enhanced,ams align,
  colback=gray!30!white,colframe=white]
%\begin{equation}
 %    \boxed{ 
 \varepsilon_{\alpha \beta} \ = \ \frac{3}{4 \sqrt{2}G_F}  \left(\frac{\lambda_{\alpha u}^{\star} \lambda_{\beta u}}{ m_{\rho^{-2/3}}^2} + \frac{\lambda_{\alpha d}^{\star} \lambda_{\beta d}}{ 2m_{\rho^{1/3}}^2}\right) \, .
 %} %\ 
    %  \equiv \ 3\varepsilon^u_{\alpha \beta}+\frac{3}{2}\varepsilon^d_{\alpha \beta} \, ,
      \label{eq:epsrho}
%\end{equation}
\end{tcolorbox}
This is same as the triplet contribution in Eq.~\eqref{eq:NSI-O39} and the maximum allowed values are given in Eq.~\eqref{eq:NSI-rho}.  
%%%%%%%%%%%%%%%%%%%%%%%%%%%%%%%%%%%%%%%%%%%%%%%%%%%%%%%%%%%%%%%%%%%%%%%%%%%%%%%%%%%%%%%%%%%%%%
%%%%%%%%%%%%%%%%%%%%%%%%%%%%%%%%%%%%%%%%%%%%%%%%%%%%%%%%%%%%
%%%%%%%%%%%%%%%%%%%%%%%%%%%%%%%%%%%%%%%%%%%%%%%%%%%
 %%%%%%%%%%%%%%%%%%%%%%%%%%%%%%
\subsubsection{A new extended one-loop leptoquark model} \label{subsec:1loopLQ}
%%%%%%%%%%%%%%%%%%%%%%%%%%%%%%
\begin{figure}[!t]
    \centering
    \includegraphics[scale=0.5]{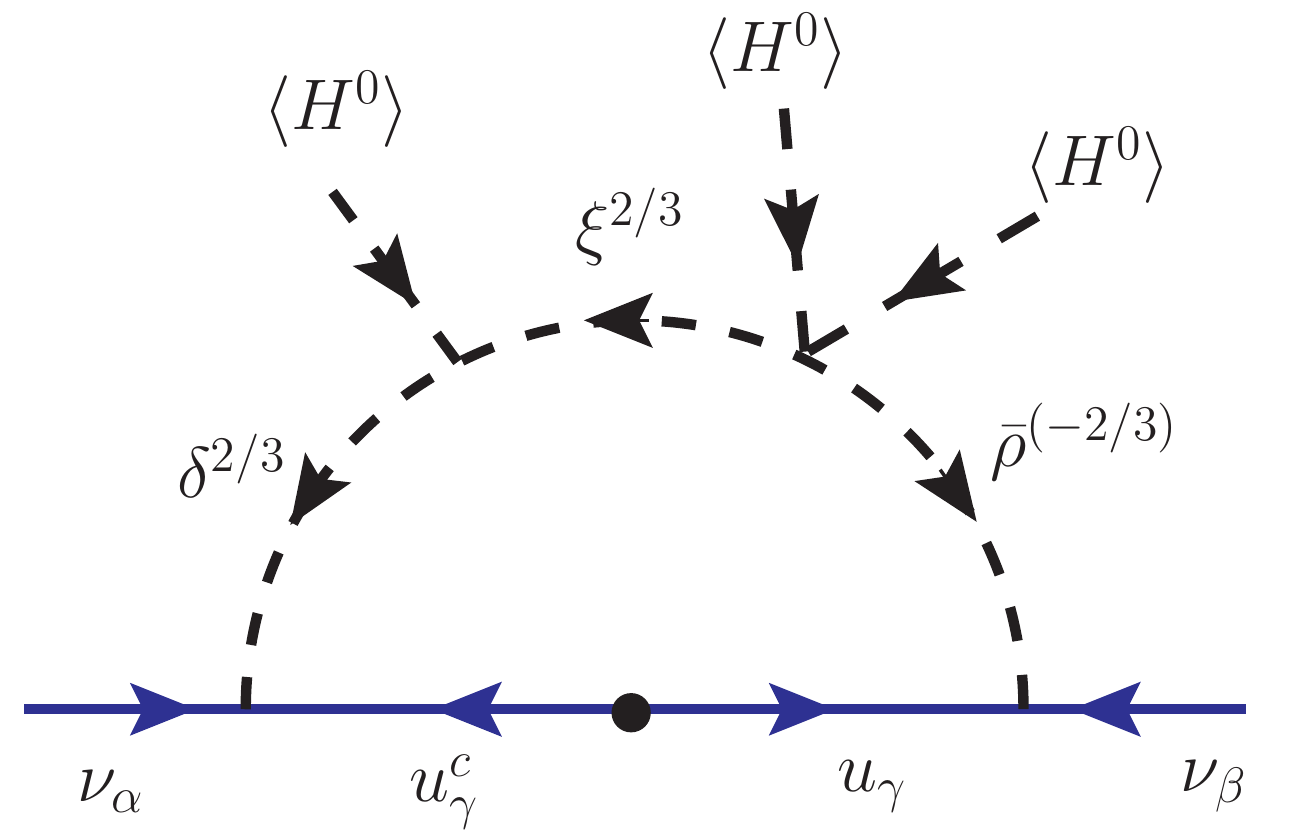}
    \caption{Feynman diagram for neutrino mass generation in the extended one-loop LQ model with up-type quark chiral suppression in the loop. The $\Delta L = 2$ effective operator is ${\cal \widetilde{O}}_1$ of Eq.~\eqref{eq:Ot1}. }
    \label{1loopLQ}
\end{figure}
%%%%%%%%%%%%%%%%%%%%%%%%%%%%%%%%%%%%%%%%%%%%%%%%%%%%%%
Here we present a  variation of the one-loop LQ model of Sec.~\ref{sec:LQ} wherein the neutrino mass is generated with up-quark chiral suppression (see Fig.~\ref{1loopLQ}), rather than down-quark mass suppression (as in Fig.~\ref{coloredzee}). The effective operator of the model is of dimension nine, given by
\begin{equation}
\mathcal{\widetilde{O}}_1 = (LQ)(Lu^c)(HH)H \, ,
\label{eq:Ot1}
\end{equation}
which may appear to be a product of $\mathcal {O}_1$ of Eq. (\ref{O1}) and the SM operator $(Q u^c H)$; but the $SU(2)_L$ contractions mix the two sub-operators.  
To realize this operator at the one-loop level, three $SU(3)_c$-triplet LQ fields are introduced:
   $\delta \left({\bf 3},{\bf 2},\frac{7}{6}\right) = 
          \left( \delta^{5/3}, \, 
           \delta^{2/3} \right), \,
    \Bar{\rho}\left(\Bar{\bf 3}, {\bf 3}, \frac{1}{3}\right) =
               \left(  \Bar{\rho}^{4/3} ,
                 \Bar{\rho}^{1/3} ,
                 \Bar{\rho}^{-2/3}\right) \, ,
    \xi\left({\bf 3},{\bf 1},\frac{2}{3}\right)$. 
Since three new fields are introduced, this model may be viewed as non-minimal, and does not fit into the classification of 
The corresponding Lagrangian for the neutrino mass generation reads as
\begin{align}
  -  \mathcal{L}_Y \  \supset \ & \lambda_{\alpha \beta} L_\alpha u_\beta^c \delta + \lambda'_{\alpha \beta} L_\alpha Q_\beta \bar{\rho} +  {\rm H.c.} 
  \  = \   \lambda_{\alpha \beta} \left(\nu_\alpha u_\beta^c \delta^{2/3} - \ell_\alpha u_\beta^c \delta^{5/3}\right) \nonumber \\
  & \qquad +\lambda'_{\alpha \beta} \left[\ell_\alpha d_\beta \Bar{\rho}^{4/3} - \frac{1}{\sqrt{2}} \left(\nu_\alpha d_\beta + \ell_\alpha u_\beta\right) \Bar{\rho}^{1/3} + \nu_\alpha u_\beta \Bar{\rho}^{-2/3}\right]+  {\rm H.c.} \, 
  \label{eq:lagLQ2}
\end{align}
%%%%%%%%%%%%%%%%%%%%%%%%%%%%%%%%%%%%%%%%%%%%%%%
Neutrino mass is generated by the diagram shown in Fig.~\ref{1loopLQ} using the Lagrangian~\eqref{eq:lagLQ2}, together with the potential terms 
\begin{eqnarray}
    V \ &\supset& \ \lambda_1 \Bar{\rho} \widetilde{H} \widetilde{H} \xi + \mu  \widetilde{H} \delta \xi^\star + {\rm H.c.} \ = \ \lambda_1 \xi^{2/3} \left(\bar{\rho}^{4/3} H^- H^- + \sqrt{2} \bar{\rho}^{1/3} H^0 H^- + \bar{\rho}^{-2/3} H^0 H^0 \right)\nonumber \\
   &+& \mu \xi^{ \star -2/3}\left(H^0 \delta^{2/3} + H^- \delta^{5/3}  \right) + {\rm H.c.}
\end{eqnarray}
%%%%%%%%%%%%%%%%%%%%%%%%%%%%%%%%%%%%%%%%%%%%%%%
where $\widetilde{H}=(H^0,\,-H^-)$ represents the SM Higgs doublet. The neutrino mass matrix can be estimated as
\begin{equation}
    M_\nu \ \sim \ \frac{1}{16 \pi^2}\frac{\mu \lambda_1 v^3}{m_1^2 m_2^2}(\lambda M_u \lambda'^T + \lambda' M_u \lambda^T)  \, ,
\end{equation}
%%%%%%%%%%%%%%%%%%%%%%%%%%%%%%%%%%%%%%%%%%%%%%%
where $m_1$ and $m_2$ are the masses of the heaviest two LQs among the $\delta$, $\bar{\rho}$ and $\xi$ fields, and $M_u$ is the diagonal mass matrix in the up-quark sector. To get small neutrino masses, we need the product $\lambda\lambda'\ll 1$. We may take $\lambda\sim {\cal O}(1)$ and $\lambda'\ll \lambda$ which is preferable to the other case of $\lambda\ll \lambda'$, since the $\lambda'$ couplings are constrained by $D$-meson decays (see Sec.~\ref{sec:Dmeson}).  

After integrating out the heavy LQ fields, Eq.~\eqref{eq:lagLQ2} leads to an effective NSI Lagrangian with up-quarks in the neutrino propagation through matter. The NSI parameters read as
\begin{tcolorbox}[enhanced,ams align,
  colback=gray!30!white,colframe=white]
%\begin{equation}
%      \boxed{
      \varepsilon_{\alpha \beta} \ = \ \frac{3}{4 \sqrt{2}G_F} \left(\frac{\lambda_{\alpha u}^{\star} \lambda_{\beta u} }{ m_\delta^2} + \frac{\lambda_{\alpha u}^{\prime\star} \lambda'_{\beta u}}{ m_{\rho^{-2/3}}^2} + \frac{\lambda_{\alpha d}^{\prime\star} \lambda'_{\beta d}}{ 2m_{\rho^{1/3}}^2}\right)  \, .
      %}
      \label{eq:uploopNSI}
%\end{equation}
\end{tcolorbox}
For $\lambda\gg \lambda'$, this expression is exactly the same as the doublet LQ contribution derived in Eq.~\eqref{nsi_coloredzee} and the corresponding maximum NSI can be read off from Table~\ref{tab:LQ} for the doublet component. For $\lambda'\gg \lambda$, Eq.~\eqref{eq:uploopNSI} is the same as Eq.~\eqref{eq:epsrho}. This latter choice maximizes NSI in this model and is summarized in Table~\ref{Table_Models}. 

%The constraints on these NSI parameters are exactly the same as those for the NSI with down-quark discussed in Sec.~\ref{sec:NSILQ} and summarized in Table~\ref{tab:LQ}. The only exception is $\varepsilon_{\mu\mu}$, where the constraint from NuTeV is a factor of five stronger~\cite{Davidson:2003ha}. The maximum NSI obatined in this model are tabulated in Table~\ref{Table_Models} and reproduced below: 
%    & \varepsilon_{ee}^{\rm max} \ = \ 0.004 \, , \qquad 
 %   \varepsilon_{\mu\mu}^{\rm max} \ = \ 0.009 \, , \qquad 
 %   \varepsilon_{\tau\tau}^{\rm max} \ = \ 0.343 \, , \nonumber \\
 %  & \varepsilon_{e\mu}^{\rm max} \ = \ 1.5\times 10^{-8} \, , \qquad 
 %   \varepsilon_{e\tau}^{\rm max} \ = \ 0.0036 \, , \qquad 
 %   \varepsilon_{\mu\tau}^{\rm max} \ = \ 0.0043 \, . 
 %   \label{eq:extendedLQ}
%\end{align}

There are other variations of one-loop LQ models with more exotic particles~\cite{Dorsner:2017wwn, AristizabalSierra:2007nf}, where the neutrino mass is proportional to up-type quark mass. The NSI predictions in these models are the same as in Eq.~\eqref{eq:uploopNSI}.  
%%%%%%%%%%%%%%%%%%%%%%%%%%%%%%%%%%%%%%%%%%%%%%%
%%%%%%%%%%%%%%%%%%%%%%%%%%%%%%%%%%%%%%%%%%%%%%%%%%%%%%%%%%%%%%%%%%%%%%%%%%%%%%%%%%%%%%%%%%%%%%
%%%%%%%%%%%%%%%%%%%%%%%%%%%%%%%%%%%%%%%%%%%%%%%%%%%%%%%%%%%%%%%%%%%%%%%%%%%%%%%%%%%%%%%%%%%%%%

%%%%%%%%%%%%%%%%%%%%%%%%%%%%%%%%%%%%%%%%%%%%%%%%%%%%%%%%%%%%%%%%%%%%%%%%%%%%%%%%%%%%%%%%%%%%%%
%%%%%%%%%%%%%%%%%%%%%%%%%%%%%%%%%%%%%%%%%%%%%%%%%%%%%%%%%%%%%%%%%%%%%%%%%%%%%%%%%%%%%%%%%%%%%%
%%%%%%%%%%%%%%%%%%%%%%%%%%%%%%%%%%%%%%%%%%%%%%%%%%%%%%%%%%%%%%%%%%%%%%%%%%%%%%%%%%%%%%%%%%%%%%

%%%%%%%%%%%%%%%%%%%%%%%%%%%%%%%%%%%%%%%%%%%%%%%
\subsection{Two-loop models} \label{sec:2loop}
%%%%%%%%%%%%%%%%%%%%%%%%%%%%%%%%%%%%%%%%%%%%%%%
%%%%%%%%%%%%%%%%%%%%%%%%%%%%%%%%%%%%%%%%%%%%%%%
\subsubsection{Zee-Babu model}\label{sec:zeebabu}
\begin{figure}[t!]
 %   \begin{minipage}[t]{7cm}
    \centering
        \includegraphics[scale=0.55]{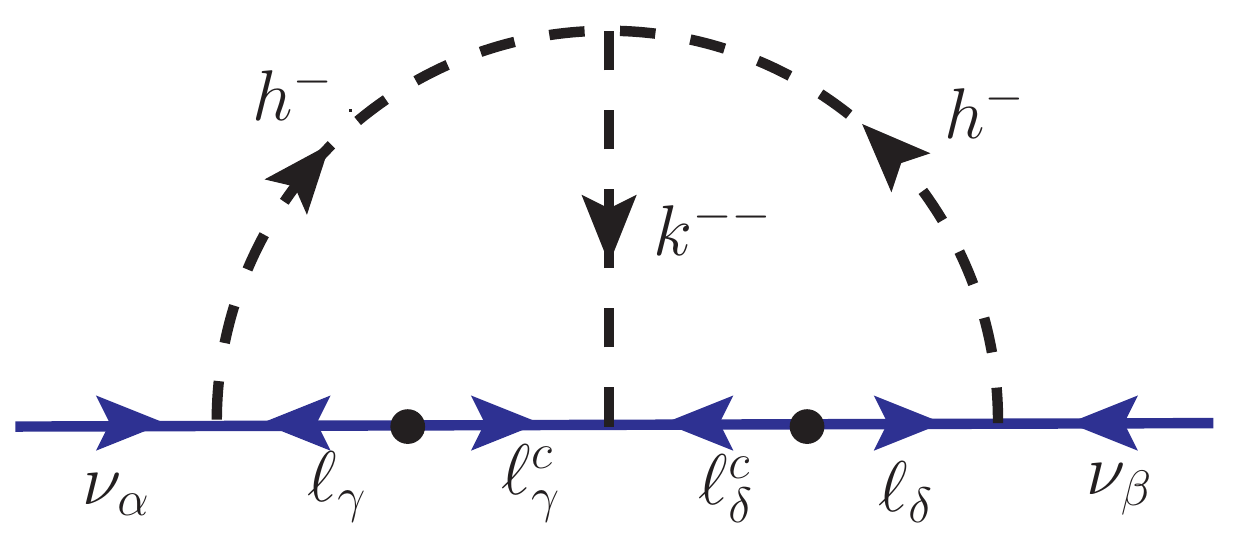}
  %   \caption*{ (a)} 
 %   \end{minipage}
  %  \hspace{5mm}
 %   \begin{minipage}[t]{7cm}
  %  \centering
  %      \includegraphics[scale=0.5]{figures/zee_babu_diag2.pdf}
  %    \caption*{(b)}
 %   \end{minipage}
 \caption{Neutrino mass generation at two-loop in the Zee-Babu model~\cite{Zee:1985id,Babu:1988ki}. This model generates operator $\mathcal{O}_9$ of Eq.~\eqref{O9}. }
 %   \caption{ (a) Zee-Babu 2 loop model (b) Diagram responsible for non-standard interaction from charged scalar singlet}
    \label{zee_babu}
 \end{figure}

This model realizes the operator $\mathcal{O}_9$ of Eq. (\ref{O9}).  
In this model~\cite{Zee:1985id,Babu:1988ki}, two $SU(2)_L$-singlet Higgs fields, $h^+({\bf 1},{\bf 1},1)$ and $k^{++}({\bf 1},{\bf 1},2)$, are introduced. The corresponding Lagrangian for the generation of neutrino mass reads:
\begin{align}
   - \mathcal{L}_Y & \ \supset \  f_{\alpha \beta} L^i_{\alpha}C L^j_{\beta} h^+ \epsilon_{ij} + h_{\alpha \beta} {\ell^T}_{\alpha} C \ell_{\beta } k^{++} +{\rm H.c.} \nonumber \\
   & \ = \ f_{\alpha \beta} (\nu^T_{\alpha} C \ell_{\beta } -\nu^T_{\beta} C \ell_{\alpha} ) h^+ + h_{\alpha \beta} \ell^T_{\alpha} C \ell_{\beta } k^{++} +{\rm H.c.} \, 
    \label{eq:zee-babu}
\end{align}
 Majorana neutrino masses are induced at two-loop as shown in Fig.~\ref{zee_babu} by the Lagrangian~\eqref{eq:zee-babu}, together with the potential term 
     \begin{align}
       V \ \supset \   - \mu h^- h^- k^{++} +{\rm H.c.} \, .
     \end{align}
 The neutrino mass matrix reads:
  \begin{equation}
 %      M_{\nu} \ \simeq \ \frac{1}{(16 \pi^2)^2} \frac{8 \mu }{M^2} \,   \, f_{\alpha c}  \, h_{cd}^{\star} \, m_c  \, m_d  \, f_{d \beta} \,  \tilde{I}\left(\frac{m_{k^{++}}^2}{m_{h^+}^2}\right) \, ,
 M_\nu \ \simeq \ \frac{1}{(16\pi^2)^2}\frac{8\mu}{M^2}fM_u h^\dag M_u f^T\cal{I} \, ,
  \end{equation}
where $M = {\rm max}(m_{k^{++}}, m_{h^+})$ and $\cal{I}$ is a dimensionless function that depends on the ratio of the masses of the two new scalars~\cite{Babu:2002uu,Nebot:2007bc,Schmidt:2014zoa,Herrero-Garcia:2014hfa}. The singly charged scalar $h^+$ induces NSI at tree-level through the $f$-type Yukawa coupling in Eq.~\eqref{eq:zee-babu}. After integrating out the heavy scalars, NSI induced in neutrino propagation through normal matter can be  written as in Eq.~\eqref{eq:nsio21}, with the replacement $m_{\eta^+}\to m_{h^+}$, and 
%\begin{equation}
 %   \boxed{\varepsilon_{\alpha \beta} \ \equiv \ \varepsilon_{\alpha \beta}^{ee} \ = \ \frac{1}{\sqrt{2}G_F}\frac{f_{ e \alpha}^{\star}f_{ e \beta }  }{m_{h^+}^2}  \, .}
%    \label{nsi_zeebabu}
%\end{equation}
%This is exactly the same as Eq.~\eqref{eq:nsio21} for which 
the maximum NSI are given by Eq.~\eqref{eq:maxZB}. These are severely constrained by cLFV searches and universality of charged currents~\cite{Nebot:2007bc} (cf.~Table~\ref{tab:zee-babu}), restricting the maximum NSI to ${\cal O}(10^{-3})$ level~\cite{Ohlsson:2009vk}. These numbers are  summarized in Table~\ref{Table_Models}.

%%%%%%%%%%%%%%%%%%%%%%%%%%%%%%%%%%%%%%%%%%%%%%%%%%%%
%%%%%%%%%%%%%%%%%%%%%%%%%%%%%%%%%%%%%%%%%%%%%%
\subsubsection{Leptoquark/diquark variant of the Zee-Babu model} \label{subsec:colorzeebabu}
\begin{figure}[t!]
    \centering
     \includegraphics[scale=0.6]{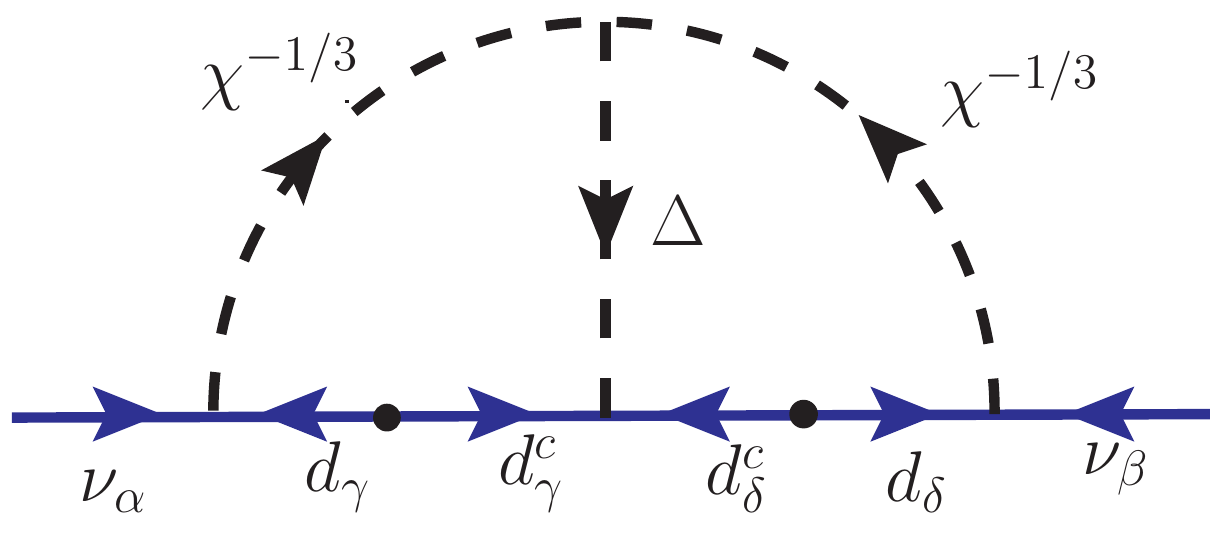}
    \caption{Neutrino mass generation at two-loop in the LQ/DQ variant of the Zee-Babu model which generates operator $\mathcal{O}_{11}$~\cite{Kohda:2012sr}, cf. Eq.~\eqref{O9}.}
    \label{other2loop}
\end{figure}

 One can also generate neutrino mass at two-loop by replacing leptons with quarks in the Zee-Babu model as shown in Fig. \ref{other2loop}. Here the effective operator is of dimension nine, given by
 \begin{equation}
     \mathcal{O}_{11} \ = \ L^i L^j Q^k d^c Q^l d^c \epsilon_{ik} \epsilon_{jl}~.
     \label{eq:O11}
 \end{equation}
 In addition to the SM fields, this model~\cite{Kohda:2012sr} employs a scalar LQ $\chi \left({\bf 3},{\bf 1},-\frac{1}{3}\right)$ and a scalar DQ  $\Delta\left({\bf 6},{\bf 1},-\frac{2}{3}\right)$. The $\chi~ (\Delta)$ field plays the role of singly (doubly)-charged scalar in the Zee-Babu model. The relevant Yukawa Lagrangian for the neutrino mass generation is written as
\begin{align}
  -  \mathcal{L}_Y \ \supset \ & \lambda_{\alpha \beta} L_{\alpha}^i Q_\beta^j \chi^{\star} \epsilon_{ij} + h_{\alpha \beta} d_\alpha^c d_\beta^c \Delta^{-2/3}  + {\rm H.c.} \, \nonumber \\
  \ = \ & \lambda_{\alpha\beta} \left(\nu_\alpha d_\beta  - \ell_\alpha u_\beta\right) 
  \chi^{\star} + h_{\alpha \beta} d_\alpha^c d_\beta^c \Delta^{-2/3}  + {\rm H.c.} \, 
  \label{eq:LQZB}
\end{align}
%%%%%%%%%%%%%%%%%%%%%%
Neutrino mass is generated at two-loop via the Lagrangian~\eqref{eq:LQZB} in combination with the potential term
\begin{align}
   V \ \supset \ - \mu \chi^{\star} \chi^{\star} \Delta^{-2/3}+{\rm H.c.} \, 
\end{align}
%%%%%%%%%%%%%%%%%%%%%%
%%%%%%%%%%%%%%%%%%%%%%%%%%
The neutrino mass matrix can be calculated as
\begin{equation}
    M_\nu \ \sim \  \frac{24 \mu}{(16 \pi^2)^2 M^2} \lambda M_{d} h^\dagger  M_{d} \lambda^T {\cal I} \, ,
\end{equation}
where $M \equiv $ max($m_\chi, m_\Delta$), $M_{d}$ is the diagonal down-type quark mass matrix, and ${\cal I}$ is a dimensionless two-loop integral defined in terms of the ratio of $m_\Delta^2$ and $m_\chi^2$~\cite{Babu:2002uu}. After integrating out the heavy scalars, the NSI parameters in this model are  given by 
%\begin{equation}
 %   \boxed{\varepsilon_{\alpha \beta} \ = \ \frac{3}{4 \sqrt{2}G_F} \frac{\lambda_{\alpha d}^{\star} \lambda_{\beta d} }{ m_\chi^2} \, .}
%\end{equation}
%This is exactly same as the singlet LQ contribution in 
Eq.~\eqref{eq:714} and the corresponding maximum NSI are given in Eq.~\eqref{eq:NSI-singlet}. 
This is also summarized in Table~\ref{Table_Models}.

There are a few variants of this LQ/DQ version of the Zee-Babu model.  First, one could replace the color sextet field $\Delta\left({\bf 6},{\bf 1},-\frac{2}{3}\right)$ by a color triplet scalar $\Delta\left({\bf 3},{\bf 1},-\frac{2}{3}\right)$ in Fig. \ref{other2loop}. The cubic term $\chi^ \star \chi^ \star \Delta$ will not be allowed by Bose symmetry in this case.  By assuming two copies of the $\chi$ field, namely, $\chi_1$ and $\chi_2$, one could restore this coupling from $\chi_1^ \star \chi_2^ \star \Delta$, in which case the diagram of Fig.~\ref{other2loop} can be connected~\cite{Klein:2019jgb}. The NSI in such a model is identical to the model described in this section.  Second, one could replace the internal down quarks of Fig.~\ref{other2loop} by up-type quarks, with a simultaneous replacement of $\chi\left({\bf 3},{\bf 1},-\frac{1}{3}\right)$ by $\rho\left({\bf 3},{\bf 3},-\frac{1}{3}\right)$ and $\Delta\left({\bf 6},{\bf 1},-\frac{2}{3}\right)$ by $\Delta\left({\bf 6},{\bf 1},\frac{4}{3}\right)$.  Neutrino NSI will then follow the $\rho$ NSI predictions as in Sec.~\ref{sec:CCSVO35}.  In this up-quark variant, one could replace the DQ $\Delta\left({\bf 6},{\bf 1},\frac{4}{3}\right)$ by a color triplet field $\Delta\left({\bf 3},{\bf 1},\frac{4}{3}\right)$ as well~\cite{Klein:2019jgb}.

%%%%%%%%%%%%%%%%%%%%%%%%%%%%%%%%%%%%%%%%%%%%%%%%%%%%
%%%%%%%%%%%%%%%%%%%%%%%%%%%%%%%%%%%%%%%%%%%%%%%%%%%%
\subsubsection{Model with \texorpdfstring{$SU(2)_L$}{su2lll}-doublet and singlet leptoquarks} \label{subsec:color2} 
Operator $\mathcal{O}_{8}^4$ of Table. (\ref{tab:O8}) does not induce neutrino mass via one-loop diagrams owing to the $SU(2)_L$ index structure.  This operator will, however, lead to generation of neutrino masses at the two-loop level.  
A simple realization of $\mathcal{O}_{8}^4$ is given in Ref.~\cite{Babu:2010vp}. This model uses the same gauge symmetry and particle content as in the LQ variant of the Zee model (cf.~Sec.~\ref{sec:LQ}),~i.e.,~$\Omega \left({\bf 3},{\bf 2}, \frac{1}{6}\right)=\left(\omega^{2/3},\omega^{-1/3}\right)$ and $\chi \left({\bf 3},{\bf 1},-\frac{1}{3}\right)$,  with $\chi$ coupling modified as follows:
\begin{align}
  -  \mathcal{L}_Y \ \supset \ & \lambda_{\alpha \beta} L^i_\alpha d_\beta^c \Omega^j\epsilon_{ij} + f_{\alpha \beta} \ell_\alpha^c u_\beta^c \chi + {\rm H.c.} \, ,\nonumber \\ 
   \ = \ & \lambda_{\alpha\beta}\left(\nu_\alpha d_\beta^c \omega^{-1/3} - \ell_\alpha d_\beta^c \omega^{2/3} \right)+ f_{\alpha \beta} \ell_\alpha^c u_\beta^c \chi + {\rm H.c.} \, 
    \label{lag_2lepto}
\end{align}
Note that these Yukawa couplings conserve both baryon and lepton number as can be seen by assigning $(B,L)$ charges of $\left(\frac{1}{3},-1\right)$ to $\Omega$ and $\left(\frac{1}{3},1\right)$ to $\chi$. The couplings $\tilde\lambda_{\alpha\beta}u^c_\alpha d^c_\beta \chi^\star$, allowed by the gauge symmetry are forbidden by $B$, and the couplings $\lambda'_{\alpha\beta}L_\alpha Q_\beta\chi^\star$ (as in Eq.~\eqref{lagLQ}), allowed by gauge symmetry as well as $B$ are forbidden by $L$.\footnote{The simultaneous presence of the $f$  and $\lambda'$ couplings will drastically alter the successful $V-A$ structure of the SM~\cite{Buchmuller:1986iq}, and therefore, the $\lambda'$ terms must be forbidden in this model by $L$ symmetry.} The $L$ symmetry is softly broken by the cubic term in the scalar potential~\eqref{pot_lepto}. 

%%%%%%%%%%%%%%%%%%%%%%%%%%%%%%%%%%%%%%
\begin{figure}[t!]
    \centering
    \subfigure[]{
        \includegraphics[scale=0.5]{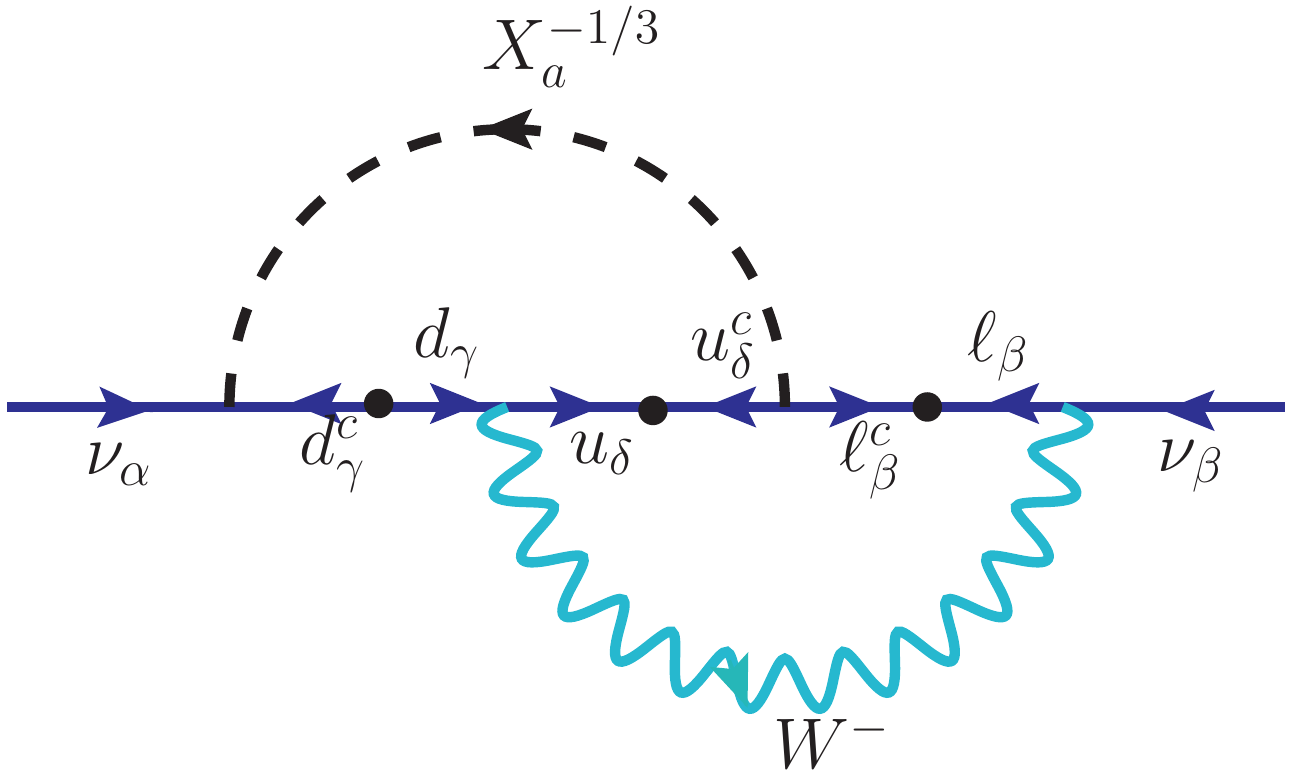}} 
        \hspace{5mm}
    \subfigure[]{
        \includegraphics[scale=0.41]{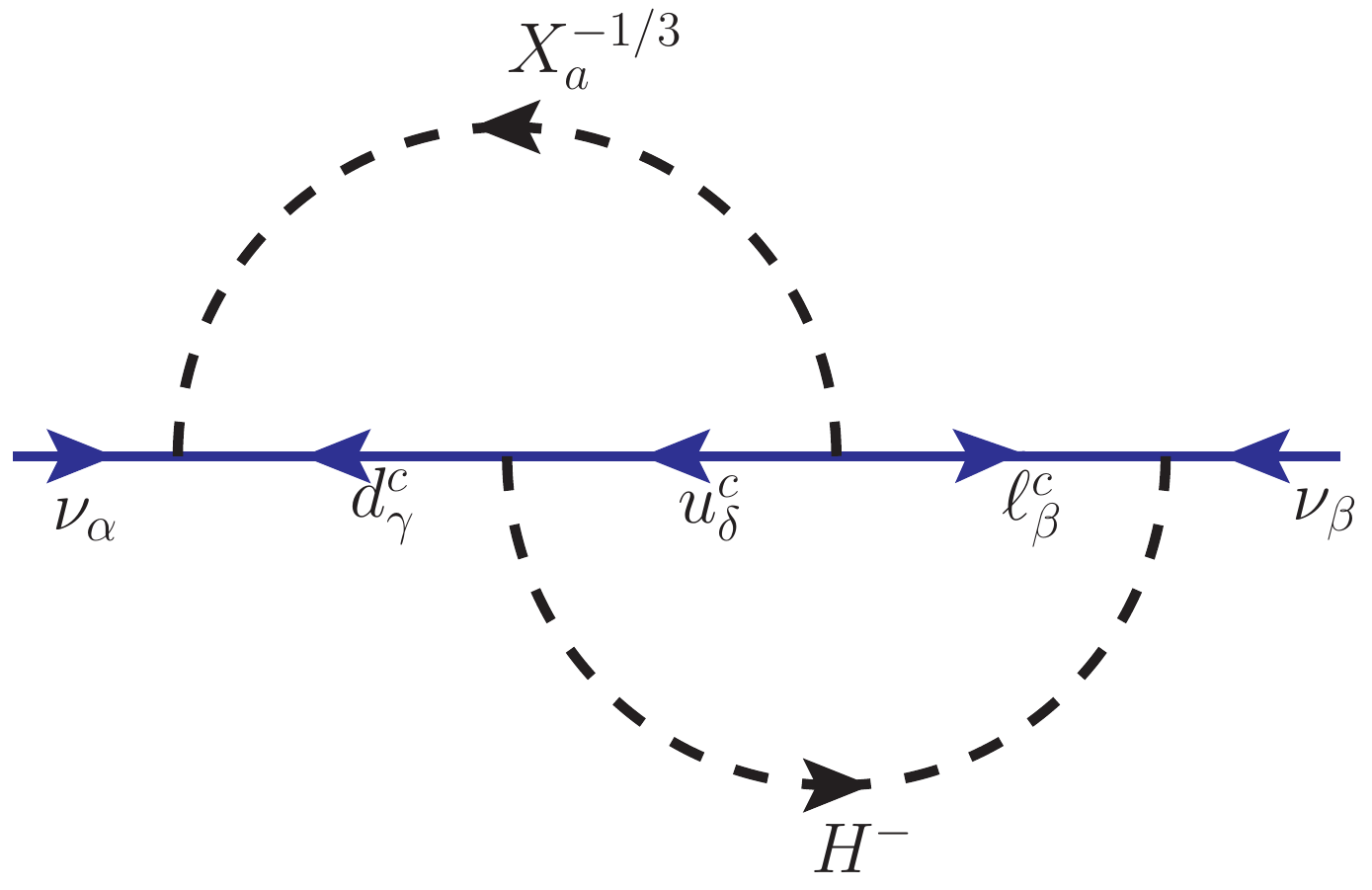}}
    \caption{Two-loop diagrams contributing to neutrino mass generation in the model of Ref.~\cite{Babu:2010vp}. The model realizes operator $\mathcal{O}_{8}^4$ of Table~\ref{tab:O8}.}
    \label{babujulio}
 \end{figure}
%%%%%%%%%%%%%%%%%%%%

The simultaneous presence of Eqs.~\eqref{lag_2lepto} and \eqref{pot_lepto} would lead to neutrino mass generation at two-loop level as shown in Fig.~\ref{babujulio}. Here $X_a$ (with $a=1,2$) are the mass eigenstates obtained from the mixture of the $\omega^{-1/3}$ and $\chi^{-1/3}$ states (cf.~Eq.~\eqref{eq:X12}).  Evaluation of the LQ-$W$ exchange diagrams in Fig.~\ref{babujulio} (a) give the neutrino mass matrix as 
\begin{equation}
 %   (M_{\nu})_{\alpha \beta} \approx \frac{C g^2 \sin 2\alpha}{(16 \pi^2)^2 M^2} \{ \lambda_{\alpha k} (M_d)_k (M_u)_l (F^\dagger )_{l \beta} (M_l)_\beta + (M_l)_\alpha (F^{\star})_{\alpha l} (M_u)_l (M_d)_k (\lambda^T)_{k \beta} \} \mathcal{I}
 M_\nu \ \sim \ \frac{3g^2 \sin 2\alpha}{(16 \pi^2)^2 M^2}\left[\lambda M_d  V^T M_u f^\dag M_\ell+ M_\ell f^\star M_u V M_d \lambda^T\right] {\cal I} \, ,
\end{equation}
where 3 is a color factor, $\alpha$ is the $\omega-\chi$ mixing angle (cf.~Eq.~\eqref{eq:tan2a}), $M_{u,d,\ell}$ are diagonal mass matrices for the up- and down-type quarks, and charged leptons, respectively, $V$ is the CKM mixing matrix, $M \equiv {\rm min}(m_1, m_2)$ (with $m_{1,2}$ given by Eq.~\eqref{eq:mX12}),  and  $\mathcal{I}$ is a dimensionless two-loop integral that depends on $m_{1,2}$, $m_W$ and $M_{u,d,\ell}$~\cite{Babu:2010vp}. 

NSI induced in this LQ model has the same features as the LQ variant of the Zee model discussed in Sec.~\ref{sec:NSILQ}. Note that the $f_{\alpha \beta}$-couplings in Eq.~\eqref{lag_2lepto}  do not lead to neutrino NSI. The expression for the NSI parameters is given by Eq.~\eqref{eq:719}, with the maximum 
%\begin{equation}
  % \boxed{ \varepsilon_{\alpha \beta} \ = \ \frac{3}{4 \sqrt{2}G_F} \frac{\lambda_{\alpha d}^{\star} \lambda_{\beta d} }{ m_\omega^2} \, .}
%   \label{nsi_eqn_LQ}
%\end{equation}
%The maximum 
allowed values given in Eq.~\eqref{eq:maxNSI_BJ} and also summarized in Table~\ref{Table_Models}.

%%%%%%%%%%%%%%%%%%%%%%%%%%%%%%%%%%%%%%
%%%%%%%%%%%%%%%%%%%%%%%%%%%%%%%%

%%%%%%%%%%%%%%%%%%%%%%%%%%
\subsubsection{Leptoquark model with \texorpdfstring{$SU(2)_L$-}{su22l}singlet vectorlike quark} \label{subsec:VQLQ}
%%%%%%%%%%%%%%%%%%%%%%%%
\begin{figure}[!t]
    \centering
    \includegraphics[scale=0.6]{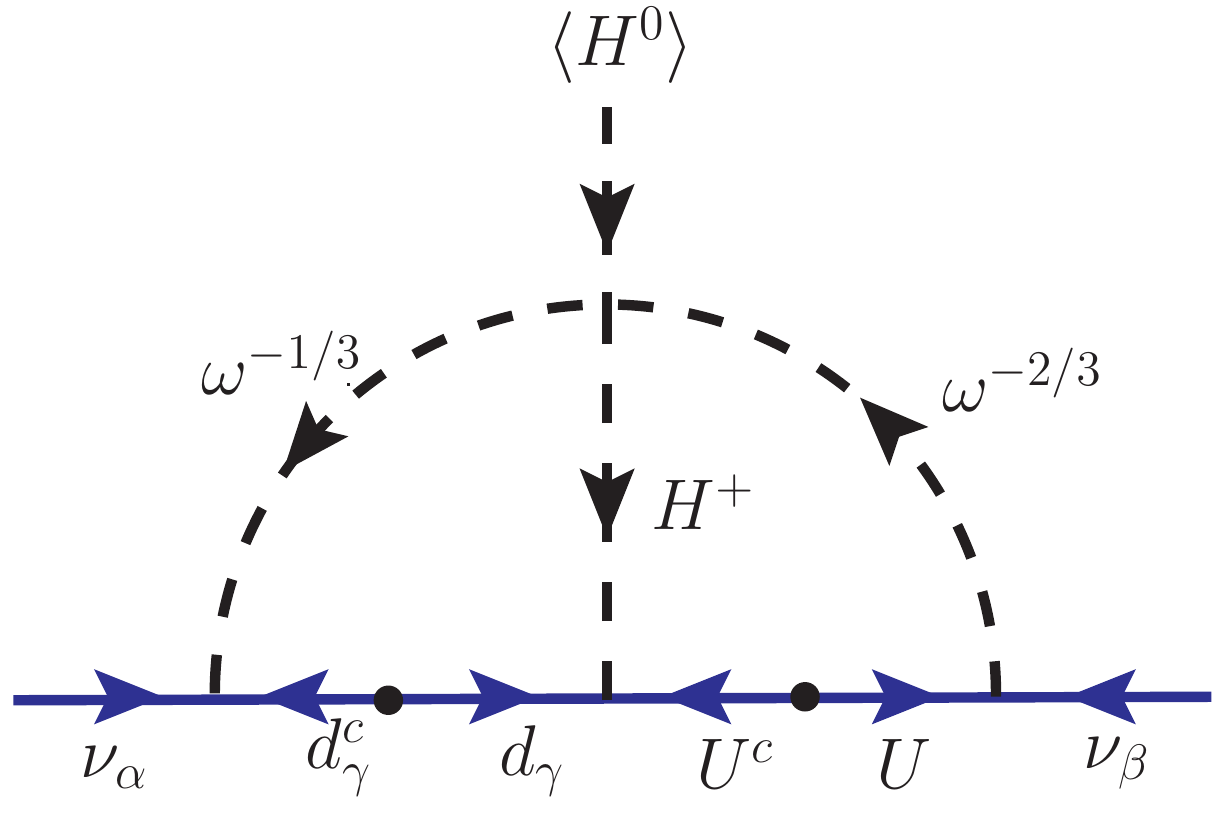}
    \caption{Two-loop neutrino mass generation in the model of Ref.~\cite{Babu:2011vb} with a LQ and a vector-like quark. This model corresponds to $\mathcal{O}_3^3$ of Table \ref{tab:O3}. }
    \label{vectorlike}
\end{figure}
This model utilizes the dimension-7 operator $L^i L^j \epsilon_{ij} Q^k H^l \epsilon_{kl} d^c$ to generate two-loop neutrino mass~\cite{Babu:2011vb}. This specific realization corresponds to the model $\mathcal{O}_3^3$ of Table \ref{tab:O3} \cite{Cai:2014kra}. In addition to the SM fields, an $SU(2)_L$-singlet vector-like quarks $U\left({\bf 3},{\bf 1},\frac{2}{3}\right)$ and $U^c \left({\bf 3}^\star,{\bf 1},-\frac{2}{3}\right)$, and a scalar doublet LQ $\Omega\left({\bf 3},{\bf 2},\frac{1}{6}\right)=\left(\omega^{2/3},\omega^{-1/3}\right)$  are added to the SM spectrum.  Addition of these fields leads to the following new Yukawa Lagrangian: 
\begin{align}
   - \mathcal{L}_Y \ \supset \ & \lambda_{\alpha \beta} L_\alpha \Omega d_\beta^c  + \lambda'_\alpha L_\alpha  \widetilde{\Omega}U + f_\alpha  Q_\alpha H U^c + {\rm H.c.}  \, , \nonumber \\
   \ = \ & \lambda_{\alpha\beta} (\nu_\alpha d_\beta^c \omega^{-1/3} - \ell_\alpha d_\beta^c \omega^{2/3} )+ \lambda_\alpha' \left[\left(\omega^{-1/3}\right)^\star \ell_\alpha U + \nu_\alpha \left(\omega^{2/3}\right)^\star U\right] \\ \nonumber
 &  + f_\alpha (u_\alpha H^0 U^c - d_\alpha H^+ U^c) + {\rm H.c.}  \, ,
    \label{eq:LQVQ1}
\end{align}
%%%%%%%%%%%%%%%%%%%%%
where $\widetilde{\Omega} \equiv i \tau_2 \Omega^{\star}$. The presence of all three Yukawa terms implies that lepton number is not conserved. Together with the quartic coupling term in the potential 
\begin{align}
    V \ \supset \ \lambda_\omega |\Omega^i H^j\epsilon_{ij}|^2 \ \supset \ - \lambda_\omega \omega^{-1/3}\omega^{-2/3}H^+H^0 +{\rm H.c.} \, , 
\end{align}
the Lagrangian~\eqref{eq:LQVQ1} leads to neutrino mass  generation at two-loop as shown in Fig.~\ref{vectorlike}. This can be estimated as
%%%%%%%%%%%%%%%%%%%%
\begin{equation}
    M_{\nu} \ \simeq \ \frac{\lambda_\omega}{(16 \pi^{2})^{2}} \frac{v}{M^2}
    (\lambda M_d f M_U \lambda'^T+ \lambda'M_U^T f^TM_d^T\lambda^T )  \, ,
\end{equation}
%%%%%%%%%%%%%%%%%%%%
where $M_d$ and $M_U$ are the diagonal down quark and vectorlike quark mass matrices respectively, and $M=$max($m_\omega,m_{U_i}$), with $m_{U_i}$ being the eigenvalues of $M_U$.

NSI in this model are induced by the $\omega^{-1/3}$ LQ and are given by Eq.~\eqref{eq:719}. 
%\begin{equation}
%    \boxed{\varepsilon_{\alpha \beta} \ = \ \frac{3}{4 \sqrt{2}G_F} \frac{\lambda_{\alpha d}^{\star} \lambda_{\beta d} }{ m_\omega^2} \, .}
%\end{equation}
%This is same as the doublet LQ contribution in Eq.~\eqref{nsi_coloredzee}. 
The maximum NSI that can be obtained in this model are given in Eq.~\eqref{eq:maxNSI_BJ} and are also summarized in Table~\ref{Table_Models}.
%%%%%%%%%%%%%%%%%%%%%%%%%%
%%%%%%%%%%%%%%%%%%%%%%%%%%%%%%%%%%%%%%%%%%%%%%%%%%%%
\subsubsection{Angelic model} \label{subsec:angelic}
\begin{figure}[t!]
    \centering
    \includegraphics[scale=0.6]{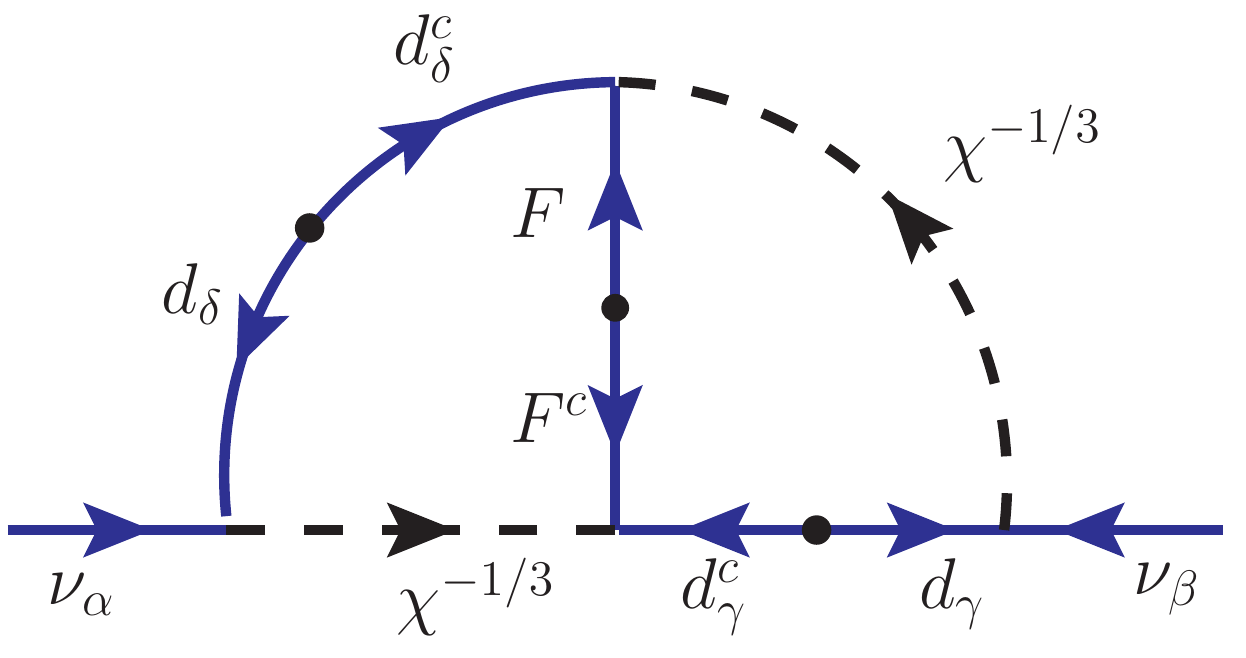}
    \caption{Two-loop neutrino mass generation in the Angelic model~\cite{Angel:2013hla}. This model induces operator $\mathcal{O}_{11}$ of Ref. \cite{Babu:2001ex}.}
    \label{angelic}
\end{figure}

This model induces operator $\mathcal{O}_{11}$ of Ref. \cite{Babu:2001ex}:
\begin{equation}
    \mathcal{O}_{11} = L^i L^j Q^k d^c Q^l d^c \epsilon_{ik} \epsilon_{jl}~.
\end{equation}
In this model~\cite{Angel:2013hla}, one adds two scalar LQs $\chi_a\left({\bf 3},{\bf 1},-\frac{1}{3}\right)$ (with $a=1,2$) and a color-octet Majorana fermion $F({\bf 8},{\bf 1},0)$. The relevant Yukawa Lagrangian is written as
\begin{equation}
 -   \mathcal{L}_Y \ \supset \ \lambda_{\alpha\beta a} L_\alpha  Q_\beta \chi_a + \lambda'_{\alpha a} d^c_\alpha F \chi_a + \lambda''_{\alpha \beta a} e^c_\alpha u_\beta \chi_a + {\rm H.c.} \, 
\end{equation}
%%%%%%%%%%%%%%%%%%%
Expanding the first term, we get 
\begin{equation}
    -  \mathcal{L}_Y \ \supset \ \lambda_{\alpha\beta 1} \left(\nu_\alpha d_\beta  - \ell_\alpha u_\beta\right) \chi_1^\star +  \ \lambda_{\alpha\beta 2} \left(\nu_\alpha d_\beta  - \ell_\alpha u_\beta\right) \chi_2^\star + {\rm H.c.} \, 
\end{equation}
Within this framework, neutrino mass is induced at two-loop level as shown in Fig. \ref{angelic} which  can be estimated as
\begin{equation}
 %   (M_\nu)_{\alpha \beta} \approx 4 \frac{m_f m_b^2 V_{tb}^2}{(16 \pi^2)^2} \sum_{k, k' = 1}^{N_S} (\lambda_{\alpha3k} \lambda_{3k}^{df}) (\mathcal{I}_{k k'}) (\lambda_{\beta3k'} \lambda_{3k'}^{df})
 M_\nu \ \sim \ \frac{4m_F}{(16\pi^2)^2M^2}(\lambda \lambda' V)(M_d {\cal I} M_d)(\lambda \lambda' V)^T \, ,
\end{equation}
where $V$ is the CKM-matrix, $M_d$ is the diagonal down-quark mass matrix, $M\equiv {\rm max}(m_F,m_{\chi_a})$, and $\mathcal{I}$ is a loop function containing $m_{\chi_a}, m_F$ and $M_d$~\cite{Angel:2013hla}. 

NSI in this model are induced by the singlet LQ $\chi$ and are given by
\begin{tcolorbox}[enhanced,ams align,
  colback=gray!30!white,colframe=white]
%\begin{equation}
 %  \boxed{ 
   \varepsilon_{\alpha \beta} \ = \  \frac{3}{4 \sqrt{2}G_F} \frac{\lambda_{\alpha d a}^{\star} \lambda_{\beta d a} }{ m_{\chi_a}^2} \, .
   %} 
%\end{equation}
\end{tcolorbox}
This is similar to the singlet LQ contribution in Eq.~\eqref{eq:714}. The maximum NSI in this model are the same as in Eq.~\eqref{eq:NSI-singlet}. This is tabulated in Table~\ref{Table_Models}.
%%%%%%%%%%%%%%%%%%%%%%%%%%
%%%%%%%%%%%%%%%%%%%%%%%%%%%%%%%%%%%%%%%%%%%%%%%%%%%%
%%%%%%%%%%%%%%%
%%%%%%%%%%%%%%%%%%%%%%%%%%%%%%%%%%%%%%%%%%%%%%%%%%%%%%%%%%%%%%%%%%
\subsubsection{Model with singlet scalar and vectorlike quark} \label{subsec:CCSVO31}
%%%%%%%%%%%%%%%%%%%%%%%%%%%%%%%%%%%%%%%%%%%%%%%
\begin{figure}
    \centering
    \includegraphics[scale=0.5]{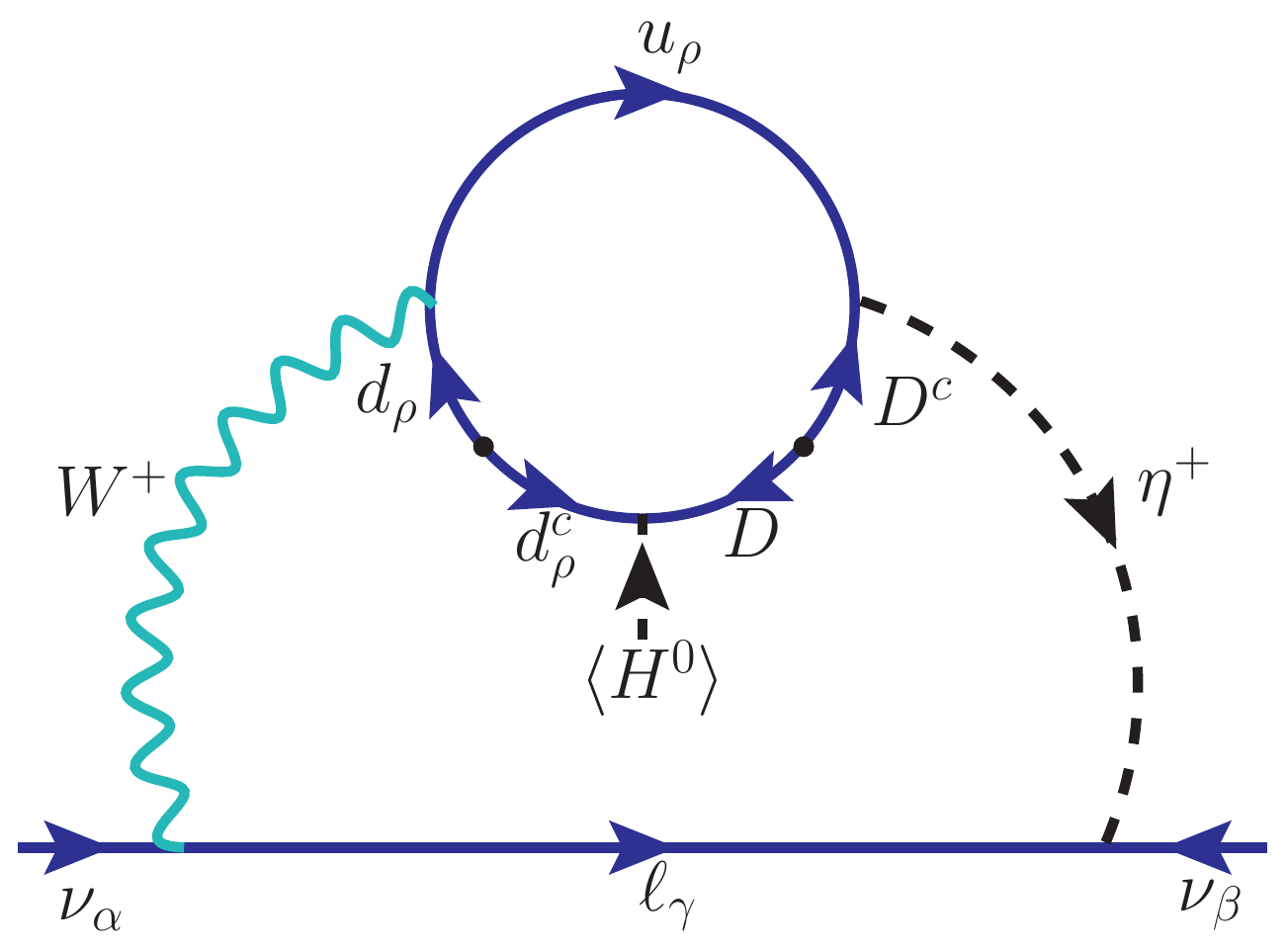}
    \caption{Two-loop neutrino mass generation with singlet scalar and vector-like quark, corresponding to $\mathcal{O}_3^1$ or Table \ref{tab:O3}~\cite{Cai:2014kra}.}
    \label{CCSVO31}
\end{figure}
%%%%%%%%%%%%%%%%%%%%%%%%%%%%%%%%%%%%%%%%%%%%%%%
This model realizes the ${\cal O}_3^1$ operator (cf.~Table~\ref{tab:O3}) by adding a singlet scalar $\eta^+({\bf 1},{\bf 1},1)$ and vectorlike quark $\mathcal{Q}\left({\bf 3},{\bf 2},-\frac{5}{6}\right)=\left(D^{-1/3},X^{-4/3}\right)$. 
Neutrino mass is generated at two-loop level as shown in the Fig.~\ref{CCSVO31}. The relevant Lagrangian for the neutrino mass generation can be read as:
\begin{eqnarray}
   -\mathcal{L}_Y &\ \supset \ & f_{\alpha \beta} L_\alpha L_\beta \eta^+ + f_\alpha^\prime \mathcal{Q}^c Q_\alpha \eta^- + Y_\alpha \mathcal{Q} d_\alpha^c H + {\rm H.c.} \nonumber\\
   & \ = \ & f_{\alpha \beta} (\nu_\alpha \ell_\beta \eta^+ - \ell_\alpha \nu_\beta \eta^+) - f_\alpha^\prime (X^c d_\alpha \eta^- + D^c u_\alpha \eta^-) \nonumber \\
   & & \qquad \qquad + Y_\alpha (D d_\alpha^c H^0 - X d_\alpha^c H^+) +{\rm H.c.} \, 
\end{eqnarray}
The neutrino mass can be estimated as 
\begin{equation}
    M_\nu \ \sim \ \frac{g^2  \sin\varphi }{(16 \pi^2)^2 m_\eta^2}\left(M_\ell^2 f + f^T M_\ell^2 \right) \, ,
    \label{eq:CCSVO31}
\end{equation}
where $\sin\varphi$ represents the mixing between $W^+$ and $\eta^+$. The role of the vectorlike quarks in this model is to achieve such a mixing, which requires lepton number violation.  Note that only the longitudinal component of $W$ mixes with $\eta^+$, which brings in two powers of lepton mass suppression in the neutrino mass estimate -- one from the Yukawa coupling of the longitudinal $W$ and the other from a required chirality-flip inside the loop.  
It is to be noted that Eq.~\eqref{eq:CCSVO31} does not fit the neutrino oscillation data as it has all diagonal entries zero, owing to the anti-symmetric nature of the $f$-couplings. 

Other operators which lead to similar inconsistency with the neutrino oscillation data are  $\mathcal{O}_3^2, \: \mathcal{O}_4^1$ and $\mathcal{O}_4^2$ (cf.~Tables~\ref{tab:O3} and~\ref{tab:O4}). Therefore, we do not discuss the NSI prospects in these models.
%%%%%%%%%%%%%%%%%%%%%%%%%%%%%%%%%%%%%%%%%%%%%%%%%%%%%%%%%%%%%%%%%%%%%%%%%%%%%%%%%%%%%%%%%%%%%%

\subsubsection{Leptoquark model with vectorlike lepton}\label{sec:CCSVO82}
%%%%%%%%%%%%%%%%%%%%%%%%%%%%%%%%%%%%%%%%%%%%%%%
\begin{figure}[t!]
    \centering
    \includegraphics[scale=0.6]{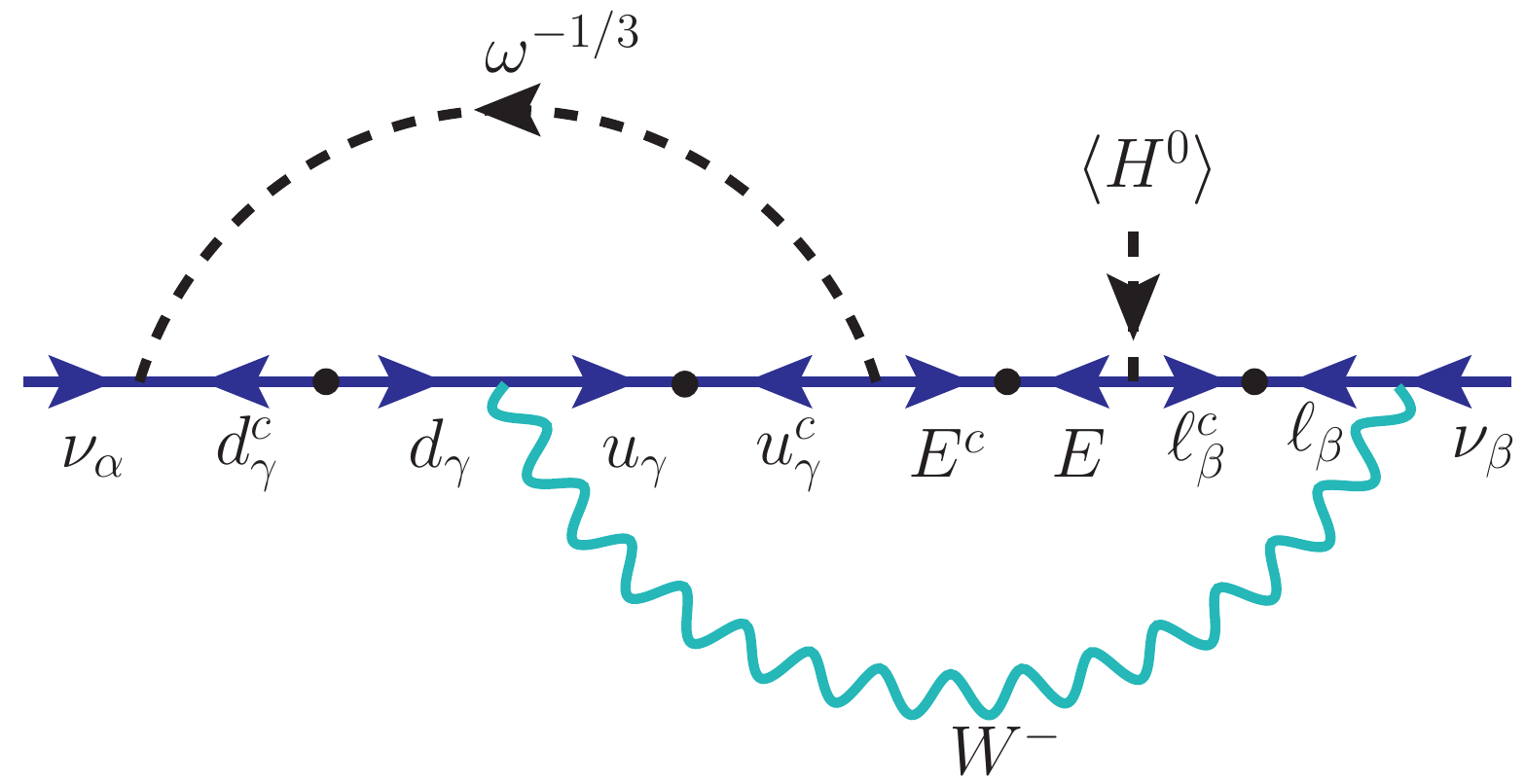}
    \caption{Two-loop neutrino mass generation with $SU(2)_L$-doublet LQ and vector-like lepton, corresponding to $\mathcal{O}_8^2$ of Table \ref{tab:O8}~\cite{Cai:2014kra}.}
    \label{CCSVO82}
\end{figure}
%%%%%%%%%%%%%%%%%%%%%%%%%%%%%%%%%%%%%%%%%%%%%%%
This model is a realization of ${\cal O}_8^2$  in Table~\ref{tab:O8}. This is achieved  by adding an $SU(2)_L$-doublet LQ $\Omega\left({\bf 3},{\bf 2},\frac{1}{6}\right)$ and a vectorlike lepton $\psi\left({\bf 1},{\bf 2},-\frac{1}{2}\right) = (N,\,E)$. 
The Lagrangian responsible for neutrino mass generation can be written as 
\begin{eqnarray}
  -\mathcal{L}_Y & \ \supset \ & m_\psi \psi \psi^c + (\lambda_{\alpha \beta} L_\alpha \Omega d_\beta^c + \lambda_\alpha^\prime \psi^c u_\alpha^c \Omega + \lambda_\alpha^{\prime \prime} \psi \ell_\alpha^c \widetilde{H} + {\rm H.c.}) \nonumber\\
  & \ = \ & m_\psi (NN^c + EE^c) + \big[\lambda_{\alpha \beta}(\nu_\alpha d_\beta^c \omega^{-1/3} - \ell_\alpha d_\beta^c \omega^{2/3}) + \lambda_\alpha^\prime (E^c \omega^{-1/3} + N^c \omega^{2/3}) u_\alpha^c \nonumber \\
  && \qquad \qquad + \lambda_\alpha^{\prime \prime} (N H^- + E \overline{H}^0) \ell_\alpha^c +{\rm H.c.})\big] \, .
\end{eqnarray}
Neutrino masses are generated at two-loop level via diagrams shown in Fig.~\ref{CCSVO82} and can be estimated as:
\begin{equation}
    M_\nu \ \sim \ \frac{g^2}{(16\pi^2)^2} \frac{v}{m_\omega^2 m_E^2} \left(\lambda M_d M_u \lambda^{\prime  \star} M_E \lambda^{\prime \prime \dagger} M_\ell + M_\ell \lambda^{\prime \prime  \star} M_E \lambda^{\prime \dagger} M_u M_d \lambda^T \right) \, ,
\end{equation}
where $M_d, \: M_u, \: M_\ell$ and $M_E$ are the diagonal mass matrices for down quark, up quark, charged leptons and vectorlike leptons, respectively, and $m_E$ is the largest eigenvalue of $M_E$. 
The NSI parameters can be written as
%\begin{equation}
 %    \boxed{ \varepsilon_{\alpha \beta} \ = \ \frac{3}{4 \sqrt{2}G_F} \frac{\lambda_{\alpha d}^{\star} \lambda_{\beta d} }{ m_\omega^2} \, .}
%\end{equation}
%This is exactly the same expression as the doublet contribution 
in Eq.~\eqref{eq:719}, with the maximum values given in Eq.~\eqref{eq:maxNSI_BJ} and also summarized in Table~\ref{Table_Models}.
%%%%%%%%%%%%%%%%%%%%%%%%%%%%%%%%%%%%%%%%%%%%%%%%%%%%%%%%%%%%%%%%%%%%%%%%%%%%%%%%%%%%%%%%%%%%%%
%%%%%%%%%%%%%%%%%%%%%%%%%%%%%%%%%%%%%%%%%%%%%%%%%%%%%%%%%%%%%%%%%%%%%%%%%%%%%%%%%%%%%%%%%%%%%%
%%%%%%%%%%%%%%%%%%%%%%%%%%%%%%%%%%%%%%%%%%%%%%%%%%%%%%%%%%%%%%%%%%%%%%%%%%%%%%%%%%%%%%%%%%%%%%
\subsubsection{Leptoquark model with \texorpdfstring{$SU(2)_L$-}{SU}-doublet vectorlike quark}\label{sec:CCSVO83}
%%%%%%%%%%%%%%%%%%%%%%%%%%%%%%%%%%%%%%%%%%%%%%%
\begin{figure}[t!]
    \centering
    \includegraphics[scale=0.6]{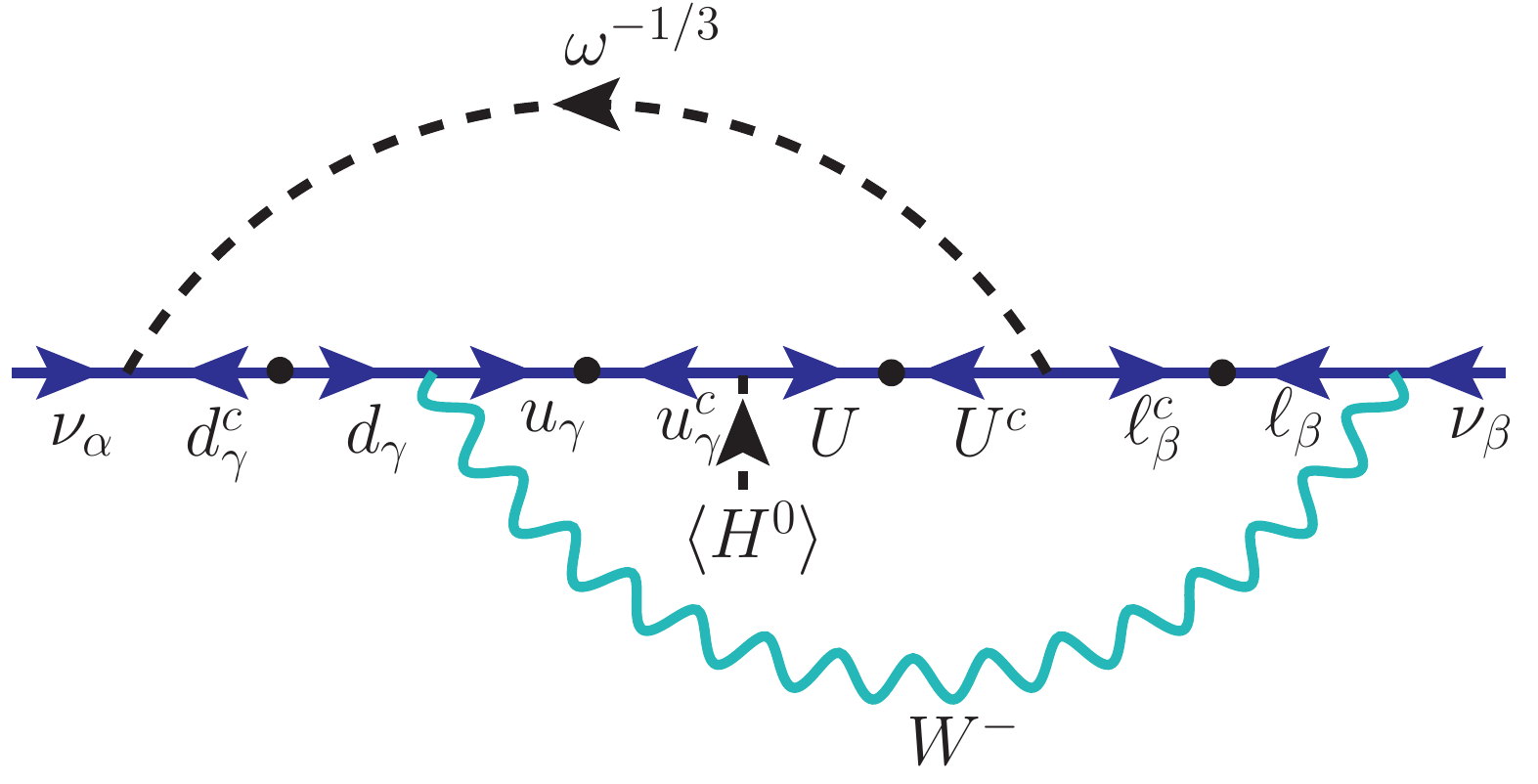}
    \caption{Two-loop neutrino mass generation with $SU(2)_L$-doublet LQ and $SU(2)_L$-doublet vectorlike quark corresponding to $\mathcal{O}_8^3$ or Table \ref{tab:O8}~\cite{Cai:2014kra}.}
    \label{CCSVO83}
\end{figure}
%%%%%%%%%%%%%%%%%%%%%%%%%%%%%%%%%%%%%%%%%%%%%%%
This model realizes the ${\cal O}_8^3$ operator (cf.~Table~\ref{tab:O8}) by adding an $SU(2)_L$-doublet LQ $\Omega\left({\bf 3},{\bf 2},\frac{1}{6}\right)$ and an $SU(2)_L$-doublet vectorlike quark $\xi\left({\bf 3},{\bf 2},\frac{7}{6}\right)=\left(V^{5/3},U^{2/3}\right)$. 
The corresponding Lagrangian for the neutrino mass generation is given by
\begin{eqnarray}
  -\mathcal{L}_Y & \ \supset \ & m_\xi \xi \xi^c + ( \lambda_{\alpha \beta} L_\alpha \Omega d_\beta^c + \lambda_\alpha^\prime \xi u_\alpha^c \widetilde{H} + \lambda_\alpha^{\prime \prime} \xi^c \ell_\alpha^c \Omega + {\rm H.c.} ) \nonumber \\
  & \ = \ & m_\xi (VV^c + UU^c ) + \big[\lambda_{\alpha \beta} (\nu_\alpha \omega^{-1/3} - \ell_\alpha \omega^{2/3}) d_\beta^c - \lambda_\alpha^\prime (VH^- + U \Bar{H}^0) u_\alpha^c \nonumber \\
  &&\qquad \qquad  + \lambda_\alpha^{\prime \prime} (U^c \omega^{-1/3} + V^c \omega^{2/3}) \ell_\alpha^c + {\rm H.c.}\big] \, . 
\end{eqnarray}
Neutrino mass is generated at two-loop level as shown in Fig.~\ref{CCSVO83} and can be estimated as
\begin{equation}
    M_\nu \ \sim \ \frac{g^2}{(16\pi^2)^2}\frac{v}{m_\omega^2 m_U^2}\left (\lambda M_d M_u \lambda^{\prime  \star} M_U \lambda^{\prime \prime \dagger} M_\ell + M_\ell \lambda^{\prime \prime  \star} M_U M_ \lambda^{\prime \dagger} M_u M_d \lambda^T \right) \, .
\end{equation}
where $M_d, \: M_u, \: M_\ell$ and $M_U$ are the diagonal mass matrices for down quark, up quark, charged leptons and vectorlike quarks, respectively, and $m_U$ is the largest eigenvalue of $M_U$. 
The NSI parameters can be written as
%\begin{equation}
 %     \varepsilon_{\alpha \beta} \ = \ \frac{3}{4 \sqrt{2}G_F} \frac{\lambda_{\alpha d}^{\star} \lambda_{\beta d} }{ m_\omega^2} \, .
%\end{equation}
%This is exactly the same expression as the doublet contribution 
in Eq.~\eqref{eq:719}, with the maximum values given in Eq.~\eqref{eq:maxNSI_BJ}.   
%%%%%%%%%%%%%%%%%%%%%%%%%%%%%%%%%%%%%%%%%%%%%%%
%%%%%%%%%%%%%%%%%%%%%%%%%%%%%%%%%%%%%%%%%%%%%%%
%%%%%%%%%%%%%%%%%%%%%%%%%%
%%%%%%%%%%%%%%%%%%%%%%%%%%%%%%%%%%%%%%%%%%%%%%%%%%%%
\subsubsection{A new two-loop leptoquark model} \label{subsec:newcolor}
\begin{figure}[!t]
    \centering
    \includegraphics[scale= 0.6]{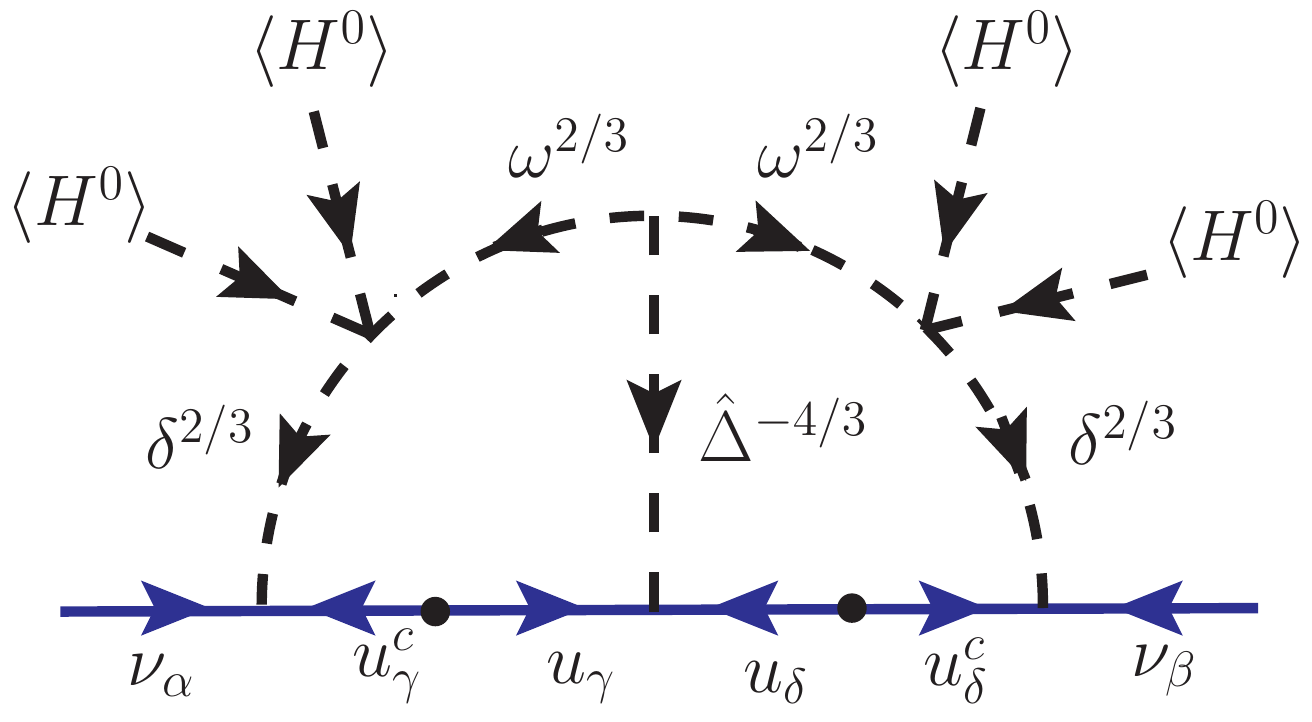}
    \caption{New two-loop scalar LQ model with up-quark loops. The operator induced in the model is ${\cal O}_{d=13}$ in Eq.~\eqref{Onew}. }
    \label{new2loopLQ}
\end{figure}
%%%%%%%%%%%%%%%%%%%%%%%%%%
Here we propose a new two-loop LQ model for neutrino mass, where one can get NSI with up-quark. The effective $\Delta L = 2$ operator is $d=13$, and is given by
\begin{equation}
\mathcal{O}_{d=13}= Q \,L \,u^c\, Q\, L\, u^c\, H\, H\, H\, H~. 
\label{Onew}
\end{equation}
This model utilizes two scalar LQs --  $\delta \left({\bf 3},{\bf 2},\frac{7}{6}\right)=\left(\delta^{5/3},\delta^{2/3}\right)$ and $\Omega \left({\bf 3},{\bf 2}, \frac{\bf 1}{6}\right)=\left(\omega^{2/3},\omega^{-1/3}\right)$, and a scalar DQ $\hat{\Delta} \left({\bf 6}^\star,{\bf 3},-\frac{1}{3}\right)=\left(\hat{\Delta}^{-4/3},\hat{\Delta}^{-1/3},\hat{\Delta}^{2/3}\right)$. The relevant Yukawa Lagrangian for the neutrino mass generation reads as
\begin{align}
   - \mathcal{L}_Y \ \supset \ & f_{\alpha \beta} L_\alpha \delta u_\beta^c  + h_{\alpha \beta} Q_\alpha \hat{\Delta} Q_\beta  + y_{\alpha \beta} Q_\alpha H u_\beta^c  + {\rm H.c.}  \nonumber \\
    \ = \ & f_{\alpha \beta} \left(\nu_\alpha u_\beta^c \delta^{2/3} - \ell_\alpha u_\beta^c \delta^{5/3}\right) + h_{\alpha \beta} \left(u_\alpha u_\beta \hat{\Delta}^{-4/3} + \sqrt{2} u_\alpha d_\beta \hat{\Delta}^{-1/3} + d_\alpha d_\beta \hat{\Delta}^{2/3}\right) \nonumber \\
  &  \qquad \qquad + y_{\alpha \beta} \left(u_\alpha H^0 u_\beta^c - d_\alpha H^+ u_\beta^c\right) + {\rm H.c.} \, 
\end{align}
%%%%%%%%%%%%%%%%%%%%%
The relevant terms in the potential that leads to neutrino mass generation read as 
\begin{align}
    V \ \supset \ \mu \Omega^2 \hat{\Delta} + \lambda \delta^\dagger \Omega H H +{\rm H.c.} \, 
\end{align}
The neutrino mass is induced at two-loop level as shown in  Fig.~\ref{new2loopLQ} and can be  estimated as
\begin{equation}
    M_\nu \ \sim \ \frac{1}{(16 \pi^2)^2} \frac{\mu v^4 \lambda^2}{m_\delta^2 m_\omega^2 m_{\hat{\Delta}}^2} f M_u h M_u f^T \, ,
\end{equation}
where $M_{u}$ is the diagonal up-type quark mass matrix. Note that $M_\nu$ is a symmetric matrix, as it should be, since $h=h^T$.  

After integrating out the heavy scalars, NSI induced in this model can be written as
\begin{tcolorbox}[enhanced,ams align,
  colback=gray!30!white,colframe=white]%\begin{equation}
 %   \boxed{
    \varepsilon_{\alpha \beta} \ = \ \frac{3}{4 \sqrt{2}G_F} \frac{f_{\alpha u}^{\star} f_{\beta u} }{ m_\delta^2} \, .
    %}
%\end{equation}
\end{tcolorbox}
This is same as the extended one-loop LQ model prediction in Eq.~\eqref{eq:uploopNSI} for $\lambda\gg \lambda'$ with the exception that $\varepsilon_{\mu\mu}$ and $\varepsilon_{\tau\tau}$ are now constrained by IceCube. The maximum allowed values are given in Eq.~\eqref{eq:maxNSI_BJ}. This is also summarized in  Table~\ref{Table_Models}.
%%%%%%%%%%%%%%%%%%%%%%%%%%
%%%%%%%%%%%%%%%%%%%%%%%%%%
%%%%%%%%%%%%%%%%%%%%%%%%%%
\subsection{Three-loop models} \label{sec:3loop}
    
\subsubsection{KNT Model} \label{sec:KNT}
\begin{figure}[t!]
    \centering
    \includegraphics[scale=0.5]{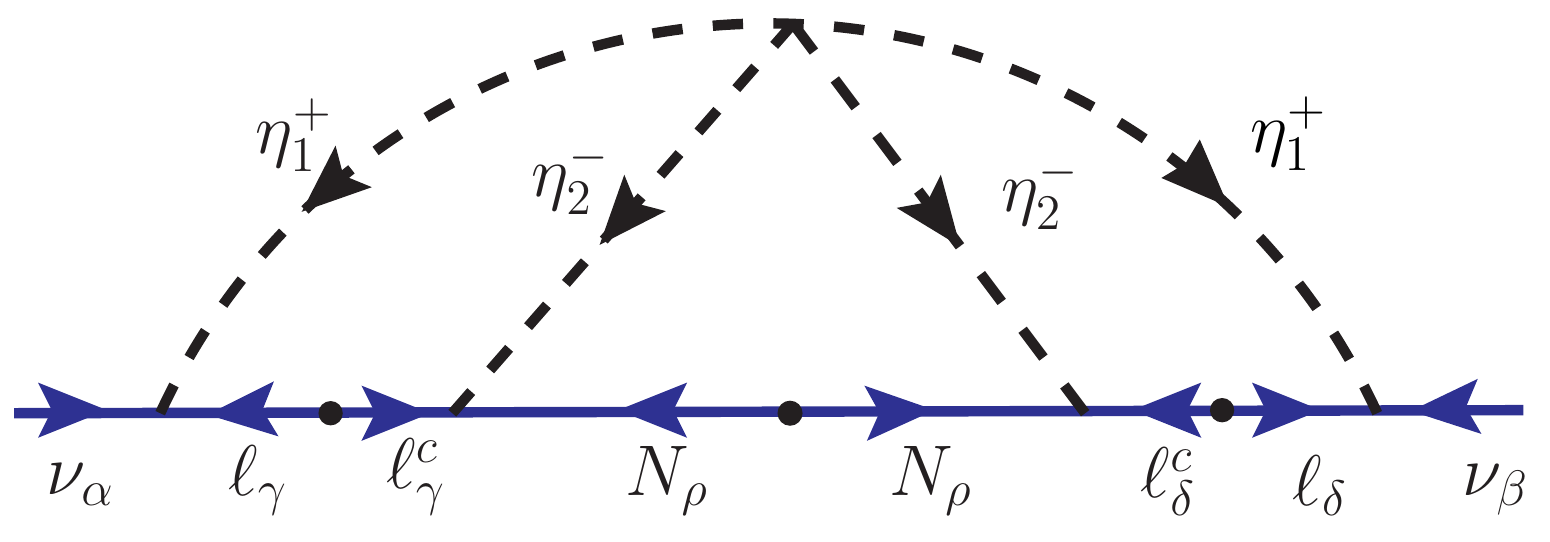}
    \caption{Three-loop neutrino mass generation in the KNT model~\cite{Krauss:2002px}. The model induces operator $\mathcal{O}_9$ of Eq. (\ref{O9}).}
    \label{KNT}
\end{figure}

The Krauss-Nasri-Trodden (KNT) model~\cite{Krauss:2002px} generates the $d=9$ operator $\mathcal{O}_9$ of Eq. (\ref{O9}).  SM-singlet fermions $N_\alpha({\bf 1},{\bf 1},0)$ and two SM-singlet scalars $\eta_1^+$ and $\eta_2^+$ with SM charges $({\bf 1},{\bf 1},1)$ are introduced. The relevant Yukawa Lagrangian is written as
\begin{equation}
  -   \mathcal{L}_Y \ \supset \ f_{\alpha \beta} \, {L}_{\alpha} L_\beta \eta_1^+ + f'_{\alpha\beta} \, \ell_\alpha^c N_\beta \eta_2^- + \frac{1}{2} (M_{N})_{\alpha\beta} N_\alpha N_\beta \, .
     \label{lagKNT}
\end{equation}
Tree level mass is prevented by imposing a $Z_2$ symmetry under which the fields $\eta_2^+$ and $N$ are odd, while the other fields are even. The Majorana mass term for $N$ as shown in Eq.~\eqref{lagKNT} explicitly breaks lepton number. 
Neutrino masses are generated at three-loop as shown in Fig.~\ref{KNT} by the Lagrangian~\eqref{lagKNT}, together with the quartic term in the potential
\begin{align}
    V \ \supset \ \lambda_s (\eta_1^+\eta_2^-)^2 \, .
\end{align}
The estimated neutrino mass matrix reads as
\begin{equation}
M_\nu \ \simeq \ -\frac{\lambda_s}{(16 \pi^2)^3}\frac{1}{M^2}fM_\ell f^{\prime\dag} M_N f^{\prime\star} M_\ell f^T {\cal I} \, ,
%    (M_\nu)_{\alpha \beta} \approx \frac{- \mu}{(16 \pi^2)^3} \frac{1}{M^2} f_{\alpha a} M_{la} g_{a i}^\dagger M_{N_i} g_{ib}^{\star} m_{l b} f_{b \beta } F(r_{N_i}, r_{\eta_1}, r_{\eta_2})
\end{equation}
where $M_\ell$ is the diagonal charged lepton mass matrix, $M_N={\rm diag}(m_{N_\alpha})$ is the diagonal Majorana mass matrix for $N_\alpha$ fermions, $M \equiv {\rm max}(m_{N_\alpha},m_{\eta_1},m_{\eta_2})$, and ${\cal I}$ is a three-loop function obtained in general by numerical integration~\cite{Cheung:2016ypw}. 

NSI in the KNT model arise from singly-charged scalar $\eta_1^+$ that has the same structure as in the Zee-Babu model (cf.~Sec.~\ref{sec:zeebabu}) and are given by Eq.~\eqref{eq:nsio21}. 
%\begin{equation}
 %\boxed{\varepsilon_{\alpha \beta} \ = \ \frac{1}{\sqrt 2 G_F}\frac{f_{ e \alpha}^{\star}f_{ e \beta }  }{m_{\eta_1}^2}  \, .}
%\end{equation}
The maximum NSI one can get in this model are same as in Eq.~\eqref{eq:maxZB} and also summarized in Table~\ref{Table_Models}.

%%%%%%%%%%%%%%%%%%%%%%%%%%%%%%%%%%%%%%%%%%
%%%%%%%%%%%%%%%%%%%%%%%%%%%%%%%%%%%%%%%%%%
\subsubsection{AKS model} \label{sec:AKS}
\begin{figure}[!t]
    \centering
    \includegraphics[scale=0.6]{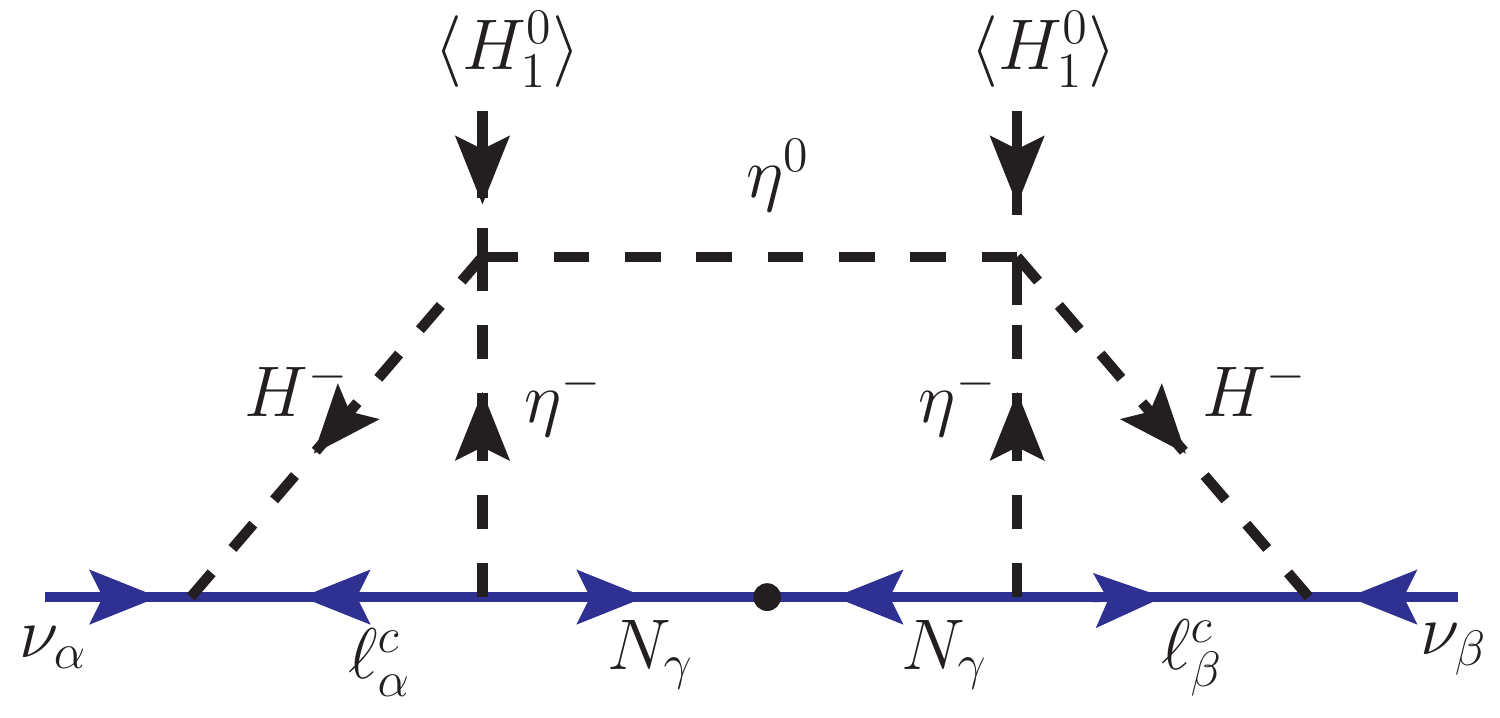}
    \caption{Three-loop neutrino mass generation in the AKS model~\cite{Aoki:2008av}. The model induces operator ${\cal O}'_3$ of Eq.~\eqref{eq:Op2}.}
    \label{AKS}
\end{figure}

In the Aoki-Kanemura-Seto (AKS) model~\cite{Aoki:2008av} an effective $\Delta L = 2$ operator of dimension 11 is induced:
\begin{equation}
{\cal O}'_3 \  = \ LLHH e^c e^c \overline{e^c} \,\overline{e^c}~.  
\label{eq:Op2}
\end{equation}
Note that there is a chiral suppression in this model unlike generic operators of type ${\cal O}_1'$ given in Eq.~\eqref{Op}.  
In addition to the SM fields, the following particles are added: an isospin doublet scalar $\Phi_2\left({\bf 1},{\bf 2},\frac{1}{2}\right)$, a singly-charged scalar singlet $\eta^+({\bf 1},{\bf 1},1)$, a real scalar singlet $\eta^0({\bf 1},{\bf 1},0)$, and two isospin-singlet right-handed neutrinos $N_{\alpha}({\bf 1},{\bf 1},0)$ (with $\alpha=1,2$). The relevant Yukawa Lagrangian for the neutrino mass generation reads as
\begin{equation}
  -  \mathcal{L}_Y \ \supset \  y_{\alpha\beta a} \widetilde{\Phi}_a L_\alpha \ell_\beta^c + h_{\alpha\beta} \ell_{\alpha}^c N_{\beta} \eta^- + \frac{1}{2} (M_N)_{\alpha\beta} N_{\alpha}  N_{\beta } +{\rm H.c.} \, ,
    \label{eq:AKS}
\end{equation}
where $\Phi_1\left({\bf 1},{\bf 2},\frac{1}{2}\right)$ is the SM Higgs doublet. 
%%%%%%%%%%%%%%%%%%%%%
Tree-level neutrino mass is forbidden by imposing a $Z_2$ symmetry under which $\eta^\pm$, $\eta^0$ and $N_{\alpha R}$ are odd, while the remaining fields are even. Neutrino masses are generated at three-loop, as shown in Fig.~\ref{AKS}, by combining Eq.~\eqref{eq:AKS} with the quartic term in the potential
\begin{align}
    V \ \supset \ \kappa \epsilon_{ab} (\Phi_a^c)^\dagger \Phi_b \eta^- \eta^0 +{\rm H.c.} \, 
\end{align}
In Fig.~\ref{AKS} $H^\pm$ are the physical charged scalars from a linear combination 
of $\Phi_1$ and $\Phi_2$. The neutrino mass matrix reads as follows: 
 %%%%%%%%%%%%%%%%%%%%%% 
\begin{equation}
M_\nu \ \simeq \ \frac{1}{(16 \pi^2)^3} \frac{\left(-m_{N} v^2\right)}{m_{N}^2 - m_{\eta^0}^2}4\kappa^2 \tan^2\beta (yh)(yh)^T{\cal I} \, ,
%    (M_\nu)_{\alpha \beta} = \frac{1}{(16 \pi^2)^3} \frac{-m_{N_R} v^2}{m_{N_R}^2 - m_{\eta^0}^2} \sum_{i = 1}^2 4 \kappa^2 \tan^2 \beta (y_{e\alpha} h_\alpha^i ) ( y_{e \beta} h_\beta^i)\, \mathcal{I}(m_{H^{\pm}}, m_{\eta^{\pm}}, m_{N_R}, m_\eta^0)
\end{equation}
where $\tan \beta\equiv \langle \Phi_2^0 \rangle/\langle \Phi_1^0 \rangle$ and $\mathcal{I}$ is a dimensionless three-loop integral function that depends on the masses present inside the loop. 

NSI in this model are induced by the charged scalar $H^-$. After integrating out the heavy scalars, the NSI expression can be written as
\begin{tcolorbox}[enhanced,ams align,
  colback=gray!30!white,colframe=white]
%\begin{equation}
 %    \boxed{
     \varepsilon_{\alpha \beta} = \frac{1}{4 \sqrt{2}G_F} \frac{y_{ e \alpha a}^{\star}y_{ e \beta a}  }{m_{H^-}^2} \, .
     %}
%\end{equation}
\end{tcolorbox}
This is similar to the heavy charged scalar contribution in Eq.~\eqref{eq:nsi2}. However, since the same Yukawa couplings $y_{e\alpha a}$ contribute to the electron mass in Eq.~\eqref{eq:AKS}, we expect 
\begin{align}
    \varepsilon_{\alpha \beta} \ \propto \ y_e^2 \tan^2\beta \ \sim 
    \ {\cal O}\left(10^{-10}\right) \, ,
\end{align}
where $y_e$ is the electron Yukawa coupling in the SM. Thus, the maximum NSI in this model are of order of ${\cal O}\left(10^{-10}\right)$, as summarized in Table~\ref{Table_Models}.
%%%%%%%%%%%%%%%%%%%%%%%%%%%%%%
%%%%%%%%%%%%%%%%%%%%%%%%%%%%%%
\subsubsection{Cocktail Model} \label{sec:cocktail}
\begin{figure}[!t]
    \centering
    \includegraphics[scale=0.6]{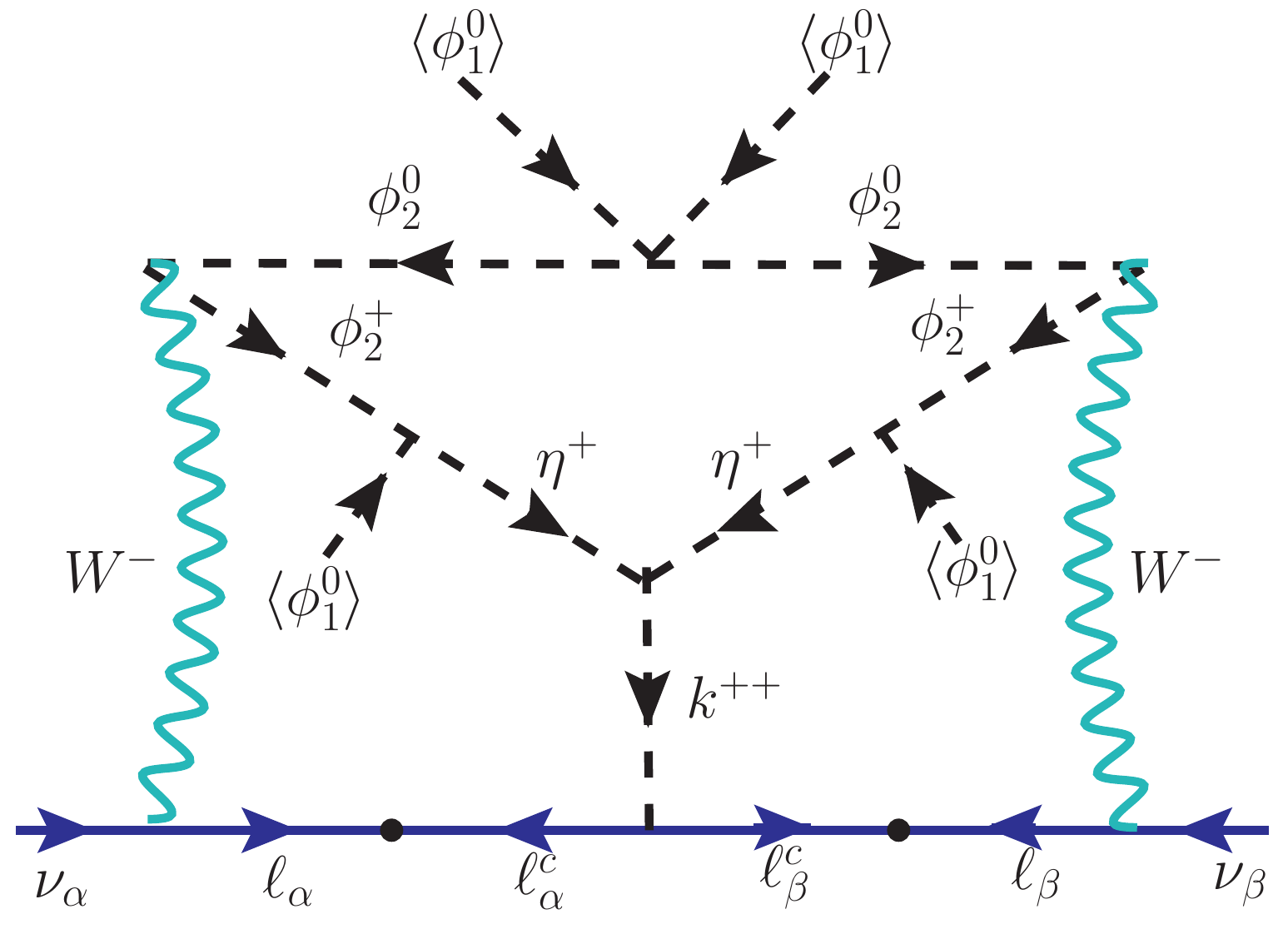}
    \caption{Three-loop neutrino mass generation in the cocktail model~\cite{Gustafsson:2012vj}. The effective operator induced is $\mathcal{O}_{d=15}$ of Eq. (\ref{O9}).}
    \label{cocktail}
\end{figure}

This model~\cite{Gustafsson:2012vj} induces operator $\mathcal{O}_{d=15}$ at the three-loop level:
\begin{equation}
    \mathcal{O}_{d=15} \ = \ LLHH (\bar{\Psi}\Psi) ( \bar{\Psi}\Psi) (H^\dagger H)^2
\end{equation}
with $\Psi = L$ or $e^c$. The model includes two $SU(2)_L$-singlet scalars $\eta^+({\bf 1},{\bf 1},1)$ and $k^{++}({\bf 1},{\bf 1},2)$, and a second scalar doublet $\Phi_2\left({\bf 1},{\bf 2},\frac{1}{2}\right)$, in addition to the SM Higgs doublet $\Phi_1\left({\bf 1},{\bf 2},\frac{1}{2}\right)$. The fields $\eta^+$ and $\Phi_2$ are odd under a $Z_2$ symmetry, while $k^{++}$ and all SM fields are even. With this particle content, the relevant term in the Lagrangian reads as 
\begin{eqnarray}
  -  \mathcal{L}_Y \ \supset \  y_{\alpha \beta} \widetilde{\Phi}_1 L_\alpha  \ell_\beta^c+Y_{\alpha \beta} \ell_{\alpha}^c \ell_{\beta} k^{++} + {\rm H.c.} \, ,
\end{eqnarray}
which breaks lepton number when combined with the following cubic and quartic terms in the potential:
\begin{align}
V \ \supset \ \frac{\lambda}{2} (\Phi_1^\dagger \Phi_2)^2 + \kappa_1 \Phi_2^T i \tau_2 \Phi_1 \eta^- + \kappa_2 k^{++} \eta^- \eta^-  + \xi \Phi_2^T i \tau_2 \Phi_1 \eta^+ k^{--} +{\rm H.c.} \, 
\end{align}
The $\Phi_2$ field is inert and does not get a VEV. After electroweak symmetry breaking, it can be written as 
\begin{align}
    \Phi_2 \ = \ \begin{pmatrix} \phi_2^+ \\ H+iA
    \end{pmatrix} \, .
\end{align}
For $\kappa_1\neq 0$, the singly-charged state $\phi_2^+$ mixes with $\eta^+$ (with mixing angle $\beta$), giving rise to two singly-charged scalar mass eigenstates: 
\begin{align}
    H_1^+ \ & = \ c_\beta \phi_2^+ + s_\beta \eta^+ \, , \nonumber \\
    H_2^+ \ & = \ -s_\beta \phi_2^++c_\beta \eta^+ \, ,
\end{align}
where $s_\beta\equiv \sin\beta$ and $c_\beta\equiv \cos\beta$. 

The neutrino mass matrix is obtained from the three-loop diagram as shown in Fig.~\ref{cocktail} and reads as~\cite{Gustafsson:2012vj} 
\begin{equation}
M_\nu \ \sim \ \frac{g^2}{(16 \pi^2)^3} M_\ell (Y+Y^T) M_\ell \, ,
\end{equation}
where $M_\ell$ stands for the diagonal charged lepton mass matrix.
%\begin{equation}
%    M_\nu \ = \ \frac{ s_{2\beta}}{(16 \pi^2)^3} M_\ell Y M_\ell %\frac{m_W^4}{v^8}\frac{\Delta m_+^2 \Delta m_0^2}{m_{k^{++}}m_0^{1/2} %m_+^{1/2}}\left[\frac{5\:\Delta m_+^2}{m_{k^{++}}^{1/2}m_+^{3/2}}(\kappa_2 s_{2\beta}+\xi v %c_{2\beta})\mathcal{I}_1 -29\:\xi v\mathcal{I}_2\right]
%\end{equation}
%where $M_\ell$ is the diagonal charged lepton mass matrix, $\Delta %m_0^2=m^2_{A^0}-m^2_{H^0}$, $\Delta m_+^2=m^2_{H_2^+}-m^2_{H_1^+}$, %$m_0^2=m^2_{A^0}+m^2_{H^0}$, $m_+^2=m^2_{H_2^+}+m^2_{H_1^+}$, and $\mathcal{I}_{1,2}$ are %dimensionless $\mathcal{O}$(1) entities that come from three-loop integrals. 

As for the NSI, since both $\Phi_2$ and $\eta^+$ are odd under $Z_2$ and the SM fields are even, there is no tree-level NSI in this model.  Note that neutrino mass generation utilizes the $W$ boson couplings, thus the neutrino matter effects in this model are the same as in the SM.

%%%%%%%%%%%%%%%%%%%%%%%%%%%%%%%%%%%%%%%%%%
%%%%%%%%%%%%%%%%%%%%%%%%%%%%%%
%%%%%%%%%%%%%%%%%%%%%%%%%%%%%%
\subsubsection{Leptoquark variant of the KNT model}  \label{sec:3loopLQ}
\begin{figure}[t!]
    \centering
    \includegraphics[scale=0.5]{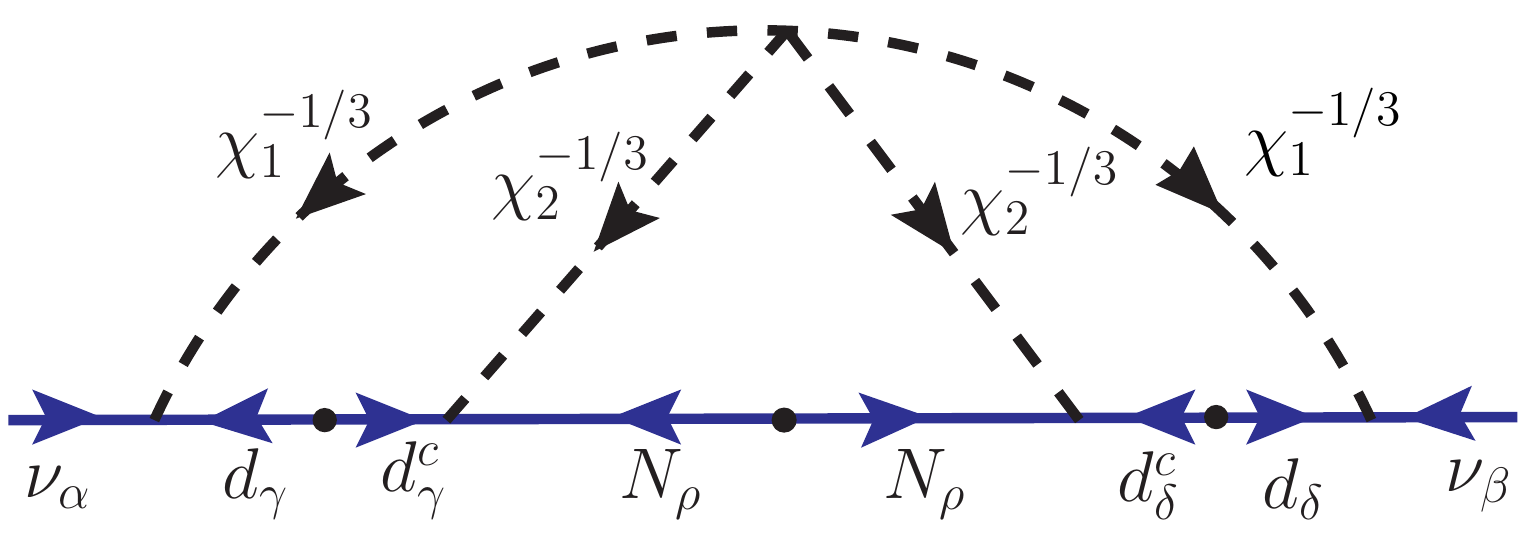}
    \caption{Three-loop neutrino mass generation in the LQ variant of the KNT model, which induces operator $\mathcal{O}_{11}$~\cite{Nomura:2016ezz}. }
    \label{knt_LQ}
\end{figure}
%%%%%%%%%%%%%%%%%%%%%%%%%%

One can replace the charged leptons in the KNT model (cf.~Sec.~\ref{sec:KNT}) by quarks, and the charged scalars by LQs. The effective operator induced in this model remains as $\mathcal{O}_{11}$ or Eq.~(\ref{eq:O11}). To achieve this, two isospin-singlet scalar LQs $\chi_a^{-1/3}\left({\bf 3},{\bf 1},-\frac{1}{3}\right)$  (with $a=1,2$) and at least two SM-singlet right-handed neutrinos $N_{\alpha}({\bf 1},{\bf 1},0)$ (with $\alpha=1,2$) are supplemented to the SM fields. A $Z_2$ symmetry is invoked under which $\chi_2^{-1/3}$ and $N$ are odd, while the rest of the fields are even. The relevant Yukawa Lagrangian is as follows:
    \begin{equation}
      - \mathcal{L}_Y \ \supset \ \lambda_{\alpha \beta} L_\alpha^i Q_\beta^j \chi_1^{\star 1/3} \epsilon_{ij}+ \lambda_{\alpha \beta}' d_\alpha^{c} N_\beta \chi_2^{\star 1/3} + \frac{1}{2} (M_{N})_{\alpha\beta} N_\alpha  N_\beta + {\rm H.c.} \, 
    \end{equation}
Here the first term expands to give
$\lambda_{\alpha\beta} \left(\nu_\alpha d_\beta  - \ell_\alpha u_\beta\right)\chi_1^{\star 1/3}$.
These interactions, along with the quartic term in the potential 
\begin{align}
    V \ \supset \ \lambda_0 \left(\chi_1^{\star 1/3}\chi_2^{-1/3}\right)^2 \, ,
\end{align}
generate neutrino masses at three-loop level, as shown in Fig.~\ref{knt_LQ}. The 
neutrino mass matrix reads as
\begin{equation}
  M_\nu \ \sim \ \frac{15 \lambda_0}{(16 \pi^2)^3 m_{\chi_1}^2 } \lambda M_d \lambda^{'\star} M_N \lambda^{'\dagger} M_d \lambda^T \, \cal{I} \, ,
\end{equation}
where the factor 15 comes from total color-degrees of freedom, $M_d$ and $M_N$ are the diagonal down-type quark and right-handed neutrino mass matrices, respectively, and $\cal{I}$ is a dimensionless three-loop integral that depends on the ratio of the masses of particles inside the loop~\cite{Nomura:2016ezz}.

NSI in this model arise from the $\chi_1^{-1/3}$ interactions with neutrinos and down-quarks. The expression for NSI parameters is given as in  
%\begin{equation}
%    \varepsilon_{\alpha \beta} \ = \ \frac{3}{4 \sqrt{2}G_F} \frac{\lambda_{\alpha d}^{\star} \lambda_{\beta d}}{ m_{\chi_1}^2} \, , 
%\end{equation}
%which is the same as the singlet contribution in 
Eq.~\eqref{eq:714}, with the replacement $m_\chi\to m_{\chi_1}$.  
The maximum NSI for this model are the same as those given in Eq.~\eqref{eq:NSI-singlet} and are summarized in Table~\ref{Table_Models}.
%%%%%%%%%%%%%%%%%%%%%%%%%%%%%%%%%%%%%%%%%%
%%%%%%%%%%%%%%%%%%%%%%%%%%%%%%
%%%%%%%%%%%%%%%%%%%%%%%%%%%%%%
\subsubsection{\texorpdfstring{$SU(2)_L$-}{SUULL}-singlet three-loop model}  \label{sec:singlet3loop}
\begin{figure}[t!]
    \centering
    \includegraphics[scale=0.5]{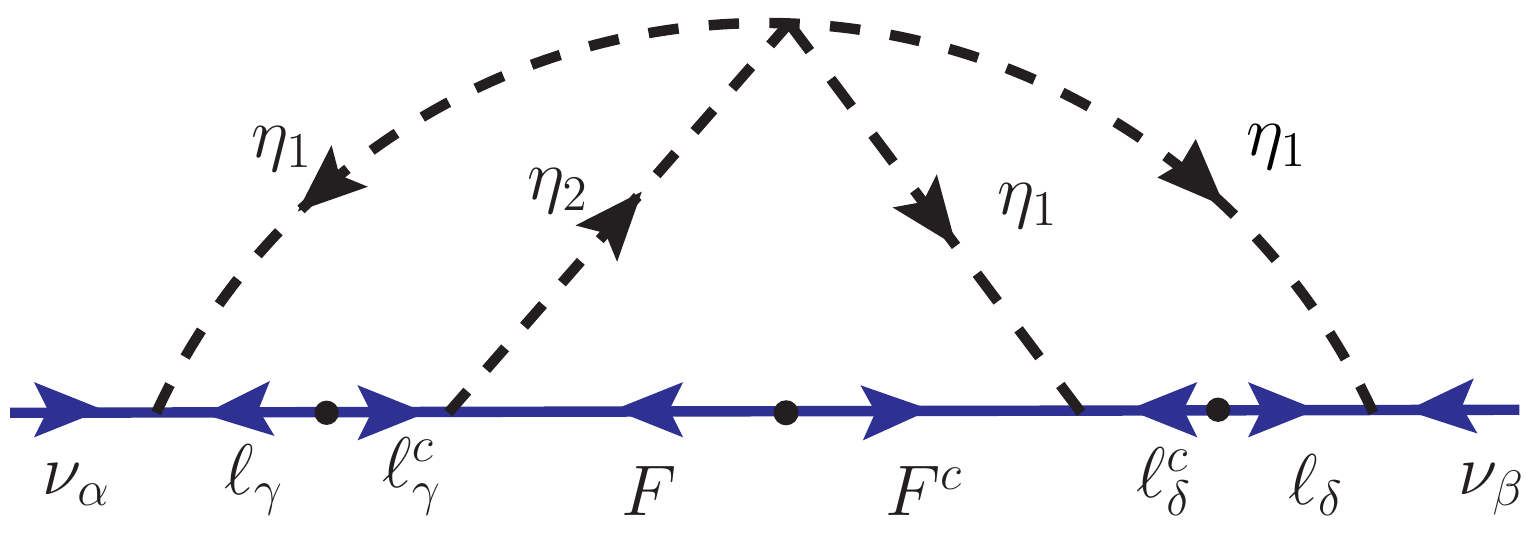}
    \caption{Three-loop neutrino mass generation with $SU(2)_L$-singlet scalar and fermion fields~\cite{Cepedello:2018rfh}, which induces operator $\mathcal{O}_9$. }
    \label{singlet3loop}
\end{figure}
%%%%%%%%%%%%%%%%%%%%%%%%%%%%%%
This model~\cite{Cepedello:2018rfh} introduces two $SU(2)_L$-singlet scalars $\eta_1 ({\bf 1},{\bf 1},1)$ and $\eta_2 ({\bf 1},{\bf 1},3)$, and a singlet fermion $F ({\bf 1},{\bf 1},2)$, in addition to the SM fields. The effective operator induced in this model is $\mathcal{O}_9$ in Eq.~(\ref{O9}). The relevant Lagrangian term for the neutrino mass generation can be read as:
\begin{equation}
    -\mathcal{L}_Y  \ \supset \ M_F FF^c + (f_{\alpha \beta} \eta_1 L_\alpha L_\beta + f_\alpha^\prime \ell_\alpha^c F \eta_2^\star + f_\alpha^{\prime \prime} \ell_\alpha^c F^c \eta_1  + {\rm H.c.} ) \, ,
    \label{eq:783}
\end{equation}
With the potential term 
\begin{align}
    V \ \supset \ \lambda \eta_1 \eta_1 \eta_1 \eta_2^\star + {\rm H.c.} \, ,
\end{align}
the Lagrangian~\eqref{eq:783} generates the neutrino mass at three-loop level, as shown in Fig.~\ref{singlet3loop}. The neutrino mass matrix can be written as
\begin{equation}
    M_\nu \ \simeq \ \frac{f M_\ell f^{\prime\dagger} M_F f^{\prime \prime \star} M_{\ell^\prime} f^T \lambda}{(16 \pi^2)^3 M^2} \, ,
\end{equation}
where $M_\ell$ is the diagonal charged lepton mass matrix and $M \equiv \text{max}(m_{F}, m_{\eta_1},  m_{\eta_2})$. NSI in this model arise from singly-charged $\eta_1$ that has the same structure as in the Zee-Babu (cf.~Sec.~\ref{sec:zeebabu}) and KNT (cf.~Sec.~\ref{sec:KNT}) models  and  and are given 
by Eq.~\eqref{eq:nsio21}. 
%\begin{equation}
% \varepsilon_{\alpha \beta} \ = \ \frac{1}{\sqrt 2 G_F}\frac{f_{ e \alpha}^{\star}f_{ e \beta }  }{m_{\eta_1}^2}  \, .
%\end{equation}
The maximum NSI one can get in this model are same as in Eq.~\eqref{eq:maxZB} and also summarized in Table~\ref{Table_Models}. Other three-loop models of this type  discussed in Ref.~\cite{Cepedello:2018rfh} will have similar NSI predictions.
%%%%%%%%%%%%%%%%%%%%%%%%%%%%%%%%%%%%%%%%%%%%%%%%%%%%%%%%%%%%%%%%%%%%%
%%%%%%%%%%%%%%%%%%%%%%%%%%%%%%%%%%%%%%%%%%%%%%%%%%%%%%%%%%%%%%%%%%%%%%%%%

\subsection{Four- and higher-loop models}\label{sec:four-loop}

As noted in the introduction, it is very unlikely that neutrino masses and mixing of the right order can be induced in type-I radiative models at four or higher loops.  The magnitude of $m_\nu$ in such models would be much smaller than needed to explain neutrino oscillation data, provided that the loop diagrams have chiral suppression proportional to  a SM fermion mass.  We illustrate below the difficulties with higher loop models with a four loop model presented in Ref. \cite{deGouvea:2019xzm}.

In Ref.  \cite{deGouvea:2019xzm} an effective $d=9$ operator involving only $SU(2)_L$-singlet fermions of the SM was studied.  The operator has the form 
\begin{equation}
{\cal O}_{\rm s} \ = \ \ell^c \ell^c u^c u^c \overline{d^c}\, \overline{d^c}~.
\end{equation}
Various UV completions are possible to induce this operator, with differing fermion contractions. All these models will induce light neutrino mass only at the four-loop level, since each fermion in ${\cal O}_{\rm s}$ has to be annihilated.  A rough (and optimistic) estimate of the four-loop induced neutrino mass is \cite{deGouvea:2019xzm}
\begin{equation}
    m_\nu \ \sim \ \frac{(y_t y_b v)^2}{(16 \pi^2)^4 \Lambda}
\end{equation}
where $\Lambda$ is the UV cut-off scale.  If the other Yukawa couplings involved are all of order one, $\Lambda = (100~ {\rm MeV} - 1 ~{\rm GeV})$ is needed to generate $m_\nu \sim 0.05$ eV.  However, such a low value of $\Lambda$ will be inconsistent with experimental data on search for new particles, since the mediators needed to induce ${\cal O}_{\rm s}$ are either colored or electrically charged, with lower limits of order TeV on their masses from collider searches.

Models with such higher dimensional operators are nevertheless very interesting, as they can lead to lepton flavor and lepton number violating processes, without being constrained by neutrino masses, as emphasized in Ref. \cite{deGouvea:2019xzm}.  For example, neutrinoless double beta decay may occur at an observable level purely from ${\cal O}_{\rm s}$, which would be unrelated to the neutrino mass.

\section{Type II radiative models} \label{sec:type2}
%%%%%%%%%%%%%%%%%%%%%%%%%%%%
 \begin{figure}[!t]
     \centering
     \includegraphics[scale=0.6]{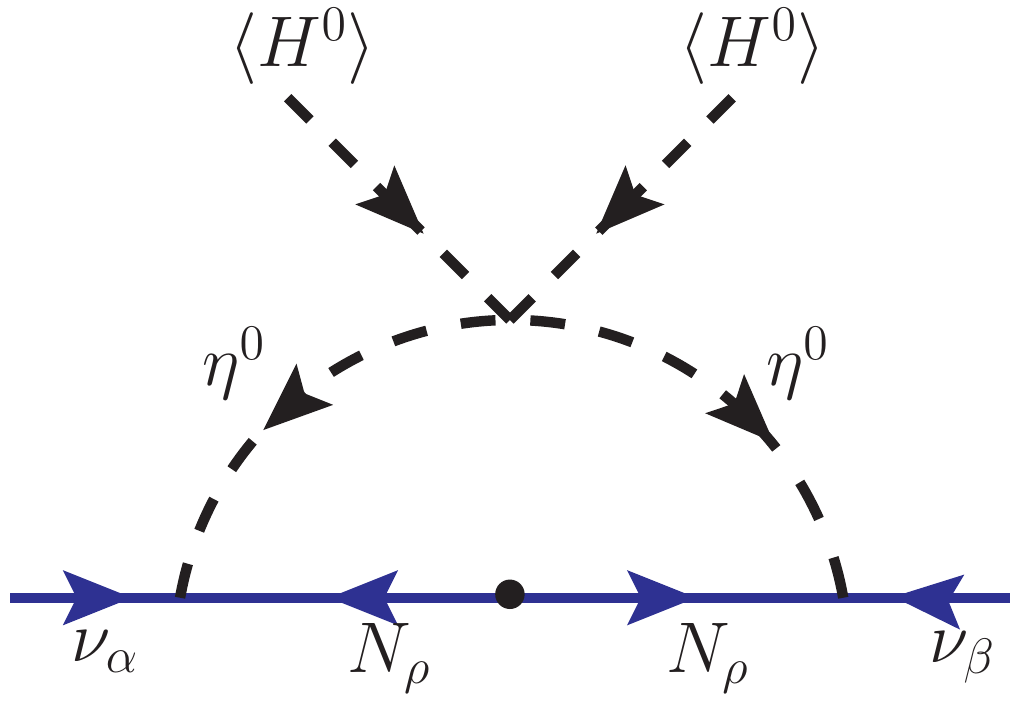}
     \caption{Neutrino mass generation at one-loop in the scotogenic model~\cite{Ma:2006km}.}
     \label{scotogenic}
 \end{figure}
%%%%%%%%%%%%%%%%%%%%%%%%%%%%%%%%%%%%%%
As discussed in the introduction (cf.~Sec.~\ref{sec:intro1}), type-II radiative neutrino mass models in our nomenclature contain no SM particle inside the loop diagrams generating $m_\nu$, and therefore, do not generally contribute to tree-level NSI, although small loop-level NSI effects are possible~\cite{Bischer:2018zbd}. To illustrate this point, let us take the scotogenic model~\cite{Ma:2006km} as a prototypical example. The new particles introduced in this model are SM-singlet fermions $N_\alpha({\bf 1},{\bf 1},0)$ (with $\alpha=1,2,3$) and an $SU(2)_L$ doublet scalar $\eta\left({\bf 1},{\bf 2},\frac{1}{2}\right):(\eta^{+}, \eta^{0})$. A $Z_2$ symmetry is imposed under which the new fields $N_\alpha$ and $\eta$ are odd, while all the SM fields are even. The new Yukawa interactions in this model are given by 
\begin{align}
    -{\cal L}_Y 
    %\ & = \  %h_{\alpha \beta} \overline{L_\alpha^c}\eta N_\beta+\frac{1}{2}(M_N)_{\alpha\beta}N_\alpha N_\beta+{\rm H.c.} \nonumber \\
    \ & \supset \ h_{\alpha\beta}(\nu_\alpha \eta^0-\ell_\alpha \eta^+)N_\beta+\frac{1}{2}(M_N)_{\alpha\beta}N_\alpha  N_\beta+{\rm H.c.} \, 
    \label{eq:lagscot}
\end{align}
Together with the scalar quartic term 
\begin{align}
    V \ \supset \ \frac{\lambda_5 }{2}(\Phi^\dag \eta)^2+{\rm H.c.} \, ,
\end{align}
where $\Phi$ is the SM Higgs doublet, the Lagrangian~\eqref{eq:lagscot} gives rise to neutrino mass at one-loop, as shown in Fig.~\ref{scotogenic}. Since this diagram does not contain any SM fields inside the loop, it cannot be cut to generate an effective higher-dimensional operator of the SM. Therefore, we label it as a type-II radiative model. The neutrino mass in this model is given by 
\begin{align}
    M_\nu \ = \ \frac{\lambda_5 v^2}{8\pi^2}\frac{hM_N h^T}{m_0^2-M_N^2}\left[1-\frac{M_N^2}{m_0^2-M_N^2}\log\left(\frac{m_0^2}{M_N^2}\right) \right] \, ,
    \label{eq:Ma}
\end{align}
where we have assumed $M_N$ to be diagonal, and $m_0^2$ is the average squared mass of the real and imaginary parts of $\eta^0$. It is clear from Eq.~\eqref{eq:Ma} that the neutrino mass is not chirally suppressed by any SM particle mass. 

%%%%%%%%%%%%%%%%%%%%%%%%%%%%%%%%%%%%%%
\begin{figure}[!t]
    \centering
    \includegraphics[scale=0.6]{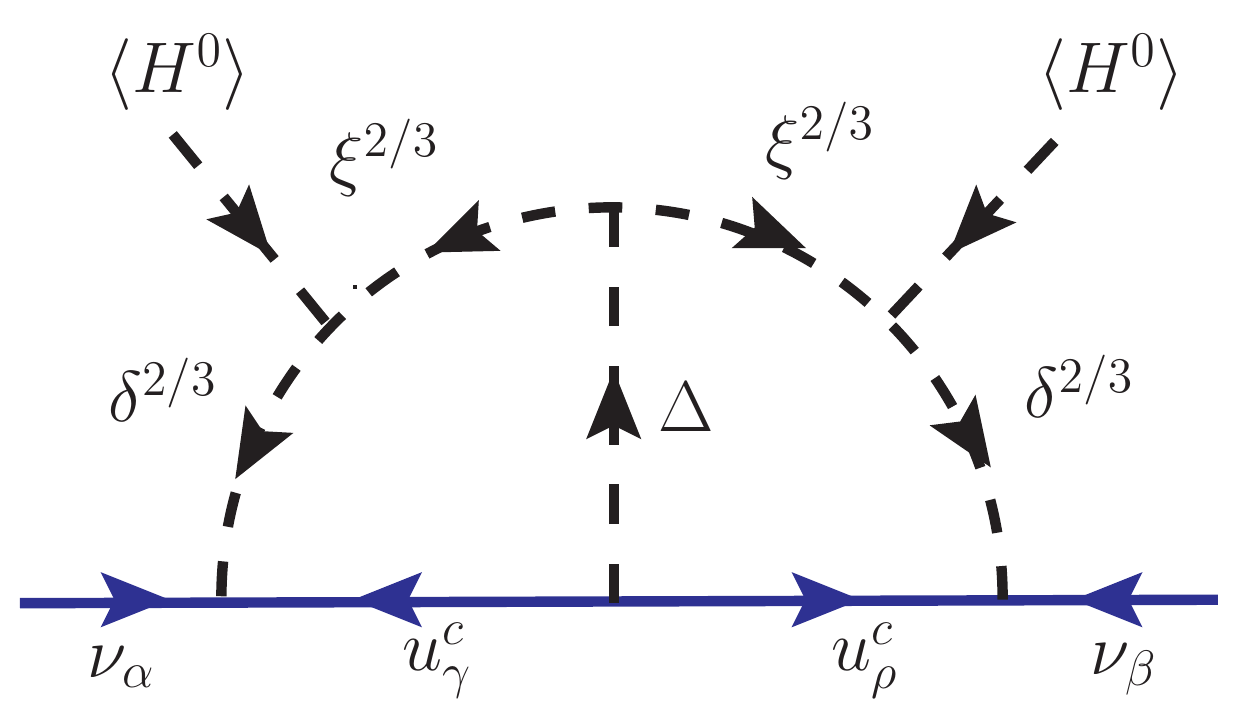}
    \caption{A new example of type-II radiative neutrino mass model.}
    \label{type0}
\end{figure}

A new example of type-II-like radiative model is shown in Fig.~\ref{type0}, where the new particles added are as follows: one color-sextet DQ $\Delta\left({\bf 6}, {\bf 1}, \frac{4}{3}\right)$, one $SU(2)_L$ doublet scalar LQ  $\delta\left({\bf 3},{\bf 2},\frac{7}{6}\right)=(\delta^{5/3},\delta^{2/3})$, and an $SU(2)_L$ singlet scalar LQ $\xi\left({\bf 3},{\bf 1},\frac{2}{3}\right)$. The relevant Yukawa Lagrangian is given by 
\begin{align}
    -{\cal L}_Y \ & \supset \ 
    %f_{\alpha\beta} \overline{L_\alpha^c} \delta u_\beta^c+\lambda_{\alpha\beta} u_\alpha^c\Delta u_\beta^c+{\rm H.c} \, \nonumber \\
 %    \ & = \ 
 f_{\alpha\beta}(\nu_\alpha  \delta^{2/3}-\ell_\alpha  \delta^{5/3})u_\beta^c+\lambda_{\alpha\beta} u_\alpha^c\Delta u_\beta^c+{\rm H.c.} \, 
 \label{eq:lagII}
\end{align}
Together with the scalar potential terms 
\begin{align}
    V \ \supset \ \mu \delta^\dag \Phi \delta+\mu' \delta^2 \Delta+{\rm H.c.} \, ,
\end{align}
where $\Phi$ is the SM Higgs doublet, the Lagrangian~\eqref{eq:lagII} gives rise to neutrino mass at two-loop level, as shown in Fig.~\ref{type0}. 
The neutrino mass can be approximated as follows:
\begin{equation}
    M_\nu \ \sim \ \frac{1}{(16 \pi^2)^2} \frac{\mu^2 \mu' v^2}{m_1^2 m_2^2} (f \lambda f^T)  \, ,
\end{equation}
where $m_1$ and $m_2$ are the masses of the heaviest two LQs among the $\delta$, $\xi$ and $\Delta$ fields that run in the loop. Thus, although this model can be described as arising from an effective $\Delta L = 2$ operator $\mathcal{O}'_1$ of Eq. \eqref{Op}, the neutrino mass has no chiral suppression here. In this sense, this can be put in the type-II radiative model category, although it leads to tree-level NSI induced by the $\delta$ LQs, as in the one-loop type-I model discussed in Sec.~\ref{subsec:1loopLQ}. A similar two-loop radiative model without the chiral suppression can be found in Ref.~\cite{Liu:2016mpf}.  

%%%%%%%%%%%%%%%%%%%%%%%%%%%%%%%%%%%%%%%%%%%%%%%
%%%%%%%%%%%%%%%%%%%%%%%%%%%%%%%%%%%%%%%%%%%%%%%
\section{Conclusion}\label{sec:con}
\begin{figure}[!t]
    \includegraphics[height=16.5cm,width=1.0\textwidth]{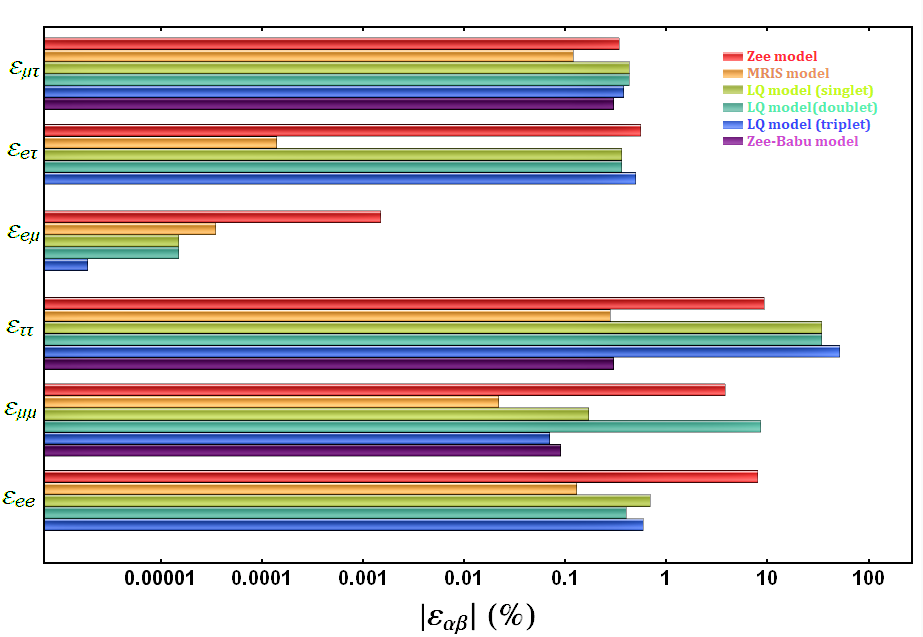}
    \caption{ Summary of maximum NSI strength  |$\varepsilon_{\alpha \beta}$| allowed in different classes of radiative neutrino mass models discussed here. Red, yellow, green, cyan,  blue and purple bars correspond to the Zee model, minimal radiative inverse seesaw model, LQ model with singlet, doublet and triplet LQs, and Zee-Babu model respectively. 
    %Light blue for $\varepsilon_{\mu\mu}$ corresponds to the singlet-induced NSI, which is smaller than the doublet-induced contribution due to $D$-meson decay constraints (cf.~Table~\ref{tab:LQ}).
    }
\label{fig:summaryplot}
\end{figure}

We have made a comprehensive analysis of neutrino non-standard interactions generated by new scalars in radiative neutrino mass models.  For this purpose, we have proposed a new nomenclature to classify radiative neutrino mass models, viz., the class of models with at least one SM particle in the loop are dubbed as {\it type-I} radiative models, whereas those models with no SM particles in the loop are called {\it type-II} radiative models. From NSI perspective, the type-I radiative models are most interesting, as the neutrino couples to a SM fermion (matter field) and a new scalar directly, thus generating NSI at tree-level, unlike type-II radiative models. After taking into account various theoretical and experimental constraints, we have derived the maximum possible NSI in all the type-I radiative models. Our results are summarized in Fig.~\ref{fig:summaryplot} and Table~\ref{Table_Models}. 

We have specifically analyzed two popular type-I radiative models, namely, the Zee model and its variant with LQs replacing the charged scalars, in great detail. In the Zee model with  $SU(2)_L$ singlet and doublet scalar fields, we find that large NSI can be obtained via the exchange of a light charged scalar, arising primarily from the $SU(2)_L$-singlet field but with some admixture of the doublet field. A light charged scalar with mass as low as $\sim$100 GeV is found to be consistent with various experimental constraints, including charged-lepton flavor violation (cf.~Sec.~\ref{sec:lfv}), monophoton constraints from LEP (cf.~Sec.~\ref{sec:monop}),  direct searches for charged scalar pair and single production at LEP (cf.~Sec.~\ref{sec:LEPZee}) and LHC (cf.~Sec.~\ref{sec:LHCZee}), Higgs physics constraints from LHC (cf.~Sec.~\ref{sec:HiggsOb}), and lepton universality in $W^\pm$ (cf.~Sec.~\ref{sec:Wuniv}) and $\tau$ (cf.~Sec.~\ref{sec:taudecay}) decays. In addition, for the Yukawa couplings and the mixing between singlet and doublet scalars, we have considered the contact interaction limits from LEP (cf.~Sec.~\ref{sec:contact}), electroweak precision constraints from $T$-parameter (cf.~Sec.~\ref{sec:ewpt}), charge-breaking minima of the Higgs potential (cf.~Sec.~\ref{sec:CBM}), as well as perturbative unitarity of Yukawa and quartic couplings. After imposing all these constraints, we find diagonal values of the NSI parameters ($\varepsilon_{ee},\,\varepsilon_{\mu\mu},\,\varepsilon_{\tau\tau})$ can be as large as $(8\%,\,3.8\%,\,9.3\%)$, while the off-diagonal NSI parameters $(\varepsilon_{e\mu},\, \varepsilon_{e\tau},\,\varepsilon_{\mu\tau})$ can be at most  $(10^{-3}\%,\,0.56\%,\,0.34\%)$, as summarized in Fig.~\ref{fig:summaryplot} and Table~\ref{tab:Zee}. Most of these NSI values are still allowed by the global-fit constraints from neutrino oscillation and scattering experiments, and some of these parameters can be probed at future long-baseline neutrino oscillation experiments, such as DUNE.

We have also analyzed in detail the LQ version of the Zee model, the results of which can be applied to other LQ models with minimal modification.  This analysis took into account the  experimental constraints from direct searches for LQ pair and single production at LHC (cf.~Sec.~\ref{sec:highconstraints}), as well as the low-energy constraints from APV (cf.~Sec.~\ref{sec:APV}), charged-lepton flavor violation (cf.~Secs.~\ref{sec:llgLQ} and \ref{sec:tauLQ}) and rare meson decays (cf.~Sec.~\ref{sec:Dmeson}), apart from the theoretical constraints from perturbative unitarity of the Yukawa couplings. Including all these constraints we found that diagonal NSI ($\varepsilon_{ee},\,\varepsilon_{\mu\mu},\,\varepsilon_{\tau\tau})$ can be as large as $(0.4\%,\,21.6\%,\,34.3\%)$, while off-diagonal NSI $(\varepsilon_{e\mu},\, \varepsilon_{e\tau}\,\varepsilon_{\mu\tau})$ can only be as large as $(10^{-5}\%,\,0.36\%,\,0.43\%)$, as summarized in Fig.~\ref{fig:summaryplot} and Table~\ref{tab:LQ}. A variant of the LQ model with triplet LQs (cf.~Sec.~\ref{sec:CCSVO39}) allows for larger $\varepsilon_{\tau\tau})$ which can be as large as $51.7\%$. Neutrino  scattering experiments are found to be the most constraining for the diagonal NSI parameters $\varepsilon_{ee}$ and $\varepsilon_{\mu\mu}$, while the cLFV searches are the most constraining for the off-diagonal NSI.  $\varepsilon_{\tau\tau}$ is the least constrained and can be probed at future long-baseline neutrino oscillation experiments, such as DUNE, whereas the other NSI parameters are constrained to be below the DUNE sensitivity reach.

%%%%%%%%%%%%%%%%%%%%%%%%%%%%%%%%%%%%%%%%%%%%%
\section*{Acknowledgments}
We thank Sanjib Agarwalla, Sabya  Chatterjee, Peter Denton, Radovan Dermi\v{s}ek, Arman Esmaili, Tao Han, Chris Kolda, Pedro Machado, Michele Maltoni, Ivan Martinez-Soler, Jordi Salvado, Yongchao Zhang and Yue Zhang for useful discussions. The work of KB, SJ, and AT was supported in part by the US Department of Energy Grant Number DE-SC 0016013. The work of BD was supported in part by the  US  Department of Energy under Grant No. DE-SC0017987 and by the MCSS. This work was also supported by the US Neutrino Theory Network Program under Grant No. DE-AC02-07CH11359. KB is supported in part by a Fermilab Distinguished Scholar program. We thank the Fermilab Theory Group for warm hospitality during the completion of this work.  In addition, BD thanks the Department of Physics at Oklahoma State University for warm hospitality during the completion of this work. SJ and AT thank the Department of Physics at Washington University in St. Louis for warm hospitality, where part of this work was done. 

%%%%%%%%%%%%%%%%%%%%%%%%%%%%%%%%%%%%%%%%%%%%%

%%%%%%%%%%%%%%%%%%%%%%%%%%%%%%%%%%%%%%%%%%%%%%%
%%%%%%%%%%%%%%%%%%%%%%%%%%%%%%%%%%%%%%%%%%%%%%%
%%%%%%%%%%%%%%%%%%%%%%%%%%%%%%%%%%%%%%%%%%%%%
%%%%%%%%%%%%%%%%%%%%%%%%%%%%%%%%%%%%%%%%%%%%%

\begin{sidewaystable}
%\hspace{-5cm}
\tiny
    \centering
%    \begin{adjustbox}{width=1\textwidth}
   \resizebox{1\textwidth}{!}{
\begin{tabularx}{\textwidth}{|X|c|c|c|c|c|c|c|c|c|c|c|}
    \hline \hline
 \textbf{Term} & $\mathcal{O}$ & \multirow{2}{*} { {\textbf{\, \, Model}}}  & {\bf Loop} & $\textcolor{blue}{\mathcal{S}}/$ & \multirow{2}{*}{{\textbf{New particles}}} & \multicolumn{6}{c|}{{\textbf{Max NSI @ tree-level}}}\\
 \cline{7-12}
 
 & & & {\bf level} & $\textcolor{blue}{\mathcal{F}}$ &  &  $|\varepsilon_{ee}|$ & $|\varepsilon_{\mu \mu}|$ & $|\varepsilon_{\tau \tau}|$ & $|\varepsilon_{e \mu}|$ & $|\varepsilon_{e \tau}|$ & $|\varepsilon_{\mu \tau}| $  \\ \hline \hline
 
\rule{0pt}{10pt} $ L \ell^c \Phi^\star $ & $\mathcal{O}_2^2$ & Zee~\cite{Zee:1980ai} & 1 & $\textcolor{blue}{\mathcal{S}}$ & $\eta^+  ({\bf 1},{\bf 1},1)$, $\Phi_2  ({\bf 1},{\bf 2},1/2)$ & 0.08 & 
0.038 & 0.093 &  $\mathcal{O}(10^{-5})$ & 0.0056 &  0.0034  \\ \cline{1-12} 

\rule{0pt}{10pt}  & $\mathcal{O}_9$ & Zee-Babu~\cite{Zee:1985id, Babu:1988ki} & 2 & $\textcolor{blue}{\mathcal{S}}$ & $h^+  ({\bf 1},{\bf 1},1)$, $k^{++}  ({\bf 1},{\bf 1},2)$ & &  &  &  &  & \\ \cline{2-6}

\rule{0pt}{10pt}  & $\mathcal{O}_9$ & KNT~\cite{Krauss:2002px} & 3 & $\textcolor{blue}{\mathcal{S}}$ & $\eta_1^+  ({\bf 1},{\bf 1},1)$, $\textcolor{red}{\eta_2^+}  ({\bf 1},{\bf 1},1)$   &  &  &  &  &  &  \\ 
\rule{0pt}{5pt}& & & &$\textcolor{blue}{\mathcal{F}}$  & $\textcolor{red}{N}  ({\bf 1},{\bf 1},0)$ & 0 & $0.0009$ & $0.003$ & 0 & 0 & 0.003 \\\cline{2-6} 

\rule{0pt}{10pt} $LL\eta$ & $\mathcal{O}_9$ & 1S-1S-1F~\cite{Cepedello:2018rfh} & 3 & $\textcolor{blue}{\mathcal{S}}$ & $\eta_1  ({\bf 1},{\bf 1},1)$, $\eta_2  ({\bf 1},{\bf 1},3)$   &  & & &  &  & \\ 
\rule{0pt}{5pt}& & & &$\textcolor{blue}{\mathcal{F}}$  & $F  ({\bf 1},{\bf 1},2)$ &  & & &  &  &  \\\cline{2-6} 

\rule{0pt}{10pt}  & $\mathcal{O}_2^1$ & 1S-2VLL~\cite{Cai:2014kra}  & 1  &$\textcolor{blue}{\mathcal{S}}$ & $\eta({\bf 1},{\bf 1}, 1)$ &   &  &  &  &  &  \\

\rule{0pt}{5pt} & & & &$\textcolor{blue}{\mathcal{F}}$ & $\Psi({\bf 1},{\bf 2},-3/2) $ &  & & &  &  &  \\ \cline{1-12}

\rule{0pt}{10pt} & ${\cal O}'_3$ & AKS~\cite{Aoki:2008av} & 3  & $\textcolor{blue}{\mathcal{S}}$ & $\Phi_2 ({\bf 1},{\bf 2},1/2) $,\,\,$\textcolor{red}{\eta^{+}} ({\bf 1},{\bf 1},1)$,\,\,$ \textcolor{red}{\eta^0} ({\bf 1},{\bf 1},0)$ & ${\cal O}(10^{-10})$ & ${\cal O}(10^{-10})$ & ${\cal O}(10^{-10})$ & ${\cal O}(10^{-10})$ & ${\cal O}(10^{-10})$ & ${\cal O}(10^{-10})$ \\ 

\rule{0pt}{5pt} $L\ell^c \phi^\star$ & & & & $\textcolor{blue}{\mathcal{F}}$ &  $\textcolor{red}{N}  ({\bf 1},{\bf 1},0) $ &  & & &  &  &  \\ \cline{1-12}

\rule{0pt}{10pt} --- & ${\cal O}_{d=15}$ & Cocktail~\cite{Gustafsson:2012vj} & 3  &$\textcolor{blue}{\mathcal{S}}$ & $\textcolor{red}{\eta^+}  ({\bf 1},{\bf 1},1)$, $k^{++}  ({\bf 1},{\bf 1},2) , \textcolor{red}{\Phi_2}  ({\bf 1},{\bf 2},1/2)$ & 0 & 0 & 0 & 0 & 0 & 0 \\

%\rule{0pt}{10pt}&  & & &  $ &  & & &  &  &  \\
\hline
 \hline 
 \rule{0pt}{10pt}\textbf{$W/Z$} & $\mathcal{O}_2'$ & MRIS~\cite{Dev:2012sg}   & 1  & $\textcolor{blue}{\mathcal{F}}$ &  $N  ({\bf 1},{\bf 1},0)$, $S  ({\bf 1},{\bf 1},0)$ & 0.0013 & ${\cal O}(10^{-4})$  & 0.0028 & ${\cal O}(10^{-5})$ & ${\cal O}(10^{-4})$ & 0.0012  \\ 
% \rule{0pt}{1pt}&  & & &  &  & & &  &  &  \\
 \hline
 \hline 
\rule{0pt}{10pt} $L\Omega d^c$ & $\mathcal{O}_3^8$ & LQ variant of Zee~\cite{AristizabalSierra:2007nf} & 1  & $\textcolor{blue}{\mathcal{S}}$ & $\Omega  ({\bf 3},{\bf 2},1/6)$, $\chi({\bf 3},{\bf 1},-1/3)$ & 0.004 & 0.216 & 0.343 & ${\cal O}(10^{-7})$ & 0.0036 & 0.0043 \\ \cline{2-6}

\rule{0pt}{10pt}$(LQ\chi^\star)$ & $\mathcal{O}_8^4$ & 2LQ-1LQ~\cite{Babu:2010vp} & 2  & $\textcolor{blue}{\mathcal{S}}$ & $\Omega  ({\bf 3},{\bf 2},1/6)$, $\chi  ({\bf 3},{\bf 1},-1/3)$  &  (0.0069) & (0.0086) &  &  &  &  \\ \cline{1-12}

\rule{0pt}{10pt}  & $\mathcal{O}_3^3$ & 2LQ-1VLQ~\cite{Babu:2011vb} & 2  & $\textcolor{blue}{\mathcal{S}}$ & $\Omega  ({\bf 3},{\bf 2},1/6)$ &   &  &  &  & & \\

\rule{0pt}{5pt}& & & & $\textcolor{blue}{\mathcal{F}}$ &  $U  ({\bf 3},{\bf 1},2/3)$  &  & & &  &  &  \\ \cline{2-6}

\rule{0pt}{10pt} $L\Omega d^c$ & $\mathcal{O}_3^6$ & 2LQ-3VLQ~\cite{Cai:2014kra}  & 1  &$\textcolor{blue}{\mathcal{S}}$ & $\Omega({\bf 3},{\bf 2}, 1/6)$ &  &  &  &  & & \\

\rule{0pt}{5pt}& & & &$\textcolor{blue}{\mathcal{F}}$ & $\Sigma({\bf 3},{\bf 3},2/3) $ & 0.004 & 0.093 & 0.093 & ${\cal O}(10^{-7})$ & 0.0036 & 0.0043   \\ \cline{2-6}

\rule{0pt}{10pt}  & $\mathcal{O}_8^2$ & 2LQ-2VLL~\cite{Cai:2014kra}  & 2  &$\textcolor{blue}{\mathcal{S}}$ & $\Omega({\bf 3},{\bf 2}, 1/6)$ &   &  &  &  &  &  \\

\rule{0pt}{5pt}& & & &$\textcolor{blue}{\mathcal{F}}$ & $\psi({\bf 1},{\bf 2},-1/2) $ &  & & &  &  &  \\ \cline{2-6}

\rule{0pt}{10pt}  & $\mathcal{O}_8^3$ & 2LQ-2VLQ~\cite{Cai:2014kra}  & 2  &$\textcolor{blue}{\mathcal{S}}$ & $\Omega({\bf 3},{\bf 2}, 1/6)$ &   &  &  &  &  &  \\

\rule{0pt}{5pt}& & & &$\textcolor{blue}{\mathcal{F}}$ & $\xi({\bf 3},{\bf 2},7/6) $ &  & & &  &  &  \\ \cline{1-12}

\rule{0pt}{10pt} $ L\Omega d^c  $ & $\mathcal{O}_3^9$ & Triplet-Doublet LQ~\cite{Cai:2014kra}  & 1  &$\textcolor{blue}{\mathcal{S}}$ & $\rho({\bf 3},{\bf 3}, -1/3)$, $\Omega({\bf 3},{\bf 2}, 1/6)$ &  0.0059 & 0.0249 & 0.517 & ${\cal O}(10^{-8})$ & 0.0050 & 0.0038 \\ 

\rule{0pt}{5pt} $ ( LQ\Bar{\rho} )$ & & & & &  &  & & &  &  &  \\ \cline{1-12}

\rule{0pt}{10pt}  & $\mathcal{O}_{11}$ & LQ/DQ variant Zee-Babu~\cite{Kohda:2012sr}   & 2  & $\textcolor{blue}{\mathcal{S}}$ & $\chi ({\bf 3},{\bf 1},-1/3)$ , $\Delta({\bf 6},{\bf 1},-2/3)$  &   &  &  &  &  &   \\ \cline{2-6}

\rule{0pt}{10pt}  & $\mathcal{O}_{11}$ & Angelic~\cite{Angel:2013hla}    & 2  & $\textcolor{blue}{\mathcal{S}}$ & $\chi({\bf 3},{\bf 1},1/3)$  &  &  &  &  &  &   \\

\rule{0pt}{5pt}& & & & $\textcolor{blue}{\mathcal{F}}$ & $F  ({\bf 8},{\bf 1},0)$ &  & & &  &  &  \\ \cline{2-6}

%\rule{0pt}{10pt}& & &  & $\hat{\Delta}  ({\bf 6}^\star, {\bf 3},-1/3)$   &  & & &  &  &  \\ 

\rule{0pt}{10pt} $LQ\chi^\star$ & $\mathcal{O}_{11}$ & LQ variant of KNT~\cite{Nomura:2016ezz}  & 3  &$\textcolor{blue}{\mathcal{S}}$ & $\chi ({\bf 3},{\bf 1},-1/3)$,\, $\textcolor{red}{\chi_2}({\bf 3},{\bf 1},-1/3)$ &  0.0069 & 0.0086 & 0.093 & ${\cal O}(10^{-7})$ & 0.0036 & 0.0043 \\

\rule{0pt}{5pt}& & & &$\textcolor{blue}{\mathcal{F}}$ & $\textcolor{red}{N}({\bf 1},{\bf 1},0) $ &  & & &  &  &  \\ \cline{2-6}

\rule{0pt}{10pt}  & $\mathcal{O}_3^4$ & 1LQ-2VLQ~\cite{Cai:2014kra}  & 1  &$\textcolor{blue}{\mathcal{S}}$ & $\chi({\bf 3},{\bf 1}, -1/3)$ &   &  &  & &  &  \\

\rule{0pt}{5pt}& & & &$\textcolor{blue}{\mathcal{F}}$ & $\mathcal{Q}({\bf 3},{\bf 2},-5/6) $ &  & & &  &  &  \\ \cline{1-12}

\rule{0pt}{10pt} $Lu^c\delta $ & $\mathcal{\Tilde{O}}_{1}$ & \textbf{3LQ-2LQ-1LQ (New)} & 1  & $\textcolor{blue}{\mathcal{S}}$ & $\Bar{\rho}  (\bar{{\bf 3}},{\bf 3},1/3)$, $\delta  ({\bf 3},{\bf 2},7/6)$, $\xi  ({\bf 3},{\bf 1},2/3)$  &  0.004 & 0.216 & 0.343 & ${\cal O}(10^{-7})$ & 0.0036 & 0.0043 \\ 

\rule{0pt}{5pt}$(LQ\Bar{\rho})$ & & &  &  &  & (0.0059) & (0.007)&(0.517)  &  &(0.005) &(0.0038) \\ 
\cline{1-12}

\rule{0pt}{10pt} $ Lu^c\delta $ & ${\cal O}_{d=13}$ & \textbf{3LQ-2LQ-2LQ(New)}   & 2  & $\textcolor{blue}{\mathcal{S}}$ & $\delta  ({\bf 3},{\bf 2},7/6)$, $\Omega  ({\bf 3},{\bf 2},1/6)$,$\hat{\Delta}  ({\bf 6}^\star, {\bf 3},-1/3)$ &  0.004 & 0.216 & 0.343 & ${\cal O}(10^{-7})$ & 0.0036 & 0.0043 \\ \cline{1-12}

\rule{0pt}{10pt} $LQ\Bar{\rho}$ & $\mathcal{O}_3^5$ & 3LQ-2VLQ~\cite{Cai:2014kra}  & 1  &$\textcolor{blue}{\mathcal{S}}$ & $\Bar{\rho}({\bf \bar{3}},{\bf 3}, -1/3)$ &   &  &  &  &  &  \\

\rule{0pt}{5pt} & & & &$\textcolor{blue}{\mathcal{F}}$ & $\mathcal{Q}({\bf 3},{\bf 2},-5/6) $ & 0.0059 & 0.0007 & 0.517 & ${\cal O}(10^{-7})$ & 0.005 & 0.0038 \\ \cline{1-12}

\rule{0pt}{15pt} &  \multicolumn{5}{c|}{{\textbf{ All Type-II Radiative models}}} & 0 & 0 & 0 & 0 & 0 & 0 \\ 

\hline \hline
\end{tabularx}
%\end{adjustbox}
}
\caption{\scriptsize A comprehensive summary of type-I radiative neutrino mass models, with the new particle content and their $(SU(3)_c,~SU(2)_L,~U(1)_Y)$ charges, and the maximum tree-level NSI allowed in each model. Red-colored exotic particles are odd under a $Z_2$ symmetry. $\textcolor{blue}{\mathcal{S}}$ and $\textcolor{blue}{\mathcal{F}}$ represent scalar and fermion fields respectively.}
 \label{Table_Models}
 %\hspace{-5cm}
\end{sidewaystable}
\clearpage
%\restoregeometry
%}

\section*{Appendix}

\appendix

\section{Analytic Expressions for Charged Higgs Cross Sections} \label{app:A}
It is instructive to write down the explicit formula for the charged-Higgs pair-production (Figs.~\ref{fig:feyn1} (a) and \ref{fig:feyn1} (b) cross section: 
\begin{align}
\sigma(e^+e^-\to h^+h^-) \ = \ & \frac{\beta^3}{48\pi s} \left[e^4 +\frac{g^4}{8c_w^4}(1-4s_w^2+8s_w^4)\left(s_w^2-\frac{1}{2}\sin^2\varphi\right)^2\frac{s^2}{(s-m_Z^2)^2+\Gamma_Z^2m_Z^2}\right. \nonumber \\ & \qquad \qquad \left. +\frac{e^2g^2}{2c_w^2} (4s_w^2-1)\left(s_w^2-\frac{1}{2}\sin^2\varphi\right)\frac{s(s-m_Z^2)}{(s-m_Z^2)^2+\Gamma_Z^2m_Z^2}\right] \nonumber \\
& +\frac{|Y_{\alpha e}|^4}{32\pi s}\left[-\beta+\frac{1}{2}(1+\beta^2)\ln\frac{1+\beta}{1-\beta}\right]\nonumber \\
& -\frac{|Y_{\alpha e}|^2}{128\pi s}\left[2\beta(1+\beta^2)-(1-\beta^2)^2\ln\frac{1+\beta}{1-\beta}\right]\nonumber \\
& \quad \times \left[e^2+\frac{g^2}{c_w^2}\left(s_w^2-\frac{1}{2}\sin^2\varphi\right)(2s_w^2-1)\frac{s(s-m_Z^2)}{(s-m_Z^2)^2+\Gamma_Z^2 m_Z^2}\right] \, ,
\end{align}
where $\beta=\sqrt{1-4m_{h^+}^2/s}$, $s$ is the squared center-of-mass energy, $e$ and $g$ are the electromagnetic and $SU(2)_L$ coupling strengths, respectively, and $c_w\equiv \cos\theta_w,~s_w\equiv \sin\theta_w$ ($\theta_w$ being the weak mixing angle). Note that the $t$-channel cross section depends on the Yukawa coupling $Y_{\alpha e}$, and it turns out there is a destructive interference between the $s$ and $t$-channel processes. 
Similarly, the differential cross section for the production of $h^\pm W^\mp$ (Fig.~\ref{fig:feyn1} (c)) is given by 
\begin{align}
\frac{d\sigma(e^+e^-\to h^\pm W^\mp)}{d\cos\theta} \ & = \ \frac{g^2|Y_{ee}|^2}{64\pi s}\lambda^{1/2}\left(1,\frac{m_{h^+}^2}{s},\frac{m_W^2}{s}\right)\nonumber \\
& \qquad \qquad \times 
\frac{A\cos^2\theta+B\cos\theta+C}{\left[1-\frac{m_{h^+}^2+m_W^2}{s}
-\lambda^{1/2}\left(1,\frac{m_{h^+}^2}{s},\frac{m_W^2}{s}\right)\cos\theta\right]^2} \, , 
\end{align}
where $\theta$ is the angle made by the outgoing $h^\pm$ with respect to the initial $e^-$-beam direction, $\lambda(x,y,z)=  x^2+y^2+z^2-2xy-2xz-2yz$, and 
\begin{align}
    A \ &= \ \frac{s}{4m_W^2}\left[1-\frac{(m_{h^+}-m_W)^2}{s}\right]\left[1-\frac{(m_{h^+}+m_W)^2}{s}\right]\left[1-\frac{2m_W^2}{s}\right]\, \\
B\ & = \ -\frac{s}{2m_W^2}\left(1-\frac{m_{h^+}^2+m_W^2}{s}\right)\lambda^{1/2}\left(1,\frac{m_{h^+}^2}{s},\frac{m_W^2}{s}\right) \, ,\\
C \ & = \ \frac{s}{4m_W^2}\left(1-\frac{2m_{h^+}^2}{s}-\frac{3m_W^4}{s^2}-\frac{2m_{h^+}^2m_W^2}{s^2}+\frac{2m_W^6}{s^3}-\frac{2m_{h^+}^2m_W^4}{s^3}%\right. \nonumber \\
%& \left. \qquad \qquad \qquad \qquad \qquad \qquad 
+\frac{m_{h^+}^4}{s^2} +\frac{m_{h^+}^4m_W^2}{s^3}\right) \, .
\end{align}

The analytic cross section formula for the single-production of charged Higgs via Drell-Yan process (Fig.~\ref{fig:feyn1} (d)) is more involved due to the three-body phase space and is not given here. 

%%%%%%%%%%%%%%%%%%%%%%%%%%%
\bibliographystyle{utphys}
\bibliography{reference}

\end{document}